\documentstyle[12pt,epsf]{article}

\def\naive{na\"{\i}ve}
\newcommand\M[2]{|{\cal{M}}^{#1}_{#2}|^2}
\newcommand\as{\alpha_{\mathrm{S}}}
\newcommand\smfrac[2]{{\textstyle\frac{#1}{#2}}}
\def\hf{\smfrac{1}{2}}
\def\ep{\epsilon}
\def\ee{$e^+e^-$}
\def\beq{\begin{equation}}
\def\eeq{\end{equation}}
\def\beeq{\begin{eqnarray}}
\def\eeeq{\end{eqnarray}}
\def\cm{{\cal M}}
\def\ket#1{|{#1}>}
\def\bra#1{<{#1}|}
\def\mket#1{|{#1}>_m}
\def\mbra#1{{}_m\!\!<{#1}|}
\def\oket#1{|{#1}>_{m+1}}
\def\obra#1{_{m+1}\!\!<{#1}|}
\def\aket#1{|{#1}>_{m+1+a}}
\def\abra#1{{}_{m+1+a}\!\!<{#1}|}
\def\amket#1{|{#1}>_{m+a}}
\def\ambra#1{{}_{m+a}\!\!<{#1}|}
\def\bom#1{{\mbox{\boldmath $#1$}}}
\def\to{\rightarrow}
\def\kper{k_{\perp}}
\def\KFS#1{K^{{#1}}_{\scriptscriptstyle\rm F\!.S\!.}}
\def\HFS#1{H_{{#1}}^{\scriptscriptstyle\rm F\!.S\!.}}
\def\Kab{\KFS{ab}}
\def\Hba{\HFS{ba}}
\def\Kcb{\KFS{cb}}
\def\Kac{\KFS{ac}}
\def\Hac{\HFS{ac}}
\def\Hcb{\HFS{cb}}
\def\Hab{\HFS{ab}}

\def\np#1#2#3{Nucl.\ Phys.\ B#1 (19#3) #2}
\def\pl#1#2#3{Phys.\ Lett.\ #1B (19#3) #2}
\def\pr#1#2#3{Phys.\ Rev.\ D #1 (19#3) #2}
\def\prep#1#2#3{Phys.\ Rep.\ #1 (19#3) #2}
\def\prl#1#2#3{Phys.\ Rev.\ Lett.\ #1 (19#3) #2}

\def\zp#1#2#3{Zeit.\ Phys.\ C#1 (19#3) #2}

\relax

\renewcommand{\theequation}{\arabic{section}.\arabic{equation}}

\begin{document}

\begin{titlepage}
\renewcommand{\thefootnote}{\fnsymbol{footnote}}
\begin{flushright}
     CERN-TH/96-29 \\ hep-ph/9605323
     \end{flushright}
\par \vspace{10mm}
\begin{center}
{\Large \bf
A General Algorithm for Calculating \\[1ex]
Jet Cross Sections in NLO QCD\footnote{Research supported in part by
EEC Programme {\it Human Capital and Mobility}, Network {\it Physics at High
Energy Colliders}, contract CHRX-CT93-0357 (DG 12 COMA).}}
\end{center}
\par \vspace{2mm}
\begin{center}
{\bf S. Catani}\\

\vspace{5mm}

{I.N.F.N., Sezione di Firenze}\\
{and Dipartimento
di Fisica, Universit\`a di Firenze}\\
{Largo E. Fermi 2, I-50125 Florence, Italy}

\vspace{5mm}

{\bf M.H. Seymour}\\

\vspace{5mm}

{Theory Division, CERN}\\
{CH-1211 Geneva 23, Switzerland}
\end{center}

\par \vspace{2mm}
\begin{center} {\large \bf Abstract} \end{center}
\begin{quote}
  We present a new general algorithm for calculating arbitrary jet cross
  sections in arbitrary scattering processes to next-to-leading accuracy in
  perturbative QCD\@. The algorithm is based on the subtraction method.  The
  key
  ingredients are new factorization formulae, called dipole formulae, which
  implement in a Lorentz covariant way both the usual soft and collinear
  approximations, smoothly interpolating the two.  The corresponding dipole
  phase space obeys exact factorization, so that the dipole contributions to
  the cross section can be exactly integrated analytically over the whole of
  phase space. We obtain explicit analytic results for any jet observable in
  any scattering or fragmentation process in lepton, lepton-hadron or
  hadron-hadron collisions. All the analytical formulae necessary to construct
  a numerical program for next-to-leading order QCD calculations are provided.
  The algorithm is straightforwardly implementable in general purpose Monte
  Carlo programs.
\end{quote}
\vspace*{\fill}
\begin{flushleft}
     CERN-TH/96-29 \\   May 1996
\end{flushleft}
\end{titlepage}

\setcounter{tocdepth}{1}
\thispagestyle{empty}
\addtocounter{page}{-1}
\enlargethispage*{8mm}
\tableofcontents

\renewcommand{\thefootnote}{\fnsymbol{footnote}}

\newpage

\setcounter{equation}{0}

\section{Introduction}
\label{int}

Most of the recent progress in the understanding of strong interaction physics
at large momentum transfer has been due to the comparison between precise
experimental data and very accurate QCD calculations to higher perturbative
orders [\ref{QCDrev}].

The perturbative QCD approach for computing hadronic cross sections is based
on the parton model picture. According to this picture, the cross section for
any hard-scattering process (i.e.\ any process involving a large transferred
momentum $Q$) can be written as a convolution of structure ($f_a(x,Q^2)$)
and fragmentation ($d_a(x,Q^2)$) functions of partons (quarks and gluons)
and a hard-cross section factor. The former are non-perturbative quantities but
are universal, that is, they are process independent. The latter is instead
dominated by momentum regions of the order of $Q$ and hence,
provided that $Q \gg \Lambda$ (where $\Lambda$ is the QCD scale), it can
be computed in QCD perturbation theory to the lowest order in the `small'
(due to asymptotic freedom) running coupling $\as(Q) \sim (\beta_0
\ln Q^2/\Lambda^2)^{-1}$.

This \naive\ parton model approach corresponds to the so-called
leading-order (LO) approximation. It is justified by the high-momentum
behaviour of the running coupling $\as(Q)$. However, just because of its
perturbative nature, the running of the QCD coupling can be hidden in
higher-order corrections by the replacement $\as(Q)= \as^{(0)} \,[1 + \,K(Q)
\,\as(Q) + \dots]$, $\as^{(0)}$ being the values of $\as$ at a fixed (and
arbitrary) momentum scale. It follows that a LO calculation
predicts only the order of magnitude of a given cross section and the rough
features of a certain observable\footnote{Typically, this poor predictivity
is quantitatively signalled by a strong dependence on the (unphysical)
renormalization and factorization scales.}. The accuracy of the perturbative
QCD expansion is instead controlled by the size
of the higher-order contributions. Any definite perturbative QCD prediction
thus requires (at least) a next-to-leading order (NLO) calculation, and NLO
definitions of $\as,$ $f_a(x,Q^2)$ and $d_a(x,Q^2)$.

This is the reason why the results of the higher-order QCD calculations which
are available at present have been
proved to be of vital importance to assess
the progress mentioned above.

These higher-order computations have been carried out over a period of
about fifteen years, often long after the accuracy of experimental data
has made them necessary, because of the difficulties in setting up a
general and straightforward calculational procedure. The physical origin of
these difficulties is in the necessity of factorizing the long- and
short-distance components of the scattering processes
and is reflected in the perturbative
calculation by the presence of divergences.

In general, when evaluating higher-order QCD cross sections,
one has to consider
real-emission contributions and virtual-loop corrections and one has to deal
with different kind of singularities. The customary {\em ultraviolet\/}
singularities, present in the virtual contributions, are removed by
renormalization. The low-momentum ({\em soft\/}) and small-angle
({\em collinear\/}) regions instead produce singularities both in the real and
in the virtual contributions. In order to handle these divergences, the
observable one is interested in has to be properly defined. It has to be a
{\em jet quantity}, that is, a hadronic observable that turns out to be
infrared safe and either collinear safe or collinear factorizable: its actual
value
has to be independent of the number of soft and collinear particles in the
final state (see Sect.~\ref{siee} for a formal definition).
In the case of jet
quantities, the coherent sum over different (real and virtual) soft and
collinear partonic configurations in the final state leads to the cancellation
of soft singularities. The left-over collinear singularities are then
factorized into the process-independent structure and fragmentation functions
of partons (parton distributions), leading to predictable scaling violations.
As a result, jet cross sections are finite (calculable) at the partonic level
order by order in perturbation theory. All the dependence on long-distance
physics is either included in the parton distributions or in non-perturbative
corrections that are suppressed by inverse powers of the (large)
transferred momentum $Q$ that controls the scattering process.

Because of this complicated pattern of singularities, the simplest quantities
that can be computed in QCD perturbation theory are fully inclusive.
In this case one considers all possible final states and integrates
the QCD matrix elements over the whole available final-state phase space.
Thus one can add real and virtual contributions before performing the relevant
momentum integrations in such a way that only ultraviolet singularities appear
at the intermediate steps of the calculation. Owing to this simplification,
powerful techniques have been set up [\ref{moscow}] to perform analytic
calculations up to next-to-next-to-leading order (NNLO), i.e.\ to relative
accuracy ${\cal O}(\as^3)$ with respect to the lowest-order approximation.

In the case of less inclusive cross sections,
QCD calculations beyond leading order (LO) are much more
involved. Owing to the complicated phase space for multi-parton
configurations, analytic calculations are in practice impossible for
all but the simplest quantities. However, the use of numerical methods is
far from trivial because real and virtual contributions have a different
number of final-state partons and thus have to be integrated
{\em separately\/} over different phase space regions. Unlike the case of fully
inclusive observables, one cannot take advantage of the cancellation of soft
and collinear divergences at the integrand level.
Soft and collinear singularities, present in the
intermediate steps, have to be first regularized, generally by analytic
continuation in a
number of space-time dimensions $d=4-2\epsilon$ different from four.
Then the real and virtual contributions should be calculated independently,
yielding equal-and-opposite poles in $\ep$. Great progress has been made in
recent years in the analytical techniques for calculating the
virtual processes [\ref{loopy}], but the
analytic continuation greatly complicates the Lorentz algebra in the
evaluation of the matrix elements and prevents a
straightforward implementation of numerical
integration techniques. Despite these difficulties, efficient computational
techniques have been set up, at least to NLO, during the last few years.

There are, broadly speaking, two types of algorithm used for NLO
calculations: one based on the phase-space {\em slicing\/} method and the
other
based on the {\em subtraction\/} method\footnote{We refer the reader to the
Introduction of Ref.~[\ref{KS}] for an elementary description of the basic
difference between the two methods.}. The main difference between these
algorithms and the standard procedures of analytic calculations is that only
a minimal part of the full calculation is treated analytically, namely only
those contributions giving rise to the singularities. Moreover, for any given
process, these contributions are computed in a manner that is independent of
the particular jet observable considered. Once every singular term has been
isolated and the cancellation/factorization of divergences achieved, one can
perform the remaining part of the calculation in four space-time dimensions.
Although, when possible, one still has the freedom of completing
the calculation analytically, at this point the use of numerical integration
techniques (typically, Monte Carlo methods) is certainly more convenient.
First of all, the numerical approach allows one to calculate any number and
any type of observable simultaneously by simply histogramming the appropriate
quantities, rather than having to make a separate analytic calculation for each
observable. Furthermore, using the numerical approach, it is easy to
implement different experimental conditions, for example, detector acceptances
and experimental cuts. In other words, the phase-space slicing and subtraction
algorithms provide the basis for setting up a general-purpose Monte
Carlo program for carrying out arbitrary NLO QCD calculations in a given
process.

Both the slicing [\ref{KL}] and the
subtraction [\ref{ERT}] methods were first used in the context of NLO
calculations of three-jet cross sections in \ee\ annihilation. Then they
have been applied to other cross sections adapting the method each time
to the particular process. Only recently has it
become clear that both algorithms are generalizable in a process-independent
manner. The key observation is that the singular parts of the QCD matrix
elements for real emission can be singled out in a general way by using
the factorization properties of soft and collinear radiation [\ref{BCM}].

At present, a general version of the slicing algorithm is available for
calculating NLO cross sections for production of {\em any\/} number of jets both
in lepton [\ref{GG}] and hadron [\ref{GGK}] collisions. To our knowledge,
fragmentation processes have been considered only in the particular cases of
prompt-photon production [\ref{BOO}] and single- and double-hadron inclusive
distributions [\ref{Aver},\ref{Chiap}].
The complete generalization of this method to include fragmentation functions
and heavy flavours is in progress [\ref{GKL}].

As for the subtraction algorithm, a general NLO formalism has been set up
for computing three-jet observables in \ee\ annihilation [\ref{ERT},\ref{KN}]
and cross sections up to two final-state jets in hadron
collisions\footnote{After completion of the present work, the method of
Ref.~[\ref{KS}] has been modified to deal with three-jet cross sections
[\ref{Frix}].  The formalism presented in Ref.~[\ref{Frix}] can be extended to
$n$-jet production both in lepton and hadron collisions. A similar method has
been presented in Ref.~[\ref{NT}].}
[\ref{KS},\ref{EKS}]. Also the treatment of massive partons has been
considered in the particular case of heavy-quark correlations in hadron
collisions [\ref{MNR}].

In this paper, we present a {\em completely general\/}
version of the  subtraction algorithm. This generality is obtained by fully
exploiting the factorization properties of soft and collinear emission and,
thus, deriving new improved factorization formulae, called {\em dipole
factorization formulae}. They allow us to introduce a set of universal
counterterms that can be used for {\em any\/} NLO QCD calculation. Therefore,
our
version of the subtraction method can be compared with those used for three-jet
observables in \ee\ annihilation [\ref{ERT}] and two-jet quantities in
hadron collisions [\ref{KS}] (although, in these known cases our treatment
turns out to differ
in many respects from the previous ones). Moreover, we are able
to consider the production of {\em any\/} number of jets in lepton and hadron
cross sections and to provide a general treatment of fragmentation processes
and multi-particle correlations. The inclusion of heavy quarks in the
algorithm can also be performed in a completely general and process-independent
manner [\ref{CS1}].
The extension of our method
to polarized scattering is not considered here but it is straightforward.

Besides discussing in detail our general formalism, we explicitly carry out
the $d$-dimen\-sional analytical part of the NLO calculation for all the
(unpolarized) scattering processes involving massless quarks and gluons.
Knowing the relevant QCD matrix elements, the results of our algorithm can be
straightforwardly implemented in NLO numerical codes without any additional
calculation.  Detailed numerical applications to \ee\
annihilation~[\ref{CSlett}] and deep inelastic lepton-hadron
scattering~[\ref{CSdis}] are presented elsewhere.

We begin in Sect.~\ref{gen} by giving a brief overview of the general method,
describing the subtraction procedure and how our dipole formulae are used to
implement it.  In Sect.~\ref{not} we establish the notation used throughout
the paper.  In Sect.~\ref{facsc} we review the factorization properties of QCD
matrix elements in the soft and collinear limits before presenting, in
Sect.~\ref{dff}, our dipole factorization formulae, which smoothly interpolate
these two limiting regions.  After briefly recalling, in Sect.~\ref{NLOxs},
the precise definitions of QCD cross sections at NLO, we go on to describe in
detail our subtraction method for evaluating these cross sections, in
Sects.~\ref{siee}--\ref{eemp}.  In Sect.~\ref{summ} we summarize and discuss
our results.  Appendix~A gives more details, and some examples, of the
necessary colour algebra.  In Appendix~B we
explicitly perform the only difficult
integral we encounter.  In Appendix~C we collect together the main formulae
needed to implement our method in specific calculations.  Finally in
Appendix~D we work through a few simple examples of applying our method to
specific cross sections.

Since the paper is quite long, readers mainly interested in understanding the
general method or in some particular application are advised to first read
Sects.~\ref{gen},
\ref{not}, \ref{facsc}, \ref{dfss} and~\ref{siee}. Here we discuss in detail our
general formalism and its use for processes with no initial-state hadrons like,
for instance, \ee\ annihilation (in this case, a brief description of
our method has already appeared [\ref{CSlett}]).
Sections~\ref{sidis}--\ref{eemp}
and the other Subsections in Sect.~\ref{dff} can then be read quite
independently from one another. The final formulae that are necessary for the
actual numerical implementation of our algorithm in  each different scattering
process are summarized in a Subsection at the end of each of
Sects.~\ref{siee}--\ref{eemp}.

\newpage

\setcounter{equation}{0}

\section{The general method}
\label{gen}

In this Section we explain the general idea behind our version of the
subtraction method by describing the subtraction procedure (Sect.~\ref{subproc})
and considering mainly the simplified case of jet cross sections
in processes with no initial-state hadrons, for instance, \ee\ annihilation
(Sect.~\ref{subee}). A brief description of our method for more complicated
scattering processes is sketched in Sect.~\ref{subpp}.

\subsection{The subtraction procedure}
\label{subproc}

Suppose we want to compute a jet cross section $\sigma$ to NLO, namely
\beq
\label{sig}
\sigma = \sigma^{LO} + \sigma^{NLO} \;.
\eeq
Here the LO cross section $\sigma^{LO}$ is obtained by integrating the fully
exclusive cross section $d\sigma^{B}$ in the Born approximation over the phase
space for the corresponding jet quantity. Suppose also that this LO calculation
involves $m$ partons in the final state. Thus, we write
\beq
\label{sLO}
\sigma^{LO} = \int_m d\sigma^{B} \;,
\eeq
where, in general, all the quantities (QCD matrix elements and
phase space) are evaluated in $d=4-2\ep$ space-time dimensions. However, by
definition, at this LO the phase space integration in Eq.~(\ref{sLO}) is finite
so that the whole calculation can be carried out (analytically or numerically)
in four dimensions.

Now we go to NLO\@. We have to consider the exclusive cross section
$d\sigma^{R}$
with $m+1$ partons in the final-state and the one-loop correction $d\sigma^{V}$
to the process with $m$ partons in the final state:
\beq
\label{sNLO}
\sigma^{NLO} \equiv \int d\sigma^{NLO} =
 \int_{m+1} d\sigma^{R} + \int_{m} d\sigma^{V} \;.
\eeq

The two integrals on the right-hand side of Eq.~(\ref{sNLO}) are separately
divergent if $d=4$, although their sum is finite. Therefore, before any
numerical calculation can be attempted, the separate pieces have to be
regularized. Using dimensional regularization, the divergences (arising out
of the integration) are replaced
by double (soft and collinear) poles $1/\ep^2$ and single (soft, collinear
or ultraviolet) poles $1/\ep$. Suppose that one has already carried out the
renormalization procedure in $d\sigma^{V}$ so that all its ultraviolet poles
have been removed.

The general idea of the subtraction method for writing a general-purpose
Monte Carlo program is to use the identity
\beq
\label{dsNLO}
d\sigma^{NLO} = \left[ d\sigma^{R} - d\sigma^{A}  \right]
+  d\sigma^{A} +  d\sigma^{V} \;\;,
\eeq
where $d\sigma^{A}$ is a proper approximation of $d\sigma^{R}$ such as to have
the same {\em pointwise\/} singular behaviour (in $d$ dimensions) as
$d\sigma^{R}$ itself.
Thus, $d\sigma^{A}$ acts as a {\em local\/} counterterm for
$d\sigma^{R}$ and, introducing the phase space integration,
\beq
\label{sNLO1}
\sigma^{NLO} = \int_{m+1} \left[ d\sigma^{R} - d\sigma^{A}  \right]
+  \int_{m+1} d\sigma^{A} +  \int_m d\sigma^{V} \;\;,
\eeq
one can safely perform the limit $\ep \to 0$ under the integral sign in the
first
term on the right-hand side of Eq.~(\ref{sNLO1}). Hence, this first term can be
integrated numerically in four dimensions.

All the singularities are now associated to the last two terms on the
right-hand side of Eq.~(\ref{sNLO1}). If one is able to carry out analytically
the integration of $d\sigma^{A}$ over the one-parton subspace leading to the
$\ep$ poles, one can combine these poles with those in $d\sigma^{V}$, thus
cancelling all the divergences, performing the limit $\ep \to 0$ and
carrying out numerically the remaining integration over the $m$-parton phase
space. The final structure of the calculation is as follows
\beq
\label{sNLO2}
\sigma^{NLO} =
\int_{m+1} \left[ \left( d\sigma^{R} \right)_{\ep=0}
- \left( d\sigma^{A} \right)_{\ep=0}  \;\right]
+  \int_m \left[ d\sigma^{V} +  \int_1 d\sigma^{A} \right]_{\ep=0} \;\;,
\eeq
and can be easily implemented in a `partonic Monte Carlo' program, which generates
appropriately weighted partonic events with $m+1$ final-state partons and
events with $m$ partons.

Note that the subtracted term $[\;d\sigma^{R} - d\sigma^{A} \;]$ in
Eq.~(\ref{sNLO2}) is integrable in four dimensions by definition. The fact
that all the divergences cancel in the second term on the right-hand side of
Eq.~(\ref{sNLO2}) is instead not a general feature of all hadronic cross
section\footnote{The presence of singularities in a QCD cross section computed
in perturbation theory does not mean that the theory itself is inconsistent.
It simply means that one is considering a cross section that cannot be
reliably
estimated using the perturbative expansion. At any energy scale, it
is affected by non-perturbative phenomena that are as big as the
perturbative ones.}. The cancellation of divergences is guaranteed only for the
hadronic observables that we are considering in this paper, namely jet
observables.

These quantities have to be experimentally (theoretically) defined in such a
way that their actual value is independent of the number of soft and
collinear hadrons (partons) produced in the final state. In particular, this
value has to be the same in a given $m$-parton configuration and in all
$m+1$-parton configurations that are kinematically degenerate with it
(i.e.\ that are obtained from the $m$-parton configuration by adding a soft
parton or replacing a parton with a pair of collinear partons carrying the same
total momentum). This property can be simply
restated in a formal way. If the function $F_J^{(n)}$ gives the value of a
certain jet observable in terms of the momenta of the $n$ final-state partons,
we should have
\beq\label{fjm}
F_J^{(m+1)} \to F_J^{(m)} \;\;,
\eeq
in any case where the $m+1$-parton and the $m$-parton configurations are
kinematically degenerate.

The Born-level cross section $d\sigma^{B}$ can be (symbolically) written as
a function of the jet-defining function $F_J^{(m)}$ in the following way
\beq\label{dsbf}
d\sigma^{B} = d\Phi^{(m)} \; |{\cal M}_m|^2 \;F_J^{(m)} \;\;,
\eeq
where $d\Phi^{(m)}$ and ${\cal M}_m$ respectively are the full phase space and
the QCD matrix element to produce $m$ final-state partons. The corresponding
expression for the real cross section $d\sigma^{R}$ is:
\beq\label{dsrf}
d\sigma^{R} = d\Phi^{(m+1)} \; |{\cal M}_{m+1}|^2 \;F_J^{(m+1)} \;\;.
\eeq

The structure of Eq.~(\ref{dsrf}) and the fundamental property (\ref{fjm}) are
essential for the feasibility of the subtraction procedure described in this
Subsection. There are obviously many ways of approximating the matrix element
${\cal M}_{m+1}$ in the neighbourhood of its soft and collinear singularities.
Correspondingly, one can approximate $F_J^{(m+1)}$ and obtain a local
counter-term $d\sigma^{A}$. The main point is that, due to the limiting
behaviour in Eq.~(\ref{fjm}), one can always find an approximation for
$F_J^{(m+1)}$ such that the one-parton subspace leading to the soft and
collinear divergences effectively decouples. Thus, one can perform the integral
$\int_1 d\sigma^{A}$ and the subtraction formula (\ref{sNLO2}) can,
{\em in principle}, always be implemented.

\subsection{Dipole formulae and universal implementation of the subtraction
procedure for jet cross sections}
\label{subee}

The key for the subtraction procedure to work {\em in practice\/} is obviously
the actual form of $d\sigma^{A}$.
One needs to find an expression for $d\sigma^{A}$ that fulfils the following
properties: $i)$~for any given process, $d\sigma^{A}$ has to be obtained in a
way
that is independent of the particular jet observable considered; $ii)$~it has
to exactly match the singular behaviour of $d\sigma^{R}$ in $d$ dimensions;
$iii)$~its form has to be
particularly convenient for Monte Carlo integration techniques;
$iv)$~it has to be exactly integrable analytically in $d$
dimensions over the single-parton subspaces leading to soft and collinear
divergences.

In Ref.~[\ref{ERT}], a suitable expression for $d\sigma^{A}$ for the process
\ee$\to 3$~jets was obtained by starting from the explicit expression (in
$d$ dimensions) of the corresponding $d\sigma^{R}$ and by performing extensive
partial fractioning of the $3+1$-parton matrix elements, so that each divergent
piece could be extracted. This is an extremely laborious and ungeneralizable
task, in the sense that having done it for \ee$\to 3$~jets does not help
us to do this for, say, \ee$\to 4$~jets or for any other process.

In Ref.~[\ref{KS}], the general properties of soft and collinear emission were
first used (in the context of the subtraction method) to construct
$d\sigma^{A}$, for one- and two-jet production in hadron collisions, in a way
that is independent of the detailed form of the corresponding $d\sigma^{R}$.

The central proposal of our version of the subtraction method is that one can
give a recipe for constructing $d\sigma^{A}$ that is completely {\em process
independent\/} (and not simply independent of the jet observable).
Starting from our physical knowledge of how the $m+1$-parton matrix elements
behave in the soft and collinear limits that produce the divergences (see
Sect.~\ref{facsc}), we derive improved factorization formulae, called
dipole formulae (see Sect.~\ref{dff}), that
allow us to write:
\beq
\label{dsA}
d\sigma^{A} = \sum_{{\rm dipoles}} \;d\sigma^{B} \otimes dV_{{\rm dipole}}
 \;\;.
\eeq
The notation in Eq.~(\ref{dsA}) is symbolic. Here $d\sigma^{B}$ denotes an
appropriate colour and spin projection of the Born-level exclusive cross
section. The symbol $\otimes$ stands for  properly defined phase
space convolutions and sums over colour and spin indices. The dipole factors
$dV_{{\rm dipole}}$ (which match the singular behaviour of $d\sigma^{R}$) are
instead universal, i.e.\ completely independent of the details of the process
and they can be computed once for all. In particular, the dependence on the
jet observable is completely embodied by the factor $d\sigma^{B}$ of
Eq.~(\ref{dsA}), in the form of Eq.~(\ref{dsbf}).

There are several dipole terms on the right-hand side of Eq.~(\ref{dsA}).
Each of them corresponds to a different kinematic configuration of $m+1$
partons. Each configuration can be thought as obtained by an effective two-step
process: an $m$-parton configuration is first produced and then one of these
partons decays into two partons. It is this two-step pseudo-process that leads
to the factorized structure on the right-hand side of Eq.~(\ref{dsA}).

The reason for having several dipoles is that each of them mimics one of the
$m+1$-parton configurations in $d\sigma^{R}$ that are kinematically degenerate
with a given $m$-parton state. Any time the $m+1$-parton state in $d\sigma^{R}$
approaches a soft and/or collinear region, there is a corresponding dipole
factor in $d\sigma^{A}$ that approaches the same region with exactly the
same probability as in $d\sigma^{R}$. In this manner $d\sigma^{A}$ acts as
a local counter-term for $d\sigma^{R}$.

Our expression for $d\sigma^{A}$ in Eq.~(\ref{dsA}) is completely defined over
the full $m+1$-parton phase space (in
particular, $d\sigma^{A}$ does not depend on any additional phase space
cut-off\footnote{This is quite a non-trivial feature of our approach. The most
\naive\ way of setting up subtraction procedures based on universal properties
in the soft and collinear limits would lead to the introduction of energy and
angular cut-offs, thus breaking Lorentz covariance at intermediate steps.
Alternative and less \naive\ methods for imposing soft and collinear cut-offs
have their own disadvantages, too. For instance, the cut-off can be related
to some kinematic invariant of the process (see Ref.~[\ref{KS}]), or one can
introduce a Lorentz covariant cut-off on parton-parton invariant masses
(see Ref.~[\ref{GG}]) rather than energies and angles. In the first case one
could spoil the universality of the subtraction procedure making it process
dependent. In the second case, it is quite difficult to arrange the cut-off
in such a way that the subtraction term is exactly integrable analytically
to any accuracy in the cut-off itself.}): momentum conservation is exactly
implemented
in each term on the right-hand side of Eq.~(\ref{dsA}) and there is a
one-to-one correspondence
between
each partonic configuration in $d\sigma^{R}$ and (each of the several)
in $d\sigma^{A}$. Therefore,
in our case, $\left[ d\sigma^{R} - d\sigma^{A} \right]$ is straightforwardly
integrable via Monte Carlo methods: one generates an $m+1$-parton event with
weight $d\sigma^{R}$ and, correspondingly, one can obtain an
$m+1$-parton counter-event with weight $d\sigma^{A}$.

Furthermore, the product structure in Eq.~(\ref{dsA}) allows us a factorizable
mapping from the $m+1$-parton phase space to an $m$-parton subspace
(that identified by the partonic variable in $d\sigma^{B}$) times a
single-parton phase space (that identified by the dipole partonic variables in
$dV_{{\rm dipole}}$). This mapping makes $dV_{{\rm dipole}}$ fully integrable
analytically. We can write (again, symbolically):
\beq
\label{dsA1}
\int_{m+1} d\sigma^{A} = \sum_{{\rm dipoles}} \;\int_m \;d\sigma^{B} \otimes
\int_1 \;dV_{{\rm dipole}} = \int_m \left[ d\sigma^{B} \otimes {\bom I}
\right] \;\;,
\eeq
where the universal factor ${\bom I}$ is defined by
\beq
\label{Ifac}
{\bom I} = \sum_{{\rm dipoles}} \;\int_1 \;dV_{{\rm dipole}} \;\;,
\eeq
 and contains all the $\ep$ poles that are necessary to cancel the (equal and
with opposite sign) poles in $d\sigma^{V}$.

The structure of the final result is given as follows in terms of two
contributions $\sigma^{NLO\,\{m+1\}}, \sigma^{NLO\,\{m\}}$ (with $m+1$-parton
and $m$-parton kinematics, respectively) which are separately finite
(and integrable) in four space-time dimensions:
\beeq
\label{sNLO3}
\sigma^{NLO} &=& \sigma^{NLO\,\{m+1\}} + \sigma^{NLO\,\{m\}}  \\
&=&\int_{m+1} \left[ \left( d\sigma^{R} \right)_{\ep=0} -
\left( \sum_{{\rm dipoles}} \;d\sigma^{B} \otimes \;dV_{{\rm dipole}}
\right)_{\ep=0} \;\right]
+  \int_m \left[ d\sigma^{V} +
d\sigma^{B} \otimes {\bom I}
\right]_{\ep=0} \;\;.\nonumber
\eeeq
Equation (\ref{sNLO3}) represents our practical implementation of the general
subtraction formula (\ref{sNLO2}).

In this paper we provide explicit expressions for both the universal factors
$dV_{{\rm dipole}}$ and~${\bom I}$.
Having
these factors at our disposal, the only other ingredients necessary for the
full NLO calculation, according to Eq.~(\ref{sNLO3}), are the following
(reading Eq.~(\ref{sNLO3}) from the right to the left):
\begin{itemize}
\item
a set
of independent colour projections\footnote{Actually, if the total number of
QCD partons involved in the LO matrix element is less than or equal to three, one
simply needs its incoherent sum over the colours (see Appendix A).}
of the matrix element
squared at the Born level, summed over parton polarizations, in $d$ dimensions;

\item
the one-loop contribution $d\sigma^{V}$ in $d$ dimensions;

\item
an additional projection of the Born level matrix element over the helicity of
each external gluon in four dimensions;

\item
the real emission  contribution $d\sigma^{R}$ in four dimensions.
\end{itemize}

\noindent These few ingredients are sufficient for writing, in a
straightforward way,
a general-purpose NLO Monte Carlo algorithm. Note in particular
that there is no need to extract a proper counter-term $d\sigma^{A}$
starting from a
cumbersome expression for $d\sigma^{R}$ in $d$ dimensions. The NLO matrix
element
contributing to $d\sigma^{R}$ can be evaluated directly in four space-time
dimensions thus leading to an extreme simplification of the Lorentz algebra,
particularly if one makes use of helicity amplitudes [\ref{HA}] to control
the rapid increase in the number of Feynman diagrams as the number of parton
grows.

\subsection{Factorization of collinear singularities and general algorithm for
processes  with identified hadrons}
\label{subpp}

The discussion in Sects.~\ref{subproc} and \ref{subee} applies to all the
processes with no initial-state hadrons (for instance, \ee\ annihilation).
However, perturbative QCD can be used also for the calculation of jet cross
sections in lepton-hadron and hadron-hadron collisions\footnote{Throughout
this paper, whenever referring to initial-state hadrons, we implicitly also
include hadron-like particles such as photons.}. The main difference
is that the presence of initial-state hadrons (partons), carrying a well
defined momentum, spoils the cancellation
of the collinear singularities arising in the perturbative treatment. The
left-over singularities can be factorized and reabsorbed into non-perturbative
and universal (process-independent) distribution functions, the parton
densities of the incoming hadron. Provided this factorization procedure is
consistently carried out, one can thus define parton-initiated jet cross
sections (see Sect.~\ref{NLOxs}) that are free from singularities and can be
computed with a subtraction procedure similar to that described in
Sect.~\ref{subproc}.

Similar features appear when the jet observable depends on the actual value of
the momentum of one or more hadrons observed in the final state (the inclusive
one-particle distribution in \ee\ annihilation is the simplest example).
Also in this case there are left-over collinear singularities that can
be reabsorbed into non-perturbative and universal distribution functions, the
fragmentation functions of the outgoing hadron. As a result, one can again
define partonic cross sections that are free from singularities and
computable in perturbation theory (see Sect.~\ref{NLOxs}).

Because of these common features, in this paper the processes with
initial-state
hadrons and those involving fragmentation functions will be referred to as
processes with identified hadrons (partons). The hadronic cross section is
obtained by convoluting partonic cross sections with non-perturbative
distribution functions. As stated above, the
NLO partonic cross sections can be evaluated using the subtraction method and,
thus, trying to implement the subtraction formula in Eq.~(\ref{sNLO2}).
There are nonetheless some additional complications with respect to the case
with no identified particles. These complications regard the construction of
the approximated cross section $d\sigma^{A}$.

Just as when there are no identified hadrons, the real cross section
$d\sigma^{R}$ is singular whenever a pair of the $m+1$ final-state partons
become collinear.  However in addition, it is also singular in the region
where one of them becomes collinear to an identified parton.
Moreover, the phase space integration has to be performed in the presence of
additional kinematic constraints, related to the fact that the momenta of the
identified partons have to be kept fixed (or, at most, rescaled by an overall
momentum fraction). As for the approximated cross section
$d\sigma^{A}$, it follows that, on one side, it should act as a local
counter-term also in the new
singular regions and, on the other side, its integral $\int_1 d\sigma^{A}$
should still be computable analytically even in the presence of the additional
phase space constraints.

The dipole formalism presented in this paper provides a simple and general
solution to these problems. Indeed, we are able to write the cross section
$d\sigma^{A}$ in the following form (see Sects.~\ref{sidis}--\ref{eemp})
\beq\label{dsAprime}
d\sigma^{A} = \sum_{{\rm dipoles}} \;d\sigma^{B} \otimes \left(
dV_{{\rm dipole}} + dV^{\prime}_{{\rm dipole}} \right) \;\;.
\eeq
Equation (\ref{dsAprime}) is completely analogous to Eq.~(\ref{dsA}). The
additional dipole terms $dV^{\prime}_{{\rm dipole}}$ on the right-hand side
match the singularities of $d\sigma^{R}$ coming from the region collinear
to the momenta of the identified partons. Moreover, these dipole terms are
still (i.e.\ even if the momenta of the identified partons are fixed) fully
integrable analytically over the one-parton subspace leading to soft and
collinear divergences.

These are peculiar features of the dipole approach. As discussed below
Eq.~(\ref{dsA}), each dipole contribution is effectively obtained by first
producing an $m$-parton configuration and then letting one parton
to decaying into two partons. The dipole formulae implement this two-step
procedure by enforcing exact momentum conservation. Actually there are
equivalent ways of doing that,  corresponding to different ways of treating
the momentum recoil in the $m$-parton configuration. This freedom allows us
to define alternative versions of the factorization formulae (Sect.~\ref{dff})
and, hence, different dipole factors (like $dV_{{\rm dipole}}$ and
$dV^{\prime}_{{\rm dipole}}$ in Eq.~(\ref{dsAprime})).  These  differences are
then used to match (and overcome) the phase-space constraints that are
encountered in the calculation of QCD cross sections with identified particles.

Having introduced the counter-term $d\sigma^A$ in Eq.~(\ref{dsAprime}), we can
proceed to its integration as in Eq.~(\ref{dsA1}). We thus obtain the singular
factor $\bom I$ in Eq.~(\ref{Ifac}) and additional singular terms, which are
reabsorbed into the non-perturbative distribution functions.

The final result of our subtraction procedure is given in terms of the NLO
partonic cross section $\sigma^{NLO}(p)$, where the dependence on the momentum
$p$ symbolically denotes the functional
dependence on the momenta of the identified partons. This cross section is
obtained by a formula that is similar to
Eq.~(\ref{sNLO3}), namely (see Sects.~\ref{sidis}--\ref{eemp})
\beeq\label{sNLO4}
\sigma^{NLO}(p) &=&
\sigma^{NLO\,\{m+1\}}(p) + \sigma^{NLO\,\{m\}}(p) + \int_0^1 dx \;
{\hat \sigma}^{NLO\,\{m\}}(x;xp) \nonumber \\
&=&
\int_{m+1} \left[ \left( d\sigma^{R}(p) \right)_{\ep=0} -
\left( \sum_{{\rm dipoles}} \;d\sigma^{B}(p) \otimes \left( \;dV_{{\rm dipole}}
+  dV^{\prime}_{{\rm dipole}} \right)
\right)_{\ep=0} \;\right]  \\
&+&  \int_m \left[ d\sigma^{V}(p) +
d\sigma^{B}(p) \otimes {\bom I}
\right]_{\ep=0} + \int_0^1 dx \;
\int_m \left[ d\sigma^{B}(xp) \otimes \left( {\bom P}
+ {\bom K} + {\bom H} \right)(x) \right]_{\ep=0} \;\;. \nonumber
\eeeq
Here, the contributions
$\sigma^{NLO\,\{m+1\}}(p)$ and $\sigma^{NLO\,\{m\}}(p)$ (with $m+1$-parton
and $m$-parton kinematics, respectively) are completely analogous to those
in Eq.~(\ref{sNLO3}).

The last term on the right-hand side of Eq.~(\ref{sNLO4}) is a finite
(in four dimensions) remainder that is left after factorization of
initial-state and final-state collinear singularities into the non-perturbative
distribution functions (parton densities and fragmentation functions). This term
involves a cross section ${\hat \sigma}^{NLO\,\{m\}}(x;xp)$ with $m$-parton
kinematics and an additional one-dimensional integration with respect to the
longitudinal momentum fraction $x$. This integration arises from the convolution
of the Born-type cross section $d\sigma^{B}(xp)$ with $x$-dependent functions
${\bom P}, {\bom K}, {\bom H}$ that are similar (but finite for $\ep \to 0$)
to the factor~${\bom I}$. The
functions ${\bom P}, {\bom K}$ and ${\bom H}$ are universal, that is,  they are
independent of the detail of the scattering process and of the jet observable:
they simply depend on the number of identified partons. Our algorithm provides
the explicit expressions for these functions. Therefore, writing a
general-purpose NLO Monte Carlo program for processes with identified particles does
not require any further conceptual or analytic effort with respect to the case
with no identified particles.

\newpage

\setcounter{equation}{0}

\section{Notation}
\label{not}

\subsection{Dimensional regularization}
\label{dimr}

In general we use dimensional regularization in $d=4-2\ep$
space-time dimensions and consider $d-2$ helicity states for gluons
and 2 helicity states for massless quarks (i.e.~fermions are four-component
spinors). This defines the usual
dimensional-regularization scheme. Other dimensional-regularization
prescriptions can be used. However, since the regularization dependence is
unphysical (i.e.\ it cancels in physical cross sections), within our formalism it
is more convenient to parametrize it in terms of simple coefficients that
enter in the one-loop contribution (see Sect.~\ref{oneloop}).

The dimensional-regularization scale, which appears in the calculation
of the matrix elements, is denoted by $\mu$. Physical cross sections do not
depend on $\mu$, although, when evaluated in fixed-order perturbation theory,
they do depend on the renormalization scale $\mu_R$
and on factorization scales $\mu_F$. In other words, the dependence on
$\mu$ cancels after having combined the matrix elements in the calculation
of physical cross sections. Therefore, in order to avoid a cumbersome notation,
we set $\mu=\mu_R$, while $\mu$ and $\mu_F$ will differ in general.

The $d$-dimensional phase space, which involves the integration over the
momenta $\{p_1, ...,$ $p_m\}$ of $m$ final-state partons, will be denoted as
follows
\beq
\label{psm}
\left[ \;\prod_{l=1}^{m} \frac{d^dp_l}{(2\pi)^{d-1}}
\,\delta_+(p_l^2) \right] \;(2\pi)^{d} \,\delta^{(d)}(p_1 + .... + p_m - Q)
\equiv d\phi_m(p_1, ...,p_m;Q) \;\;.
\eeq
When there is no ambiguity on the number of final-state partons, we shall drop
the subscript $m$ in $d\phi_m$.

\subsection{Matrix elements}
\label{me}

Let us first consider processes that involve only final-state
QCD partons (\ee-type processes). Non QCD partons
($\gamma^*, Z^0, W^{\pm}, \cdots$), carrying
a total incoming momentum $Q_\mu$, are always understood.

The ({\em tree-level\/}) matrix element with $m$ QCD partons in the final state
has the following general structure
\beq
\cm_m^{c_1, ...,c_m; s_1, ...,s_m}(p_1, ...,p_m)
\eeq
where $\{c_1, ...,c_m\}$, $\{s_1, ...,s_m\}$ and
$\{p_1, ...,p_m\}$ are respectively colour indices
($a= 1, ...,$ $N_c^2-1$ different colours for each gluon,
$\alpha=1, ..,N_c$ different colours for each quark or antiquark),
spin indices ($\mu=1,...,d$ for gluons, $s=1,2$ for massless fermions)
and momenta.

It is useful to introduce a basis
$\{ \ket{c_1,...,c_m} \otimes \ket{s_1,...,s_m} \}$
in colour + helicity space in such a way that
\beq\label{cmmdef}
\cm_m^{c_1, ...,c_m; s_1, ...,s_m}(p_1, ...,p_m) \equiv
\Bigl( \bra{c_1,...,c_m} \otimes \bra{s_1,...,s_m} \Bigr) \mket{1, ...., m} \;.
\eeq
Thus $\mket{1, ...., m}$ is a vector in colour + helicity space.

According to this notation, the matrix element squared (summed
over final-state colours and spins) $\M{}{m}$ can be written as
\beq
\M{}{m} = {}_m\!\!<{1, ...., m}|{1, ...., m}>_m \;.
\eeq

In the following we shall always consider matrix elements squared summed
over final-state spins (the generalization to fixed-helicity
amplitudes is feasible).

As for the colour structure\footnote{Within our formalism, there is no need
to consider the decomposition of the matrix elements into colour
subamplitudes [\ref{MP}], as in Ref.~[\ref{GG}].}, it is convenient to
associate a colour charge ${\bf T}_i$
with the emission of a gluon from each parton $i$. If the emitted gluon
has colour index $c$, the colour-charge operator is:
\beq
{\bom T}_i \equiv T_i^c \ket{c}
\eeq
and its action onto the colour space is defined by
\beq
\bra{c_1, ..., c_i, .. c_m, c} {\bom T}_i
\ket{b_1, ..., b_i, .. b_m} = \delta_{c_1 b_1} ....
T_{c_i b_i}^c ...\delta_{c_m b_m} \;\;,
\eeq
where $T_{c b}^a \equiv i f_{cab}$ (colour-charge matrix
in the adjoint representation)  if the emitting particle $i$
is a gluon and $T_{\alpha \beta}^a \equiv t^a_{\alpha \beta}$
(colour-charge matrix in the fundamental representation)
if the emitting particle $i$ is a quark (in the case of an emitting
antiquark $T_{\alpha \beta}^a \equiv {\bar t}^a_{\alpha \beta}
= - t^a_{\beta \alpha }$).

The colour-charge algebra is:
\beq
{\bom T}_i \cdot {\bom T}_j ={\bom T}_j \cdot {\bom T}_i \;\;\;\;{\rm if}
\;\;i \neq j; \;\;\;\;\;\;{\bom T}_i^2= C_i,
\eeq
where $C_i$ is the Casimir operator, that is,
$C_i=C_A=N_c$ if $i$ is a gluon and $C_i=C_F=(N_c^2-1)/2N_c$ if $i$ is a quark
or antiquark.

Note that by definition, each vector $\mket{1, ...., m}$ is a
colour-singlet state. Therefore colour conservation is simply
\beq \label{cocon}
\sum_{i=1}^m {\bom T}_i \;\mket{1, ...., m} = 0 \;.
\eeq

Using this notation, we also define the square of colour-correlated
tree-amplitudes, $|{\cal{M}}_{m}^{i,k}|^2$, as follows
\beeq
\label{colam}
|{\cal{M}}_{m}^{i,k}|^2 &\equiv&
{}_m\!\!<{1, ..., m}| \,{\bom T}_i \cdot {\bom T}_k \,|{1, ..., m}>_m
\nonumber \\
&=&
\left[ {\cal M}_m^{a_1.. b_i ... b_k ... a_m}(p_1,...,p_m) \right]^*
\; T_{b_ia_i}^c \, T_{b_ka_k}^c
\; {\cal M}_m^{a_1.. a_i ... a_k ... a_m}(p_1,...,p_m) \;.
\eeeq

In the case of hard processes with QCD partons in the initial state,
in addition to $m$ final-state partons, the
relevant matrix element is:
\beq
\label{mma}
\cm_{m,a...}^{c_1, ...,c_m,c_a, ...; s_1, ...,s_m,s_a, ...}(p_1, ...,p_m;
p_a, ...) \;\;,
\eeq
and the corresponding vector in colour + helicity space will be denoted
in the following way
\beeq
\label{ketin}
\ket{1, ...., m;a, ...}_{m,a...} &\equiv& \frac{1}{\sqrt {n_c(a) \dots}}
\left( \ket{c_1,..,c_m,c_a,..} \otimes \ket{s_1,..,s_m,s_a,..} \right)
\nonumber \\
&\cdot&
\cm_{m,a...}^{c_1, ...,c_m,c_a, ...; s_1, ...,s_m,s_a, ...}(p_1, ...,p_m;
p_a, ...) \;\;.
\eeeq
Here the labels $a, ...$ refer to the initial-state partons. The normalization
of the state vector in Eq.~(\ref{ketin}) is fixed by including a factor of
$1/\sqrt {n_c(a)}$ for each initial-state parton carrying $n_c(a)$ colours.

Note that the
colour-charge operator of an initial-state parton $a$ is defined by crossing
symmetry, that is, $({\bom T}_a)^c_{bd} = i f_{bcd}$ if $a$ is a gluon and
$({\bom T}_a)^c_{\alpha \beta} = {\bar t}^c_{\alpha \beta} =
- t^c_{\beta \alpha}$ if $a$ is a quark
(if $a$ is an antiquark, $({\bom T}_a)_{\alpha \beta}^c
= t^c_{\alpha \beta}$).
The analogue of the colour-conservation condition (\ref{cocon}) is:
\beq
\left( \sum_{i=1}^m {\bom T}_i + {\bom T}_a + ... \right)
\ket{1, ...., m;a, ...}_{m,a..} = 0 \;.
\eeq

Owing to the normalization of the state vector in Eq.~(\ref{ketin}), the
square of colour-correlated tree-amplitudes is:
\beeq
\label{ccamp}
|{\cal{M}}_{m,a...}^{I,J}|^2 &\equiv&
{}_{m,a...}\!\!<{1, ..., m; a,...}| \,{\bom T}_I \cdot {\bom T}_J \,|{1, ...,
  m; a,...}>_{m,a...}
\nonumber \\
&=&
\frac1{n_c(a) \dots}
\left[ {\cal M}_{m,a...}^{a_1.. b_I ...b_J...}(p_1,...,p_m;p_a,...) \right]^*
\; T_{b_Ia_I}^c \, T_{b_Ja_J}^c
\nonumber\\&&\phantom{\frac1{n_c(a) \dots}}
\; {\cal M}_{m,a...}^{a_1.. a_I ... a_J ...}(p_1,...,p_m;p_a,...) \;,
\eeeq
where the indices $I,J$ refer either to final-state or initial-state partons.

We refer to Appendix A for more details of the colour algebra.

\subsection{One-loop matrix elements and scheme (in)dependence}
\label{oneloop}

We denote by $| \cm_{m,a...}^{(bare)}(p_1, ...,p_m;p_a, ...)|^2_{(1-loop)}$
the one-loop correction to the square of the tree-level matrix element in
Eq.~(\ref{mma}). This term enters in the computation of the virtual
contribution $d\sigma^V$ to the NLO QCD cross section. Actually, $d\sigma^V$
is proportional to the renormalized one-loop correction
$| \cm_{m,a...}(p_1, ...,p_m;p_a, ...)|^2_{(1-loop)}$ and the latter is
obtained from the corresponding bare quantity by simply adding an ultraviolet
counterterm. More precisely, if the (tree-level) matrix element squared
$| \cm_{m,a...}(p_1, ...,p_m;p_a, ...)|^2$ is proportional to the $n$-th power
of the QCD coupling $\as$, the renormalized one-loop correction is given by
\beeq
\label{mren}
\!\!| \cm_{m,a...}(p_1, ...,p_m;p_a, ...)|^2_{(1-loop)} \!\!&=&\!\!
| \cm_{m,a...}^{(bare)}(p_1, ...,p_m;p_a, ...)|^2_{(1-loop)} \\
\!\!\!\!&-&\!\! n \;\frac{\as}{2\pi} \frac{(4\pi)^\ep}{\Gamma(1-\ep)}
\left( \frac{\beta_0}{\ep} + {\tilde \beta}_0 \right)
\;| \cm_{m,a...}(p_1, ...,p_m;p_a, ...)|^2 \;\;, \nonumber
\eeeq
where $\beta_0= (11 C_A - 4 T_R N_f)/6$ is the first coefficient of the QCD
$\beta$-function ($N_f$ is the number of quark flavours and ${\rm Tr} (t^at^b) =
\delta^{ab} T_R$, i.e.\ $T_R=1/2$)
and ${\tilde \beta}_0$ is an $\ep$-independent and process-independent
coefficient that defines the renormalization scheme.
In Eq.~(\ref{mren}) and in the rest of the paper, $\as$ stands for $\as(\mu)$,
the NLO QCD running coupling evaluated at the renormalization scale $\mu$.
The actual value of the QCD coupling $\as(\mu)$ depends on the
renormalization scheme. The customary ${\overline {\rm MS}}$ scheme is
obtained by setting ${\tilde \beta}_0=0$ in Eq.~(\ref{mren}).

The detailed expression of
$| \cm_{m,a...}^{(bare)}(p_1, ...,p_m;p_a, ...)|^2_{(1-loop)}$ (and, hence,
of $| \cm |^2_{(1-loop)}$) depends on the dimensional regularization
procedure used for evaluating the loop integral. Since we are using conventional
dimensional regularization, we need the result for
$| \cm_{m,a...}^{(bare)}|^2_{(1-loop)}$ within this regularization scheme. If
the one-loop correction is known in a different scheme, we have to introduce
a correction factor proportional to the corresponding tree-level amplitude. More
precisely, we have to perform the following replacement in Eq.~(\ref{mren}):
\beeq
\label{mscheme}
| \cm_{m,a...}(p_1, ...,p_m;p_a, ...)|^2_{(1-loop)} &\!\to\!&
| \cm_{m,a...}(p_1, ...,p_m;p_a, ...)|^2_{(1-loop)}  \\
&\!\!&- \;\frac{\as}{2\pi}
\left[\; \sum_{i=1}^m  {\tilde \gamma}_i +  {\tilde \gamma}_a + \dots \right]
\;| \cm_{m,a...}(p_1, ...,p_m;p_a, ...)|^2 \;\;, \nonumber
\eeeq
where the universal coefficients ${\tilde \gamma}_i, {\tilde \gamma}_a, \dots$
depend only on the flavour of the QCD partons. The actual values of these
coefficients for several different regularization schemes can be found, for
instance, in Ref.~[\ref{KST}].

\newpage

\setcounter{equation}{0}

\section{Factorization in the soft and collinear limits}
\label{facsc}

\subsection{Soft and collinear singularities in tree-level amplitudes}
\label{softcoll}

\begin{figure}[b]
\centerline{\epsfbox{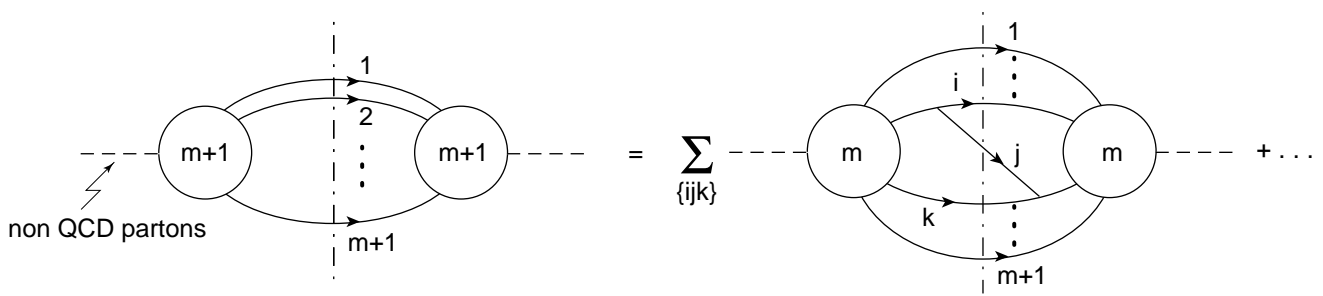}}
\caption{Diagrammatic representation of the external-leg insertion rule.
The blobs denote the tree-level matrix elements and their complex conjugate.
The dots on the right-hand side stand for non-singular terms both in the soft
and collinear limits.}
\end{figure}

Let us consider a generic tree-level matrix element $\cm_{m+1,a...}$
with $m+1$ QCD partons (Fig.~1). The dependence of $|\cm_{m+1,a...}|^2$
on the momentum $p_j$ of a final-state parton $j$ is singular in two different
phase-space regions: a) in the {\em soft\/} region, defined by the limit
$p_j = \lambda q , \;\;\lambda \to 0$ (where $q$ is an arbitrary four
momentum),
$|\cm_{m+1,a...}|^2$ behaves as $1/\lambda^2$; in the {\em collinear\/} region,
defined by the limit $p_j \to (1-z)p_i/z$ (where $p_i$ is the momentum of another
QCD parton in $\cm_{m+1,a...}$), $|\cm_{m+1,a...}|^2$ behaves as $1/(p_ip_j)$.
This singular behaviour of $|\cm_{m+1,a...}|^2$ leads to the soft and
collinear
divergences of the NLO contribution $\int_{m+1} d\sigma^R$ in Eq.~(\ref{sNLO})
if the phase-space integration over $p_j$ is performed in four dimensions.

The starting point of our method for constructing the counter-term $d\sigma^A$
in Eq.~(\ref{sNLO1}) is the observation that the singular behaviour of
$|\cm_{m+1,a...}|^2$ is universal, that is, it is not dependent on the very
detailed structure of $\cm_{m+1,a...}$ itself. The origin of this universality
is in the fact that, for its singular terms with respect to the momentum
$p_j$, the tree amplitude $\cm_{m+1,a...}$  can always be
considered as being
obtained by the insertion of the parton $j$ over all the possible external
legs of a tree-level amplitude  $\cm_{m,a...}$  with $m$ QCD partons (Fig.~1).
Thus, the singular behaviour of $\cm_{m+1,a...}$ is essentially factorizable
with respect to $\cm_{m,a...}$ and the singular factor only depends
on the momenta and quantum numbers of the QCD partons in $\cm_{m,a...}$.
Actually, according to the external-leg insertion sketched in Fig.~1, it is
evident that the singular factor we are looking for is {\em quasi-local}, in
the
sense that it only depends on the momenta and quantum numbers of three partons:
the parton $j$ that is inserted onto $\cm_{m,a...}$ and the partons $i$
and $k$
in $\cm_{m,a...}$.
\begin{figure}
\centerline{\epsfbox{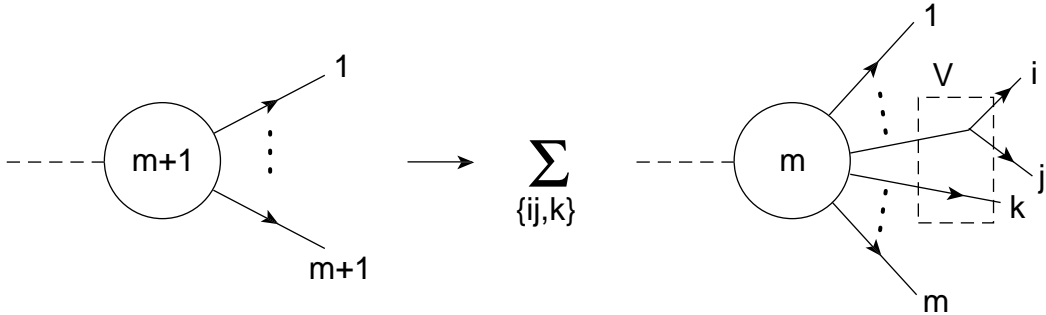}}
\caption{Pictorial representation of the dipole factorization procedure.
When the partons $i$ and $j$ become soft and/or collinear, the singularities
are
factorized into the term ${\bf V}_{ij,k}$ (the dashed box on the right-hand
side) which embodies correlations with a single additional parton $k$.}
\end{figure}

This feature of the soft and collinear singularities will be used to obtain
factorization formulae with the following symbolic structure (Fig.~2)
\beq
\label{Vsim}
|\cm_{m+1,a...}|^2 \to |\cm_{m,a...}|^2 \otimes  {\bom V}_{ij,k} \;\;.
\eeq
Here ${\bom V}_{ij,k}$ is the singular factor, which depends on the momenta and
quantum numbers of the three partons $i,j,k$. As explained in detail in
Sect.~\ref{dff},
two of these partons (e.g. $i$ and $j$) will play the role of `emitter' and the
third parton (e.g. $k$) that of `spectator'. Because of this structure, the
factorization formulae described in Sect.~\ref{dff} will be called {\em dipole
factorization formulae}. In order to explicitly construct these formulae, we
should first recall the known properties [\ref{BCM},\ref{AP}] of the tree-level
QCD matrix elements in the soft and collinear limits.

\subsection{Soft limit}
\label{softlim}

Let us consider the matrix element $\cm_{m+1,a...}$ and the corresponding
vector (in colour + helicity space) $\ket{1, ...., m+1;a, ...}_{m+1,a..}$
defined in Eq.~(\ref{ketin}).
Let us denote by $p_j$ the momentum of a final-state gluon in $\cm_{m+1,a...}$.
In the soft limit, which we parametrize in terms of an arbitrary four
vector $q^\mu$ and a scale parameter $\lambda$:
\beq
\label{slim}
p_j^\mu = \lambda q^\mu \;, \;\;\;\lambda \to 0 \;\;,
\eeq
the matrix element squared behaves as follows [\ref{BCM}]
\beeq
\label{sfac}
&&{}_{m+1,a..}\!\!\bra{1, ...., m+1;a, ...}
\ket{1, ...., m+1;a, ...}_{m+1,a..} \to \nonumber \\
&&- \frac{1}{\lambda^2} \; 4 \pi \mu^{2\ep} \as \;
{}_{m,a..}\!\!\bra{1,..., m+1;a,..}
\left[ {\bom J}^{\mu}(q) \right]^\dagger {\bom J}_{\mu}(q)
\ket{1, ...., m+1;a, ...}_{m,a..} \;\;.
\eeeq
Here we have neglected all the contributions
less singular than $1/\lambda^2$.
The $m$-parton matrix element on the right-hand side of Eq.~(\ref{sfac})
is obtained from $\cm_{m+1,a...}$ by simply removing the soft gluon $p_j$.
The term ${\bom J}_{\mu}(q)$ is the eikonal current for the emission of the
soft gluon $q$. Its explicit expression is given in terms of the momenta and
colour charges of the partons in $\ket{1, ...., m+1;a, ...}_{m,a..}$:
\beq
\label{eikonal}
{\bom J}^{\mu}(q) = \sum_i {\bom T}_i \,\frac{p_i^\mu}{p_i \cdot q}
+ {\bom T}_a \, \frac{p_a^\mu}{p_a \cdot q} + \dots \;\;.
\eeq

The formula in Eq.~(\ref{sfac}) is well known. Here we limit ourselves to
recalling a
few properties of Eq.~(\ref{sfac}), which are relevant for understanding the
structure of the improved
factorization formulae that we shall introduce in Sect.~\ref{dff}.

Although first derived in the four-dimensional case, Eq.~(\ref{sfac}) is
actually valid in any number $d=4-2\ep$ of space-time dimensions. The only
dependence on $d$ is in the overall scale factor $\mu^{2\ep}$ on the right-hand
side.

The $m$-parton matrix element is not exactly factorized. Since the eikonal
current in Eq.~(\ref{eikonal}) depends on the colour charges of the hard
partons, it leads to {\em colour correlations\/} on the right-hand of
Eq.~(\ref{sfac}).

In the actual calculation of cross sections, Eq.~(\ref{sfac}) cannot be used
as a true factorization formula not only
because of these colour correlations. In fact, the tree-level matrix elements
are unambiguously defined only when momentum conservation is fulfilled exactly.
Since, in general, the $m+1$- parton phase space does not factorize into an
$m$-parton times a single-parton phase space, the right-hand side of
Eq.~(\ref{sfac}) is unequivocally defined only  in the strict soft
limit $\lambda=0$.
Away from the point $\lambda=0$, care has to be taken in implementing momentum
conservation.

The form of the eikonal current in Eq.~(\ref{eikonal}) is actually valid
both for massless and massive partons. Squaring the current as in
Eq.~(\ref{sfac}), and taking the massless limit, one obtains:
\beq
\label{eikonal2}
\left[ {\bom J}^{\mu}(q) \right]^\dagger {\bom J}_{\mu}(q) =
\sum_{k,i} \;{\bom T}_k \cdot {\bom T}_i \;\frac{p_k p_i}{(p_k q) (p_iq)}
+ \left( 2 \sum_i {\bom T}_a \cdot {\bom T}_i \;\frac{p_a p_i}{(p_a q) (p_iq)}
+ \dots \right)  \;\;.
\eeq
Each term $p_kp_i/(p_kq)(p_iq)$ on the right-hand side of Eq.~(\ref{eikonal2})
leads to collinear
singularities when the soft momentum $q$ is parallel either to $p_i$ or
to $p_k$ or $p_a$.
These collinear singularities can be disentangled by using the following
identity
\beq
\label{eikid}
\frac{p_k p_i}{(p_k q) (p_iq)} = \frac{p_k p_i}{p_k\cdot q \;(p_i+p_k)\cdot q}
+ \;\frac{p_k p_i}{p_i \cdot q \;(p_i+p_k)\cdot q} \;\;,
\eeq
and likewise for the terms $p_ap_i/(p_aq)(p_iq) \;, \dots $. Thus,
Eq.~(\ref{sfac}) can be rewritten as follows
\beeq
\label{sfacdip}
\!\!\!\!\!\!&&\!\!\!\!\!\!{}_{m+1,a..}\!\!\bra{1, ...., m+1;a, ...}
\ket{1, ...., m+1;a, ...}_{m+1,a..} \to
- \frac{1}{\lambda^2} \; 8 \pi \mu^{2\ep} \as\\
\!\!\!\!&&\!\!\!\!\!\!\cdot \;\sum_i
\;\frac{1}{p_i q} \;\sum_{k \neq i} \;
{}_{m,a..}\!\!\bra{1,..., m+1;a,..}
\frac{{\bom T}_k \cdot {\bom T}_i \;p_k p_i}{(p_i+p_k) q}
\ket{1, ...., m+1;a, ...}_{m,a..}  + \dots \;\;,
\nonumber
\eeeq
where the dots stand for similar contributions that involve the initial-state
parton $a , \dots $.
The dipole structure mentioned at the end of Sect.~\ref{softcoll}
starts to emerge from Eq.~(\ref{sfacdip}). Each term on the right-hand side
of Eq.~(\ref{sfacdip}) depends on the radiated soft momentum $q$, on the
`emitter' momentum $p_i$ (whose direction signals the presence of a collinear
singularity) and on the `spectator' parton $k$ (which accounts for colour
correlations).

\subsection{Collinear limit}
\label{collim}

Let us consider the momenta $p_i$ and $p_j$ of two final-state partons in
$\cm_{m+1,a...}$. The limit where $p_i$ and $p_j$ become collinear is precisely
defined as follows
\beeq
\label{clim}
&&p_i^\mu = z p^\mu + k_\perp^\mu - \frac{k_\perp^2}{z} \frac{n^\mu}{2 pn} \;\;,
\;\;\; p_j^\mu =
(1-z) p^\mu - k_\perp^\mu - \frac{k_\perp^2}{1-z} \frac{n^\mu}{2 pn}\;\;,
\nonumber \\
&&2 p_i p_j = - \frac{k_\perp^2}{z(1-z)} \;\;,
\;\;\;\;\;\;\;\; k_\perp \to 0 \;\;.
\eeeq
In Eq.~(\ref{clim}) the light-like ($p^2=0$) vector $p^\mu$ denotes the
collinear direction, while $n^\mu$ is an auxiliary light-like vector which
is necessary to specify the transverse component $k_\perp$ ($k_\perp^2<0$)
($k_\perp p = k_\perp n = 0$) or, equivalently, how the collinear direction
is approached.
In the small-$k_\perp$-limit (i.e.\ neglecting terms that are less singular than
$1/k_\perp^2$), the $m+1$-parton matrix element behaves as follows [\ref{AP}]
\beeq
\label{cfac}
\!\!&&{}_{m+1,a..}\!\!\bra{1, ...., m+1;a, ...}
\ket{1, ...., m+1;a, ...}_{m+1,a..} \to \nonumber \\
\!\!&&\frac{1}{p_i p_j} \; 4 \pi \mu^{2\ep} \as \;
{}_{m,a..}\!\!\bra{1,..., m+1;a,..} \;{\hat P}_{(ij),i}(z,\kper;\ep) \;
\ket{1, ...., m+1;a, ...}_{m,a..} \;\;.
\eeeq
The $m$-parton matrix element on the right-hand side of Eq.~(\ref{cfac}) is
obtained by replacing the partons $i$ and $j$ in $\cm_{m+1,a...}$ with a
single parton denoted by $ij$. This parton carries the quantum numbers of the
pair $i+j$ in the collinear limit. In other words, its momentum is $p^\mu$ and
its other quantum numbers (flavour, colour) are obtained according to the
following rule: anything~+~gluon gives anything  and quark~+~antiquark gives
gluon.

The kernel ${\hat P}_{(ij),i}$ in Eq.~(\ref{cfac}) is the $d$-dimensional
Altarelli-Parisi splitting function. It depends not only on the momentum
fraction $z$ involved in the collinear splitting $ij \to i + j$, but also on
the transverse momentum $\kper$ and on the helicity of the parton $ij$ in the
$m$-parton matrix element. More precisely, ${\hat P}_{(ij),i}$ is a matrix
acting on the spin indices of the parton $ij$ in
${}_{m,a..}\!\!\bra{1,..., m+1;a,..}$ and $\ket{1, ...., m+1;a, ...}_{m,a..}$.
Because of these {\em spin correlations}, the square of the $m$-parton matrix
element cannot be simply factorized on the right-hand side of Eq.~(\ref{cfac}).

The explicit expressions of ${\hat P}_{ab}(z,\kper;\ep)$ for the splitting
processes
\beq
\label{sppro}
a(p) \to b(zp + \kper + {\cal O}(\kper^2)) +
c((1-z) p - \kper + {\cal O}(\kper^2)) \;\;
\eeq
are as follows
\beeq
\label{hpqqep}
\bra{s} {\hat P}_{qq}(z,\kper;\ep) \ket{s'} = \delta_{ss'} \;C_F
\;\left[ \frac{1 + z^2}{1-z} - \ep (1-z) \right] \;\;,
\eeeq
\beeq
\label{hpqgep}
\bra{s} {\hat P}_{qg}(z,\kper;\ep)\ket{s'} = \delta_{ss'} \;C_F
\;\left[ \frac{1 + (1-z)^2}{z} - \ep z \right] \;\;,
\eeeq
\beeq
\label{hpgqep}
\bra{\mu} {\hat P}_{gq}(z,\kper;\ep) \ket{\nu} = T_R
\left[ - g^{\mu \nu} + 4 z(1-z) \frac{\kper^{\mu} \kper^{\nu}}{\kper^2}
\right] \;\;,
\eeeq
\beq
\label{hpggep}
\bra{\mu} {\hat P}_{gg}(z,\kper;\ep) \ket{\nu} = 2C_A
\;\left[ - g^{\mu \nu} \left( \frac{z}{1-z} + \frac{1-z}{z} \right)
- 2 (1-\ep) z(1-z) \frac{\kper^{\mu} \kper^{\nu}}{\kper^2}
\right] \;\;,
\eeq
where the spin indices of the parent parton $a$ have been denoted by $s,s'$
if $a$ is a fermion and $\mu,\nu$ if $a$ is a gluon.

Equations (\ref{hpqqep})--(\ref{hpggep}) lead to the more familiar form of the
$d$-dimensional splitting functions only after average over the polarizations
of the parton $a$. The $d$-dimensional average is obtained by means of the
factors
\beq
\frac{1}{2} \;\delta_{ss'}
\eeq
for a fermion,
and (the gauge terms are proportional either to $p^\mu$ or to $p^\nu$)
\beq
\frac{1}{d-2} d_{\mu \nu}(p) = \frac{1}{2(1-\ep)}
(-g_{\mu \nu} + {\rm gauge \; terms} ) \;,
\eeq
with
\beq
-g^{\mu \nu} \,d_{\mu \nu}(p) = d-2 \;, \;\;\;\;
p^\mu \,d_{\mu \nu}(p) = 0 \;\;,
\eeq
for a gluon with on-shell momentum $p$. Denoting by $<{\hat P}_{ab}>$
the average of ${\hat P}_{ab}$ over the polarizations of the parton
$a$, we have:
\beeq
\label{avhpqq}
< {\hat P}_{qq}(z;\ep) > = C_F
\;\left[ \frac{1 + z^2}{1-z} - \ep (1-z) \right] \;\;,
\eeeq
\beeq
\label{avhpqg}
< {\hat P}_{qg}(z;\ep) > = C_F
\;\left[ \frac{1 + (1-z)^2}{z} - \ep z \right] \;\;,
\eeeq
\beeq
\label{avhpgq}
< {\hat P}_{gq}(z;\ep) > = T_R
\left[ 1 - \frac{2 z(1-z)}{1-\ep} \right] \;\;,
\eeeq
\beq
\label{avhpgg}
< {\hat P}_{gg}(z;\ep) > = 2C_A
\;\left[ \frac{z}{1-z} + \frac{1-z}{z}
+ z(1-z) \right] \;\;,
\eeq

So far, we have considered the case in which two final-state partons in
$\cm_{m+1,a...}$ become collinear. In general,
one has to deal also with the case in which a final-state parton $i$
becomes collinear to an initial-state parton $a$. Here the collinear limit
is defined as follows
\beeq
\label{calim}
&&p_i^\mu = (1-x) p_a^\mu + k_\perp^\mu - \frac{k_\perp^2}{1-x}
\frac{n^\mu}{2 p_a n} \;\;,
\nonumber \\
&&2 p_i p_a = - \frac{k_\perp^2}{1-x} \;\;,
\;\;\;\;\;\;\;\; k_\perp \to 0 \;\;,
\eeeq
and the corresponding splitting process $a \to ai + i$ involves the transition
from the initial-state parton $a$ to the initial-state parton $ai$ with the
associated emission of the final-state parton $i$. The quantum numbers of the
parton $ai$ are assigned according to their conservation at the QCD tree-level
vertices:
if $a$ and $i$ are partons of the same species, then $ai$ is
a gluon; if $a$ is a fermion (gluon) and $i$ is a gluon (fermion) then
$ai$ is a fermion (antifermion).

The analogue of Eq.~(\ref{cfac}) in the collinear limit (\ref{calim}) is the
following
\beeq
\label{cafac}
\!\!\!\!&&\!\!\!\!{}_{m+1,a..}\!\!\bra{1, ...., m+1;a, ...}
\ket{1, ...., m+1;a, ...}_{m+1,a..} \to  \\
\!\!\!\!&&\!\!\!\!\frac{1}{x}\, \frac{1}{p_i p_a} \; 4 \pi \mu^{2\ep} \as \;
{}_{m,ai..}\!\!\bra{1,..., m+1;ai,..} \;{\hat P}_{a,(ai)}(x,\kper;\ep) \;
\ket{1, ...., m+1;ai, ...}_{m,ai..} \;\;. \nonumber
\eeeq
Now, the $m$-parton matrix element on the right-hand side is obtained from
$\cm_{m+1,a...}$ by removing the final-state parton $i$ and replacing the
initial-state parton $a$ with the parton $ai$. Note two main differences with
respect to Eq.~(\ref{cfac}): on the right-hand side of Eq.~(\ref{cafac})
there is an additional factor of $1/x$ and the initial-state parton $ai$ carries
the momentum $x p_a^\mu$.

As in the case of Eq.~(\ref{sfac}), Eqs.~(\ref{cfac}) and (\ref{cafac}) have to
be regarded as limiting formulae rather than factorization formulae. Their
implementation in the calculation of QCD cross sections indeed requires a
careful treatment of momentum conservation away from the collinear limit.

Note that the splitting functions in Eqs.~(\ref{hpqqep}--\ref{hpggep}) are
divergent for $z \to 0,1$. These divergences are the soft singularities already
discussed in Sect.~\ref{softlim}. When using Eqs.~(\ref{sfacdip}) and
(\ref{cfac})
to approximate the singular behaviour of $\cm_{m+1,a...}$ care has to be
taken in order to avoid double counting the soft and collinear divergences in
their overlapping region.

Note also that Eqs.~(\ref{cfac}) and (\ref{cafac}) do not depend solely on the
collinear momenta $p_i, p_j$ and $p_i, p_a$. In fact, the Altarelli-Parisi
splitting functions produce spin correlations with respect to the directions
of the other momenta in the matrix element. The dipole structure of these
limiting formulae is thus hidden in the azimuthal dependence of these
correlations.

\newpage

\setcounter{equation}{0}

\section{Dipole factorization formulae}
\label{dff}

In this Section we present in detail our basic formalism to construct the local
counter-term $d\sigma^A$ for the NLO cross section in Eq.~(\ref{dsNLO}).
We introduce
improved factorization formulae for the QCD matrix elements. These formulae
are based on a dipole structure with respect to {\em colour\/} and {\em spin\/}
indices and have the following
main features. Our dipole factorization formulae coincide with
Eqs.~(\ref{sfacdip})
and (\ref{cfac}) (or (\ref{cafac}) ) respectively in the soft and collinear
limit. These limits are approached smoothly, thus avoiding double counting
of overlapping soft and collinear singularities. This smooth transition is
possible because the dipole formulae fulfil exact momentum conservation.
Actually, we present several alternative versions of the factorization
formulae  that differ from one another in the implementation of momentum
conservation away from the soft and collinear limits. These differences are
then used to match the phase-space
constraints that are encountered in the calculation of different kinds of
QCD cross sections. In this manner, for any QCD process, we achieve the
analytical integrability of the counter-term $d\sigma^A$ over the single-parton
subspace leading to soft and collinear divergences.

We start by considering the case of final-state singularities for matrix elements
without (Sect.~\ref{dfss}) or with (Sect.~\ref{dfssa}) initial-state partons.
Sections \ref{diss}
and \ref{dissa} deal with initial-state singularities in the case of one or
two initial-state partons, respectively. Cross sections with identified hadrons
in the final state require the introduction of fragmentation functions.
Factorization formulae suitable for these fragmentation processes are presented
in Sects.~\ref{dfrag} and~\ref{dcorr}.

The main properties of the dipole factorization formulae are considered in
detail
in Sect.~\ref{dfss}. In the following Subsections we limit ourselves to
writing down the formalism and emphasizing the relevant differences with
respect to the case discussed in Sect.~\ref{dfss}.

\subsection{Final-state singularities with no initial-state partons}
\label{dfss}
\setcounter{footnote}{0}

The dipole factorization formula in the limit $p_i \cdot p_j \rightarrow 0$
for the matrix elements with no partons in the initial state is the following
\beeq \label{ff}
&&\obra{1, ...., m+1} \oket{1, ...., m+1} = \sum_{k \neq i,j}
{\cal D}_{ij,k}(p_1, ...,p_{m+1}) + \dots
\eeeq
where $\dots$ stands for terms that are not singular in the
limit $p_i \cdot p_j \rightarrow 0$ and the dipole contribution
${\cal D}_{ij,k}$ is given by
\beeq \label{dipff}
\!\!\!&{\cal D}_{ij,k}&\!\!\!\!(p_1, ...,p_{m+1}) =
- \frac{1}{2 p_i \cdot p_j}  \\
&&\cdot \,
\mbra{1, .., {\widetilde {ij}},.., {\widetilde k},.., m+1}
\,\frac{{\bom T}_k \cdot {\bom T}_{ij}}{{\bom T}_{ij}^2} \; {\bom V}_{ij,k} \,
\mket{1, .., {\widetilde {ij}},.., {\widetilde k},.., m+1} \;.
\nonumber
\eeeq

The $m$-parton matrix element on the right-hand side of Eq.~(\ref{dipff}) is
obtained from the original $m+1$-parton matrix element by replacing $a)$ the
partons $i$ and $j$ with a single parton ${\widetilde {ij}}$
({\em the emitter\/})
and $b)$ the parton $k$ with the parton $\widetilde k$
({\em the spectator\/})\footnote{In general, we use the following notation in
the dipole formulae (Fig.~3). A pair of indices like ${\widetilde {ij}}$
denotes the
emitter parton and a single index like $\widetilde k$ denotes the spectator
parton. If these indices appear as subscripts or superscripts, they respectively
indicate final-state or initial-state partons.}.
\begin{figure}
\centerline{\epsfbox{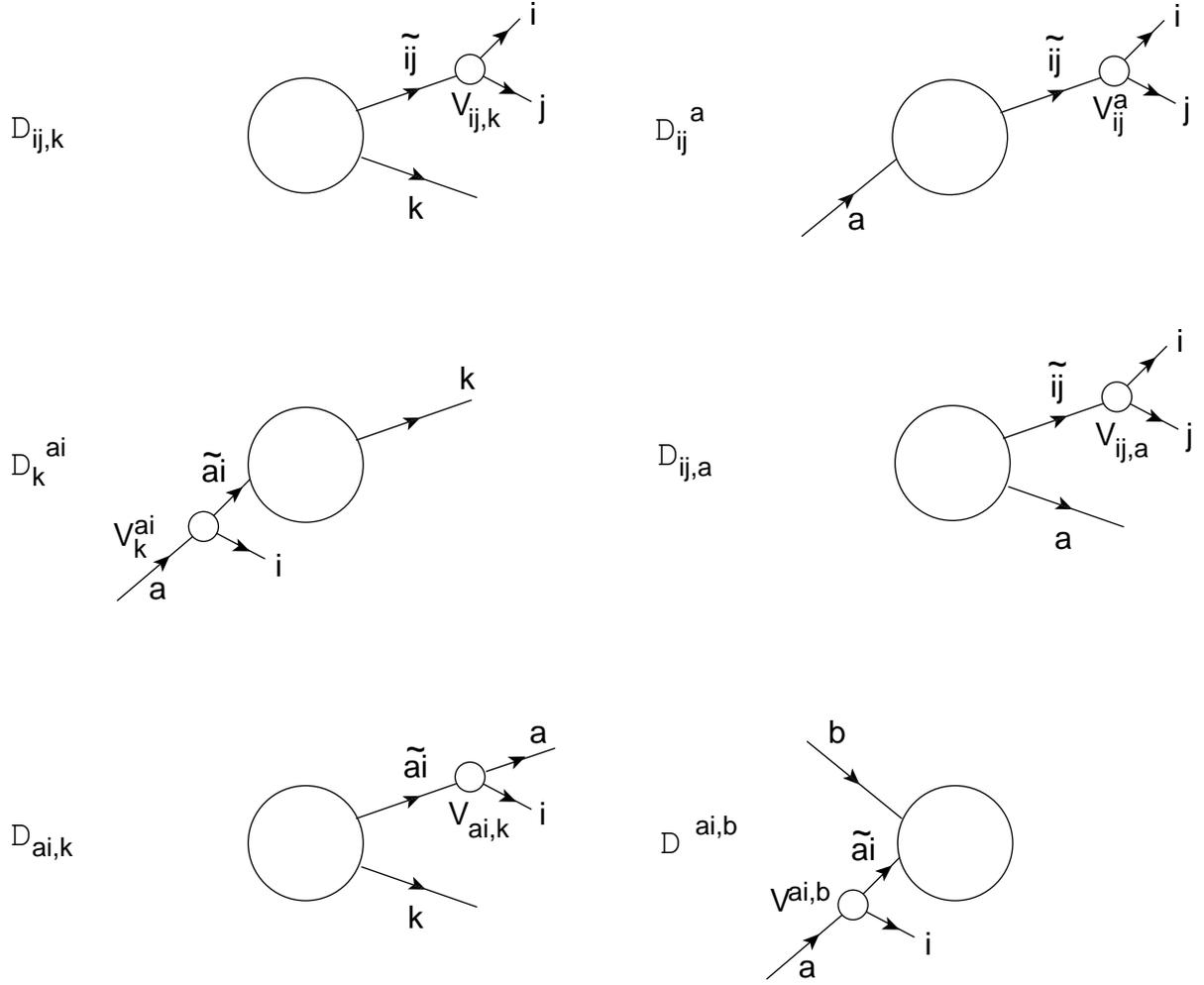}}
\caption{Effective diagrams for the different dipole formulae introduced
in Sect.~\protect\ref{dff}. The blobs denote the $m$-parton matrix element.
Incoming and
outgoing lines respectively stand for initial-state and final-state partons.}
\end{figure}

All the quantum numbers except momenta are assigned as follows. The spectator
parton $\widetilde k$ has the same quantum numbers as $k$. The quantum numbers
of the emitter parton ${\widetilde {ij}}$ are obtained according to their
conservation in the collinear splitting process ${\widetilde {ij}} \to i + j$
(cfr. Sect.~\ref{collim}). This  rule applies to Eq.~(\ref{dipff}) as well
as to all the dipole formulae we shall introduce in the following Subsections.

The momenta of the emitter and the spectator are defined in different ways in
different dipole formulae. In Eq.~(\ref{dipff}) we have
\beq \label{pk}
{\widetilde p}_k^\mu = \frac{1}{1-y_{ij,k}} \,p_k^\mu \;\;,
\;\;\;\;{\widetilde p}_{ij}^\mu =
p_i^\mu  + p_j^\mu  - \frac{y_{ij,k}}{1-y_{ij,k}} \,p_k^\mu \;\;,
\eeq
where the dimensionless variable $y_{ij,k}$ is given by
\beq \label{yijk}
y_{ij,k} = \frac{p_ip_j}{p_ip_j+p_jp_k+p_kp_i} \;.
\eeq

Note that both the emitter and the spectator are on-shell
$({\widetilde p}_{ij}^2 = {\widetilde p}_k^2 = 0)$ and that, performing the
replacement $\{i,j,k\} \to \{{\widetilde {ij}},{\widetilde k}\}$, momentum
conservation is implemented exactly:
\beq \label{momcon}
p_i^\mu  + p_j^\mu + p_k^\mu = {\widetilde p}_{ij}^\mu +
{\widetilde p}_k^\mu \;\;.
\eeq
These are common features of all the dipole formulae in the paper.

In the bra-ket on the right-hand side of Eq.~(\ref{dipff}),
${\bom T}_k$ and ${\bom T}_{ij}$ are the colour charges of the emitter
and the spectator and ${\bom V}_{ij,k}$ are matrices in the helicity space
of the emitter. These splitting matrices, which depend on $y_{ij,k}$ and
on the kinematic variables ${\tilde z}_i, {\tilde z}_j$:
\beeq
\label{tilz}
{\tilde z}_i
= \frac{p_ip_k}{p_jp_k+p_ip_k} =
\frac{p_i{\widetilde p}_k}{{\widetilde p}_{ij}{\widetilde p}_k}
\;, \;\;\;
{\tilde z}_j = \frac{p_jp_k}{p_jp_k+p_ip_k}=
\frac{p_j{\widetilde p}_k}{{\widetilde p}_{ij}{\widetilde p}_k}=
1-{\tilde z}_i  \;,
\eeeq
are related to the $d$-dimensional Altarelli-Parisi splitting functions
in Eqs.~(\ref{hpqqep})--(\ref{hpggep}).

For fermion + gluon splitting we have
($s$ and $s'$ are the spin indices of the fermion ${\widetilde {ij}}$
in $\bra{..,{\widetilde {ij}}..}$ and $\ket{..,{\widetilde {ij}},..}$
respectively)
\beeq
\label{Vqigjk}
\bra{s}
{\bom V}_{q_ig_j,k}({\tilde z}_i;y_{ij,k}) \ket{s'}
&=& 8\pi \mu^{2\ep} \as\; C_F\,
\left[ \frac{2}{1-{\tilde z}_i(1-y_{ij,k})}
- \frac{}{} (1+{\tilde z}_i)
-\ep (1-{\tilde z}_i)\right] \;\delta_{ss'} \nonumber \\
&\equiv& V_{q_ig_j,k} \;\delta_{ss'}\;.
\eeeq

For quark + antiquark and gluon + gluon splitting we have
($\mu$ and $\nu$ are the spin indices of the gluon ${\widetilde {ij}}$
in $\bra{..,{\widetilde {ij}}..}$ and $\ket{..,{\widetilde {ij}},..}$
respectively)
\beq \label{vqbqk}
\bra{\mu}
{\bom V}_{q_i{\bar q}_j,k}({\tilde z}_i)
\ket{\nu}
= 8\pi \mu^{2\epsilon} \as \;
T_R \;
\left[ -g^{\mu \nu} - \frac{2}{p_ip_j} \,
( {\tilde z}_i p_i^{\mu} - {\tilde z}_j p_j^{\mu} )
\,( {\tilde z}_i p_i^{\nu} - {\tilde z}_j p_j^{\nu} ) \,\right]
\equiv V_{q_i{\bar q}_j,k}^{\mu \nu} \,,
\eeq
\beeq \label{vggk}
&\!&\!\!\!\!\!\!\!\!\!\!\bra{\mu}{\bom V}_{g_ig_j,k}({\tilde z}_i;y_{ij,k})
\ket{\nu} = 16\pi
\mu^{2\epsilon} \as \;C_A\,
\left[ -g^{\mu \nu} \left(
\frac{1}{1-{\tilde z}_i(1-y_{ij,k})} \right. \right.\nonumber \\
&\!&\!\!\!\!\!\!\!\!\!\!+ \left. \left.
\frac{1}{1-{\tilde z}_j(1-y_{ij,k})} - 2 \right)
+ (1-\ep) \frac{1}{p_ip_j} \,
( {\tilde z}_i p_i^{\mu} - {\tilde z}_j p_j^{\mu} )
\,( {\tilde z}_i p_i^{\nu} - {\tilde z}_j p_j^{\nu} ) \,\right]
\equiv V_{g_ig_j,k}^{\mu \nu} \;.
\eeeq

\vspace{.5cm}
\noindent {\it Soft and collinear limits}
\vspace{.1cm}

Note that the matrices ${\bom V}_{ij,k}$ do not lead to two-particle
singularities
in any of the limits $p_i \cdot p_j, p_i\cdot p_k, p_j \cdot p_k  \to 0$.
This is because the only non-polynomial dependence on ${\tilde z}_i$ is in the
factors
\beq \label{sofac}
\frac{1}{1-{\tilde z}_i(1-y_{ij,k})} =
\frac{p_ip_j+p_jp_k+p_kp_i}{(p_i + p_k)p_j} \;\;.
\eeq
Therefore the dipole term on the right-hand side of Eq.~(\ref{ff})
contains only collinear and soft divergences for $p_i \cdot p_j \to 0$.
\pagebreak[3]

In the collinear limit of Eq.~(\ref{clim}), the dipole variables become:
\beeq \label{dcoll}
y_{ij,k} &\to& - \;\frac{\kper^2}{2z(1-z) pp_k} \;, \;\;\;
{\tilde z}_i = 1 - {\tilde z}_j \to z \;\;, \nonumber \\
{\widetilde p}_k^\mu  &\to& p_k^\mu \;, \;\;\;\;\;\;
{\widetilde p}_{ij}^\mu \to p^\mu \;\;.
\eeeq
Thus, the $m$-parton matrix element in Eq.~(\ref{dipff}) coincides with that
in Eq.~(\ref{cfac}). Moreover, the vector
${\tilde z}_i p_i^{\mu} - {\tilde z}_j p_j^{\mu}$ in
Eqs.~(\ref{vqbqk},\ref{vggk}) becomes
$[{\tilde z}_i z - {\tilde z}_j (1-z)] p^{\mu} + \kper^{\mu}$ and,
since its $p^{\mu}$-component gives a vanishing contribution to
Eq.~(\ref{dipff}) ($p^{\mu} \bra{\mu}  \mket{.., p,..} = 0$ because of
gauge invariance), the matrices ${\bom V}_{ij,k}$ become proportional to
Altarelli-Parisi splitting functions:
\beq \label{vcoll}
{\bom V}_{ij,k} \to 8 \pi \mu^{2\ep} \as \;{\hat P}_{(ij),i}(z,\kper;\ep) \;\;.
\eeq
In particular, the only dependence on $k$ that survives in
Eq.~(\ref{dipff}) is that on ${\bom T}_k$. Therefore, one can perform
the sum over the colour charges on the right-hand side of Eq.~(\ref{ff}) and,
using charge conservation $(\sum_{k \neq i,j} {\bom T}_k = - {\bom T}_{ij})$,
one can check that this equation reproduces the collinear behaviour in
Eq.~(\ref{cfac}).

In the soft limit of Eq.~(\ref{slim}), we have
\beeq \label{dsoft}
y_{ij,k} &\to& 0 \;, \;\;\;{\tilde z}_i \to 1 \;, \;\;\;{\tilde z}_j \to 0
\;\;, \nonumber \\
{\widetilde p}_k^\mu  &\to& p_k^\mu \;, \;\;\;\;\;\;
{\widetilde p}_{ij}^\mu \to p_i^\mu \;\;.
\eeeq
Thus the $m$-parton matrix element in Eq.~(\ref{dipff}) coincides with that in
Eq.~(\ref{sfacdip}). Moreover, the only singular factor in ${\bom V}_{ij,k}$
is due to the term in Eq.~(\ref{sofac}):
\beq \label{ssofac}
\frac{1}{1-{\tilde z}_i(1-y_{ij,k})} \to \frac{1}{\lambda} \;
\frac{p_kp_i}{(p_i + p_k)q} \;\;,
\eeq
which gives
\beq \label{vsoft}
\lambda {\bom V}_{ig_j,k} \to 16 \pi \mu^{2\ep} \as \;{\bom T}_{ij}^{\,2}
\;\frac{p_kp_i}{(p_i + p_k)q} \;\;.
\eeq
Inserting this expression into Eqs.~(\ref{dipff}), we see that the dipole term
in Eq.~(\ref{ff}) reproduces the soft limit in Eq.~(\ref{sfacdip}).

This discussion proves that the dipole formula (\ref{ff}) provides a point-wise
approximation of the $m+1$-parton matrix element in the singular region
$p_i \cdot p_j \to 0$. Note that to achieve this, the helicity dependence
of the splitting kernels ${\bom V}_{ij,k}$ is essential. The azimuthal
correlations due to this dependence cancel after integration over $p_i$
(see Eqs.~(\ref{intdip},\ref{vijep})) and hence, provided that the counting
of helicity states is consistently performed in $d$ dimensions, they are not
relevant for reproducing the correct
poles in $1/\ep$ in the contribution $\int_{m+1} d\sigma^A$ to the NLO
cross section in Eq.~(\ref{sNLO1}). Nonetheless, these correlations
have to be properly taken into
account in constructing the {\em local\/} counter-term $d\sigma^A$, which makes
the
contribution $[ d\sigma^R - d\sigma^A ]$ integrable in four dimensions.
Indeed, the parton azimuthal correlations due to this dependence are not only
essential in the most general case when the jet cross section explicitly
depends on them, but even when it does not\footnote{In this case the
  evaluation of $\int_{m+1} d\sigma^R$ in four dimensions usually involves
  double angular integrals of the type $\int_{-1}^{+1} d\cos\theta
  \int_{0}^{2\pi} d\varphi \;\cos\varphi/(1-\cos\theta),$ where $\varphi$ is
  the azimuthal angle. These integrals are mathematically ill-defined. If
  their numerical integration is attempted, one can obtain any answer
  whatsoever, depending on the detail of the integration procedure.
  Performing the integral analytically before going to 4 dimensions, one
  obtains $\int_{-1}^{+1} d\cos\theta \int_{0}^{2\pi} d\varphi
  \;\cos\varphi/(1-\cos\theta)\sin^{-2\ep}\theta\sin^{-2\ep}\varphi=0$.}.

\vspace{.5cm}
\noindent {\it Phase space factorization}
\vspace{.1cm}

The definition (\ref{pk}) of the dipole momenta is particularly useful because
it allows us to exactly factorize the phase space of the partons $i,j,k$
into the dipole phase space times a single-parton contribution. Indeed, let us
consider the following 3-particle contribution to the final-state
phase space:
\beq
d\phi(p_i,p_j,p_k;Q) =
\frac{d^{d}p_i}{(2\pi)^{d-1}} \,\delta_+(p_i^2) \;
\frac{d^{d}p_j}{(2\pi)^{d-1}} \,\delta_+(p_j^2) \;
\frac{d^{d}p_k}{(2\pi)^{d-1}} \,\delta_+(p_k^2) \;
\;(2\pi)^d\,\delta^{(d)}(Q-p_i-p_j-p_k) \;.
\eeq

In terms of the momenta ${\widetilde p}_{ij}, \,
{\widetilde p}_k$ and $p_i$, this phase-space
contribution takes the factorized form:
\beeq \label{psfac}
d\phi(p_i,p_j,p_k;Q) = d\phi({\widetilde p}_{ij},
{\widetilde p}_k;Q) \;\left[ dp_i({\widetilde p}_{ij},
{\widetilde p}_k) \right] \;\;,
\eeeq
where
\beeq
\label{dpi}
\left[ dp_i({\widetilde p}_{ij},{\widetilde p}_k) \right]
= \frac{d^{d}p_i}{(2\pi)^{d-1}} \,\delta_+(p_i^2) \;
{\cal J}(p_i;{\widetilde p}_{ij},{\widetilde p}_k) \;,
\eeeq
and the Jacobian factor is
\beq
\label{Jac}
{\cal J}(p_i;{\widetilde p}_{ij},{\widetilde p}_k)
= \Theta(1- {\tilde z}_i) \,\Theta(1-y_{ij,k}) \;
\frac{(1-y_{ij,k})^{d-3}}{1- {\tilde z}_i} \;.
\eeq
In terms of the kinematic variables defined earlier, we have
\beeq
\label{exdpi}
\left[ dp_i({\widetilde p}_{ij},{\widetilde p}_k) \right]
&=& \frac{(2{\widetilde p}_{ij}{\widetilde p}_k)^{1-\ep}}{16\pi^2}
\;\frac{d\Omega^{(d-3)}}{(2\pi)^{1-2\ep}}
\;d{\tilde z}_i \;dy_{ij,k}
\;\Theta({\tilde z}_i(1-{\tilde z}_i))
\;\Theta(y_{ij,k}(1-y_{ij,k})) \nonumber \\
&\cdot&
\;\left({\tilde z}_i(1-{\tilde z}_i) \right)^{-\ep}
\;\left( 1-y_{ij,k} \right)^{1-2\ep} y_{ij,k}^{-\ep} \;,
\eeeq
where $d\Omega^{(d-3)}$ is an element of solid angle perpendicular to
${\widetilde p}_{ij}$ and ${\widetilde p}_k$ and thus
\beq
\int d\Omega^{(d-3)} = \frac{2\pi}{\pi^\ep\Gamma(1-\ep)} \;\;.
\eeq

\vspace{.5cm}
\noindent {\it Integration of the splitting functions
${\bom V}_{ij,k}$}
\vspace{.1cm}

To evaluate the integral $\int_{m+1} d\sigma^A$ in Eq.~(\ref{sNLO1}), we can
compute the integral of the splitting functions ${\bom V}_{ij,k}$ over
$[dp_i({\widetilde p}_{ij},{\widetilde p}_k)]$ once and for all. The only
non-trivial point involved in this integration regards azimuthal correlations.

Note, however, that the spin correlation tensor
\beq\label{spinten}
\frac{1}{p_ip_j} \,
( {\tilde z}_i p_i^{\mu} - {\tilde z}_j p_j^{\mu} )
\,( {\tilde z}_i p_i^{\nu} - {\tilde z}_j p_j^{\nu} )
\eeq
in Eqs.~(\ref{vqbqk},\ref{vggk})
is orthogonal
to both ${\widetilde p}_{ij}^\mu$ and ${\widetilde p}_{ij}^\nu$.
Using this property and Lorentz covariance (the integral may depend only
on ${\widetilde p}_{ij}$ and ${\widetilde p}_{k}$), it follows that the
azimuthal integration of the spin correlation tensor gives a contribution of
the type
\beq
K_{\mu \nu} = A \left[ -g_{\mu \nu} +
\frac{{\widetilde p}_{ij}^\mu \,{\widetilde p}_k^\nu
+\,{\widetilde p}_k^\mu \,
{\widetilde p}_{ij}^\nu}{{\widetilde p}_{ij} \cdot
{\widetilde p}_k} \right] + B \,{\widetilde p}_{ij}^\mu
\,{\widetilde p}_{ij}^\nu \;.
\eeq
On the other hand, the (on-shell) matrix element
$\mket{1, .., {\widetilde {ij}},.., {\widetilde k},.., m+1}$ is conserved
(gauge invariance), that is,
\beq
{\widetilde p}_{ij}^\mu \,
\bra{\mu}
\mket{1, .., {\widetilde {ij}},.., {\widetilde k},.., m+1} = 0 \;.
\eeq
Thus, only the term $- A \,g_{\mu \nu}$ in
$K_{\mu \nu}$ contributes to the dipole formula
after integration over
$[dp_i({\widetilde p}_{ij},{\widetilde p}_k)]$. This term
can be singled out by taking the $d$-dimensional average over the polarizations
of the emitter:
\beq
\frac{1}{d-2} d^{\mu \nu}({\widetilde p}_{ij}) \,K_{\mu \nu} =
\frac{1}{d-2} d^{\mu \nu}({\widetilde p}_{ij}) \, [ -g_{\mu \nu} \,A] \;.
\eeq
In conclusion, after integration of the dipole
${\cal D}_{ij,k}(p_1, ...,p_{m+1})$
over $[dp_i({\widetilde p}_{ij},{\widetilde p}_k)]$, only
colour correlations survive, in the form:
\beeq
\label{intdip}
\!\!\!\!\!\!\!\!&&\!\!\!\!\!\! \int
\left[dp_i({\widetilde p}_{ij},{\widetilde p}_k)\right] \;
{\cal D}_{ij,k}(p_1, ...,p_{m+1}) \nonumber \\
\!\!\!\!\!\!\!\!&=&\!\!\!\!- \;{\cal V}_{ij,k} \;\;
\mbra{1, .., {\widetilde {ij}},.., {\widetilde k},.., m+1}
\,\frac{{\bom T}_k \cdot {\bom T}_{ij}}{{\bom T}_{ij}^2} \;
\mket{1, .., {\widetilde {ij}},.., {\widetilde k},.., m+1} \;,
\eeeq
where
\beq
\label{azcor}
\label{vijep}
{\cal V}_{ij,k} =
\int \left[dp_i({\widetilde p}_{ij},{\widetilde p}_k)\right]
\; \frac{1}{2 p_i \cdot p_j} \; <{\bom V}_{ij,k}>
\;\equiv \;\frac{\as}{2\pi}
\frac{1}{\Gamma(1-\ep)} \left(
\frac{4\pi \mu^2}{2 {\widetilde p}_{ij}{\widetilde p}_k} \right)^{\ep}
{\cal V}_{ij}(\ep) \;,
\eeq
and $<{\bom V}_{ij,k}>$ denotes the average of
${\bom V}_{ij,k}$ over the polarizations of the emitter parton
${\widetilde {ij}}$.

The function ${\cal V}_{ij}(\ep)$ depends only on the flavour indices $i$
and $j$.
Using Eq.~(\ref{exdpi}), from the definition
of ${\cal V}_{ij}(\ep)$ in Eq.~(\ref{vijep}) we obtain
\beq
\label{intvij}
{\cal V}_{ij}(\ep) =
\int_0^1 d{\tilde z}_i \,
\left({\tilde z}_i (1-{\tilde z}_i) \right)^{-\ep}
\int_0^1 \frac{dy}{y} \, \left( 1-y \right)^{1-2\ep} y^{-\ep} \;
\frac{<{\bom V}_{ij,k}({\tilde z}_i;y)>}{8 {\pi} \as
\mu^{2\ep}} \;,
\eeq
where the spin-averaged splitting functions are:
\beeq
\label{avVqgk}
\frac{<{\bom V}_{qg,k}({\tilde z};y)>}{8 {\pi} \as
\mu^{2\ep}} = C_F\,
\left[ \frac{2}{1-{\tilde z}(1-y)}
-  (1+{\tilde z})
-\ep (1-{\tilde z})\right]  \;,
\eeeq
\beeq
\label{avVqqk}
\frac{<{\bom V}_{q{\bar q},k}({\tilde z};y)>}{8 {\pi} \as
\mu^{2\ep}} =
T_R \;
\left[ 1 - \frac{2{\tilde z}(1-{\tilde z})}
{1-\ep} \,\right] \;,
\eeeq
\beq
\label{avVggk}
\frac{<{\bom V}_{gg,k}({\tilde z};y)>}{8 {\pi} \as
\mu^{2\ep}} =
2 C_A \,\left[
\frac{1}{1-{\tilde z}(1-y)} +
\frac{1}{1-(1-{\tilde z})(1-y)} - 2
+ {\tilde z} (1 - {\tilde z}) \right] \;.
\eeq
Performing the integration in Eq.~(\ref{intvij}), we find
\beq
\label{vqgep}
{\cal V}_{qg}(\ep) = \frac{\Gamma^3(1-\ep)}{\Gamma(1-3\ep)}
\; C_F \left[ \frac{1}{\ep^2} + \frac{1}{\ep} \; \frac{3+\ep}{2(1-3\ep)}
\right]
= C_F \left[ \frac{1}{\ep^2} + \frac{3}{2\ep} + 5 - \frac{\pi^2}{2}
+ {\cal O}(\ep)  \right] \;\;,
\eeq
\beq
\label{vqqep}
{\cal V}_{q{\bar q}}(\ep) = \frac{\Gamma^3(1-\ep)}{\Gamma(1-3\ep)}
\; T_R \left[ - \frac{1}{\ep} \; \frac{2(1-\ep)}{(1-3\ep)(3-2\ep)} \right]
= T_R \left[ - \frac{2}{3\ep} - \frac{16}{9}
+ {\cal O}(\ep) \right] \;\;,
\eeq
\beq
\label{vggep}
{\cal V}_{gg}(\ep) = \frac{\Gamma^3(1-\ep)}{\Gamma(1-3\ep)}
\; 2 C_A \left[ \frac{1}{\ep^2} + \frac{1}{\ep}
\; \frac{11-7\ep}{2(1-3\ep)(3-2\ep)} \right]
= 2 C_A \left[ \frac{1}{\ep^2} + \frac{11}{6\ep} + \frac{50}{9}
 - \frac{\pi^2}{2}
+ {\cal O}(\ep)  \right] \;\;.
\eeq
Note that in each case, the first result is exact in any number of
dimensions, $d=4-2\ep,$ while the latter is valid for $\ep\to0$.

\subsection{Final-state singularities with initial-state partons}
\label{dfssa}

In the presence of initial-state partons $a, \dots$, the $m+1$-parton
matrix element has both final-state $(p_i \cdot p_j \to 0)$ and
initial-state $(p_a \cdot p_j \to 0)$ singularities. Let us first
consider the dipole factorization formula for final-state
singularities. Neglecting terms that are not singular when
$p_i \cdot p_j \to 0$, we have:
\beeq \label{fi}
&&{}_{m+1,a..}\!\!\bra{1, ...., m+1;a, ...}
\ket{1, ...., m+1;a, ...}_{m+1,a..} = \sum_{k \neq i,j}
{\cal D}_{ij,k}(p_1, ...,p_{m+1};p_a,..) \nonumber \\
&&+ \left[ \;{\cal D}_{ij}^a(p_1, ...,p_{m+1};p_a,..) + \dots \right] +
\dots \;\;.
\eeeq
The first term on the right-hand side is the same contribution as in
Eqs.~(\ref{ff},\ref{dipff}), while the terms in the square bracket represent
additional dipole contributions in which the emitter is the final-state parton
${\widetilde {ij}}$ and the spectators are the initial-state partons
${\widetilde {a}}, \dots$ . These dipole terms are explicitly given by
\beeq \label{dipfi}
&&{\cal D}_{ij}^a(p_1, ...,p_{m+1};p_a,..) = - \frac{1}{2 p_i \cdot p_j} \;
\frac{1}{x_{ij,a}} \\
&&\cdot \;\,
{}_{m,a..}\!\!\bra{1,.., {\widetilde {ij}},.., m+1;{\widetilde a},..}
\,\frac{{\bom T}_a \cdot {\bom T}_{ij}}{{\bom T}_{ij}^2} \; {\bom V}_{ij}^a \,
\ket{1, .., {\widetilde {ij}}, .., m+1;{\widetilde a}, ...}_{m,a..}
\;\;. \nonumber
\eeeq

In Eq.~(\ref{dipfi}) the momenta of the spectator ${\widetilde {a}}$ and the
emitter ${\widetilde {ij}}$ are defined as follows
\beeq
\label{wpa}
{\widetilde p}_a^\mu = x_{ij,a} \,p_a^\mu  \;, \;\;\;\;
{\widetilde p}_{ij}^\mu  = p_i^\mu  + p_j^\mu  - (1 - x_{ij,a} ) \,p_a^\mu  \;,
\eeeq
\beq\label{wpax}
x_{ij,a} = \frac{p_ip_a+p_jp_a-p_ip_j}{(p_i+p_j)p_a} \;,
\eeq
and the corresponding splitting functions are:
\beeq
\label{Vqigja}
\bra{s} {\bom V}_{q_ig_j}^a({\tilde z}_i;x_{ij,a}) \ket{s'}
&=& 8\pi \mu^{2\ep} \as\; C_F\;
\left[ \frac{2}{1-{\tilde z}_i + (1 - x_{ij,a})}
- \frac{}{} (1 + {\tilde z}_i)
-\ep (1-{\tilde z}_i) \right]
\;\delta_{ss'} \;, \nonumber \\
&\equiv& V_{q_ig_j}^a \;\delta_{ss'} \;\;,
\eeeq
\beeq
&\bra{\mu}&\!\!\!\!{\bom V}_{g_ig_j}^a({\tilde z}_i;x_{ij,a})
\ket{\nu} = 16\pi
\mu^{2\epsilon} \as \;C_A\;
\left[ -g^{\mu \nu} \left(
\frac{1}{1-{\tilde z}_i+(1-x_{ij,a})} \right. \right.
\nonumber \\
&\!\!\!\!\!\!\!\!+\!\!\!\!\!\!\!\!&\!\!\!\!\!\!\!\!\left.
\frac{1}{1-{\tilde z}_j+(1-x_{ij,a})} - 2
\right) + \left. (1-\ep) \frac{1}{p_ip_j} \,
( {\tilde z}_i p_i^{\mu} - {\tilde z}_j p_j^{\mu} )
\,( {\tilde z}_i p_i^{\nu} - {\tilde z}_j p_j^{\nu} ) \,\right] \;,
\nonumber \\
&&\,
\eeeq
\beq\label{Vqibqja}
\bra{\mu}
{\bom V}_{q_i{\bar q}_j}^a({\tilde z}_i)
\ket{\nu}
= 8\pi \mu^{2\epsilon} \as \;
T_R \;
\left[ -g^{\mu \nu} - \frac{2}{p_ip_j} \,
( {\tilde z}_i p_i^{\mu} - {\tilde z}_j p_j^{\mu} )
\,( {\tilde z}_i p_i^{\nu} - {\tilde z}_j p_j^{\nu} ) \,\right] \,,
\eeq
where
\beeq\label{tilzi}
{\tilde z}_i
= \frac{p_ip_a}{p_ip_a+p_jp_a} =
\frac{p_i{\widetilde p}_a}{{\widetilde p}_{ij}{\widetilde p}_a}
\;, \;\;\;
{\tilde z}_j = \frac{p_jp_a}{p_jp_a+p_ip_a}=
\frac{p_j{\widetilde p}_a}{{\widetilde p}_{ij}{\widetilde p}_a}=
1-{\tilde z}_i  \;.
\eeeq

In the definition of the splitting functions ${\bom V}_{ij}^a$, the kinematic
variable $x_{ij,a}$ plays a role similar to that of $y_{ij,k}$ in
${\bom V}_{ij,k}$: it provides a smooth interpolation between the soft and
collinear limits. Following the same argument as in Sect.~\ref{dfss}, it is
straightforward to check that Eqs.~(\ref{fi},\ref{dipfi}) reproduce the
correct soft and collinear behaviour of the $m+1$-parton matrix element in the
limit $p_i \cdot p_j \to 0$.

Note that, comparing Eqs.~(\ref{Vqigja}--\ref{Vqibqja}) with the equivalent
expressions when the spectator is a final-state parton,
Eqs.~(\ref{Vqigjk}--\ref{vggk}), they are not symmetric under crossing
symmetry, $p_k\to-p_a$.  While the non-soft terms do become identical in the
collinear limit, the soft terms do not in the soft limit.  This is because the
crossing-symmetric eikonal term is split between the two dipole terms that,
separately, are not crossing symmetric.  In fact,
after adding Eqs.~(\ref{Vqigja}--\ref{Vqibqja}) and the equivalent expressions
for initial-state singularities given below in
Eqs.~(\ref{Vqagik}--\ref{Vgagik}),
the soft terms for $p_j \to 0$ do obey crossing:
\beeq
\label{cs1}
\!\!\!\!\!\!\!\!\!\!\!\!  \mbox{Final-state parton} \;p_k :~ \hspace{1em}
  \frac{p_kp_i}{p_ip_j(p_ip_j+p_kp_j)}
  + \frac{p_kp_i}{p_kp_j(p_ip_j+p_kp_j)}
  &=& \frac{p_kp_i}{(p_ip_j)(p_kp_j)} \;, \\
\label{cs2}
\!\!\!\!\!\!\!\!\!\!\!\!
  \mbox{Initial-state parton} \;p_a : \hspace{1em}
  \frac{p_ap_i}{p_ip_j(p_ip_j+p_ap_j)}
  + \frac{p_ap_i}{p_ap_j(p_ip_j+p_ap_j)}
  &=& \frac{p_ap_i}{(p_ip_j)(p_ap_j)} \;.
\eeeq
If, instead, \naive\ crossing, $p_k\to-p_a$, had been used on the individual
terms on the left-hand side of
Eqs.~(\ref{cs1},\ref{cs2}), spurious singularities would develop at
$p_ip_j=p_ap_j$.
Similar
comments concerning crossing apply to the formulae in all later Subsections.

Considering the three-parton phase space
\beq
d\phi(p_i,p_j;Q+p_a) =
\frac{d^{d}p_i}{(2\pi)^{d-1}} \,\delta_+(p_i^2) \;
\frac{d^{d}p_j}{(2\pi)^{d-1}} \,\delta_+(p_j^2) \;
\;(2\pi)^d\,\delta^{(d)}(Q+p_a-p_i-p_j) \;,
\eeq
the analogue of the phase-space {\em factorization\/} in Eq.~(\ref{psfac})
is the following phase-space {\em convolution\/}
\beq \label{psconv}
d\phi(p_i,p_j;Q+p_a) =  \int_0^1 dx \; d\phi({\widetilde p}_{ij};
Q + x p_a) \;\left[ dp_i({\widetilde p}_{ij};
p_a, x) \right] \;,
\eeq
where
\beq \label{px}
\left[ dp_i({\widetilde p}_{ij}; p_a, x) \right] =
\frac{d^{d}p_i}{(2\pi)^{d-1}} \,\delta_+(p_i^2) \;
\Theta(x) \Theta(1-x) \;
\delta(x - x_{ij,a}) \frac{1}{1-{\tilde z}_i} \;\;,
\eeq
or, more explicitly, using the kinematic variables in
Eqs.~(\ref{wpax},\ref{tilzi}):
\beeq
\label{expx}
\left[ dp_i({\widetilde p}_{ij};p_a, x) \right]
&=& \frac{(2{\widetilde p}_{ij}p_a)^{1-\ep}}{16\pi^2}
\;\frac{d\Omega^{(d-3)}}{(2\pi)^{1-2\ep}}
\;d{\tilde z}_i \;dx_{ij,a}
\;\Theta({\tilde z}_i(1-{\tilde z}_i))
\;\Theta(x(1-x)) \nonumber \\
&\cdot&
\;\left({\tilde z}_i(1-{\tilde z}_i) \right)^{-\ep}
\;\delta(x-x_{ij,a}) \;\left( 1-x \right)^{-\ep} \;,
\eeeq
where $d\Omega^{(d-3)}$ is an element of solid angle perpendicular to
${\widetilde p}_{ij}$ and $p_a$.
Note that we are interested in using Eqs.~(\ref{psconv},\ref{px}) for a NLO
calculation whose leading-order kinematic is $Q+p_a=p$ with $p^2 =0,
p_0 \geq 0$. Thus the physical constraint $Q^2 \leq 0$ is always understood
in Eqs.~(\ref{psconv},\ref{px}).

The integration of the splitting function ${\bom V}_{ij}^a$ over the phase
space in Eq.~(\ref{px}) defines the functions ${\cal V}_{ij}(x;\ep)$
(the treatment of the azimuthal correlations is similar to that
in Sect.~\ref{dfss}):
\beq
\frac{\as}{2\pi} \frac{1}{\Gamma(1-\ep)} \left(
\frac{4\pi \mu^2}{2 {\widetilde p}_{ij}p_a} \right)^{\ep}
{\cal V}_{ij}(x;\ep)
\equiv  \int \left[ dp_i({\widetilde p}_{ij}; p_a, x) \right]
\frac{1}{2p_ip_j} <{\bom V}_{ij}^a({\tilde z}_i;x_{ij,a})> \;,
\eeq
that is,
\beq
\label{vijx}
{\cal V}_{ij}(x;\ep) = \Theta(x) \Theta(1-x) \left(
\frac{1}{1-x} \right)^{1+\ep} \;
\int_0^1 d{\tilde z}_i \,
\left({\tilde z}_i (1-{\tilde z}_i) \right)^{-\ep}
\frac{<{\bom V}_{ij}^a({\tilde z}_i;x)>}{8 {\pi} \as
\mu^{2\ep}} \;.
\eeq
As usual, $<{\bom V}_{ij}^a>$ denotes the average of
${\bom V}_{ij}^a$ over the polarizations of the parton
${\widetilde {ij}}$.

The $\ep$-dependence of ${\cal V}_{ij}(x;\ep)$ has to be interpreted with
care. For $\ep=0$, ${\cal V}_{ij}(x;\ep)$ is well-defined everywhere except
the point $x=1$ where it is singular as $\frac{\ln (1-x)}{1-x}$.
In order to avoid this singularity, the limit $\ep \to 0$ has to be taken
uniformly in $x$. In this manner, ${\cal V}_{ij}(x;\ep)$ defines an
$x$-distribution whose coefficients contain poles in $1/\ep$. More precisely,
we have:
\beq
\label{vpd}
{\cal V}_{ij}(x;\ep) = \left[ \,{\cal V}_{ij}(x;\ep) \,\right]_{+} +
\delta(1-x) \int_0^1 dz \;{\cal V}_{ij}(z;\ep) \;\;,
\eeq
\beq \label{vplus}
\left[ \,{\cal V}_{ij}(x;\ep) \,\right]_{+} =
\left[ \,{\cal V}_{ij}(x;\ep=0) \,\right]_{+} + {\cal O}(\ep) \;\;,
\eeq
where the `$+$'-distribution is defined, as usual, by its action on a generic
test function $g(x)$:
\beq
\int_0^1 dx \, g(x) \left[ \,{\cal V}(x) \,\right]_{+} \equiv
\int_0^1 dx \, [ g(x) - g(1) ] \,{\cal V}(x)  \;\;.
\eeq
Note that in Eq.~(\ref{vplus}), the expansion around $\ep=0$ is well defined,
i.e.\ the ${\cal O}(\ep)$ term on the right-hand side is integrable in $x$.

The explicit form of the spin-averaged splitting functions
$<{\bom V}_{ij}^a>$ is the following:
\beeq
\label{vqga}
\frac{<{\bom V}_{qg}^a({\tilde z};x)>}{8 {\pi} \as
\mu^{2\ep}} = C_F\,
\left[ \frac{2}{1-{\tilde z} + (1-x)}
-  (1+{\tilde z})
-\ep (1-{\tilde z})\right]  \;,
\eeeq
\beeq
\label{vqqa}
\frac{<{\bom V}_{q{\bar q}}^a({\tilde z};x)>}{8 {\pi} \as
\mu^{2\ep}} =
T_R \;
\left[ 1 - \frac{2{\tilde z}(1-{\tilde z})}
{1-\ep} \,\right] \;,
\eeeq
\beq
\label{vgga}
\frac{<{\bom V}_{gg}^a({\tilde z};x)>}{8 {\pi} \as
\mu^{2\ep}} =
2 C_A \,\left[
\frac{1}{1-{\tilde z} + (1-x)} +
\frac{1}{{\tilde z} + (1-x)} - 2
+ {\tilde z} (1 - {\tilde z}) \right] \;.
\eeq
Inserting Eqs.~(\ref{vqga}--\ref{vgga}) into Eqs.~(\ref{vijx}--\ref{vplus})
we obtain:
\beeq
\label{vqgxep}
{\cal V}_{qg}(x;\ep) &=& C_F \left[ \left( \frac{2}{1-x} \ln \frac{1}{1-x}
\right)_+ - \frac{3}{2} \left( \frac{1}{1-x} \right)_+ +
\frac{2}{1-x} \ln (2-x) \right]
\nonumber \\
&+& \delta(1-x) \left[
{\cal V}_{qg}(\ep) - \frac{3}{2} \,C_F\right] + {\cal O}(\ep) \;\;,
\eeeq
\beeq
\label{vqqxep}
{\cal V}_{q{\bar q}}(x;\ep) = \frac{2}{3} \;T_R
\left( \frac{1}{1-x} \right)_+ + \delta(1-x) \left[
{\cal V}_{q{\bar q}}(\ep) + \frac{2}{3} \;T_R \right] + {\cal O}(\ep) \;\;,
\eeeq
\beeq
\label{vggxep}
{\cal V}_{gg}(x;\ep) &=& 2 C_A \left[ \left( \frac{2}{1-x} \ln \frac{1}{1-x}
\right)_+ - \frac{11}{6} \left( \frac{1}{1-x} \right)_+ +
\frac{2}{1-x} \ln (2-x) \right] \nonumber \\
&+& \delta(1-x) \left[
{\cal V}_{gg}(\ep) - \frac{11}{3} \, C_A \right] + {\cal O}(\ep) \;\;,
\eeeq
where ${\cal V}_{ij}(\ep)$ are the functions in Eqs.~(\ref{vqgep}--\ref{vggep}).
Note that the terms $\delta(1-x) \,{\cal V}_{ij}(\ep)$ account for all the
$\ep$-poles of ${\cal V}_{ij}(x;\ep)$, i.e.\ ${\cal V}_{ij}(x;\ep) -
\delta(1-x) \,{\cal V}_{ij}(\ep)$ are finite for $\ep \to 0$.

\subsection{Initial-state singularities: one initial-state parton}
\label{diss}

In the limit $p_a \cdot p_i \to 0$, the dipole factorization formula for the
$m+1$-parton matrix element with a single initial-state parton $a$
is the following
\beq
\label{if}
{}_{m+1,a}\!\!\bra{1, ...., m+1;a}
\ket{1, ...., m+1;a}_{m+1,a} = \sum_{k \neq i}
{\cal D}^{ai}_k(p_1,...,p_{m+1};p_a) + \dots \;\;,
\eeq
where the dots stand for non-singular contributions and the first term
is the sum of the dipole contributions
in which the emitter is the initial-state parton ${\widetilde {ai}}$
and the spectator is the final-state parton $k$. These dipoles are given by
\beeq \label{dipif}
&&{\cal D}^{ai}_k(p_1,...,p_{m+1};p_a) =
- \frac{1}{2 p_a \cdot p_i} \;\frac{1}{x_{ik,a}} \nonumber \\
&&\cdot
{}_{m,a}\!\!\bra{1,.., {\widetilde k},., m+1;{\widetilde {ai}}}
\,\frac{{\bom T}_k \cdot {\bom T}_{ai}}{{\bom T}_{ai}^2} \; {\bom V}^{ai}_k \,
\ket{1,..,{\widetilde k},.., m+1;{\widetilde {ai}}}_{m,a}  \;\;.
\eeeq
Note that, unlike the case in which the emitter is
a final state ${\widetilde {ij}}$, now the dipole is not symmetric under
$a \leftrightarrow i$. In fact, the momentum of the emitter parton
$\widetilde {ai}$ is parallel to $p_a$:
\beq\label{wpaia}
{\widetilde p}_{ai}^\mu = x_{ik,a} \,p_a^\mu \;,
\eeq
where
\beq\label{wpaix}
x_{ik,a} = \frac{p_kp_a+p_ip_a-p_ip_k}{(p_k+p_i)p_a} \;.
\eeq
On the other hand, the momentum of the spectator $\widetilde k$
is not parallel to $p_k$:
\beq\label{wpk}
{\widetilde p}_k^\mu = p_k^\mu  + p_i^\mu  - (1 - x_{ik,a} ) \,p_a^\mu  \;,
\eeq

The splitting functions
${\bom V}^{ai}_k$ are matrices in the helicity space of the parton
$\widetilde {ai}$. Their explicit expression is the following
\beeq\label{Vqagik}
\bra{s}
{\bom V}^{q_ag_i}_k(x_{ik,a};u_i) \ket{s'}
&=& 8\pi \mu^{2\ep} \as\; C_F\;\delta_{ss'} \left[ \frac{2}{1-x_{ik,a}+u_i}
-  \frac{}{} (1+ x_{ik,a}) -\ep (1-x_{ik,a}) \right]
\;, \nonumber \\
&\equiv& V^{q_ag_i}_k \;\delta_{ss'} \;\;,
\eeeq
\beeq\label{Vqagik1}
\bra{s}
{\bom V}^{g_a{\bar q}_i}_k(x_{ik,a}) \ket{s'}
= 8\pi \mu^{2\ep} \as\; T_R\;
\left[ 1 - \ep -2 x_{ik,a}(1-x_{ik,a}) \right]
\;\delta_{ss'} \equiv V^{g_a{\bar q}_i}_k \;\delta_{ss'} \;,
\eeeq
\beeq
&\bra{\mu}&\!\!\!
{\bom V}^{q_aq_i}_k(x_{ik,a};u_i)
\ket{\nu}
= 8\pi \mu^{2\epsilon} \as \; C_F \;
\left[ -g^{\mu \nu}x_{ik,a} \frac{}{} \right. \nonumber \\
&+&\!\!\! \left. \frac{1-x_{ik,a}}{x_{ik,a}} \;
\frac{2u_i(1-u_i)}{p_ip_k} \,
\left( \frac{p_i^{\mu}}{u_i} - \frac{p_k^{\mu}}{1-u_i} \right)
\left( \frac{p_i^{\nu}}{u_i} - \frac{p_k^{\nu}}{1-u_i} \right)
\,\right] \,,
\eeeq
\beeq\label{Vgagik}
&\bra{\mu}&\!\!\!\!{\bom V}^{g_ag_i}_k(x_{ik,a};u_i)
\ket{\nu} = 16\pi
\mu^{2\epsilon} \as \;C_A\;
\left[ -g^{\mu \nu} \left(
\frac{1}{1-x_{ik,a} +u_{i}} -1
+ x_{ik,a}(1-x_{ik,a}) \right) \right. \nonumber \\
&+&\!\!\!\!\!\! \left.
(1-\ep) \;\frac{1-x_{ik,a}}{x_{ik,a}} \;\frac{u_i(1-u_i)}{p_ip_k} \,
\left( \frac{p_i^{\mu}}{u_i} - \frac{p_k^{\mu}}{1-u_i} \right)
\left( \frac{p_i^{\nu}}{u_i} - \frac{p_k^{\nu}}{1-u_i} \right)
\,\right] \;,
\eeeq
where
\beeq\label{dui}
u_i=  \frac{p_ip_a}{p_ip_a+p_kp_a} \;\;.
\eeeq

Note that the factor $1/x_{ik,a}$ on the right-hand side of Eq.~(\ref{dipif})
is relevant for reproducing the collinear limit in Eq.~(\ref{cafac}).

The analogue of the phase-space convolution in Eq.~(\ref{psconv})
is the following identity (similarly to Eqs.~(\ref{psconv},\ref{px}),
we have $Q^2 \leq 0$ in the physical region
of the phase space we are interested in)
\beq
\label{psconva}
d\phi(p_i,p_k;Q+p_a) =  \int_0^1 dx \; d\phi({\widetilde p}_{k};
Q + x p_a) \;\left[ dp_i({\widetilde p}_{k};
p_a, x) \right] \;,
\eeq
where
\beq
\label{dpix}
\left[ dp_i({\widetilde p}_{k}; p_a, x) \right] =
\frac{d^{d}p_i}{(2\pi)^{d-1}} \,\delta_+(p_i^2) \;
\Theta(x) \Theta(1-x) \;
\delta(x - x_{ik,a}) \frac{1}{1-u_i} \;\;.
\eeq
Using the kinematic variables in Eqs.~(\ref{wpaix},\ref{dui}), the phase space
in Eq.~(\ref{dpix}) can be written as follows
\beeq
\label{exdpix}
\left[ dp_i({\widetilde p}_{k};p_a, x) \right]
&=& \frac{(2{\widetilde p}_{k}p_a)^{1-\ep}}{16\pi^2}
\;\frac{d\Omega^{(d-3)}}{(2\pi)^{1-2\ep}}
\;du_i \;dx_{ik,a}
\;\Theta(u_i(1-u_i))
\;\Theta(x(1-x)) \nonumber \\
&\cdot&
\;\left(u_i(1-u_i) \right)^{-\ep}
\;\delta(x-x_{ik,a}) \;\left( 1-x \right)^{-\ep} \;,
\eeeq
where $d\Omega^{(d-3)}$ is an element of solid angle perpendicular to
${\widetilde p}_{k}$ and $p_a$.

It is also useful to introduce the following integral of the
splitting function ${\bom V}^{ai}_k$:
\beq
\label{intVAik}
\frac{\as}{2\pi} \frac{1}{\Gamma(1-\ep)} \left(
\frac{4\pi \mu^2}{2 {\widetilde p}_{k}p_a} \right)^{\ep}
{\cal V}^{a,ai}(x;\ep)
\equiv \int \left[ dp_i({\widetilde p}_{k}; p_a, x) \right]
\frac{1}{2p_ap_i} \frac{n_s({\widetilde {ai}})}{n_s(a)}
\; <{\bom V}^{ai}_k(x_{ik,a};u_i)> \;,
\eeq
where $n_s(a) \;(n_s({\widetilde {ai}}))$ is the number of polarizations
of the
parton $a \;({\widetilde {ai}})$ ($n_s=2$ for fermions and $n_s=d-2=2(1-\ep)$
for gluons) and $<{\bom V}^{ai}_k>$ denotes the average of
${\bom V}^{ai}_k$ over the polarizations of the emitter parton
${\widetilde {ai}}$.

{}From the definition in Eq.~(\ref{intVAik}), we have
\beq
\label{Vaix}
{\cal V}^{a,ai}(x;\ep) = \Theta(x) \Theta(1-x) \left(
\frac{1}{1-x} \right)^{\ep} \;
\int_0^1 \frac{du_i}{u_i} \,
\left(u_i (1-u_i) \right)^{-\ep}
\frac{n_s({\widetilde {ai}})}{n_s(a)}
\frac{<{\bom V}^{ai}_k(x;u_i)>}{8 {\pi} \as
\mu^{2\ep}} \;.
\eeq

The integration over $u_i$ in Eq.~(\ref{Vaix}) leads to $\ep$-poles
of the type $f(x)/\ep$, where $f(x)$ can be either an $x$-integrable function
for $x=1$ or $(1-x)^{-1}$. In order to get a series in $\ep$ whose
coefficients
are well-defined $x$-distributions, we write:
\beq
\label{vaai}
{\cal V}^{a,ai}(x;\ep) = \frac{1}{\ep} \left\{ \frac{1}{x}
\left[ \,\ep x {\cal V}^{a,ai}(x;\ep) \,\right]_{+} + \ep
\delta(1-x) \int_0^1 dz \; z\, {\cal V}^{a,ai}(z;\ep) \right\} \;\;.
\eeq
The two terms in the curly bracket can separately be expanded in $\ep$ and
lead to contributions with the following structure
\beq
P^{a,{\widetilde {ai}}}(x) + {\mbox {\rm const.}} \left( \frac{1}{\ep} +
{\cal O}(1) \right)
\delta(1-x)
+ {\cal O}(\ep) \;\;,
\eeq
where $P^{a,{\widetilde {ai}}}(x)$ is the
{\em four-dimensional\/} regularized (real + virtual)
Altarelli-Parisi probability for the splitting process
$a(p_a) \to {\widetilde {ai}}(xp_a) + i((1-x)p_a)$.

In order to obtain this result, we first compute the
spin-averaged splitting functions $<{\bom V}^{ai}_k(x;u)>$:
\beeq
\label{vqgu}
\frac{n_s({\widetilde q})}{n_s(q)}
\frac{<{\bom V}^{qg}_k(x;u)>}{8 {\pi} \as
\mu^{2\ep}} = C_F\,
\left[ \frac{2}{1- x + u}
-  (1+x)
-\ep (1-x)\right]  \;,
\eeeq
\beeq
\label{vgqu}
\frac{n_s({\widetilde q})}{n_s(g)}
\frac{<{\bom V}^{g{\bar q}}_k(x)>}{8 {\pi} \as
\mu^{2\ep}} =
T_R \;
\left[ 1 - \frac{2x(1-x)}{1-\ep} \,\right] \;,
\eeeq
\beq
\label{vggu}
\frac{n_s({\widetilde g})}{n_s(g)}
\frac{<{\bom V}^{gg}_k(x;u)>}{8 {\pi} \as
\mu^{2\ep}} =
2 C_A \,\left[
\frac{1}{1- x + u} +
\frac{1-x}{x} - 1
+ x (1 - x) \right] \;.
\eeq
\beeq
\label{vqqu}
\frac{n_s({\widetilde g})}{n_s(q)}
\frac{<{\bom V}^{qq}_k(x)>}{8 {\pi} \as
\mu^{2\ep}} =
C_F \;
\left[ (1 - \ep) x + 2 \frac{1-x}{x} \,\right] \;.
\eeeq
Then, inserting Eqs.~(\ref{vqgu}--\ref{vqqu}) into Eqs.~(\ref{Vaix},\ref{vaai}),
we find:
\beeq
\label{cvqgxep}
{\cal V}^{q,g}(x;\ep) = - \frac{1}{\ep} \;P^{qg}(x) + P^{qg}(x) \ln(1-x)
+ C_F \;x + {\cal O}(\ep) \;\;,
\eeeq
\beeq
\label{cvgqxep}
{\cal V}^{g,q}(x;\ep) = - \frac{1}{\ep} \;P^{gq}(x) + P^{gq}(x) \ln(1-x)
+ T_R \;2x(1-x) + {\cal O}(\ep) \;\;,
\eeeq
\beeq
\label{cvqqxep}
{\cal V}^{q,q}(x;\ep) &=& - \frac{1}{\ep} \;P^{qq}(x) + \delta(1-x) \left[
{\cal V}_{qg}(\ep) + \left( \frac{2}{3} \pi^2 - 5 \right) C_F \right]
\nonumber \\
&+& C_F \left[ - \left(\frac{4}{1-x} \ln \frac{1}{1-x} \right)_+ -
\frac{2}{1-x} \ln(2-x) \right. \nonumber \\
&+& \left. \frac{}{} (1-x) - (1+x) \ln(1-x) \right] + {\cal O}(\ep) \;\;,
\eeeq
\beeq
\label{cvggxep}
{\cal V}^{g,g}(x;\ep) &=& - \frac{1}{\ep} \;P^{gg}(x) + \delta(1-x) \left[
\frac{1}{2} {\cal V}_{gg}(\ep) + N_f {\cal V}_{q{\bar q}}(\ep) +
\left( \frac{2}{3} \pi^2 - \frac{50}{9} \right)
C_A \right. \nonumber \\
&+&\left. \frac{16}{9} N_f T_R \right]
+ C_A \left[ - \left(\frac{4}{1-x} \ln \frac{1}{1-x} \right)_+ -
\frac{2}{1-x} \ln(2-x) \right. \nonumber \\
&+& \left. 2 \left( -1 +x(1-x) + \frac{1-x}{x} \right)
\ln(1-x) \right] + {\cal O}(\ep) \;\;,
\eeeq
where the functions ${\cal V}_{ij}(\ep)$ are given in
Eqs.~(\ref{vqgep}--\ref{vggep}) and the (customary) regularized Altarelli-Parisi
probabilities are
\beeq \label{pqg4}
P^{qg}(x) = C_F \;\frac{1 + (1-x)^2}{x} \;\;,
\eeeq
\beeq \label{pgq4}
P^{gq}(x) = T_R \left[ x^2 + (1-x)^2 \right] \;\;,
\eeeq
\beeq \label{pqq4}
P^{qq}(x) = C_F \;\left( \frac{1 + x^2}{1-x} \right)_+ \;\;,
\eeeq
\beq \label{pgg4}
P^{gg}(x) = 2C_A \;\left[ \left( \frac{1}{1-x} \right)_+ + \frac{1-x}{x}
-1 + x(1-x) \right]
+ \delta(1-x) \left( \frac{11}{6} C_A - \frac{2}{3}
N_f T_R \right) \;\;.
\eeq

To simplify our notation in the following Subsections, it is
also convenient to introduce the regular (i.e.\ not singular for $x \to 1$)
parts $P^{ab}_{{\rm reg}}(x)$ of the Altarelli-Parisi probabilities as follows
\beeq
\label{preg}
P^{ab}_{{\rm reg}}(x) \equiv P^{ab}(x) - \delta^{ab} \left[
2 \;{\bom T}_a^2 \left( \frac{1}{1-x} \right)_+ + \gamma_a \;\delta(1-x) \right]
\;\;,
\eeeq
\beeq
\label{cgamma}
\gamma_{a=q,{\bar q}} = \frac{3}{2} C_F \;, \;\;\;\;
\gamma_{a=g} = \frac{11}{6} C_A - \frac{2}{3} T_R N_f \;,
\eeeq
that is,
\beeq
\label{prex}
&&P^{ab}_{{\rm reg}}(x) = P^{ab}(x) \;\; \;\;\;\;\;\;\;\;\;\;\;\;\;{\rm if}
\;\;a \neq b \;\;,
\nonumber \\
&&P^{qq}_{{\rm reg}}(x) = - C_F \,(1 + x) \;\;, \;\;\;
P^{gg}_{{\rm reg}}(x) = 2\, C_A \left[ \frac{1-x}{x}  - 1 + x(1-x) \right] \;\;.
\eeeq

\subsection{Dipole formulae with one identified parton in the final state}
\label{dfrag}

The perturbative QCD treatment of cross sections that, in addition to jets,
have identified particles in the final state requires the introduction
of fragmentation functions. In order to deal with the collinear singularities
related to the fragmentation process,
it is convenient to consider dipole factorization formulae that treat the
final-state partons in an unsymmetric manner. Thus, the factorization formulae
we are going to present below  rescale the momenta of identified and
non-identified partons in a different way.

Let us consider the matrix element $\aket{a,1, ...., m+1}$,
where $a$ denotes a parton whose momentum $p_a$ is identified in the final
state. In this case we can get singularities from the regions where
$p_i \cdot p_j \rightarrow 0$ or $p_i \cdot p_a \rightarrow 0$. In the first
region we write the following dipole formula
\beeq \label{ffa}
&&\abra{a,1, ...., m+1} \aket{a,1, ...., m+1} \nonumber \\
&&= \sum_{k \neq i.j} {\cal D}_{ij,k}(p_a,p_1,...,p_{m+1}) +
{\cal D}_{ij,a}(p_a,p_1,...,p_{m+1}) + \dots \;\;
\eeeq
where $\dots$ stands for terms that are not singular in the
limit $p_i \cdot p_j \rightarrow 0$.

The first term on the right-hand side of Eq.~(\ref{ffa}) is analogous to that
in Eq.~(\ref{ff}), i.e.\ the dipole ${\cal D}_{ij,k}(p_a,p_1,...,p_{m+1})$ is:
\beeq
&&{\cal D}_{ij,k}(p_a,p_1,...,p_{m+1}) = - \frac{1}{2 p_i \cdot p_j}
\nonumber \\
&&\ambra{a,1, .., {\widetilde {ij}},.., {\widetilde k},.., m+1}
\,\frac{{\bom T}_k \cdot {\bom T}_{ij}}{{\bom T}_{ij}^2} \; {\bom V}_{ij,k} \,
\amket{a,1, .., {\widetilde {ij}},.., {\widetilde k},.., m+1}  \;\;.
\eeeq

The second term is similar to the first one, that is,
\beeq \label{dipffa}
&&{\cal D}_{ij,a}(p_a,p_1,...,p_{m+1}) = - \frac{1}{2 p_i \cdot p_j}
\nonumber \\
&&\ambra{{\widetilde a},1, .., {\widetilde {ij}},.., m+1}
\,\frac{{\bom T}_a \cdot {\bom T}_{ij}}{{\bom T}_{ij}^2} \; {\bom V}_{ij,a} \,
\amket{{\widetilde a},1, .., {\widetilde {ij}},.., m+1} \;\;,
\eeeq
but now the spectator is the identified particle $\widetilde a$.
It is thus convenient to rewrite Eqs.~(\ref{pk})--(\ref{vggk}) in terms of the
following variable
\beq\label{dzija}
z_{ij,a}= 1-y_{ij,a} = \frac{(p_i+p_j)p_a}{p_ip_j+p_jp_a+p_ap_i} \;.
\eeq
The dipole momenta ${\widetilde p}_a$ and ${\widetilde p}_{ij}$ are:
\beq
{\widetilde p}_a^\mu = \frac{1}{z_{ij,a}} \,p_a^\mu \;, \;\;\;\;
{\widetilde p}_{ij}^\mu = p_i^\mu + p_j^\mu -
\frac{1-z_{ij,a}}{z_{ij,a}} \,p_a^\mu \;,
\eeq
and the splitting functions are given in terms of $z_{ij,a}$ and the variables
${\tilde z}_i$, ${\tilde z}_j$,
\beq\label{tzia}
{\tilde z}_i
= \frac{p_ip_a}{(p_j+p_i)p_a} =
\frac{p_i{\widetilde p}_a}{{\widetilde p}_{ij}{\widetilde p}_a} \;, \;\;\;
{\tilde z}_j = \frac{p_jp_a}{(p_j+p_i)p_a}=
\frac{p_j{\widetilde p}_a}{{\widetilde p}_{ij}{\widetilde p}_a}=
1-{\tilde z}_i  \;,
\eeq
by the following expressions
\beq\label{Vqigjaa}
\bra{s}
{\bom V}_{q_ig_j,a}({\tilde z}_i;1-z_{ij,a}) \ket{s'}
= 8\pi \mu^{2\ep} \as\; C_F\,
\left[ \frac{2}{1-{\tilde z}_i \,z_{ij,a}}
- \frac{}{} (1+{\tilde z}_i)
-\ep (1-{\tilde z}_i)\right] \;\delta_{ss'} \;,
\eeq
\beq
\bra{\mu}
{\bom V}_{q_i{\bar q}_j,a}({\tilde z}_i)
\ket{\nu}
= 8\pi \mu^{2\epsilon} \as \;
T_R \;
\left[ -g^{\mu \nu} - \frac{2}{p_ip_j} \,
( {\tilde z}_i p_i^{\mu} - {\tilde z}_j p_j^{\mu} )
\,( {\tilde z}_i p_i^{\nu} - {\tilde z}_j p_j^{\nu} ) \,\right] \,,
\eeq
\beeq\label{Vgigjaa}
&\!&\!\!\!\!\!\!\!\!\!\!\bra{\mu}{\bom V}_{g_ig_j,a}({\tilde z}_i;1-z_{ij,a})
\ket{\nu} = 16\pi
\mu^{2\epsilon} \as \;C_A\,
\left[ -g^{\mu \nu} \left(
\frac{1}{1-{\tilde z}_i \,z_{ij,a}} \right. \right.\nonumber \\
&\!&\!\!\!\!\!\!\!\!\!\!+ \left. \left.
\frac{1}{1-{\tilde z}_j \,z_{ij,a}} - 2 \right)
+ (1-\ep) \frac{1}{p_ip_j} \,
( {\tilde z}_i p_i^{\mu} - {\tilde z}_j p_j^{\mu} )
\,( {\tilde z}_i p_i^{\nu} - {\tilde z}_j p_j^{\nu} ) \,\right] \;.
\eeeq

The only difference in the treatment of the two dipole contributions in
Eq.~(\ref{ffa}) is that $a$ is identified in the final state and, hence, in the
physical cross section we do not have to integrate over its momentum $p_a$.
Thus we are led to consider the following convolution formula for the phase
space (the factor $1/z^{2-2\ep}$ is introduced for later convenience: see
Eq.~(\ref{saCinc}) and Sect.~\ref{sifra})
\beq
\label{psconvfr}
d\phi(p_i,p_j;Q-p_a) =  \int_0^1 \frac{dz}{z^{2-2\ep}}
\; d\phi({\widetilde p}_{ij};
Q - p_a/z) \;\left[ dp_i({\widetilde p}_{ij},
p_a;z) \right] \;,
\eeq
where
\beq
\label{dpiz}
\left[ dp_i({\widetilde p}_{ij},p_a;z) \right] =
\frac{d^{d}p_i}{(2\pi)^{d-1}} \,\delta_+(p_i^2) \;
\Theta(z) \Theta(1-z) \;
\delta(z - z_{ij,a}) \frac{z^{2-2\ep}}{1-{\tilde z_i}} \;\;.
\eeq
In terms of the kinematic variables in Eqs.~(\ref{dzija},\ref{tzia}),
Eq.~(\ref{dpiz}) is
\beeq
\label{exdpiz}
\left[ dp_i({\widetilde p}_{ij},p_a;z) \right]
&=& \frac{(2{\widetilde p}_{ij}p_a)^{1-\ep}}{16\pi^2}
\;\frac{d\Omega^{(d-3)}}{(2\pi)^{1-2\ep}}
\;d{\tilde z}_i \;dz_{ij,a}
\;\Theta({\tilde z}_i(1-{\tilde z}_i))
\;\Theta(z(1-z)) \nonumber \\
&\cdot&
\;\left({\tilde z}_i(1-{\tilde z}_i) \right)^{-\ep}
\;\delta(z-z_{ij,a}) \;\left( z(1-z) \right)^{-\ep} \;,
\eeeq
where $d\Omega^{(d-3)}$ is an element of solid angle perpendicular to
${\widetilde p}_{ij}$ and $p_a$.

The integration of the splitting functions ${\bom V}_{ij,a}$ over the phase
space in Eq.~(\ref{exdpiz}) defines the functions
${\overline {\cal V}}_{ij}(z;\ep)$:
\beq
\frac{\as}{2\pi} \frac{1}{\Gamma(1-\ep)} \left(
\frac{4\pi \mu^2}{2 {\widetilde p}_{ij}p_a} \right)^{\ep}
{\overline {\cal V}}_{ij}(z;\ep)
\equiv \int \left[ dp_i({\widetilde p}_{ij}, p_a; z) \right]
\frac{1}{2p_ip_j} <{\bom V}_{ij,a}({\tilde z}_i;1-z_{ij,a})> \;,
\eeq
that is,
\beq
\label{vbarijz}
{\overline {\cal V}}_{ij}(z;\ep) = \Theta(z) \Theta(1-z) \;\
\frac{z^{1-\ep}}{(1-z)^{1+\ep}} \;
\int_0^1 d{\tilde z}_i \,
\left({\tilde z}_i (1-{\tilde z}_i) \right)^{-\ep}
\frac{<{\bom V}_{ij,a}({\tilde z}_i;1-z)>}{8 {\pi} \as
\mu^{2\ep}} \;.
\eeq
As in the case of the function ${\cal V}_{ij}(x;\ep)$ in Eq.~(\ref{vijx}),
${\overline {\cal V}}_{ij}(z;\ep)$ is a $z$-distribution with $\ep$-dependent
coefficients:
\beq
\label{barvplus}
{\overline {\cal V}}_{ij}(z;\ep) =
\left[ \,{\overline {\cal V}}_{ij}(z;\ep=0) \,\right]_{+} +
\delta(1-z) \int_0^1 dz'
\;{\overline {\cal V}}_{ij}(z';\ep) + {\cal O}(\ep)  \;\;.
\eeq

The explicit form of the spin-averaged splitting functions
$<{\bom V}_{ij,a}>$ is the same as in Eqs.~(\ref{avVqgk}--\ref{avVggk}),
apart from the replacements $k \to a \;, \;\;1-y \to z$:
\beeq
\label{barvqga}
\frac{<{\bom V}_{qg,a}({\tilde z};1-z)>}{8 {\pi} \as
\mu^{2\ep}} = C_F\,
\left[ \frac{2}{1-{\tilde z} z}
-  (1+{\tilde z})
-\ep (1-{\tilde z})\right]  \;,
\eeeq
\beeq
\label{barvqqa}
\frac{<{\bom V}_{q{\bar q},a}({\tilde z})>}{8 {\pi} \as
\mu^{2\ep}} =
T_R \;
\left[ 1 - \frac{2{\tilde z}(1-{\tilde z})}
{1-\ep} \,\right] \;,
\eeeq
\beq
\label{barvgga}
\frac{<{\bom V}_{gg,a}({\tilde z};1-z)>}{8 {\pi} \as
\mu^{2\ep}} =
2 C_A \,\left[
\frac{1}{1-{\tilde z}z} +
\frac{1}{1-(1-{\tilde z})z} - 2
+ {\tilde z} (1 - {\tilde z}) \right] \;.
\eeq
Inserting Eqs.~(\ref{barvqga})--(\ref{barvgga}) into Eq.~(\ref{vbarijz}),
performing the $\tilde z$-integration and expanding in $\ep$ as in
Eq.~(\ref{barvplus}), we find
\beq
\label{vqgzep}
{\overline {\cal V}}_{qg}(z;\ep) = C_F \left[ \left( \frac{2}{1-z}
\ln \frac{1}{1-z}
\right)_+ - \frac{3}{2} \left( \frac{z}{1-z} \right)_+  \right]
+ \delta(1-z) \;\;{\cal V}_{qg}(\ep) + {\cal O}(\ep) \;\;,
\eeq
\beeq
\label{vqqzep}
{\overline {\cal V}}_{q{\bar q}}(z;\ep) = \frac{2}{3} \;T_R
\left( \frac{z}{1-z} \right)_+ + \delta(1-z) \;\;
{\cal V}_{q{\bar q}}(\ep) + {\cal O}(\ep) \;\;,
\eeeq
\beq
\label{vggzep}
{\overline {\cal V}}_{gg}(z;\ep) = 2 C_A \left[ \left( \frac{2}{1-z}
\ln \frac{1}{1-z}
\right)_+ - \frac{11}{6} \left( \frac{z}{1-z} \right)_+
\right]
+ \delta(1-z) \;\;{\cal V}_{gg}(\ep) + {\cal O}(\ep) \;\;,
\eeq
where ${\cal V}_{ij}(\ep)$ are the functions in Eqs.~(\ref{vqgep}--\ref{vggep}).

Let us now consider the singularities from the region where
$p_i \cdot p_a \rightarrow 0$. In this
case we write:
\beq \label{aff}
\abra{a,1, ...., m+1} \aket{a,1, ...., m+1} =
\sum_{k \neq i} {\cal D}_{ai,k}(p_a,p_1,...,p_{m+1}) + \dots \;\;,
\eeq
\beeq \label{dipaff}
&&{\cal D}_{ai,k}(p_a,p_1,...,p_{m+1}) =
- \frac{1}{2 p_i \cdot p_a}  \nonumber \\
&&\cdot
\ambra{{\widetilde {ai}},1, ... {\widetilde k},.., m+1}
\,\frac{{\bom T}_k \cdot {\bom T}_{ai}}{{\bom T}_{ai}^2} \; {\bom V}_{ai,k} \,
\amket{{\widetilde {ai}},1, ...{\widetilde k},.., m+1}  \;\;.
\eeeq

The parton momenta in the dipole (\ref{dipaff}) are:
\beq\label{wpaimu}
{\widetilde p}_{ai}^\mu = \frac{1}{z_{ik,a}} \,p_a^\mu  \;, \;\;\;\;
{\widetilde p}_{k}^\mu  = p_i^\mu  + p_k^\mu  -
\frac{1-z_{ik,a}}{z_{ik,a}} \,p_a^\mu  \;,
\eeq
where
\beq\label{wpaimuz}
z_{ik,a}= \frac{(p_i+p_k)p_a}{p_ip_k+p_kp_a+p_ap_i} \;.
\eeq
Note that, unlike the case of the final-state dipole
$\{ {\widetilde {ij}},{\widetilde k}\}$ with no
identified partons, here the momentum of $\widetilde {ai}$ ($\widetilde k$)
is (is not) parallel to $a$ ($k$).

The splitting functions ${\bom V}_{ai,k}$ are:
\beq\label{Vqagika}
\bra{s}
{\bom V}_{q_ag_i,k}(z_{ik,a};u_i) \ket{s'}
= 8\pi \mu^{2\ep} \as\; C_F\;\delta_{ss'} \left[ \frac{2}{1-z_{ik,a}(1-u_i)}
- \frac{}{} (1+ z_{ik,a}) -\ep (1-z_{ik,a})) \right]
\;,
\eeq
\beeq
\bra{s}
{\bom V}_{g_a{\bar q}_i,k}(z_{ik,a}) \ket{s'}
= 8\pi \mu^{2\ep} \as\; C_F\;\delta_{ss'} \left[
\frac{1+(1-z_{ik,a})^2}{z_{ik,a}} -\ep \;z_{ik,a} \right] \;\;,
\eeeq
\beeq
&\bra{\mu}&\!\!\!
{\bom V}_{q_a{\bar q}_i,k}(z_{ik,a};u_i)
\ket{\nu}
= 8\pi \mu^{2\epsilon} \as \; T_R\; \\
&\cdot&\!\!\! \left[ -g^{\mu \nu} \frac{}{}
- 2 z_{ik,a}(1-z_{ik,a})
\frac{u_i(1-u_i)}{p_ip_k} \,
\left( \frac{p_i^{\mu}}{u_i} - \frac{p_k^{\mu}}{1-u_i} \right)
\left( \frac{p_i^{\nu}}{u_i} - \frac{p_k^{\nu}}{1-u_i} \right)
\,\right] \,, \nonumber
\eeeq
\beeq\label{Vgagika}
&\bra{\mu}&\!\!\!\!{\bom V}_{g_ag_i,k}(z_{ik,a};u_i)
\ket{\nu} = 16\pi
\mu^{2\epsilon} \as \;C_A\;
\left[ -g^{\mu \nu} \left(
\frac{1}{1-z_{ik,a}(1-u_{i})} -2 + \frac{1}{z_{ik,a}} \right)
 \right. \nonumber \\
&+&\!\!\!\!\!\left.
(1-\ep) (1-z_{ik,a})z_{ik,a} \;
\frac{u_i(1-u_i)}{p_ip_k} \,
\left( \frac{p_i^{\mu}}{u_i} - \frac{p_k^{\mu}}{1-u_i} \right)
\left( \frac{p_i^{\nu}}{u_i} - \frac{p_k^{\nu}}{1-u_i} \right)
\,\right] \;,
\nonumber \\
&&\,
\eeeq
where the variable $u_i$ is defined as follows
\beq\label{duia}
u_i= \frac{p_ip_a}{(p_i+p_k)p_a} \;\;.
\eeq

The phase-space for the dipole $\{ai,k\}$ can be written in terms of the
rescaled momenta ${\widetilde p}_{k}, {\widetilde p}_{ai}$ by using the
following convolution formula
\beq
\label{psconfrag}
d\phi(p_i,p_k;Q-p_a) =  \int_0^1 \frac{dz}{z^{2-2\ep}}
\; d\phi({\widetilde p}_{k};
Q - p_a/z) \;\left[ dp_i({\widetilde p}_{k},
p_a;z) \right] \;,
\eeq
where
\beq
\label{psaik}
\left[ dp_i({\widetilde p}_{k},p_a;z) \right] =
\frac{d^{d}p_i}{(2\pi)^{d-1}} \,\delta_+(p_i^2) \;
\Theta(z) \Theta(1-z) \;
\delta(z - z_{ik,a}) \frac{z^{2-2\ep}}{1-u_i} \;\;.
\eeq
Using the kinematic variables in Eqs.~(\ref{wpaimuz},\ref{duia}),
Eq.~(\ref{psaik}) can be written as
\beeq
\label{expsaik}
\left[ dp_i({\widetilde p}_{k},p_a;z) \right]
&=& \frac{(2{\widetilde p}_{k}p_a)^{1-\ep}}{16\pi^2}
\;\frac{d\Omega^{(d-3)}}{(2\pi)^{1-2\ep}}
\;du_i \;dz_{ik,a}
\;\Theta(u_i(1-u_i))
\;\Theta(z(1-z)) \nonumber \\
&\cdot&
\;\left(u_i(1-u_i) \right)^{-\ep}
\;\delta(z-z_{ik,a}) \; \left( z(1-z) \right)^{-\ep} \;,
\eeeq
where $d\Omega^{(d-3)}$ is an element of solid angle perpendicular to
${\widetilde p}_{k}$ and $p_a$.

The integration of the splitting functions ${\bom V}_{ai,k}$ over the phase
space in Eq.~(\ref{expsaik}) defines the functions ${\cal V}_{ai,a}(z;\ep)$:
\beq
\label{intVAikt}
\frac{\as}{2\pi} \frac{1}{\Gamma(1-\ep)} \left(
\frac{4\pi \mu^2}{2 {\widetilde p}_{k}p_a} \right)^{\ep}
{\cal V}_{ai,a}(z;\ep)
\equiv \int \left[ dp_i({\widetilde p}_{k}, p_a; z) \right]
\frac{1}{2p_ap_i}
\; <{\bom V}_{ai,k}(z_{ik,a};u_i)> \;,
\eeq
that is,
\beq
\label{Vaiz}
{\cal V}_{ai,a}(z;\ep) = \Theta(z) \Theta(1-z) \;\left[
z(1-z) \right]^{-\ep} \;
\int_0^1 \frac{du_i}{u_i} \,
\left(u_i (1-u_i) \right)^{-\ep}
\frac{<{\bom V}_{ai,k}(z;u_i)>}{8 {\pi} \as
\mu^{2\ep}} \;.
\eeq

The spin-averaged splitting functions $<{\bom V}_{ai,k}(z;u)>$ are:
\beeq
\label{vqgzu}
\frac{<{\bom V}_{q_ag_i,k}(z;u)>}{8 {\pi} \as
\mu^{2\ep}} = C_F\,
\left[ \frac{2}{1- z (1-u)}
-  (1+z)
-\ep (1-z)\right]  \;,
\eeeq
\beeq
\label{vgqzu}
\frac{<{\bom V}_{g_aq_i,k}(z)>}{8 {\pi} \as
\mu^{2\ep}} =
C_F \;
\left[ \frac{1 + (1-z)^2}{z} - \ep z \,\right] \;,
\eeeq
\beeq
\label{vqqzu}
\frac{<{\bom V}_{q_a{\bar q_i},k}(z)>}{8 {\pi} \as
\mu^{2\ep}} =
T_R \;
\left[ 1 - \frac{2z(1-z)}{1-\ep} \,\right] \;,
\eeeq
\beq
\label{vggzu}
\frac{<{\bom V}_{g_ag_i,k}(z;u)>}{8 {\pi} \as
\mu^{2\ep}} =
2 C_A \,\left[
\frac{1}{1- z(1- u)} - 2
+ \frac{1}{z}
+ z (1 - z) \right] \;.
\eeq
Inserting Eqs.~(\ref{vqgzu}--\ref{vggzu}) into Eq.~(\ref{Vaiz}) and expanding
in $\ep$, we obtain the following $z$-distributions
\beeq
\label{cvqgzep}
{\cal V}_{q,g}(z;\ep) = - \frac{1}{\ep} \;P_{qg}(z) + P_{qg}(z) \;\ln[z(1-z)]
+ C_F \;z + {\cal O}(\ep) \;\;,
\eeeq
\beeq
\label{cvgqzep}
{\cal V}_{g,q}(z;\ep) = - \frac{1}{\ep} \;P_{gq}(z) + P_{gq}(z) \;\ln[z(1-z)]
+ T_R \;2z(1-z) + {\cal O}(\ep) \;\;,
\eeeq
\beeq
\label{cvqqzep}
{\cal V}_{q,q}(z;\ep) &=& - \frac{1}{\ep} \;P_{qq}(z) + \delta(1-z) \left[
{\cal V}_{qg}(\ep) + \left( \frac{2}{3} \pi^2 - 5 \right) C_F \right]
\nonumber \\
&+& C_F \left[ - \left(\frac{4}{1-z} \ln \frac{1}{1-z} \right)_+ +
\frac{2}{1-z} \ln z \right. \nonumber \\
&+& \left. \frac{}{} (1-z) - (1+z) \;\ln[z(1-z)] \;\right] + {\cal O}(\ep) \;\;,
\eeeq
\beeq
\label{cvggzep}
{\cal V}_{g,g}(z;\ep) &=& - \frac{1}{\ep} \;P_{gg}(z) + \delta(1-z) \left[
\frac{1}{2} {\cal V}_{gg}(\ep) + N_f {\cal V}_{q{\bar q}}(\ep) +
\left( \frac{2}{3} \pi^2 - \frac{50}{9} \right)
C_A \right. \nonumber \\
&+&\left. \frac{16}{9} N_f T_R \right]
+ C_A \left[ - \left(\frac{4}{1-z} \ln \frac{1}{1-z} \right)_+ +
\frac{2}{1-z} \ln z \right. \nonumber \\
&+& \left. 2 \left( -2 +z(1-z) + \frac{1}{z} \right) \;
\ln[z(1-z)] \;\right] + {\cal O}(\ep) \;\;,
\eeeq
where the functions ${\cal V}_{ij}(\ep)$ are given in
Eqs.~(\ref{vqgep}--\ref{vggep}) and $P_{ab}(z) = P^{ab}(z)$
are the regularized Altarelli-Parisi probabilities in
Eqs.~(\ref{pqg4}--\ref{pgg4}).

\subsection{Initial-state singularities: two initial-state partons}
\label{dissa}

Let us come back to the treatment of initial-state singularities
in the case with two partons $a$ and $b$ in the initial state.
In the limit $p_a \cdot p_i \to 0$, the dipole factorization
formula is
\beeq
\label{iif}
&&{}_{m+1,ab}\!\!\bra{1, ...., m+1;a,b}
\ket{1, ...., m+1;a,b}_{m+1,ab} \nonumber \\
&&= \sum_{k \neq i} {\cal D}^{ai}_k(p_1,...,p_{m+1};p_a,p_b)
+ {\cal D}^{ai,b}(p_1,...,p_{m+1};p_a,p_b) + \dots \;\;.
\eeeq
This equation generalizes Eq.~(\ref{if}).

The
second term on the right-hand side represents a new dipole contribution in
which the emitter is the initial-state parton $\widetilde {ai}$ and
the spectator is the other initial-state parton~$b$:
\beeq
\label{dipiif}
&&{\cal D}^{ai,b}(p_1,...,p_{m+1};p_a,p_b) =
- \frac{1}{2 p_a \cdot p_i}  \;\frac{1}{x_{i,ab}} \nonumber \\
&&{}_{m,ab}\!\!\bra{{\widetilde 1},.., {\widetilde {m+1}};{\widetilde {ai}},b}
\,\frac{{\bom T}_b \cdot {\bom T}_{ai}}{{\bom T}_{ai}^2} \;
{\bom V}^{ai,b} \,
\ket{{\widetilde 1},..,{\widetilde {m+1}};{\widetilde {ai}},b}_{m,ab}
\;\;.
\eeeq
This dipole contribution differs from that in which
the spectator is a final-state parton because, when computing the
cross section (see Sect.~\ref{sipp}), it is convenient to leave
the momentum $p_b$ {\em unchanged}. Thus the matrix element
$\ket{{\widetilde 1},..,{\widetilde {m+1}};{\widetilde {ai}},b}_{m,ab}$
involves an initial-state parton ${\widetilde {ai}}$ with momentum parallel
to $p_a$ :
\beeq\label{wpab}
{\widetilde p}_{ai}^\mu &=& x_{i,ab} \,p_a^\mu  \;, \\
\label{wxiab}
x_{i,ab} &=& \frac{p_ap_b-p_ip_a-p_ip_b}{p_ap_b} \;,
\eeeq
and {\em all\/} other final-state momenta $k_j$ (and {\em not only\/} the
momenta $p_j$ of the QCD partons!) rescaled as follows
\beq
\label{ktilde}
{\widetilde k_j}^{\mu} = k_j^{\mu} - \frac{2 k_j \cdot (K+{\widetilde K})}
{(K+{\widetilde K})^2} \;(K+{\widetilde K})^{\mu} + \frac{2 k_j \cdot K}
{K^2} \;{\widetilde K}^{\mu}\;,
\eeq
where the momenta $K^{\mu}$ and ${\widetilde K}^\mu$ are defined by
\beq\label{dqmu}
\begin{array}{rcl}
K^{\mu} &=& p_a^\mu + p_b^\mu - p_i^\mu \;\;, \\
{\widetilde K}^{\mu} &=& {\widetilde p_{ai}}^\mu + p_b^\mu \;\;.
\end{array}
\eeq

Note that the momentum conservation constraint in the $m+1$-parton matrix
element is
\beq
p_a^{\mu} + p_b^{\mu} - \sum_j k_j^{\mu} - p_i^\mu = 0 \;\;.
\eeq
Therefore we have $2 \sum_j k_j \cdot K = 2K^2$ and $2 \sum_j k_j \cdot
(K+{\widetilde K}) = 2K^2+2K\cdot{\widetilde K} = (K+{\widetilde K})^2$
and it is straightforward to
check that momentum conservation is exactly implemented in the $m$-parton
matrix element on the right-hand side of 
Eq.~(\ref{dipiif}), that is,
\beq
{\widetilde p}_{ai}^{\mu} + p_b^{\mu} - \sum_j {\widetilde k}_j^{\mu} = 0 \;\;.
\eeq

Note also that Eq.~(\ref{ktilde}) can be rewritten in the following way:
\beeq
\label{klor}
{\widetilde k_j}^{\mu} &=& \Lambda^{\mu}_{\;\; \nu}(K,{\widetilde K})
\;k_j^{\nu} \;\;, \\
\label{klor2}
\Lambda^{\mu}_{\;\; \nu}(K,{\widetilde K}) &=& g^{\mu}_{\;\; \nu} -
\frac{2 (K+{\widetilde K})^\mu (K+{\widetilde K})_\nu}{(K+{\widetilde K})^2}
+\frac{2 {\widetilde K}^\mu K_\nu}{K^2}\;\;,
\eeeq
and thus the matrix $\Lambda^{\mu}_{\;\; \nu}(K,{\widetilde K})$
generates a Lorentz transformation (actually, a {\em proper\/} Lorentz
transformation) on all the final-state momenta.

The splitting functions
${\bom V}^{ai,b}$  in Eq.~(\ref{iif})
are as follows
\beq\label{Vqagib}
\bra{s}
{\bom V}^{q_ag_i,b}(x_{i,ab}) \ket{s'}
= 8\pi \mu^{2\ep} \as\; C_F\;\delta_{ss'} \left[ \frac{2}{1-x_{i,ab}}
- \frac{}{} (1+ x_{i,ab}) -\ep (1-x_{i,ab})) \right]
\;,
\eeq
\beeq
\bra{s}
{\bom V}^{g_a{\bar q}_i,b}(x_{i,ab}) \ket{s'}
= 8\pi \mu^{2\ep} \as\; T_R\;
\left[ 1 - \ep -2 x_{i,ab}(1-x_{i,ab}) \right]
\;\delta_{ss'} \;,
\eeeq
\beeq
&\bra{\mu}&\!\!\!
{\bom V}^{q_aq_i,b}(x_{i,ab})
\ket{\nu}
= 8\pi \mu^{2\epsilon} \as \; C_F \;
\left[ -g^{\mu \nu}x_{i,ab} \frac{}{} \right. \nonumber \\
&+&\!\!\! \left. \frac{1-x_{i,ab}}{x_{i,ab}}
\frac{2p_a \cdot p_b}{p_i \cdot p_a \;p_i \cdot p_b} \,
\left( p_i^{\mu} - \frac{p_ip_a}{p_bp_a} p_b^{\mu} \right)
\left( p_i^{\nu} - \frac{p_ip_a}{p_bp_a} p_b^{\nu} \right)
\,\right] \,,
\eeeq
\beeq\label{Vgagib}
&\bra{\mu}&\!\!\!\!{\bom V}^{g_ag_i,b}(x_{i,ab})
\ket{\nu} = 16\pi
\mu^{2\epsilon} \as \;C_A\;
\left[ -g^{\mu \nu} \left(
\frac{x_{i,ab}}{1-x_{i,ab}} + x_{i,ab}(1-x_{i,ab}) \right)
\right. \nonumber \\
&+&\!\!\!\!\!\left.
(1-\ep) \;\frac{1-x_{i,ab}}{x_{i,ab}} \;
\frac{p_a \cdot p_b}{p_i \cdot p_a \;p_i \cdot p_b} \,
\left( p_i^{\mu} - \frac{p_ip_a}{p_bp_a} p_b^{\mu} \right)
\left( p_i^{\nu} - \frac{p_ip_a}{p_bp_a} p_b^{\nu} \right)
\,\right] \;.
\eeeq

The Jacobian factor associated with the Lorentz transformation (\ref{klor})
acting on
the final-state momenta is equal to unity. Therefore the phase space
for the dipole $\{ {\widetilde {ai}},b\}$ has a trivial convolution structure:
\beq
\label{psconvab}
d\phi(p_i, k_1,...;p_a+p_b) =  \int_0^1 dx
\; d\phi({\widetilde k}_1,...;x p_a+p_b)
\;\left[ dp_i(
p_a,p_b, x) \right] \;,
\eeq
where
\beq
\label{dpiabx}
\left[ dp_i(p_a,p_b, x) \right] =
\frac{d^{d}p_i}{(2\pi)^{d-1}} \,\delta_+(p_i^2) \;
\Theta(x) \Theta(1-x) \;
\delta(x - x_{i,ab}) \;\;.
\eeq
The phase space in Eq.~(\ref{dpiabx}) can be written as follows
\beeq
\label{exdpiabx}
\left[ dp_i(p_a,p_b, x) \right]
&=& \frac{(2p_a p_b)^{1-\ep}}{16\pi^2}
\;\frac{d\Omega^{(d-3)}}{(2\pi)^{1-2\ep}}
\;d{\tilde v}_i \;dx_{i,ab}
\;\Theta(x(1-x))
\;\Theta({\tilde v}_i) \;\Theta\!\left(1-\frac{{\tilde v}_i}{1-x} \right)
\nonumber \\
&\cdot&
\;\left( 1-x\right)^{-2\ep} \;\delta(x-x_{i,ab})
\;\left[\frac{{\tilde v}_i}{1-x}
\left(1-\frac{{\tilde v}_i}{1-x}\right) \right]^{-\ep} \;,
\eeeq
where $x_{i,ab}$ is defined in Eq.~(\ref{wxiab}),
${\tilde v}_i=p_ap_i/p_ap_b$ and
$d\Omega^{(d-3)}$ is an element of solid angle perpendicular to $p_a$ and
$p_b$.

The following integral of the
splitting function ${\bom V}^{ai,b}$ defines the $x$-distribution
${\widetilde {\cal V}}^{a,ai}$ :
\beq
\label{intVAib}
\frac{\as}{2\pi} \frac{1}{\Gamma(1-\ep)} \left(
\frac{4\pi \mu^2}{2p_ap_b} \right)^{\ep}
{\widetilde {\cal V}}^{a,ai}(x;\ep)
\equiv \int \left[ dp_i(p_a,p_b, x) \right]
\frac{1}{2p_ap_i} \frac{n_s({\widetilde {ai}})}{n_s(a)}
\; <{\bom V}^{ai,b}(x_{i,ab})> \;.
\eeq
Using the phase space in Eq.~(\ref{exdpiabx}), we thus have:
\beq
\label{calvt}
{\widetilde {\cal V}}^{a,ai}(x;\ep) = - \frac{1}{\ep}
\frac{\Gamma^2(1-\ep)}{\Gamma(1-2\ep)} \Theta(x) \Theta(1-x)
\;(1-x)^{-2\ep}\;
\frac{n_s({\widetilde {ai}})}{n_s(a)}
\frac{<{\bom V}^{ai,b}(x)>}{8 {\pi} \as \mu^{2\ep}} \;,
\eeq
where the spin averages $<{\bom V}^{ai,b}(x)>$
are exactly proportional to the spin average of the $d$-dimensional
Altarelli-Parisi splitting functions in Eqs.~(\ref{avhpqq}--\ref{avhpgg}):
\beq
\label{calvtav}
\frac{n_s({\widetilde {ai}})}{n_s(a)}
\frac{<{\bom V}^{ai,b}(x)>}{8 {\pi} \as \mu^{2\ep}}
= <{\hat P}_{a,{\widetilde {ai}}}(x;\ep)> \;\;.
\eeq
Performing the $\ep$-expansion in Eqs.(\ref{calvt},\ref{calvtav}) according
to the procedure in Eqs.~(\ref{vpd}) and (\ref{vplus}),
we find
\beq
\label{wtcalv}
{\widetilde {\cal V}}^{a,b}(x;\ep) = {\cal V}^{a,b}(x;\ep) +
\delta^{ab} \,{\bom T}_a^2 \left[ \left( \frac{2}{1-x} \ln\frac{1}{1-x}
\right)_+ + \frac{2}{1-x} \ln (2-x) \right]
+ {\widetilde K}^{ab}(x) + {\cal O}(\ep) \;\;,
\eeq
where ${\cal V}^{a,b}(x;\ep)$ are the functions defined in
Eqs.~(\ref{cvqgxep}--\ref{cvggxep}) and ${\widetilde K}^{ab}(x)$ are
given in terms of the regular part (see Eq.~(\ref{preg})) of the
Altarelli-Parisi probabilities as follows
\beeq
\label{wkdef}
&&{\widetilde K}^{ab}(x) = P^{ab}_{{\rm reg}}(x) \;\ln(1-x) \nonumber \\
&&+ \;\delta^{ab} \,{\bom T}_a^2 \left[ \left( \frac{2}{1-x} \ln (1-x)
\right)_+ - \frac{\pi^2}{3} \delta(1-x) \right] \;\;.
\eeeq

\subsection{Factorization formulae with many identified partons}
\label{dcorr}

In the most general case one deals with QCD cross sections involving
initial-state partons and many identified partons in the final state.
Here the NLO matrix element has collinear singularities
when $p_i\cdot p_j \to 0$ ($i$ and $j$ being unidentified final-state partons)
and when $p_i\cdot p_a \to 0$ ($a$ being either an initial-state parton or an
identified final-state parton). The former singularities can be factorized in
terms of the dipoles ${\cal D}_{ij,k}, {\cal D}_{ij}^a, {\cal D}_{ij,a}$
in Eqs.~(\ref{dipff},\ref{dipfi},\ref{dipffa}). As for the latter
singularities, one should consider two
different possibilities.
If the spectator is an unidentified final-state parton $k$, one can factorize
in terms of the dipoles ${\cal D}^{ai}_k, {\cal D}_{ai,k}$ in
Eqs.~(\ref{dipif},\ref{dipaff}).
If the spectator $b$ is an
initial-state parton or an identified final-state parton, it is convenient to
introduce new dipoles ${\cal D}_{ai,b}^{(n)}, {\cal D}_{ai}^{(n)\,b},
{\cal D}^{(n)\, ai}_b, {\cal D}^{(n)\, ai,b}$ in which the momentum of the
spectator is left {\em unchanged}. Actually, with respect to
the momentum dependence,
these objects are `pseudo-dipoles' rather than dipoles, in the sense that they
depend on the momentum ${\widetilde p}_{ai}$ of the emitter, on
the momentum $p_b$
of the spectator and on an additional momentum $n$. This momentum $n$ is:
\beq
\label{ndef}
n^\mu = p^\mu_{\rm {in}} - \sum_{a
\in {\rm {final \,state}}} p_a^\mu \;\;,
\eeq
where $p^\mu_{\rm {in}}$ is the total incoming momentum and the second term on
the right-hand side is the sum of all the momenta of the identified partons
in the final state\footnote{Note that, in general, one can change
$p^\mu_{\rm {in}}$ by adding some momentum transfer that does not involve
QCD partons. For instance, in deep-inelastic lepton-hadron collisions
$l(k) + h(p) \to l^\prime(k') + \dots$ one can replace
$p^\mu_{\rm {in}} = k^\mu + p^\mu$ with
$p^\mu_{\rm {in}} = Q^\mu + p^\mu$ where $ Q_\mu = k_\mu - k^{\prime}_\mu$.}.
Note that, by momentum conservation, $n^\mu$ is equal to the sum
of the momenta of the final-state unidentified particles (QCD partons or not).
Therefore the momentum $n^\mu$ is time-like (and with positive definite energy).
Furthermore, since we only consider non-trivial quantities in which the
lowest order process has at least one unidentified parton, it cannot be
light-like.

As we shall see, the dipole ${\cal D}^{ai,b}$ considered in the previous
Subsection corresponds to ${\cal D}^{(n)\,ai,b}$ for the particular case
with no identified partons in the final-state (i.e.\ when $n=p_a+p_b$).

Let us first consider the singularities $p_i \cdot p_a \to 0$ when $a$ is an
initial-state parton. In this case the factorization formula is:
\beeq
\label{ainin}
&&{}_{m+1 ...,a...}\!\!\bra{1, ...., m+1,...;a..}
\ket{1, ...., m+1,...;a..}_{m+1 ...,a...} \nonumber \\
&&= \sum_{k \neq i} {\cal D}^{ai}_k(p_1,...,p_{m+1},... ;p_a,..)
+ \sum_{b \neq a} {\cal D}^{(n) \,ai}_b(p_1,...,p_{m+1},... ;p_a,..)
+ \dots \;\;.
\eeeq
The first term on the right-hand side of Eq.~(\ref{ainin}) is the same as that
in Eq.~(\ref{iif}).

The new dipole contributions are given by the second term on the right-hand
side of Eq.~(\ref{ainin}).
Note that the sum over the spectators $b$ refers to all initial-state partons
as well as to all
identified partons in the final state and no distinction is made between these
two cases. The explicit expression for these dipole terms is:
\beeq
\label{dipainin}
&&{\cal D}^{(n) \,ai}_b(p_1,...,p_{m+1},... ;p_a,..) =
- \frac{1}{2 p_a \cdot p_i} \;\frac{1}{x_{ain}} \nonumber \\
&&{}_{m ...,a...}\!\!\bra{{\widetilde 1},.., {\widetilde {m+1}}, ...;
{\widetilde {ai}}...}
\,\frac{{\bom T}_b \cdot {\bom T}_{ai}}{{\bom T}_{ai}^2} \;
{\bom V}^{(n)\, ai}_b \,
\ket{{\widetilde 1},..,{\widetilde {m+1}}, ..;{\widetilde {ai}}..}_{m..,a..}
\;\;.
\eeeq

The assignment of momenta in the matrix element
$\ket{{\widetilde 1},..,{\widetilde {m+1}}, ..;{\widetilde {ai}}..}_{m..,a..}$
is the following.
The momentum of the emitter is parallel to $p_a$:
\beeq
\label{wpai}
{\widetilde p}_{ai}^\mu &=& x_{ain} \,p_a^\mu  \;, \\
\label{xaindef}
x_{ain} &=& \frac{(p_a-p_i)\cdot n}{p_a\cdot n} \;,
\eeeq
and all other initial-state and identified final-state momenta are left
{\em unchanged}.
{\em All\/} other final-state momenta $k_j$ (QCD partons or not)
are transformed according to the following Lorentz transformation
\beeq
\label{klorn}
{\widetilde k_j}^{\mu} &=& \Lambda^{\mu}_{\;\; \nu}(K,{\widetilde K})
\;k_j^{\nu} \;\;, \\
\Lambda^{\mu}_{\;\; \nu}(K,{\widetilde K}) &=& g^{\mu}_{\;\; \nu} -
\frac{2 (K+{\widetilde K})^\mu (K+{\widetilde K})_\nu}{(K+{\widetilde K})^2}
+\frac{2 {\widetilde K}^\mu K_\nu}{K^2}\;\;,
\eeeq
where the momenta $K^{\mu}$ and ${\widetilde K}^\mu$ are defined by
\beq
\label{qdef}
\begin{array}{rcl}
K^{\mu} &=& n^\mu - p_i^\mu \;\;, \\
{\widetilde K}^{\mu} &=& n^\mu - (1-x_{ain}) p_a^\mu  \;\;.
\end{array}
\eeq
Note that Eq.~(\ref{klorn}) actually defines a proper Lorentz
transformation.

The splitting functions
${\bom V}^{(n) \,ai}_b$ in Eq.~(\ref{ainin})
are as follows
\beq\label{Vqagin}
\bra{s}
{\bom V}^{(n) \,q_ag_i}_b(x_{ain};v_{i,ab}) \ket{s'}
= 8\pi \mu^{2\ep} \as\; C_F\;\delta_{ss'} \left[
2 v_{i,ab}
- (1 + x_{ain}) -\ep (1-x_{ain}) \right]
\;,
\eeq
\beeq
\bra{s}
{\bom V}^{(n)\,g_a{\bar q}_i}_b(x_{ain}) \ket{s'}
= 8\pi \mu^{2\ep} \as\; T_R\;
\left[ 1 - \ep -2 x_{ain}(1-x_{ain}) \right]
\;\delta_{ss'} \;,
\eeeq
\beeq
&\bra{\mu}&\!\!\!
{\bom V}^{(n)\,q_aq_i}_b(x_{ain})
\ket{\nu}
= 8\pi \mu^{2\epsilon} \as \; C_F \;
\left[ -g^{\mu \nu}x_{ain} \frac{}{}
+ \frac{1-x_{ain}}{x_{ain}}
\right. \nonumber \\
&\cdot&\!\!\! \left.
\frac{4p_i \cdot p_a}{2 (p_a \cdot n) (p_i \cdot n) -  n^2 \;p_i \cdot p_a } \,
\left( \frac{n p_a}{p_i p_a} p_i^{\mu} - n^{\mu} \right)
\left( \frac{n p_a}{p_i p_a} p_i^{\nu} - n^{\nu} \right)
\,\right] \,,
\eeeq
\beeq\label{Vgagin}
&\bra{\mu}&\!\!\!\!{\bom V}^{(n)\,g_ag_i}_b(x_{ain};v_{i,ab})
\ket{\nu} = 16\pi
\mu^{2\epsilon} \as \;C_A\;
\left[\; -g^{\mu \nu} \left( \frac{}{} v_{i,ab} -1 + x_{ain}(1-x_{ain}) \right)
\right. \nonumber \\
&+&\!\!\!\!\!\left.
(1-\ep) \;\frac{1-x_{ain}}{x_{ain}} \;
\frac{2p_i \cdot p_a}{2 (p_a \cdot n) (p_i \cdot n) -  n^2 \;p_i \cdot p_a } \,
\left( \frac{n p_a}{p_i p_a} p_i^{\mu} - n^{\mu} \right)
\left( \frac{n p_a}{p_i p_a} p_i^{\nu} - n^{\nu} \right)
\,\right] \;,
\nonumber \\
&&\,
\eeeq
where we have defined
\beq
\label{viabdef}
v_{i,ab} = \frac{p_ap_b}{p_i (p_a + p_b)} \;\;\;.
\eeq

Since all the momenta ${\widetilde k}_j$ are obtained by means
of the Lorentz transformation (\ref{klorn}), the `pseudo-dipole'
phase space has a trivial convolution structure
(note that in the physical region of interest, $(\sum_j k_j)^2 - Q^2 \geq 0$),
namely
\beq
\label{psconvn}
d\phi(p_i, k_1,...;p_a + Q ) =  \int_0^1 dx
\; d\phi({\widetilde k}_1,...;x p_a+Q)
\;\left[ dp_i(n=Q+p_a,
p_a, x) \right] \;,
\eeq
where
\beq
\label{dpinx}
\left[ dp_i(n,p_a, x) \right] =
\frac{d^{d}p_i}{(2\pi)^{d-1}} \,\delta_+(p_i^2) \;
\Theta(x) \Theta(1-x) \;
\delta(x - x_{ain}) \;\;.
\eeq
The phase space in Eq.~(\ref{dpinx}) can explicitly be written
in terms of the kinematic variables $x_{ain}$ in Eq.~(\ref{xaindef})
and ${\tilde v}_i=p_ap_i/p_an$ :
\beeq
\label{exdpinx}
\left[ dp_i(n,p_a, x) \right]
&=& \frac{(2p_a n)^{1-\ep}}{16\pi^2}
\;\frac{d\Omega^{(d-3)}}{(2\pi)^{1-2\ep}}
\;d{\tilde v}_i \;dx_{ain}
\;\Theta(x(1-x))
\;\Theta({\tilde v}_i)
\;\Theta\!\left(1-\frac{n^2{\tilde v}_i}{2(1-x)p_an} \right)
\nonumber \\
&\cdot&
\;\left( 1-x\right)^{-2\ep} \;\delta(x-x_{ain})
\;\left[\frac{{\tilde v}_i}{1-x}
\left(1-\frac{n^2{\tilde v}_i}{2(1-x)p_an} \right) \right]^{-\ep} \;,
\eeeq
where $d\Omega^{(d-3)}$ is an element of solid angle perpendicular to the
light-like momenta
${\bar n}^{\mu} = n^{\mu} - n^2 p_a^{\mu}/(2p_an)$
and $p_a^{\mu}$.

Performing the integration of the splitting kernels ${\bom V}^{(n)\, ai}_b$
over the phase space in Eq.~(\ref{exdpinx}), we introduce the functions
${\widetilde {\cal V}}^{a,ai}(x;\ep;p_a,p_b,n)$ :
\beeq
\label{intVainin}
\frac{\as}{2\pi} \frac{1}{\Gamma(1-\ep)} \left(
\frac{4\pi \mu^2}{2p_ap_b} \right)^{\ep}
{\widetilde {\cal V}}^{a,ai}(x;\ep;p_a,p_b,n)
&\!\equiv\!& \int \left[ dp_i(n,p_a, x) \right]
\frac{1}{2p_ap_i} \frac{n_s({\widetilde {ai}})}{n_s(a)} \nonumber \\
&\!\cdot\!& <{\bom V}^{(n)\,ai}_b(x_{ain};v_{i,ab})> \;.
\eeeq

The spin averages of the kernels ${\bom V}^{(n)\, ai}_b$ are related
in a simple way to the corresponding averages of the ($d$-dimensional)
Altarelli-Parisi splitting functions, that is,
\beq
\label{calvtavn}
\frac{n_s({\widetilde {ai}})}{n_s(a)}
\frac{<{\bom V}^{(n) \,ai}_b(x; v_{i,ab})>}{8 {\pi} \as \mu^{2\ep}}
= \;<{\hat P}_{a,{\widetilde {ai}}}(x;\ep)> + 2 \;\delta^{a,{\widetilde {ai}}}
\;{\bom T}_a^2 \left[ v_{i,ab} - \frac{1}{1-x} \right] \;\;.
\eeq
Therefore, the only non-trivial integration in Eq.~(\ref{intVainin}) is
that which involves the term $v_{i,ab}$ (see Appendix~B) and leads
to the following result
\beeq
\label{wtcalvn}
{\widetilde {\cal V}}^{a,b}(x;\ep;p_a,p_b,n) =
{\widetilde {\cal V}}^{a,b}(x;\ep) + {\cal L}^{a,b}(x;p_a,p_b,n) +
{\cal O}(\ep) \;\;.
\eeeq
The first term on the right-hand side is given in Eq.~(\ref{wtcalv})
and the second term is defined by
\beeq
\label{calldef}
\!\!\!{\cal L}^{a,b}(x;p_a,p_b,n) &=& \delta^{ab} \;
\delta(1-x) \,2 \,{\bom T}_a^2 \left[ \,
{\rm Li}_2\!\left(1- \frac{(1+v)}{2} \frac{(p_a+p_b)\cdot n}{p_a\cdot n}
\right) \right.
\nonumber \\
&+& \left. {\rm Li}_2\!\left(1- \frac{(1-v)}{2}
\frac{(p_a+p_b)\cdot n}{p_a \cdot n} \right)
\right]
- P^{ab}_{{\rm reg}}(x) \ln \frac{n^2(p_a \cdot p_b)}{2 (p_a \cdot n)^2} \;\;,
\eeeq
\beq
\label{vmpdef}
v= \sqrt { 1 - \frac{ n^2 (p_a+p_b)^2 }{ [ (p_a+p_b)\cdot n ]^2}} \;\;,
\eeq
$P^{ab}_{{\rm reg}}(x)$ being the regular
part of the Altarelli-Parisi probabilities in Eq.~(\ref{preg})
and ${\rm Li}_2(x)$ is the dilogarithm function:
\beq
\label{dilogdef}
{\rm Li}_2(x) = - \int_0^x \frac{dz}{z} \;\ln (1-z) \;\;.
\eeq

Note that unlike the $\cal V$-functions considered in the previous
Subsections, Eq.~(\ref{wtcalvn}) depends not only on the momentum fraction
$x,$ but also on the momenta $p_a,p_b,n$. This momentum dependence is entirely
accounted for by the ${\cal L}^{a,b}$ function on the right-hand side of
Eq.~(\ref{wtcalvn}).  In the case with no final-state identified partons,
$p_b$ is the momentum of an incoming parton, $n=p_a+p_b,$ so ${\cal L}^{a,b}$
vanishes, thus recovering the results already discussed in Sect.~\ref{dissa}.

In order to deal with the singularities for $p_i \cdot p_a \to 0$, when $a$
is an identified parton in the final state, we introduce the following
factorization formula
\beeq
\label{ainf}
&&\!\!\!\!\!\!\!\!\!\!{}_{m+1 a..,...}\!\!\bra{1, ...., m+1, a...;...}
\ket{1, ...., m+1, a..;..}_{m+1 a..,...} \nonumber \\
&&\!\!\!\!\!\!\!\!\!\!=  \sum_{k \neq i}
{\cal D}_{ai,k}(p_1,...,p_{m+1},p_a,..;...) +
\sum_{b \neq a}
{\cal D}^{(n)}_{ai,b}(p_1,...,p_{m+1},p_a,..;...) + \dots \;\;,
\eeeq
where ${\cal D}_{ai,k}$ is the dipole in Eq.~(\ref{dipaff}) and the new dipole
contribution is given by
\beeq
\label{dipainf}
&&{\cal D}^{(n)}_{ai,b}(p_1,...,p_{m+1},p_a,..;...) =
- \frac{1}{2 p_a \cdot p_i}  \nonumber \\
&&\cdot \;{}_{m a..,...}\!\!\bra{{\widetilde 1},.., {\widetilde {m+1}},
{\widetilde {ai}}..;...}
\,\frac{{\bom T}_b \cdot {\bom T}_{ai}}{{\bom T}_{ai}^2} \;
{\bom V}^{(n)}_{ai,b} \,
\ket{{\widetilde 1},..,{\widetilde {m+1}}, {\widetilde {ai}}..;..}_{m a..,...}
\;.
\eeeq
As in Eq.~(\ref{ainin}), the sum over the spectators $b$ in
Eq.~(\ref{ainf}) refers to all initial-state partons as well as to all
identified partons in the final state with no distinction between these
two cases.

The assignment of momenta in the matrix element
$\ket{{\widetilde 1},..,{\widetilde {m+1}}, {\widetilde {ai}}..;..}_{m a..,...}$
is the following.
The momentum of the emitter is parallel to $p_a$:
\beeq
\label{wpaif}
{\widetilde p}_{ai}^\mu &=& \frac{1}{z_{ain}} \,p_a^\mu \;, \\
\label{zaindef}
z_{ain} &=& \frac{p_a\cdot n}{(p_a+p_i)\cdot n} \;,
\eeeq
and all other initial-state and identified final-state momenta are left
unchanged.
{\em All\/} other final-state momenta $k_j$ (QCD partons or not)
are transformed according to the Lorentz transformation in
Eqs.~(\ref{klorn}--\ref{qdef}).

The splitting functions ${\bom V}^{(n)}_{ai,b}$ in Eq.~(\ref{ainf}) are given
in terms of the variable $v_{i,ab}$ in Eq.~(\ref{viabdef}) as follows
\beq\label{Vqaginb}
\bra{s}
{\bom V}^{(n)}_{q_ag_i,b}(z_{ain};v_{i,ab}) \ket{s'}
= 8\pi \mu^{2\ep} \as\; C_F\;\delta_{ss'} \left[
2 \,\frac{v_{i,ab}}{z_{ain}}
- (1 + z_{ain}) -\ep \;(1-z_{ain}) \right]
\;,
\eeq
\beeq
\bra{s}
{\bom V}^{(n)}_{g_a{\bar q}_i,b}(z_{ain}) \ket{s'}
= 8\pi \mu^{2\ep} \as\; C_F\;
\left[ \frac{1 + (1-z_{ain})^2}{z_{ain}} -\ep \,z_{ain} \right]
\;\delta_{ss'} \;,
\eeeq
\beeq
&\bra{\mu}&\!\!\!
{\bom V}^{(n)}_{q_a{\bar q}_i,b}(z_{ain})
\ket{\nu}
= 8\pi \mu^{2\epsilon} \as \; T_R \;
\left[ -g^{\mu \nu}
- 4 z_{ain}(1-z_{ain}) \frac{}{}
\right. \nonumber \\
&\cdot&\!\!\!\!\! \left.
\frac{p_i \cdot p_a}{2 (p_a \cdot n) (p_i \cdot n) -  n^2 \;p_i \cdot p_a } \,
\left( \frac{n p_a}{p_i p_a} p_i^{\mu} - n^{\mu} \right)
\left( \frac{n p_a}{p_i p_a} p_i^{\nu} - n^{\nu} \right)
\,\right] \,,
\eeeq
\beeq\label{Vgaginb}
&\bra{\mu}&\!\!\!\!{\bom V}^{(n)}_{g_ag_i,b}(z_{ain};v_{i,ab})
\ket{\nu} = 16\pi
\mu^{2\epsilon} \as \;C_A\;
\left[\; -g^{\mu \nu} \left( \frac{v_{i,ab}}{z_{ain}} - 1 +
\frac{1-z_{ain}}{z_{ain}}
\right) \right. \nonumber \\
&+&\!\!\!\!\! \left. \frac{}{} 2(1-\ep) z_{ain}(1-z_{ain})
\frac{p_i \cdot p_a}{2 (p_a \cdot n) (p_i \cdot n) -  n^2 \;p_i \cdot p_a } \,
\left( \frac{n p_a}{p_i p_a} p_i^{\mu} - n^{\mu} \right)
\left( \frac{n p_a}{p_i p_a} p_i^{\nu} - n^{\nu} \right)
\,\right] \;.
\nonumber \\
&&\,
\eeeq

The phase space for the dipole (\ref{dipainf}) has the following convolution
form
\beq
\label{psconvnz}
d\phi(p_i, k_1,...; Q-p_a ) =  \int_0^1 \frac{dz}{z^{2-2\ep}}
\; d\phi({\widetilde k}_1,...; Q- p_a/z)
\;\left[ dp_i(n=Q-p_a;
p_a, z) \right] \;,
\eeq
where
\beq
\label{dpinz}
\left[ dp_i(n;p_a, z) \right] =
\frac{d^{d}p_i}{(2\pi)^{d-1}} \,\delta_+(p_i^2) \;
\Theta(z) \Theta(1-z) \;
\delta(z - z_{ain}) \;z^{2-2\ep} \;\;.
\eeq
Introducing the kinematic variables $z_{ain}$ in Eq.~(\ref{zaindef}) and
${\tilde v}_i=p_ap_i/p_an$, Eq.~(\ref{dpinz}) can be written
as follows
\beeq
\label{exdpinz}
\left[ dp_i(n;p_a, z) \right]
&=& \frac{(2p_a n)^{1-\ep}}{16\pi^2}
\;\frac{d\Omega^{(d-3)}}{(2\pi)^{1-2\ep}}
\;d{\tilde v}_i \;dz_{ain}
\;\Theta(z(1-z))
\;\Theta({\tilde v}_i)
\;\Theta\!\left(1-\frac{n^2{\tilde v}_iz}{2(1-z)p_an} \right)
\nonumber \\
&\cdot&
\;\left( 1-z\right)^{-2\ep} \;\delta(z-z_{ain})
\;\left[\frac{{\tilde v}_iz}{1-z}
\left(1-\frac{n^2{\tilde v}_iz}{2(1-z)p_an} \right) \right]^{-\ep} \;,
\eeeq
where $d\Omega^{(d-3)}$ is an element of solid angle perpendicular to the
light-like momenta ${\bar n}^{\mu} = n^{\mu} - n^2 p_a^{\mu}/(2p_an)$
and $p_a^{\mu}$.

Thus, the integration of the splitting kernels ${\bom V}^{(n)}_{ai,b}$ defines
the functions
${\overline {\cal V}}_{ai,a}(z;\ep;p_a,p_b,n)$:
\beq
\label{intVainf}
\frac{\as}{2\pi} \frac{1}{\Gamma(1-\ep)} \left(
\frac{4\pi \mu^2}{2p_ap_b} \right)^{\ep}
{\overline {\cal V}}_{ai,a}(z;\ep;p_a,p_b,n)
\equiv \int \left[ dp_i(n;p_a, z) \right]
\frac{1}{2p_ap_i} \;
 <{\bom V}^{(n)}_{ai,b}(z_{ain};v_{i,ab})> \;.
\eeq
As in the case of initial-state singularities, the spin averages
$ <{\bom V}^{(n)}_{ai,b}>$ are simply related to the ($d$-dimensional)
Altarelli-Parisi splitting functions:
\beq
\label{calvtavnf}
\frac{<{\bom V}^{(n)}_{ai,b}(z; v_{i,ab})>}{8 {\pi} \as \mu^{2\ep}}
= \;<{\hat P}_{{\widetilde {ai}},a}(z;\ep)> + 2 \;\delta^{{\widetilde {ai}},a}
\;{\bom T}_a^2 \left[ \frac{v_{i,ab}}{z} - \frac{1}{1-z} \right] \;\;,
\eeq
and, explicitly performing the integration of the $v_{i,ab}$ contributions in
Eq.~(\ref{calvtavnf}) (see Appendix~B), we find that the functions
${\overline {\cal V}}_{a,b}$ are equal to the analogous functions
${\widetilde {\cal V}}^{a,b}$ in Eq.~(\ref{wtcalvn}):
\beq
{\overline {\cal V}}_{a,b}(z;\ep;p_a,p_b,n) =
{\widetilde {\cal V}}^{a,b}(z;\ep;p_a,p_b,n) \;\;.
\eeq

\newpage

\setcounter{equation}{0}

\section{QCD cross sections at NLO}
\label{NLOxs}

In the following Sections we describe in detail our subtraction method for
evaluating QCD cross sections. To this end, it is useful to recall the general
and precise definitions of the NLO cross sections.

In the case of processes with no initial-state hadrons,
for instance in \ee\ annihilation, the partonic cross section
is\footnote{We are using the same notation as in Sect.~\ref{gen}. Thus, $m$
is the number of unobserved final-state partons for the leading-order process.}
\beq
\label{eexs}
\sigma = \sigma^{LO} + \sigma^{NLO} \;,
\eeq
\beq
\label{eeBRV}
\sigma^{LO} = \int_m d\sigma^{B} \;,
\;\;\;\sigma^{NLO} = \int_{m+1} d\sigma^{R} + \int_m d\sigma^{V} \;,
\eeq
where $d\sigma^{B}, d\sigma^{R}, d\sigma^{V}$ are the
cross sections in the Born approximation and to one-loop order ($R$: real
emission; $V$: virtual correction). If $\sigma$ is a jet cross section (no
final-state hadrons observed), hadron-level and parton-level cross sections
coincide.

In the case of processes with one initial-state hadron carrying momentum
$p^\mu$ (for instance, deep inelastic lepton-hadron scattering), the
calculation of the QCD cross section requires the introduction of parton
distributions. If we denote by $f_a(\eta, \mu_F^2)$ the density of partons
of type $a$ in the incoming hadron, the hadronic cross section is given by
\beq
\label{1hxs}
\sigma(p) = \sum_a \int_0^1 d\eta \, f_a(\eta, \mu_F^2)
\; \left[ \sigma_a^{LO}(\eta p) + \sigma_a^{NLO}(\eta p;\mu_F^2) \right] \;,
\eeq
and the corresponding parton-level cross sections are:
\beeq
\label{aLO}
\sigma_a^{LO}(p) &=& \int_m d\sigma_a^{B}(p) \;, \\
\label{aNLO}
\sigma_a^{NLO}(p;\mu_F^2) &=& \int_{m+1} d\sigma_a^{R}(p) +
\int_m d\sigma_a^{V}(p) +
\int_m d\sigma_a^{C}(p;\mu_F^2) \;.
\eeeq
The notation $B$, $R$, $V$ is as in Eq.~(\ref{eeBRV}). The contribution
$d\sigma_a^{C}$ represents the collinear-subtraction counterterm and is
explicitly given by the following expression
\beq
\label{saC}
d\sigma_a^{C}(p;\mu_F^2) = - \frac{\as}{2 \pi} \;\frac{1}{\Gamma(1-\ep)} \sum_b
\int_0^1 dz \,
\left[ - \frac{1}{\ep}
\left( \frac{4 \pi \mu^2}{\mu_F^2} \right)^{\ep} P^{ab}(z) + \Kab(z) \right]
\,d\sigma_b^{B}(zp) \;.
\eeq

The partonic contributions on the right-hand side of Eq.~(\ref{aNLO}) are
separately
divergent for $\ep \to 0$. Their sum  $\sigma_a^{NLO}$ is finite for
$\ep \to 0$
but depends on the {\em factorization scale\/} and on the {\em factorization
scheme\/} of collinear singularities. Both dependences are
contained in the definition of $d\sigma_a^{C}$: $\mu_F$ is the factorization
scale and the actual form of the kernel $\Kab(z)$ specifies the
factorization scheme. Setting $\Kab(z) = 0$
defines the $\overline {\rm MS}$ subtraction scheme.
The functions
$P^{ab}(z)$ in Eq.~(\ref{saC}) are the {\em four dimensional\/}
Altarelli Parisi probabilities in Eqs.~(\ref{pqg4}--\ref{pgg4}).
The parton densities
$f_a(\eta, \mu_F^2)$ are also scale/scheme dependent, so that this dependence
cancels in the hadronic cross section of Eq.~(\ref{1hxs}).

Note that
\beq
\sum_b \int_0^1 dx \;x \;P^{ab}(x) = 1 \;\;,
\eeq
and that, in the $\overline {\rm MS}$ scheme, momentum conservation reads as
follows
\beq
\label{mom}
\sum_{a} \int_0^1 dx \;x \;f_a(x, \mu_F^2) = 1 \;\;.
\eeq
In other factorization schemes the generalization of Eq.~(\ref{mom}) is:
\beq
\sum_{a,b} \int_0^1 dx \;x \;f_b(x, \mu_F^2) \left[ \delta^{ba} -
\frac{\as}{2\pi}
\int_0^1 dz \;z \;\KFS{ba}(z) \right] = 1 \;\;.
\eeq

The extension of Eq.~(\ref{1hxs}) to processes with two initial-state
hadrons is straightforward. Denoting by $f_a$ and
${\bar f}_b$ the parton densities of the two incoming hadrons, we have
\beq
\label{2hxs}
\sigma(p,{\bar p}) = \sum_{a,b} \int_0^1 d\eta \, f_a(\eta, \mu_F^2)
\;\int_0^1 d{\bar \eta} \, {\bar f}_b({\bar \eta}, \mu_F^2)
\; \left[ \sigma_{ab}^{LO}(\eta p,{\bar \eta}{\bar p})
+ \sigma_{ab}^{NLO}(\eta p,{\bar \eta}{\bar p};\mu_F^2) \right] \;,
\eeq
\beeq
\label{abLO}
\sigma_{ab}^{LO}(p,{\bar p}) &=& \int_m d\sigma_{ab}^{B}(p,{\bar p}) \;, \\
\label{abNLO}
\sigma_{ab}^{NLO}(p,{\bar p};\mu_F^2) &=& \int_{m+1}
d\sigma_{ab}^{R}(p,{\bar p}) + \int_m d\sigma_{ab}^{V}(p,{\bar p}) +
\int_m d\sigma_{ab}^{C}(p,{\bar p};\mu_F^2) \;,
\eeeq
where the collinear counterterm is:
\beeq
\label{sabC}
d\sigma_{ab}^{C}(p,{\bar p};\mu_F^2) &= &
- \frac{\as}{2 \pi} \;\frac{1}{\Gamma(1-\ep)} \sum_{cd}
\int_0^1 dz \, \int_0^1 d{\bar z} \;\;d\sigma_{cd}^{B}(zp,{\bar z}{\bar p})
\nonumber \\
&\cdot& \left\{ \delta_{bd} \delta(1-{\bar z})
\left[ - \frac{1}{\ep}
\left( \frac{4 \pi \mu^2}{\mu_F^2} \right)^{\ep} P^{ac}(z) + \Kac(z) \right]
\right. \nonumber \\
&+&  \left. \delta_{ac} \delta(1- z)
\left[ - \frac{1}{\ep}
\left( \frac{4 \pi \mu^2}{\mu_F^2} \right)^{\ep} P^{bd}({\bar z}) +
\KFS{bd}({\bar z}) \right] \right\} \;.
\eeeq
It is completely trivial to generalize the resulting formulae to the case
in which one introduces different factorization scales for the two hadrons,
as one might in photoproduction for example. The replacement
${\bar f}_b({\bar \eta}, \mu_F^2) \to {\bar f}_b({\bar \eta}, {\bar \mu}_F^2)$
in Eq.~(\ref{2hxs}) is simply accompanied by $\mu_F \to {\bar \mu}_F$ in the
second term in the curly bracket on the right-hand side of Eq.~(\ref{sabC}).

Let us now consider fragmentation processes.
The one-hadron inclusive cross section $\sigma_{(incl)}(p)$ in the case with
no initial-state hadrons is:
\beq
\label{1fxs}
\sigma_{(incl)}(p) = \sum_a \int_0^1 \frac{d\eta}{\eta^2} \, d^a(\eta, \mu_F^2)
\; \left[ \sigma_{(incl) \,a}^{LO}(p/\eta ) +
\sigma_{(incl) \,a}^{NLO}(p/\eta;\mu_F^2 ) \right] \;,
\eeq
where $d^a(\eta, \mu_F^2)$ is the fragmentation function of the parton $a$ into
the observed hadron and the partonic cross sections are:
\beeq
\label{afLO}
\sigma_{(incl) \,a}^{LO}(p) &=& \int_m d\sigma_{(incl) \,a}^{B}(p) \;, \\
\label{afNLO}
\sigma_{(incl) \,a}^{NLO}(p) &=& \int_{m+1} d\sigma_{(incl) \,a}^{R}(p) +
\int_m d\sigma_{(incl) \,a}^{V}(p) +
\int_m d\sigma_{(incl) \,a}^{C}(p;\mu_F^2) \;,
\eeeq
\beq
\label{saCinc}
d\sigma_{(incl) \,a}^{C}(p;\mu_F^2)
= - \frac{\as}{2 \pi} \;\frac{1}{\Gamma(1-\ep)}
\sum_b
\int_0^1 \frac{dz}{z^{2-2\ep}} \,
\left[ - \frac{1}{\ep}
\left( \frac{4 \pi \mu^2}{\mu_F^2} \right)^{\ep} P_{ba}(z) + \Hba(z) \right]
\; d\sigma_{(incl) \,b}^{B}(p/z) \;,
\eeq
where the Altarelli-Parisi probability $P_{ab}(z)$ describes the time-like
splitting process $a(p) \to b(zp)$. Note that $P_{ab}(z) = P^{ab}(z)$. The
kernel $\Hba(z)$ in Eq.~(\ref{saCinc}) defines the factorization scheme
($\Hba(z) = 0$ in the $\overline {\rm MS}$ subtraction scheme).

Note that the one-particle inclusive cross section $\sigma_{(incl)}(p)$ in
Eq.~(\ref{1fxs}) is defined without integrating over any component of the
momentum $p^\mu$ of the observed hadron (the unusual convolution measures
$d\eta/\eta^2$ and $dz/z^{2-2\ep}$ in Eqs.~(\ref{1fxs}) and (\ref{saCinc})
follow from that).
Thus the following integral
\beq
\int \frac{d^4p}{(2\pi)^3} \;\delta_+(p^2) \;\sigma_{(incl)}(p)
\eeq
is equal to the associated multiplicity times the total cross section. The
corresponding associated multiplicities at partonic level (i.e.\ in $d$
dimensions) are $(I=B,R,V,C,)$:
\beq
\int \frac{d^dp}{(2\pi)^{d-1}} \;\delta_+(p^2) \;d\sigma_{(incl) \,a}^I(p)
\eeq

In the most general case, one should consider multi-particle correlations,
that is, one deals with processes of the type
\beq
\label{mpkin}
p + {\bar p} \to q_1 + \dots q_n + X \;\;,
\eeq
where $p, {\bar p}$ are the momenta of two incoming hadrons, $q_1, \dots q_n$
are the momenta of $n$ hadrons detected in the final state and $X$ stands
for unobserved final-state particles or jets. Note that, by definition, the
momenta $q_1,...,q_n$ are supposed not to be parallel to each other or to
the incoming momenta $p$ and ${\bar p}$.
The hadronic cross section is:
\beeq
\label{1mxs}
&&\sigma_{(incl)}(p,{\bar p};q_1,..,q_n) = \sum_{a,b} \int_0^1 d\eta
\;f_a(\eta, \mu_F^2)  \int_0^1 d{{\bar \eta}}
\;{\bar f}_b({\bar \eta}, \mu_F^2)
\nonumber \\
&&\cdot
\sum_{a_1,...,a_n}
\int_0^1 \frac{d\eta_1}{\eta_1^2}  \;.... \int_0^1 \frac{d\eta_n}{\eta_n^2}
\; d^{a_1}(\eta_1, \mu_1^2) \;.... \,d^{a_n}(\eta_n, \mu_n^2) \nonumber \\
&&\cdot
\; \left[ \sigma_{ab,(incl) a_1,..,a_n}^{LO}(\eta p,{\bar \eta}{\bar p};
q_1/\eta _1,...,q_n/\eta _n) \right. \nonumber \\
&&+ \left.
\sigma_{ab,(incl) a_1,..,a_n}^{NLO}(\eta p,{\bar \eta}{\bar p};
q_1/\eta _1,...,q_n/\eta _n;\mu_F^2, \mu_1^2,..,\mu_n^2) \right] \;,
\eeeq
where we have introduced a single factorization scale $\mu_F^2$ for the parton
densities and $n$ different factorization scales $\mu_1^2, ...,\mu_n^2$
for the fragmentation functions.

The corresponding cross sections at parton level are the following
\beq
\sigma_{ab,(incl) a_1,..,a_n}^{LO}(p,{\bar p};q_1,...,q_n) =
\int_m d\sigma_{ab,(incl) a_1,..,a_n}^{B}(p,{\bar p};q_1,...,q_n) \;\;,
\eeq
\beeq
\label{mpNLO}
\!\!\!&&\sigma_{ab,(incl) a_1,..,a_n}^{NLO}(p,{\bar p};
q_1,...,q_n;\mu_F^2, \mu_1^2,..,\mu_n^2) =
\int_{m+1} d\sigma_{ab,(incl) a_1,..,a_n}^{R}(p,{\bar p};q_1,...,q_n)
 \\
\!\!\!&&+ \int_m d\sigma_{ab,(incl) a_1,..,a_n}^{V}(p,{\bar p};q_1,...,q_n)
+ \int_m d\sigma_{ab,(incl)
a_1,..,a_n}^{C}(p,{\bar p};q_1,...,q_n;\mu_F^2, \mu_1^2,..,\mu_n^2) \;\;,
\nonumber
\eeeq
where the collinear counterterm $d\sigma^C$ is given by
\beeq
\!\!&&d\sigma_{ab,(incl)\,a_1,..,a_n}^{C}(p,{\bar p};q_1,...,q_n;\mu_F^2,
\mu_1^2,..,\mu_n^2) =  - \frac{\as}{2 \pi} \;\frac{1}{\Gamma(1-\ep)}
\nonumber \\
\!\!&&\cdot \left\{ \sum_{a'} \int_0^1 dz
\left[ - \frac{1}{\ep}
\left( \frac{4 \pi \mu^2}{\mu_F^2} \right)^{\ep} P^{aa'}(z) + \KFS{aa'}(z)
\right] d\sigma_{a'b,(incl) \,a_1,..,a_n}^{B}(zp,{\bar p};q_1,...,q_n)
\right. \nonumber \\
\!\!&&+ \left. \sum_{b'} \int_0^1 dz
\left[ - \frac{1}{\ep}
\left( \frac{4 \pi \mu^2}{\mu_F^2} \right)^{\ep} P^{bb'}(z)
+ \KFS{bb'}(z)
\right] d\sigma_{ab',(incl) \,a_1,..,a_n}^{B}(p,z{\bar p};q_1,...,q_n)
\right. \nonumber \\
\!\!&&+ \left. \sum_{i=1}^n \sum_{a_i^\prime}
\int_0^1 \frac{dz}{z^{2-2\ep}} \,
d\sigma_{ab,(incl)
\,a_1,..,a_i^\prime,..,a_n}^{B}(p,{\bar p};q_1,..,q_i/z,..,q_n) \right.
\nonumber \\
\!\!&&\cdot \left. \left[ - \frac{1}{\ep}
\left( \frac{4 \pi \mu^2}{\mu_i^2} \right)^{\ep} P_{a_i^\prime a_i}(z)
+ \HFS{a_i^\prime a_i}(z)
\right] \right\} \;\;.
\eeeq

Note that, computing the hadronic cross sections in
Eqs.~(\ref{1hxs},\ref{2hxs},\ref{1fxs},\ref{1mxs}),
the factorization-scale evolution of the parton distribution functions has
to be consistently carried out at NLO\@. For the parton densities
$f_a(\eta, \mu_F^2)$ and the fragmentation functions $d^a(\eta, \mu_F^2)$
we have
\beeq
\frac{d \,f_a(\eta, \mu_F^2)}{d\ln \mu_F^2} = \sum_b \int_\eta^1 \frac{dz}{z}
\;f_b(\eta/z, \mu_F^2) \; \frac{\as(\mu_F^2)}{2\pi} \left[ P^{ba}(z) +
\frac{\as(\mu_F^2)}{2\pi} \;P^{(1) \,ba}(z) \right] \;\;,
\eeeq
\beeq
\frac{d \,d^a(\eta, \mu_F^2)}{d\ln \mu_F^2} = \sum_b \int_\eta^1 \frac{dz}{z}
\; \frac{\as(\mu_F^2)}{2\pi} \left[ P_{ab}(z) +
\frac{\as(\mu_F^2)}{2\pi} \;P^{(1)}_{ab}(z) \right] \;d^b(\eta/z, \mu_F^2)\;\;,
\eeeq
where $P^{(1) \,ab}(x)$ and $P^{(1)}_{ab}(x)$ respectively are the
space-like and time-like NLO Altarelli-Parisi probabilities. They depend on
the factorization scheme and this dependence is given in terms of the flavour
kernels $\Kab$ and $\Hba$ as follows
\beq
P^{(1) \,ab}(x) = P^{(1) \,ab}_{{\overline {\scriptscriptstyle \rm MS}}}(x)
+ \sum_c \int_x^1 \frac{dz}{z} \left[ P^{ac}(z) \,\Kcb(x/z)
- \Kac(x/z) \,P^{cb}(z) \right] - 2 \pi \beta_0 \,\Kab(x) \;\;,
\eeq
\beq
P^{(1)}_{ab}(x) = P^{(1) \;{\overline {\scriptscriptstyle \rm MS}}}_{ab}(x)
+ \sum_c \int_x^1 \frac{dz}{z} \left[ \Hac(x/z) \;P_{cb}(z)
- P_{ac}(z) \;\Hcb(x/z) \right] - 2 \pi \beta_0 \,\Hab(x) \;\;,
\eeq
where $P^{(1) \,ab}_{{\overline {\scriptscriptstyle \rm MS}}}(x)$ and
$P^{(1) \,{\overline {\scriptscriptstyle \rm MS}}}_{ab}(x)$ are
the corresponding probabilities
evaluated in the $\overline {\rm MS}$ subtraction scheme [\ref{CFP}].

\newpage

\setcounter{equation}{0}

\section{Jet cross sections with no initial-state hadrons}
\label{siee}

In processes with no initial-state hadrons, the QCD cross section for jet
observables is
given by Eqs.~(\ref{eexs},\ref{eeBRV}).
In terms of the QCD matrix elements, the
Born-level cross section in $d$ dimensions is the following
\beq
\label{dsmb}
d\sigma^{B} = {\cal N}_{in} \sum_{\{ m \} } \, d\phi_m(p_1, ...,p_m;Q)
\;\frac{1}{S_{\{ m \} }} \;|{\cal{M}}_{m}(p_1, ...,p_m) |^2
\;F_J^{(m)}(p_1, ...,p_m) \;\;,
\eeq
where ${\cal N}_{in}$ includes all the factors that are QCD independent,
$\sum_{\{ m \} }$ denotes the sum over all the configurations with $m$
partons, $d\phi_m$ is the partonic phase space in Eq.~(\ref{psm}),
$S_{\{ m \} }$ is the Bose symmetry factor for identical partons in the final
state and
${\cal{M}}_{m}$ is the tree-level matrix element.

The phase space function $F_J^{(m)}(p_1, ...,p_m)$ defines the jet observable
in terms of the momenta of the $m$ final-state partons. In general $F_J$ may
contain $\theta$-functions
(thus, Eq.~(\ref{dsmb}) defines precisely a cross section), $\delta$-functions
(Eq.~(\ref{dsmb}) defines a differential cross section), numerical and
kinematic factors (Eq.~(\ref{dsmb}) refers to an inclusive observable), or
any combination of these. The essential property of $F_J^{(m)}$ is that the
jet observable we are interested in has to be infrared
and collinear safe. From a formal viewpoint this implies that $F_J$ fulfils the
following properties
\beq
\label{fjsoft}
F_J^{(n+1)}(p_1,..,p_j=\lambda q,..,p_{n+1} ) \to
F_J^{(n)}(p_1,...,p_{n+1} )  \;\;\;\;\;\ {\rm if} \;\;\lambda \to 0 \;\;,
\eeq
\beq
\label{fjcoll}
F_J^{(n+1)}(p_1,..,p_i,..,p_j,..,p_{n+1} ) \to
F_J^{(n)}(p_1,..,p,..,p_{n+1} ) \;\;\;\;\;\ {\rm if} \;\;p_i \to zp \;,\;\;
p_j \to (1-z)p
\eeq
for all $n \geq m$, and
\beq
\label{fjLO}
F_J^{(m)}(p_1,...,p_m ) \to 0 \;\;\;\;\;\ {\rm if} \;\;p_i \cdot p_j
\to 0 \;\;.
\eeq

Equations (\ref{fjsoft}) and (\ref{fjcoll}) respectively guarantee that the
jet observable is infrared and collinear safe for {\em any\/} number $n$ of
final-state partons, i.e.\ to {\em any\/} order in QCD perturbation theory.
The $n$-parton jet function $F_J^{(n)}$ on the right-hand side of
Eq.~(\ref{fjsoft}) is obtained from the original $F_J^{(n+1)}$ by removing the
soft parton $p_j$, and that on the right-hand side of Eq.~(\ref{fjcoll}) by
replacing the collinear partons $\{p_i,p_j\}$ by $p = p_i + p_j$.

Equation (\ref{fjLO}) defines the leading-order cross section, that is, it
ensures that the Born-level cross section $d\sigma^{B}$ in Eq.~(\ref{dsmb})
is well-defined (i.e.\ finite after integration) in $d=4$ dimensions.

The cross section $d\sigma^{R}$ has the same expression as $d\sigma^{B}$,
apart from the replacement $m \to m + 1$.

\subsection{Subtraction term}

In order to compute $\sigma^{NLO}$ we write the following identity
\beq
\label{xsid}
\sigma^{NLO} =  \int_{m+1} \left( d\sigma^{R} - d\sigma^{A} \right)
+ \left[ \int_{m+1} d\sigma^{A} + \int_m d\sigma^{V} \right] \;\;,
\eeq
where $d\sigma^{A}$ is a local counterterm for $d\sigma^{R}$, i.e.\
$d\sigma^{A}$ has the same ({\em unintegrated\/}) singular behaviour as
$d\sigma^{R}$. An explicit and general form of $d\sigma^{A}$ is provided
by the dipole factorization formula introduced in Sect.~\ref{dfss}. Thus we
can define:
\beeq
\label{dsadef}
d\sigma^{A} &=& {\cal N}_{in} \sum_{\{ m+1 \} } \,
d\phi_{m+1}(p_1, ...,p_{m+1};Q)
\; \frac{1}{S_{\{ m+1 \} }}  \\
&\cdot& \sum_{\mathrm{pairs}\atop i,j}
\;\sum_{k\not=i,j} {\cal D}_{ij,k}(p_1, ...,p_{m+1}) \;
F_J^{(m)}(p_1, .. {\widetilde p}_{ij}, {\widetilde p}_k, ..,p_{m+1}) \;\;,
\nonumber
\eeeq
where ${\cal D}_{ij,k}(p_1, ...,p_{m+1})$ is the dipole contribution
in Eq.~(\ref{dipff}), namely
\beeq
\label{dipnew}
\!\!\!&{\cal D}_{ij,k}&\!\!\!(p_1, ...,p_{m+1}) =
- \frac{1}{2 p_i \cdot p_j} \\
&\cdot& \mbra{1, .., {\widetilde {ij}},.., {\widetilde k},.., m+1}
\,\frac{{\bom T}_k \cdot {\bom T}_{ij}}{{\bom T}_{ij}^2} \; {\bom V}_{ij,k} \,
\mket{1, .., {\widetilde {ij}},.., {\widetilde k},.., m+1} \;, \nonumber
\eeeq
and $F_J^{(m)}(p_1, .. {\widetilde p}_{ij}, {\widetilde p}_k, ..,p_{m+1})$
is the jet function for the corresponding $m$-parton state
$\{ p_1, .. {\widetilde p}_{ij}, {\widetilde p}_k, ..,p_{m+1} \}$.

We can check that the definition (\ref{dsadef}) makes the difference
$( d\sigma^{R} - d\sigma^{A} )$ integrable in $d=4$ dimensions. Indeed, its
explicit expression is
\beeq
\label{dsra}
d\sigma^{R} - d\sigma^{A} &=& {\cal N}_{in} \sum_{\{ m+1 \} } \,
d\phi_{m+1}(p_1, ...,p_{m+1};Q)
\; \frac{1}{S_{\{ m+1 \} }}  \nonumber \\
&\cdot& \Bigl\{  |{\cal{M}}_{m+1}(p_1, ...,p_{m+1}) |^2
\;F_J^{(m+1)}(p_1, ...,p_{m+1})  \nonumber \\
&-&
\sum_{\mathrm{pairs}\atop i,j}
\;\sum_{k\not=i,j} {\cal D}_{ij,k}(p_1, ...,p_{m+1}) \;
F_J^{(m)}(p_1, .. {\widetilde p}_{ij}, {\widetilde p}_k, ..,p_{m+1}) \Bigr\}
\;\;.
\eeeq
Each term in the curly bracket is separately singular in the soft and
collinear regions. However, as discussed in Sect.~\ref{dfss}, in each of these
regions both the matrix element ${\cal{M}}_{m+1}$ and the phase space
for the $m+1$-parton configuration behave as the corresponding dipole
contribution and dipole phase space:
\beeq
|{\cal{M}}_{m+1}(p_1, ...,p_{m+1}) |^2 &\to& {\cal D}_{ij,k}(p_1, ...,p_{m+1})
\;\;, \\
\{ p_1, .. p_i,..p_j,..p_k, ..,p_{m+1} \} &\to&
\{ p_1, .. {\widetilde p}_{ij}, {\widetilde p}_k, ..,p_{m+1} \} \;\;.
\eeeq
Thus, because of Eqs.~(\ref{fjsoft}) and (\ref{fjcoll}), the singularities
of the first term in the curly bracket are cancelled by similar singularities
due to the second term. On the other hand, each dipole ${\cal D}_{ij,k}$ in
Eq.~(\ref{dipnew}) has no other singularities but those due to the $m$-parton
matrix element $\mket{1, .., {\widetilde {ij}},.., {\widetilde k},.., m+1}$.
However, because of Eq.~(\ref{fjLO}), these singularities are screened
(regularized) by the jet
function $F_J^{(m)}(p_1, .. {\widetilde p}_{ij}, {\widetilde p}_k, ..,p_{m+1})$
in the curly bracket of Eq.~(\ref{dsra}).

Note that this cancellation mechanism is completely independent of the actual
form of the jet defining function but it is essential that $d\sigma^R$ and
$d\sigma^A$ are proportional to $F_J^{(m+1)}$ and $F_J^{(m)}$ respectively.
Nonetheless, both $d\sigma^R$ and $d\sigma^A$ live on the same $m+1$-parton
phase space $d\phi_{m+1}$. Thus the numerical integration (in $d=4$
dimensions) of Eq.~(\ref{dsra}) via Monte Carlo techniques is straightforward.
One
simply generates an $m+1$-parton configuration and gives it a positive
($+ \,|{\cal{M}}_{m+1}|^2$) or negative ($- \sum_{k\not=i,j}\,
{\cal D}_{ij,k}$) weight.
The role
of the two different jet functions $F_J^{(m+1)}$ and $F_J^{(m)}$
is that of binning these weighted events into different bins of the jet
observable. Any time that the generated configuration approaches a singular
region, these two bins coincide and the cancellation of the large  positive
and negative weights takes place.

\subsection{Integral of the subtraction term}
\label{intsub}

Having discussed the four-dimensional integrability of
$( d\sigma^{R} - d\sigma^{A} )$, the only other step we have to consider
is the $d$-dimensional analytical integrability of $d\sigma^{A}$ over the
one-parton subspace leading to soft and collinear divergences.

We start by noting that the dipole contribution in Eq.~(\ref{dipnew}) can be
written as follows
\beq
\label{Dijk}
{\cal D}_{ij,k}(p_1, ...,p_{m+1}) =
- \left[ \frac{{\bom V}_{ij,k}}{2 p_i \cdot p_j} \,\frac{1}{{\bom T}_{ij}^2} \,
|{\cal{M}}_{m}^{ij,k}(p_1,
..{\widetilde p}_{ij}, {\widetilde p}_k, ...,p_{m+1}) |^2 \right]_h
\eeq
where ${\cal{M}}_{m}^{ij,k}$ is a colour-correlated $m$-parton amplitude
(see Eq.~(\ref{colam}))
depending only on $p_1, .. {\widetilde p}_{ij}, {\widetilde p}_k, ..,p_{m+1}$
while ${\bom V}_{ij,k}/p_i \cdot p_j$ depends only on $p_i,p_j,p_k$ or,
equivalently, $p_i,{\widetilde p}_{ij},{\widetilde p}_k$ (the subscript
$h$ in the square bracket of Eq.~(\ref{Dijk}) means that
${\bom V}_{ij,k}$ and ${\cal{M}}_{m}^{ij,k}$ are still coupled in helicity
space).
Using the phase space factorization property in Eq.~(\ref{psfac}), we can thus
completely factorize the $p_i$ dependence in the following form
\beeq
\int_{m+1} d\sigma^{A} &=& - \int_m \;{\cal N}_{in} \sum_{\{ m+1 \} }
\,\sum_{\mathrm{pairs}\atop i,j} \;\sum_{k\not=i,j}
d\phi_{m}(p_1,..{\widetilde p}_{ij},{\widetilde p}_k,.,p_{m+1};Q)
\; \frac{1}{S_{\{ m+1 \} }} \nonumber \\
&\cdot&
F_J^{(m)}(p_1, .. {\widetilde p}_{ij}, {\widetilde p}_k, ..,p_{m+1})
 \\
&\cdot&
\left[ \frac{1}{{\bom T}_{ij}^2} \; |{\cal{M}}_{m}^{ij,k}(p_1,
..{\widetilde p}_{ij}, {\widetilde p}_k, ...,p_{m+1}) |^2 \;\int_1 \,
\frac{{\bom V}_{ij,k}}{2 p_i \cdot p_j} \right]_h
\left[ dp_i({\widetilde p}_{ij},{\widetilde p}_k) \right] \;\;, \nonumber
\eeeq
and we can perform the integration over $p_i$. In particular,
according to the discussion in Sect.~\ref{dfss}, the azimuthal correlation
between
${\bom V}_{ij,k}$ and $|{\cal{M}}_{m}^{ij,k}|^2$ vanishes after
integration over $\left[ dp_i({\widetilde p}_{ij},{\widetilde p}_k) \right]$
and we get (see Eq.~(\ref{azcor}))
\beeq
\label{idsa}
\int_{m+1} d\sigma^{A} &=& - \int_m \,{\cal N}_{in} \sum_{\{ m+1 \} }
\,\sum_{\mathrm{pairs}\atop i,j} \;\sum_{k\not=i,j}
d\phi_{m}(p_1,..{\widetilde p}_{ij},{\widetilde p}_k,.,p_{m+1};Q)
\; \frac{1}{S_{\{ m+1 \} }} \nonumber \\
&\cdot&
F_J^{(m)}(p_1, .. {\widetilde p}_{ij}, {\widetilde p}_k, ..,p_{m+1})
\, |{\cal{M}}_{m}^{ij,k}(p_1,
..{\widetilde p}_{ij}, {\widetilde p}_k, ...,p_{m+1}) |^2
\nonumber \\
&\cdot&
\frac{\as}{2\pi}
\frac{1}{\Gamma(1-\ep)} \left(
\frac{4\pi \mu^2}{2 {\widetilde p}_{ij}{\widetilde p}_k} \right)^{\ep}
\frac{1}{{\bom T}_{ij}^2} \;{\cal V}_{ij}(\ep) \;.
\eeeq

In order to rewrite Eq.~(\ref{idsa}) in terms of an $m$-parton contribution
times a factor, we have to perform the counting of the symmetry factors for
going from $m$ partons to $m+1$ partons.

Consider a Born-level $m$-parton
configuration with $m_f$ quarks of flavour $f$, ${\overline m}_f$ antiquarks
of flavour $f$, and $m_g$ gluons. From this parton configuration we can
obtain an $(m+1)$-parton configuration by changing
\beq
{\rm a)} \;m_g \to m_g+1 \;\;\;\;\;{\rm or} \;\;\;\;\;
{\rm b)} \;m_f \to m_f+1, \; {\overline m}_f \to {\overline m}_f+1, \;
m_g \to m_g-1 \;\;.
\eeq
The corresponding ratios of the symmetry factors for identical particles are
\beeq
\frac{S^{(a)}_{\{ m \} }}{S_{\{ m+1 \} }} \!&=&\!
\frac{.... \;m_g!}{.... \;(m_g+1)!}
= \frac{1}{m_g+1} \;, \\
\frac{S^{(b)}_{\{ m \} }}{S_{\{ m+1 \} }} \!&=&\!
\frac{.... \;m_f! \;{\overline m}_f! \;m_g!}
{.... \;(m_f+1)! \;({\overline m}_f+1)! \;(m_g-1)!} =
\frac{m_g}{(m_f+1)({\overline m}_f+1)} \;\;. \nonumber
\eeeq
Thus we write
\beeq
\sum_{\{ m+1 \} } \frac{1}{S_{\{ m+1 \} }}
\sum_{\mathrm{pairs}\atop i,j} \cdots &=&
{\sum_{\{ m+1 \} }}^{\!\!\!(a)} \frac{1}{S_{\{ m \} }}
\frac{1}{m_g+1} \left( \sum_{\mathrm{pairs}\atop i,j=q_f,g} \cdots
+ \sum_{\mathrm{pairs}\atop i,j={\bar q}_f,g} \cdots +
\sum_{\mathrm{pairs}\atop i,j=g,g} \cdots \right)
\nonumber  \\
&+&
{\sum_{\{ m+1 \} }}^{\!\!\!(b)} \frac{1}{S_{\{ m \} }}
\frac{m_g}{(m_f+1)({\overline m}_f+1)}
\sum_{\mathrm{pairs}\atop i,j=q_f,{\bar q}_f} \cdots  \;\;,
\eeeq
and then
\beq
\sum_{\mathrm{pairs}\atop i,j} \cdots =
\frac{ \# (i,j)_{m+1}}{ \# ({\widetilde {ij}})_m}
\sum_{\widetilde {ij}} \cdots \;\;,
\eeq
where $\# (i,j)_{m+1}$ denotes the number of pairs $i,j$ in the
configuration with
$m+1$ partons and $\# ({\widetilde {ij}})_m$ denotes the number of partons with
flavour ${\widetilde {ij}}$ in the corresponding $m$-parton configuration.
Since we have
\beeq
\frac{ \# (q_f,g)_{m+1}}{ \# (q_f)_m} &=& \frac{m_f (m_g+1)}{m_f} \,,
\;\;
\frac{ \# ({\bar q}_f,g)_{m+1}}{ \# ({\bar q}_f)_m} =
\frac{{\overline m}_f (m_g+1)}{{\overline m}_f} \,, \\
\frac{ \# (g,g)_{m+1}}{ \# (g)_m} &=& \frac{m_g(m_g+1)/2}{m_g} \,,
\;\;
\frac{ \# (q_f,{\bar q}_f)_{m+1}}{ \# (g)_m} = \frac{(m_f+1)
({\overline m}_f+1)}{m_g} \,, \nonumber
\eeeq
we end up with
\beeq
\label{sumsym}
\sum_{\{ m+1 \} } \frac{1}{S_{\{ m+1 \} }}
\,\sum_{\mathrm{pairs}\atop i,j} \cdots &=&
{\sum_{\{ m \} }}^{\!(a)} \frac{1}{S_{\{ m \} }}
\left( \sum_{{\widetilde {ij}}=q_f} \cdots
+ \sum_{{\widetilde {ij}}={\bar q}_f} \cdots
+ \frac{1}{2} \sum_{{\widetilde {ij}}=g} \cdots \right)
\nonumber \\
&+& {\sum_{\{ m \}} }^{\!(b)} \frac{1}{S_{\{ m \} }} \; N_f
\sum_{{\widetilde {ij}}=g} \cdots \;\;.
\eeeq
Inserting Eq.~(\ref{sumsym}) into Eq.~(\ref{idsa}), we obtain:
\beeq
\label{dsafin}
&&\int_{m+1} d\sigma^{A} = - \int_m {\cal N}_{in} \sum_{\{ m \} }
d\phi_{m}(p_1,..,p_m;Q)
\; \frac{1}{S_{\{ m \} }} \;F_J^{(m)}(p_1, ...,p_m) \nonumber \\
&&\cdot \sum_i \sum_{k \neq i}
|{\cal{M}}_{m}^{i,k}(p_1,...,p_m) |^2  \;\,
\frac{\as}{2\pi}
\frac{1}{\Gamma(1-\ep)} \left(
\frac{4\pi \mu^2}{2 p_i\cdot p_k} \right)^{\ep}
\;\frac{1}{{\bom T}_{i}^2} \;{\cal V}_i(\ep) \;\;,
\eeeq
where we have defined
\beeq
\label{vfep}
{\cal V}_i(\ep) &\equiv& {\cal V}_{qg}(\ep) \;,
\;\;\;\;\;\;\;\;\;\;\;\;\;\;\;\;\;\;\;\;\;\;\;\; {\rm if}
\;\;i=q,{\bar q}
\;\;, \\
\label{vgep}
{\cal V}_i(\ep) &\equiv& \frac{1}{2} {\cal V}_{gg}(\ep) + N_f \,
{\cal V}_{q{\bar q}}(\ep)\;, \;\;\; {\rm if} \;\; i=g \;\;.
\eeeq

Equation (\ref{dsafin}) explicitly shows that the subtraction contribution
$d\sigma^{A}$ can be integrated in closed analytic form over the subspace
leading to soft and collinear divergences. These divergences are indeed
collected in terms of $\ep$ poles into the factors ${\cal V}_i(\ep)$.

The final result for $\int_{m+1} d\sigma^{A}$ in Eq.~(\ref{dsafin})
can be written as follows
\beq
\label{dsaf}
\int_{m+1} d\sigma^{A} = \int_m \left[ d\sigma^{B} \cdot
{\bom I}(\ep) \right] \;\;.
\eeq
Comparing Eq.~(\ref{dsafin}) with Eq.~(\ref{dsmb}), we see that
the notation $\left[ d\sigma^{B} \cdot {\bom I}(\ep) \right]$ on the
right-hand side means that one has to write down the expression for
$d\sigma^{B}$ and then replace the corresponding matrix element squared
at the Born level
\beq
\M{}{m} = {}_m\!\!<{1, ...., m}|{1, ...., m}>_m \;,
\eeq
by
\beeq
\label{insop}
{}_m\!\!<{1, ...., m}| \;{\bom I}(\ep) \;|{1, ...., m}>_m \;,
\eeeq
where the insertion operator ${\bom I}(\ep)$ depends on the colour charges
and momenta of the $m$ final-state partons in $d\sigma^{B}$:
\beeq
\label{iee}
{\bom I}(p_1,...,p_m;\ep) = -
\frac{\as}{2\pi}
\frac{1}{\Gamma(1-\ep)} \sum_i \;\frac{1}{{\bom T}_{i}^2} \;{\cal V}_i(\ep)
\; \sum_{k \neq i} {\bom T}_i \cdot {\bom T}_k
\; \left( \frac{4\pi \mu^2}{2 p_i\cdot p_k} \right)^{\ep}
 \;\;.
\eeeq
The singular factors ${\cal V}_i(\ep)$, defined in Eqs.~(\ref{vfep},\ref{vgep}),
are given by (see Eqs.~(\ref{vqgep}--\ref{vggep}))
\beeq
\label{calvexp}
{\cal V}_{i}(\ep) = {\bom T}_{i}^2 \left( \frac{1}{\ep^2} -
\frac{\pi^2}{3} \right) + \gamma_i \;\frac{1}{\ep}
+ \gamma_i + K_i + {\cal O}(\ep) \;\;,
\eeeq
where $\gamma_i$ is defined in Eq.~(\ref{cgamma}) and we have introduced the
following constants
\beeq
\label{kcondef}
K_{i=q,{\bar q}} = \left( \frac{7}{2} - \frac{\pi^2}{6} \right) C_F \;,
\;\;\;\;
K_{i=g} = \left( \frac{67}{18} - \frac{\pi^2}{6} \right) C_A - \frac{10}{9}
T_R N_f \;.
\eeeq

Note that all the terms in ${\cal V}_{i}(\ep)$ have a simple interpretation.
The coefficient of the double pole $1/\ep^2$ is the square of the colour
charge of the parton $i$, that of the single pole $1/\ep$ is related to
the integral of the non-soft part of its four-dimensional Altarelli-Parisi
splitting function,
and the $\pi^2$-term is a customary phase space factor. The constant $K_g$
typically appears in the resummation program of higher-order logarithmic
corrections of Sudakov type [\ref{Sud}] if one uses dimensional regularization
and the ${\overline {\rm MS}}$ renormalization scheme (we do not know of an
analogous role for the constant $K_q$). Actually, in the context of
these calculations the constant $K_g$ can be absorbed by a redefinition of
the renormalized coupling, thus introducing a `more physical' renormalization
scheme, accidentally called the `Monte Carlo' scheme in Ref.~[\ref{CMW}].

\subsection{One-loop corrections}
\label{1lc}

The NLO QCD cross section in Eq.~(\ref{xsid}) is finite by definition, i.e.\
because of the infrared and collinear safety of the jet observable. Since the
first term on the right-hand side is finite, the second term
\beq
\label{dsav}
\int_{m+1} d\sigma^{A} + \int_m d\sigma^{V} = \int_m \left\{
d\sigma^{V} + \left[ d\sigma^{B} \cdot
{\bom I}(\ep) \right] \right\}
\eeq
has to be finite as well. Thus all the $\ep$ poles in $d\sigma^{V}$ must be
cancelled by those in $d\sigma^{B} \cdot {\bom I}(\ep)$. The virtual
contribution $d\sigma^{V}$ has the following expression in terms of the
(renormalized) one-loop matrix element
\beq
\label{dsv1}
d\sigma^{V} = {\cal N}_{in} \sum_{\{ m \} } \, d\phi_m(p_1, ...,p_m;Q)
\;\frac{1}{S_{\{ m \} }} \;|{\cal{M}}_{m}(p_1, ...,p_m) |^2_{(1-loop)}
\;F_J^{(m)}(p_1, ...,p_m) \;\;.
\eeq
Comparing Eq.~(\ref{dsv1}) with Eqs.~(\ref{dsafin},\ref{dsaf}), we find that
the singular terms of the one-loop matrix element have the following universal
structure
\beeq
\label{1lsing}
&&| \cm_{m}(p_1, ...,p_m)|^2_{(1-loop)} = - \;
{}_m\!\!<{1, ...., m}| \;{\bom I}(\ep) \;|{1, ...., m}>_m  + \dots \nonumber \\
&&= \frac{\as}{2\pi}
\frac{1}{\Gamma(1-\ep)}  \sum_i
\;\frac{1}{{\bom T}_{i}^2} \;{\cal V}_i(\ep) \;\sum_{k\not=i}
\;\left(\frac{4\pi \mu^2}{2 p_i\cdot p_k} \right)^{\ep}
\;|{\cal{M}}_{m}^{i,k}(p_1,...,p_m) |^2  + \dots \;\;,
\eeeq
where $|{\cal{M}}_{m}^{i,k}|^2 =
{}_m\!\!<{1, ...., m}| \,{\bom T}_i \cdot {\bom T}_k \,|{1, ...., m}>_m $
is the square of the colour-correlated tree-amplitude in Eq.~(\ref{colam})
and the dots stand
for contributions that do not contain any $\ep$ poles. Thus, using the
finiteness property of the NLO cross section, we have obtained as by-product
of our algorithm the general expression (\ref{1lsing}) for the singular terms
of the one-loop QCD amplitudes.

Alternatively, we can use the results of Ref.~[\ref{KST2}] (see also
Ref.~[\ref{GG}]) on the singular behaviour of the one-loop amplitudes to prove
that our algorithm correctly produces the cancellation of all the $\ep$-poles
in Eq.~(\ref{dsav}), thus leading to a finite NLO cross section. As a matter of
fact, using Eqs.~(\ref{calvexp}) and keeping only the $\ep$ poles, we can
rewrite
Eq.~(\ref{1lsing}) as follows
\beeq
\label{1lep}
&&| \cm_{m}(p_1, ...,p_m)|^2_{(1-loop)} =
\frac{\as}{2\pi} \frac{1}{\Gamma(1-\ep)} \sum_i \sum_{k \neq i}
\;\left[ \frac{1}{\ep^2} \left(
\frac{4\pi \mu^2}{2 p_i\cdot p_k} \right)^{\ep} + \frac{1}{\ep} \;
\frac{\gamma_i}{{\bom T}_i^2} \right] \nonumber \\
&&\cdot \;\;
{}_m\!\!<{1, ...., m}| \;{\bom T}_i \cdot {\bom T}_k \;|{1, ...., m}>_m
+ \dots \nonumber \\
&&= \frac{\as}{2\pi} \frac{1}{\Gamma(1-\ep)} \left[ \;
\sum_{i,k \atop i \neq k}
\frac{1}{\ep^2} \left( \frac{4\pi \mu^2}{2 p_i\cdot p_k} \right)^{\ep}
\; {}_m\!\!<{1, ...., m}| \;{\bom T}_i \cdot {\bom T}_k \;|{1, ...., m}>_m
\right. \nonumber \\
&&- \left. \left( \sum_i \gamma_i \right) \frac{1}{\ep}
\; {}_m\!\!<{1, ...., m}|{1, ...., m}>_m \right] + \dots \;\;,
\eeeq
where in the last expression we have used colour-charge conservation
($\sum_{k\neq i} {\bom T}_i \cdot {\bom T}_k = - {\bom T}_i^2$). Equation
(\ref{1lep}) is completely equivalent to Eqs.~(2.3) and (2.9) in
Ref.~[\ref{KST2}]\footnote{Note a misprint in Eq.~(2.9) of Ref.~[\ref{KST2}]:
the overall sign on the right-hand side should be plus instead of minus.}.

\subsection{Final formulae}
\label{finfo}

The final results of the application of our algorithm to the calculation
of jet cross sections with no hadrons in the initial state are summarized
below.

The full QCD cross section in Eq.~(\ref{eexs}) contains a LO and a NLO
component.
Assuming that the LO calculation involves $m$ final-state partons, the LO
cross section is given by
\beq
\label{LOeefin}
\sigma^{LO} = \int_m d\sigma^{B} = \int d\Phi^{(m)}
\;| \cm_{m}(p_1, ...,p_m)|^2 \;F_J^{(m)}(p_1, ...,p_m) \;\;,
\eeq
where $\cm_{m}$ is the tree-level QCD matrix element to produce $m$ partons
in the final state and the function $F_J^{(m)}$ defines the particular jet
observable we are interested in (see Eqs.~(\ref{fjsoft}--\ref{fjLO})
for the general
properties that $F_J^{(m)}$ has to fulfil). The factor $d\Phi^{(m)}$ collects
all the relevant phase space factors, i.e.\ all the remaining terms
on the right-hand side of Eq.~(\ref{dsmb}). The whole calculation (phase space
integration and evaluation of the matrix element) has to be carried out in
four space-time dimensions.

According to the notation in Eq.~(\ref{sNLO3}), the NLO cross section
is split into two terms with $m+1$-parton and  $m$-parton kinematics,
respectively:
\beq
\sigma^{NLO} = \sigma^{NLO\,\{m+1\}} + \sigma^{NLO\,\{m\}} \;\;.
\eeq
The contribution with $m+1$-parton kinematics is the following
\beeq
\label{m1eefin}
\sigma^{NLO\,\{m+1\}}
&&\!\!\! = \int_{m+1} \left[ \left( d\sigma^{R} \right)_{\ep=0} -
\left( \sum_{{\rm dipoles}} \;d\sigma^{B} \otimes \;dV_{{\rm dipole}}
\right)_{\ep=0} \;\right] \nonumber \\
&&\!\!\! = \int d\Phi^{(m+1)}  \left\{ \frac{}{}
\;| \cm_{m+1}(p_1, ...,p_{m+1})|^2 \;F_J^{(m+1)}(p_1, ...,p_{m+1}) \right. \\
&&\left. - \sum_{\mathrm{pairs}\atop i,j}
\;\sum_{k\not=i,j} {\cal D}_{ij,k}(p_1, ...,p_{m+1}) \;
F_J^{(m)}(p_1, .. {\widetilde p}_{ij}, {\widetilde p}_k, ..,p_{m+1}) \right.
\left. \frac{}{} \right\}
\;\;, \nonumber
\eeeq
where the term in the curly bracket is exactly the same as that in
Eq.~(\ref{dsra}): $\cm_{m+1}$ is the tree-level matrix element,
${\cal D}_{ij,k}$ is the dipole factor in Eq.~(\ref{dipff}) and $F_J^{(m)}$
is the
jet defining function for the corresponding $m$-parton state (note, again,
the difference between the two jet functions $F_J^{(m+1)}$ and $F_J^{(m)}$
in the curly bracket). Despite their original $d$-dimensional
definition, at this stage the full calculation is carried out in four
dimensions.

The NLO contribution with $m$-parton kinematics is given by
\beeq
\label{meefin}
&&\!\!\!\!\!\!\!\sigma^{NLO\,\{m\}} =
\int_{m} \left[ d\sigma^{V}  + d\sigma^{B} \otimes {\bom I}
 \right]_{\ep=0}  \\
&&\!\!\!\!\!\!\!
=  \int d\Phi^{(m)} \left\{ | \cm_{m}(p_1, ...,p_{m})|^2_{(1-loop)}
+ {}_m\!\!<{1, ...., m}| \;{\bom I}(\ep) \;|{1, ...., m}>_m \right\}_{\ep=0}
 F_J^{(m)}(p_1, ...,p_{m}) \;\;.\nonumber
\eeeq
The first term in the curly bracket is the one-loop\footnote{Remember that,
according to our calculation of the insertion operator $\bom I$, the one-loop
matrix element in Eq.~(\ref{meefin}) has to be evaluated by using conventional
dimensional regularization. We refer to the discussion in Sect.~\ref{oneloop}
for the use of different regularization schemes.}
{\em renormalized\/} matrix
element square to produce $m$ final-state partons. The second term is
obtained by inserting the colour-charge operator of Eq.~(\ref{iee}) into the
tree-level matrix element to produce $m$ partons as in Eq.~(\ref{insop})
(see also Appendix~C). These two terms have to be first evaluated in $d=4-2\ep$
dimensions. Then one has to carry out their expansion in $\ep$-poles (the
expansion for the singular factors ${\cal V}_i(\ep)$ is recalled in Appendix~C),
cancel analytically (by trivial addition) the poles and perform the limit
$\ep \to 0$. At this point the phase-space integration is carried out in four
space-time dimensions.

\newpage

\setcounter{equation}{0}

\section{Jet cross sections with one initial-state hadron}
\label{sidis}

Sections \ref{sidis}--\ref{eemp} are devoted to the generalization of the
results of the
previous Section to processes with identified hadrons (cfr. Sect.~\ref{subpp}).
In each of these Sections, we first describe the implementation of our
subtraction procedure by following closely (although with less detail) the
steps of
Sect.~\ref{siee}. We start by defining the jet cross sections for each class of
processes, then we introduce the explicit expression for our subtraction
term $d\sigma^A$ and, finally, we perform its integration,
calculate the appropriate combinatorial factors,
and show how the
ensuing contribution can be combined with the virtual term $d\sigma^V$ and
the collinear counter-term $d\sigma^C$ to provide a finite NLO partonic
cross section. The final formulae of our algorithm are summarized at the end
of each Section.

Let us start by considering hard-scattering processes with a single incoming
hadron (cfr. Eqs.~(\ref{1hxs}--\ref{saC})) like, for instance, deep-inelastic
lepton-hadron scattering. In the case of unpolarized scattering, the
Born-level partonic cross section with one parton of flavour $a$ and momentum
$p_a$ in the initial state
has the following
expression in terms of the QCD matrix elements
\beeq
\label{dsab}
d\sigma_a^{B}(p_a) &=& {\cal N}_{in} \frac{1}{n_s(a) n_c(a) \Phi(p_a)}
\sum_{\{ m \} } \, d\phi_m(p_1, ...,p_m;p_a+Q)
\;\frac{1}{S_{\{ m \} }} \nonumber \\
&\cdot&|{\cal{M}}_{m,a}(p_1, ...,p_m;p_a) |^2
\;F_J^{(m)}(p_1, ...,p_m;p_a) \;\;,
\eeeq
Here the factor $1/(n_s(a) n_c(a))$ accounts for the average over the number of
initial-state polarizations and colours and $\Phi(p_a)$ is the flux factor.
Since
$p_a^2=0$, the flux factor fulfils the following scaling property
\beq
\label{fluxe}
\Phi( \eta p_a) = \eta \,\Phi(p_a) \;\;.
\eeq

The jet function $F_J^{(m)}(p_1, ...,p_m;p_a)$ is infrared and collinear safe
(see Eqs.~(\ref{fjsoft}--\ref{fjLO})) and, moreover, it fulfils the property
of factorizability of initial-state collinear singularities. From a formal
viewpoint this implies that
\beq
\label{fjdiscoll}
F_J^{(n+1)}(p_1,..,p_i,..,p_{n+1};p_a) \to F_J^{(n)}(p_1, ...,p_{n+1};xp_a)
\;\;,
\;\;\;\;\;\;\; {\rm if } \;\;\;p_i \to (1-x)p_a
\eeq
for any number $n$ of final-state partons (the $n$-parton jet function on
the right-hand side is obtained from the $n+1$-parton function on the left-hand
side by removing the parton $i$) and
\beq
\label{fjdis}
F_J^{(m)}(p_1, ...,p_{m};p_a) \to 0 \;\;,
\;\;\;\;\;\; {\rm if } \;\;\;p_i \cdot p_a \to 0
\;\;
\eeq
for the leading-order process (i.e.\ $n=m$).

All the other factors in Eq.~(\ref{dsab}) are analogous to those in
Eq.~(\ref{dsmb}).

\subsection{Implementation of the subtraction procedure}
\label{impdis}

In order to compute the NLO cross section in Eq.~(\ref{aNLO}),
we write the following identity
\beeq
\label{dsradis}
\sigma_a^{NLO}(p_a;\mu_F^2) &=&  \int_{m+1} \left( d\sigma_a^{R}(p_a) -
d\sigma_a^{A}(p_a) \right)  \nonumber \\
&+& \left[ \int_{m+1} d\sigma_a^{A}(p_a) + \int_m d\sigma_a^{V}(p_a) +
\int_m d\sigma_a^{C}(p_a;\mu_F^2) \right] \;\;,
\eeeq
where, according to the dipole formulae  in Sects.~\ref{dfss}--\ref{diss},
a local
counterterm $d\sigma_a^{A}(p_a)$ is provided by:
\beeq
\label{dsapa}
d\sigma_a^{A}(p_a) &=& {\cal N}_{in} \frac{1}{n_s(a) \Phi(p_a)}
\sum_{\{ m+1 \} } \,
d\phi_{m+1}(p_1, ...,p_{m+1};p_a+Q)
\frac{1}{S_{\{ m+1 \} }}  \nonumber \\
&\cdot& \left\{ \sum_{\mathrm{pairs}\atop i,j}
\;\sum_{k\not=i,j} {\cal D}_{ij,k}(p_1, ...,p_{m+1};p_a) \;
F_J^{(m)}(p_1, .. {\widetilde p}_{ij}, {\widetilde p}_k, ..,p_{m+1};p_a)
\right. \;\; \nonumber \\
&+& \left. \sum_{\mathrm{pairs}\atop i,j} \;
{\cal D}_{ij}^{a}(p_1, ...,p_{m+1};p_a) \;
F_J^{(m)}(p_1, .. {\widetilde p}_{ij}, ..,p_{m+1};{\widetilde p}_a)
\right. \;\; \nonumber \\
&+& \left. \sum_i \;\sum_{k\not=i}
{\cal D}_{k}^{ai}(p_1, ...,p_{m+1};p_a) \;
F_J^{(m)}(p_1, .. {\widetilde p}_{k}, ..,p_{m+1};{\widetilde p}_{ai})
\right\} \;\;.
\eeeq
Here ${\cal D}_{ij,k}(p_1, ...,p_{m+1})$,
${\cal D}_{ij}^{a}(p_1, ...,p_{m+1};p_a)$ and
${\cal D}_{k}^{ai}(p_1, ...,p_{m+1};p_a)$ are respectively
the dipoles in  Eqs.~(\ref{dipff}), (\ref{dipfi}) and (\ref{dipif})
and
$F_J^{(m)}(p_1, .. {\widetilde p}_{ij}, {\widetilde p}_k, ..,p_{m+1};p_a)$,
$F_J^{(m)}(p_1, .. {\widetilde p}_{ij}, ..,$ $p_{m+1}; {\widetilde p}_a)$ and
$F_J^{(m)}(p_1, .. {\widetilde p}_{k}, .., p_{m+1}; {\widetilde p}_{ai})$
are the jet defining functions for the corresponding $m$-parton states
$\{ p_1, .. {\widetilde p}_{ij}, {\widetilde p}_k, .., p_{m+1}; p_a \}$,
$\{ p_1, .. {\widetilde p}_{ij}, .., p_{m+1}; {\widetilde p}_a \}$ and
$\{ p_1, .. {\widetilde p}_{k}, .., p_{m+1}; {\widetilde p}_{ai} \}$.

Since the dipole contributions exactly match the soft and collinear divergences
of the square of the matrix element $|{\cal{M}}_{m+1,a}|^2 $, the
subtracted expression $( d\sigma_a^{R}(p_a) - d\sigma_a^{A}(p_a) )$ in
Eq.~(\ref{dsradis}) is integrable in $d=4$ dimensions.

In order to compute the contribution in the square bracket of
Eq.~(\ref{dsradis}), we write $d\sigma_a^{A}(p_a)$ as follows
\beq
\label{ds123}
d\sigma_a^{A}(p_a)= d\sigma_a^{A \prime}(p_a) +
d\sigma_a^{A \prime \prime}(p_a) +
d\sigma_a^{A \prime \prime \prime}(p_a) \;\;,
\eeq
where the three terms on the right-hand side of Eq.~(\ref{ds123}) are in
one-to-one correspondence with those in the curly bracket on the
right-hand side of Eq.~(\ref{dsapa}).

The integration of $d\sigma_a^{A \prime}(p_a)$ can be performed analogously
to that of $d\sigma^{A}$ in the previous Section, thus leading to the
following result
\beeq
\label{dsap1}
\!\!\!&&\int_{m+1} d\sigma_a^{A \prime}(p_a) = - \int_m {\cal N}_{in}
\frac{1}{n_s(a) \Phi(p_a)}
\sum_{\{ m \} }
d\phi_{m}(p_1,..,p_m;p_a+Q)
\; \frac{1}{S_{\{ m \} }}  \\
\!\!\!&&\cdot
\; F_J^{(m)}(p_1, ...,p_m;p_a) \sum_i \sum_{k \neq i}
|{\cal{M}}_{m,a}^{i,k}(p_1,...,p_m;p_a) |^2
\; \frac{\as}{2\pi}
\frac{1}{\Gamma(1-\ep)} \left(
\frac{4\pi \mu^2}{2 p_i\cdot p_k} \right)^{\ep}
\;\frac{1}{{\bom T}_{i}^2} \;{\cal V}_i(\ep) \;\;, \nonumber
\eeeq
where the functions ${\cal V}_i(\ep)$ are defined in
Eqs.~(\ref{vfep},\ref{vgep}).

Let us now consider the integration of $d\sigma_a^{A \prime \prime}(p_a)$.
We first rewrite Eq.~(\ref{dipfi}) as follows
\beq
{\cal D}_{ij}^{a}(p_1, ...,p_{m+1};p_a) =
- \left[ \frac{{\bom V}_{ij}^{a}}{2 x_{ij,a} \,p_i \cdot p_j}
\,\frac{1}{{\bom T}_{ij}^2}
\, |{\cal{M}}_{m,a}^{ij,a}(p_1,
..{\widetilde p}_{ij},..,p_{m+1};{\widetilde p}_a) |^2 \right]_h \;\;,
\eeq
where the notation is similar to that in Eq.~(\ref{Dijk}), and then, using
the phase space convolution properties in Eqs.~(\ref{psconv},\ref{px}),
we obtain:
\beeq
\int_{m+1} d\sigma_a^{A \prime \prime}(p_a) &=& - \int_m {\cal N}_{in}
\sum_{\{ m+1 \} }
\,\sum_{\mathrm{pairs}\atop i,j} \;\int_0^1 dx \;\;
d\phi_{m}(p_1,..{\widetilde p}_{ij},..,p_{m+1};xp_a + Q)
\; \frac{1}{S_{\{ m+1 \} }}
 \nonumber \\
&\cdot&
F_J^{(m)}(p_1, .. {\widetilde p}_{ij},..,p_{m+1};xp_a)
\; \frac{1}{n_s(a) \Phi(p_a)}
 \\
&\cdot& \frac{1}{x}
\left[ \frac{1}{{\bom T}_{ij}^2} \; |{\cal{M}}_{m,a}^{ij,a}(p_1,
..{\widetilde p}_{ij},..,p_{m+1};xp_a) |^2 \;\int_1
\frac{{\bom V}_{ij}^{a}}{2 p_i \cdot p_j} \right]_h
\left[ dp_i({\widetilde p}_{ij}; p_a,x) \right] \;\;. \nonumber
\eeeq
As in the case of the splitting functions ${\bom V}_{ij,k}$, the azimuthal
correlations due
to ${\bom V}_{ij}^{a}$ vanish after integration over $p_i$ (keeping
${\widetilde p}_{ij}$ and $x$ fixed) and we find
\beeq
\label{dsapp}
&&\int_{m+1} d\sigma_a^{A \prime \prime}(p_a) = - \int_m {\cal N}_{in}
\sum_{\{ m+1 \} }
\,\sum_{\mathrm{pairs}\atop i,j} \;\int_0^1 dx \;
\frac{1}{n_s(a) \Phi(xp_a)} \nonumber \\
&&\cdot d\phi_{m}(p_1,..{\widetilde p}_{ij},..,p_{m+1};xp_a + Q)
\; \frac{1}{S_{\{ m+1 \} }}
\; F_J^{(m)}(p_1, .. {\widetilde p}_{ij},..,p_{m+1};xp_a)
\; \nonumber \\
&&\cdot \;|{\cal{M}}_{m,a}^{ij,a}(p_1,
..{\widetilde p}_{ij},..,p_{m+1};xp_a) |^2
\; \frac{\as}{2\pi}
\frac{1}{\Gamma(1-\ep)} \left(
\frac{4\pi \mu^2}{2 {\widetilde p}_{ij}\cdot p_a} \right)^{\ep}
\;\frac{1}{{\bom T}_{ij}^2} \;{\cal V}_{ij}(x;\ep) \;\;,
\eeeq
where the functions ${\cal V}_{ij}(x;\ep)$ are given in
Eqs.~(\ref{vqgxep}--\ref{vggxep}) and we have used Eq.~(\ref{fluxe}) to replace
$x\Phi(p_a)$ with $\Phi(xp_a)$.

Equation (\ref{dsapp}) is similar to Eq.~(\ref{idsa}) in Sect.~\ref{siee},
apart from the replacement $k \to a$ as spectator parton.
In order to rewrite Eq.~(\ref{dsapp}) in terms of an $m$-parton contribution
times a factor, we have to perform the counting of the symmetry factors for
going from $m$ partons to $m+1$ partons in the final state. Since, as shown
in Sect.~\ref{siee}, this counting is independent of the spectator parton,
we immediately get
\beeq
\label{dsap2}
\int_{m+1} d\sigma_a^{A \prime \prime}(p_a) &=&
 - \int_m {\cal N}_{in} \int_0^1 dx \;
\frac{1}{n_s(a) \Phi(xp_a)} \sum_{\{ m \} }
d\phi_{m}(p_1,..,p_m;xp_a + Q)  \nonumber \\
&\cdot& \frac{1}{S_{\{ m \} }} \;
F_J^{(m)}(p_1, ..,p_m;xp_a)
\sum_i \; |{\cal{M}}_{m,a}^{i,a}(p_1,
..,p_m;xp_a) |^2 \nonumber \\
&\cdot&
\frac{\as}{2\pi}
\frac{1}{\Gamma(1-\ep)} \left(
\frac{4\pi \mu^2}{2 p_i\cdot p_a} \right)^{\ep}
\;\frac{1}{{\bom T}_i^2} \;{\cal V}_{i}(x;\ep) \;\;,
\eeeq
where we have defined
\beeq
\label{vixep}
{\cal V}_i(x;\ep) &\equiv& {\cal V}_{qg}(x;\ep) \;,
\;\;\;\;\;\;\;\;\;\;\;\;\;\;\;\;\;\;\;\;\;\;\;\;\;\;\;\;\; {\rm if}
\;\;i=q,{\bar q}
\;\;, \\
\label{vgxep}
{\cal V}_i(x;\ep) &\equiv& \frac{1}{2} \;{\cal V}_{gg}(x;\ep) + N_f \,
{\cal V}_{q{\bar q}}(x;\ep)\;, \;\;\; {\rm if} \;\; i=g \;\;.
\eeeq
Recall that all the $\ep$-poles of ${\cal V}_i(x;\ep)$ are accounted for by
$\delta(1-x){\cal V}_i(\ep)$ terms (see Eqs.~(\ref{vqgxep}--\ref{vggxep})),
where ${\cal V}_i(\ep)$ are the functions
defined in Eqs.~(\ref{vfep},\ref{vgep}).

Let us now consider the integration of
$d\sigma_a^{A \prime \prime \prime}(p_a)$. We first rewrite the corresponding
dipole contribution (see Eq.~(\ref{dipif})) as follows
\beq
{\cal D}_{k}^{ai}(p_1, ...,p_{m+1};p_a) =
 - \left[ \frac{{\bom V}_{k}^{ai}}{2 x_{ik,a} \,p_i \cdot p_a}
\,\frac{1}{{\bom T}_{ai}^2}
\, |{\cal{M}}_{m,{\widetilde {ai}}}^{k,ai}(p_1,
..{\widetilde p}_{k},..,p_{m+1};{\widetilde p}_{ai}) |^2 \right]_h \;\;.
\eeq
Thus, using the phase space convolution in Eqs.~(\ref{psconva},\ref{dpix}) and
performing the integration over $p_i$ (keeping
${\widetilde p}_{k}$ and $x$ fixed), the azimuthal correlation due to
${\bom V}_{k}^{ai}$ vanishes  and we
obtain
\beeq
\label{dsa3}
&&\int_{m+1} d\sigma_a^{A \prime \prime \prime}(p_a) =
- \int_m {\cal N}_{in} \int_0^1 dx \;
\sum_{\{ m+1 \} }
\sum_i \sum_{k \neq i} \frac{1}{n_s({\widetilde {ai}}) \Phi(xp_a)}
\nonumber \\
&&\cdot \;d\phi_{m}(p_1,..{\widetilde p}_k,..,p_m;xp_a + Q)
\; \frac{1}{S_{\{ m+1 \} }}
\; F_J^{(m)}(p_1,..,{\widetilde p}_k,..,p_m;xp_a) \nonumber \\
&&\cdot
\; |{\cal{M}}_{m,{\widetilde {ai}}}^{ai,k}(p_1,
..{\widetilde p}_k,..,p_m;xp_a) |^2
\; \frac{\as}{2\pi}
\frac{1}{\Gamma(1-\ep)} \left(
\frac{4\pi \mu^2}{2 {\widetilde p}_k\cdot p_a} \right)^{\ep}
\;\frac{1}{{\bom T}_{ai}^2} \;{\cal V}^{a,ai}(x;\ep) \;\;,
\eeeq
where the functions ${\cal V}^{a,ai}(x;\ep)$ are given in
Eqs.~(\ref{cvqgxep}--\ref{cvggxep}).

The right-hand side of Eq.~(\ref{dsa3}) can easily be rewritten in terms of a
sum over $m$-parton configurations. To this end, we have to perform the
corresponding counting of symmetry factors, which, nonetheless, is trivial in
this case. If the Born-level $m$-parton configuration has $m_i$ partons of type
$i$, the corresponding $m+1$-parton configuration has $m_i+1$ partons of
the same type, so that:
\beeq
\label{sm1}
\sum_{\{ m+1 \} } \frac{1}{S_{\{ m+1 \} }} \sum_i \dots &=&
\sum_{\{ m+1 \} } \sum_i \frac{... \; m_i!}{...(m_i+1)!} \;
\frac{1}{S_{\{ m \} }}  \dots  \;\;.
\eeeq
However, there are $m_i+1$ possible ways of choosing the parton $i$ in the
$m+1$-parton configuration and, hence, we obtain
\beeq
\label{sfi}
\sum_{\{ m+1 \} } \frac{1}{S_{\{ m+1 \} }} \sum_i \dots &=&
\sum_i \frac{... \; m_i!}{...(m_i+1)!} \;(m_i+1)
\sum_{\{ m \} } \frac{1}{S_{\{ m \} }} \dots \nonumber \\
&=& \sum_i \sum_{\{ m \} } \frac{1}{S_{\{ m \} }} \dots \;\;,
\eeeq
where the $\sum_i$ in the last line of Eq.~(\ref{sfi}) simply denotes the
sum over
the flavours $i$ in the $m$-parton configuration. We can thus rewrite this
sum as a sum over the flavours $b={\widetilde {ai}}$ in the initial state
and Eq.~(\ref{dsa3}) becomes:
\beeq
\label{dsap3}
&&\int_{m+1} d\sigma_a^{A \prime \prime \prime}(p_a) =
- \int_{m} {\cal N}_{in} \sum_{\{ m \} } \int_0^1 dx \;\;
d\phi_{m}(p_1,..,p_m;xp_a + Q) \; \frac{1}{S_{\{ m \} }} \nonumber \\
&\cdot& F_J^{(m)}(p_1,..,p_m;xp_a)
\; \sum_{k} \sum_{b}
\frac{1}{n_s(b) \Phi(xp_a)}
\; |{\cal{M}}_{m,b}^{b,k}(p_1,
...,p_m;xp_a) |^2 \nonumber \\
&&\cdot \;
\frac{\as}{2\pi}
\frac{1}{\Gamma(1-\ep)} \left(
\frac{4\pi \mu^2}{2 p_k\cdot p_a} \right)^{\ep}
\;\frac{1}{{\bom T}_{b}^2} \;{\cal V}^{a,b}(x;\ep) \;\;.
\eeeq
Note that, in addition to $\delta^{ab}\delta(1-x){\cal V}_b(\ep)$ terms like
those in Eqs.~(\ref{vixep},\ref{vgxep}), ${\cal V}^{a,b}(x;\ep)$ contains
$\ep$-poles in terms of the form $P^{ab}(x)/\ep$ (see
Eqs.~(\ref{cvqgxep}--\ref{cvggxep})).

Collecting Eqs.~(\ref{dsap1},\ref{dsap2},\ref{dsap3})
and adding Eq.~(\ref{saC}), we find
\beeq
&&\!\!\!\!\!\!\!\!\int_{m+1} d\sigma_a^{A}(p_a) +  \int_m d\sigma_a^{C}(p_a) =
- \sum_{b} \; \int_0^1 dx \;\;
\int_m {\cal N}_{in} \sum_{\{ m \} }
d\phi_{m}(p_1,..,p_m;xp_a + Q) \nonumber \\
&\cdot& \frac{1}{S_{\{ m \} }} \;F_J^{(m)}(p_1,..,p_m;xp_a)
\; \frac{1}{n_s(b) \Phi(xp_a)} \; \frac{\as}{2\pi}
\frac{1}{\Gamma(1-\ep)}
\nonumber \\
&\cdot& \left\{ \delta^{ab} \;\delta(1-x) \sum_i
\sum_{k \neq i}
\; |{\cal{M}}_{m,b}^{i,k}(p_1,
...,p_m;xp_a) |^2 \;\left(
\frac{4\pi \mu^2}{2 p_i\cdot p_k} \right)^{\ep}
\;\frac{1}{{\bom T}_{i}^2} \;{\cal V}_{i}(\ep) \right.
\nonumber \\
&+& \left. \delta^{ab} \sum_{i}
\; |{\cal{M}}_{m,b}^{i,b}(p_1,
...,p_m;xp_a) |^2 \;\left(
\frac{4\pi \mu^2}{2 p_i\cdot p_a} \right)^{\ep}
\;\frac{1}{{\bom T}_{i}^2} \;{\cal V}_{i}(x;\ep) \right.
\nonumber \\
&+& \left. \sum_{i}
\; |{\cal{M}}_{m,b}^{i,b}(p_1,
...,p_m;xp_a) |^2 \;\left(
\frac{4\pi \mu^2}{2 p_i\cdot p_a} \right)^{\ep}
\;\frac{1}{{\bom T}_{b}^2} \;{\cal V}^{a,b}(x;\ep) \right.
\nonumber \\
&+& \left. \;\frac{1}{n_c(b)}
\; |{\cal{M}}_{m,b}(p_1,
...,p_m;xp_a) |^2 \;\left[ - \frac{1}{\ep}
\left(
\frac{4\pi \mu^2}{\mu_F^2} \right)^{\ep}
\;P^{ab}(x)  + \Kab(x) \right] \right\} \;\;.
\eeeq

We see that $\int_{m+1} d\sigma_a^{A}(p_a) +  \int_m d\sigma_a^{C}(p_a)$
is obtained
from the leading-order expression $\int_m d\sigma_a^{B}(xp_a)$ by
replacing the Born-level matrix element squared
\beq
\frac{1}{n_s(a)} \;\; {}_{m,a}\!\!<{...;p_a}|{...;p_a}>_{m,a} \;,
\eeq
by
\beeq
\sum_b \frac{1}{n_s(b)}
\;\;{}_{m,b}\!\!<{...;xp_a}| \;{\bom I}^{a,b}(x;\ep) \;|{...;xp_a}>_{m,b} \;,
\eeeq
and performing the $x$-integration. Here
the insertion operator ${\bom I}(x;\ep)$ depends on the colour charges,
momenta and flavours of the QCD partons
\beeq
\label{idis1}
\!\!\!\!\!\!\!\!\!&&{\bom I}^{a,b}(p_1,...,p_m;p_a,x;\ep;\mu_F^2) = -
\frac{\as}{2\pi}
\frac{1}{\Gamma(1-\ep)} \nonumber \\
\!\!\!\!\!\!\!\!\!&&\cdot \left\{ \delta^{ab} \;\delta(1-x)
\sum_i \sum_{k \neq i} {\bom T}_i \cdot {\bom T}_k
\;\left(
\frac{4\pi \mu^2}{2 p_i\cdot p_k} \right)^{\ep}
\;\frac{1}{{\bom T}_{i}^2} \;{\cal V}_{i}(\ep)
+  \delta^{ab} \sum_{i} {\bom T}_i \cdot {\bom T}_b
\;\left(
\frac{4\pi \mu^2}{2 p_i\cdot p_a} \right)^{\ep}
\;\frac{1}{{\bom T}_{i}^2} \;{\cal V}_{i}(x;\ep) \right.
\nonumber \\
\!\!\!\!\!\!\!\!\!&&+ \left. \sum_{i} {\bom T}_i \cdot {\bom T}_b
\left(
\frac{4\pi \mu^2}{2 p_i\cdot p_a} \right)^{\ep}
\;\frac{1}{{\bom T}_{b}^2} \;{\cal V}^{a,b}(x;\ep)
-   \frac{1}{\ep}
\left(
\frac{4\pi \mu^2}{\mu_F^2} \right)^{\ep}
\;P^{ab}(x)  + \Kab(x)  \right\} \;\;.
\eeeq

This form of the insertion operator can be simplified in the limit $\ep \to 0$.
We start by rewriting Eq.~(\ref{idis1}) as follows
\beeq
\label{idis2}
&{\bom I}^{a,b}&\!\!\!\!(p_1,...,p_m;p_a,x;\ep;\mu_F^2)
= \delta^{ab} \;\delta(1-x)
\; {\bom I}(p_1,...,p_m,p_a;\ep) \nonumber \\
&+&\!\!\!\! \delta^{ab} \;
{\bom I}_{(1)}(p_1,...,p_m;p_a,x;\ep) +
{\bom I}_{(2)}^{a,b}(p_1,...,p_m;p_a,x;\ep;\mu_F^2) \;\;,
\eeeq
where
\beeq
\label{i0dis}
&&{\bom I}(p_1,...,p_m,p_a;\ep) =  - \;\frac{\as}{2\pi}
\frac{1}{\Gamma(1-\ep)} \left\{ \sum_i
\frac{1}{{\bom T}_{i}^2} \;{\cal V}_{i}(\ep)
\left[ \;\sum_{k \neq i} {\bom T}_i \cdot {\bom T}_k
\;\left(
\frac{4\pi \mu^2}{2 p_i\cdot p_k} \right)^{\ep} \right. \right.
\nonumber \\
&& + \left. \left. {\bom T}_i \cdot {\bom T}_a
\;\left(
\frac{4\pi \mu^2}{2 p_i\cdot p_a} \right)^{\ep} \;\right] +
\frac{1}{{\bom T}_{a}^2} \;{\cal V}_{a}(\ep)
\sum_{i} {\bom T}_i \cdot {\bom T}_a
\left(
\frac{4\pi \mu^2}{2 p_i\cdot p_a} \right)^{\ep} \;\right\} \;\;,
\eeeq
\beeq
\label{i1dis}
{\bom I}_{(1)}(p_1,...,p_m;p_a,x;\ep) &=&
- \;\frac{\as}{2\pi} \frac{1}{\Gamma(1-\ep)}
\sum_{i} {\bom T}_i \cdot {\bom T}_a
\;\left(
\frac{4\pi \mu^2}{2 p_i\cdot p_a} \right)^{\ep}
\frac{1}{{\bom T}_{i}^2} \nonumber \\
&\cdot& \Bigl[ {\cal V}_{i}(x;\ep) - \delta(1-x)
\; {\cal V}_{i}(\ep) \Bigr] \;\;,
\eeeq
\beeq
\label{i2dis}
\!\!\!\!\!\!\!\!\!\!\!\!\!\!\!&&\!\!\!\!\!\!{\bom I}_{(2)}^{a,b}
(p_1,...,p_m;p_a,x;\ep;\mu_F^2) =
- \;\frac{\as}{2\pi} \frac{1}{\Gamma(1-\ep)}
\left\{ \sum_{i} {\bom T}_i \cdot {\bom T}_b
\;\left(
\frac{4\pi \mu^2}{2 p_i\cdot p_a} \right)^{\ep}
\;\frac{1}{{\bom T}_{b}^2} \right. \nonumber \\
\!\!\!\!\!\!\!\!\!\!\!\!\!\!\!&\cdot&\!\!\!\! \left.
\Bigl[ \;{\cal V}^{a,b}(x;\ep)
- \delta^{ab} \delta(1-x) \;{\cal V}_{a}(\ep) \Bigr]
- \frac{1}{\ep}
\;\left(
\frac{4\pi \mu^2}{\mu_F^2} \right)^{\ep}
\;P^{ab}(x)  + \Kab(x)  \right\} \;\;.
\eeeq

The operator ${\bom I}(p_1,...,p_m,p_a;\ep)$ in Eq.~(\ref{i0dis}) is fully
symmetric with respect to the dependence on colour charges and momenta of the
$m+1$ partons $\{ p_1,...,p_m,p_a \}$. In other words, this operator
is identical (apart from depending on the additional initial-state parton $a$)
to that in Eq.~(\ref{iee}). Since crossing the momentum of partons from the
final to the initial state does not change the singular terms in the
one-loop QCD amplitudes, it follows that the insertion operator (\ref{i0dis})
cancels all the singularities in
the virtual contribution $\int_m d\sigma_a^V(p_a)$.
As a consequence, the two other operators
${\bom I}_{(1)}$ and ${\bom I}_{(2)}$ should contribute as finite
counterterms. Actually, ${\bom I}_{(1)}$ and ${\bom I}_{(2)}$ are separately
finite in the limit $\ep \to 0$.

In order to show that, we can use Eqs.~(\ref{vqgxep}--\ref{vggxep}),
(\ref{vixep},\ref{vgxep}) and (\ref{vfep},\ref{vgep}) and thus obtain
\beeq
\label{vixmvi}
{\cal V}_{i}(x;\ep) - \delta(1-x)
\; {\cal V}_{i}(\ep) &=& {\bom T}_i^2 \left[
\left( \frac{2}{1-x} \ln\frac{1}{1-x} \right)_+ + \frac{2}{1-x} \ln(2-x) \right]
\nonumber \\
&-& \gamma_i \left[ \left( \frac{1}{1-x} \right)_+ + \delta(1-x) \right]
+ {\cal O}(\ep) \;\;,
\eeeq
where $\gamma_i$ are given in Eq.~(\ref{cgamma}).
Then we can write:
\beeq
\label{i1dis1}
&&{\bom I}_{(1)}(p_1,...,p_m;p_a,x;\ep) =
- \;\frac{\as}{2\pi}
\sum_{i} {\bom T}_i \cdot {\bom T}_a
\left\{ \left( \frac{2}{1-x} \ln\frac{1}{1-x} \right)_+ \right.
\nonumber \\
&+& \left. \frac{2}{1-x} \ln(2-x) - \frac{\gamma_i}{{\bom T}_{i}^2}
\left[ \left( \frac{1}{1-x} \right)_+ + \delta(1-x) \right]
\right\} + {\cal O}(\ep) \;\;.
\eeeq

Coming to ${\bom I}_{(2)}$, let us rewrite Eq.~(\ref{i2dis}) in the following
form
\beeq
\label{i2dis1}
&&\!\!\!\!{\bom I}_{(2)}^{a,b}(p_1,...,p_m;p_a,x;\ep;\mu_F^2) =
- \;\frac{\as}{2\pi} \frac{1}{\Gamma(1-\ep)}
\left\{ \sum_{i} {\bom T}_i \cdot {\bom T}_b
\;\left(
\frac{4\pi \mu^2}{2 p_i\cdot p_a} \right)^{\ep}
\;\frac{1}{{\bom T}_{b}^2} \right. \nonumber \\
&\cdot& \left. \left[ \;{\cal V}^{a,b}(x;\ep)
+ \frac{1}{\ep} \;P^{ab}(x)
- \delta^{ab} \delta(1-x) \;{\cal V}_{a}(\ep) \right] \right.
\nonumber \\
&& - \left. \left[ \sum_{i} {\bom T}_i \cdot {\bom T}_b
\;\left(
\frac{4\pi \mu^2}{2 p_i\cdot p_a} \right)^{\ep}
\;\frac{1}{{\bom T}_{b}^2} + \;\left(
\frac{4\pi \mu^2}{\mu_F^2} \right)^{\ep} \right]
\frac{1}{\ep}
\;P^{ab}(x)  + \Kab(x)  \right\} \;\;.
\eeeq
Using Eqs.~(\ref{cvqgxep}--\ref{cvggxep}), we see that the first
square bracket contribution on the right-hand side of Eq.~(\ref{i2dis1})
is finite for $\ep \to 0$ and given by
\beeq
\label{vabpab}
&&\!\!\!\!\!\!\!\!{\cal V}^{a,b}(x;\ep)
+ \frac{1}{\ep} \;P^{ab}(x)
- \delta^{ab} \delta(1-x) \;{\cal V}_{a}(\ep) =
{\overline K}^{ab}(x) + P^{ab}(x) \ln x \nonumber \\
&-& \delta^{ab} \; {\bom T}_{a}^2
\left[
\left( \frac{2}{1-x} \ln\frac{1}{1-x} \right)_+ + \frac{2}{1-x} \ln(2-x) \right]
+ {\cal O}(\ep) \;\;,
\eeeq
where the ${\overline K}^{ab}(x)$ functions are defined so as to simplify
the final formulae,
\beeq
\label{kbarqg}
{\overline K}^{qg}(x) &=& P^{qg}(x) \ln\frac{1-x}{x} + C_F \;x \;\;, \\
\label{kbargq}
{\overline K}^{gq}(x) &=& P^{gq}(x) \ln\frac{1-x}{x} + T_R \;2x(1-x) \;\;, \\
\label{kbarqq}
{\overline K}^{qq}(x) &=& C_F \left[
\left( \frac{2}{1-x} \ln\frac{1-x}{x} \right)_+
- (1+x) \ln\frac{1-x}{x} + (1-x) \right] \nonumber\\
&-& \delta(1-x) \left( 5 - \pi^2 \right) C_F  \;\;, \\
\label{kbargg}
{\overline K}^{gg}(x) &=& 2 C_A \left[
\left( \frac{1}{1-x} \ln\frac{1-x}{x} \right)_+
+ \left( \frac{1-x}{x} - 1 + x(1-x) \right) \ln\frac{1-x}{x} \right] \nonumber\\
&-& \delta(1-x) \left[ \left( \frac{50}{9} - \pi^2 \right) C_A
- \frac{16}{9} T_R N_f \right]
\;\;.
\eeeq

As for the second square bracket contribution on the right-hand side
of Eq.~(\ref{i2dis1}),
using colour conservation ($\sum_{i} {\bom T}_i = - {\bom T}_b$) and
expanding in $\ep$, we get an ${\cal O}(\ep)$ term:
\beeq
\label{oep}
\sum_{i} {\bom T}_i \cdot {\bom T}_b
\left(
\frac{4\pi \mu^2}{2 p_i\cdot p_a} \right)^{\ep}
\frac{1}{{\bom T}_{b}^2} +
\left(
\frac{4\pi \mu^2}{\mu_F^2} \right)^{\ep} \!\!&=&\!\!
\sum_{i} {\bom T}_i \cdot {\bom T}_b
\; \frac{1}{{\bom T}_{b}^2} \left[
\left(
\frac{4\pi \mu^2}{2 p_i\cdot p_a} \right)^{\ep}
- \left(
\frac{4\pi \mu^2}{\mu_F^2} \right)^{\ep} \;\right] \nonumber \\
\!\!&=&\!\! \ep \;\sum_{i} {\bom T}_i \cdot {\bom T}_b
\;\frac{1}{{\bom T}_{b}^2} \;\ln \frac{\mu_F^2}{2 p_i \cdot p_a}
+ {\cal O}(\ep^2) \;\;.
\eeeq

Inserting Eqs.~(\ref{vabpab},\ref{oep}) into Eq.~(\ref{i2dis1}), adding the
${\bom I}_{(1)}$ contribution in Eq.~(\ref{i1dis1}) and again using colour
charge conservation, we end up with
\beeq
\label{i12dis}
\!\!\!\!&&\!\!\!\!\!\!\delta^{ab} \;
{\bom I}_{(1)}(p_1,...,p_m;p_a,x;\ep) +
{\bom I}_{(2)}^{a,b}(p_1,...,p_m;p_a,x;\ep;\mu_F^2) \nonumber \\
&=& {\bom K}^{a,b}(x) +
{\bom P}^{a,b}(p_1,...,p_m;xp_a,x;\mu_F^2) +
{\cal O}(\ep) \;\;,
\eeeq
where we have defined
\beq
\label{kdef}
{\bom K}^{a,b}(x)
= \frac{\as}{2\pi}
\left\{ \frac{}{} {\overline K}^{ab}(x) - \Kab(x)
+ \delta^{ab} \; \sum_{i} {\bom T}_i \cdot {\bom T}_b
\frac{\gamma_i}{{\bom T}_i^2}
\left[ \left( \frac{1}{1-x} \right)_+ + \delta(1-x) \right] \right\}
\;,
\eeq
\beeq
\label{pdef}
{\bom P}^{a,b}(p_1,...,p_m;xp_a,x;\mu_F^2) =
\frac{\as}{2\pi} \;P^{ab}(x) \;\frac{1}{{\bom T}_{b}^2}
\sum_{i} {\bom T}_i \cdot {\bom T}_b
\;\ln \frac{\mu_F^2}{2 x p_a \cdot p_i} \;\;.
\eeeq

The final result for ${\bom I}^{a,b}(x;\ep)$ is the following
\beeq
\label{iabf}
&&{\bom I}^{a,b}(p_1,...,p_m;p_a,x;\ep;\mu_F^2) =  \delta^{ab} \;
\delta(1-x) \;{\bom I}(p_1,...,p_m,p_a;\ep) \nonumber \\
&& + {\bom K}^{a,b}(x) +
{\bom P}^{a,b}(p_1,...,p_m;xp_a,x;\mu_F^2) +
{\cal O}(\ep) \;\;,
\eeeq
where the insertion operators ${\bom I}, {\bom K}^{a,b}$ and ${\bom P}^{a,b}$
are given in Eqs.~(\ref{i0dis}), (\ref{kdef}) and (\ref{pdef}).
Therefore, using a notation similar to that in Eq.(\ref{dsaf}), we can write
\beeq
\label{dsadis}
&&\int_{m+1} d\sigma_a^{A}(p) + \int_m d\sigma_a^{C}(p;\mu_F^2) =
\int_m \left[ d\sigma_a^{B}(p) \cdot {\bom I}(\ep) \right]  \\
&&+ \sum_b \int_0^1 dx \;
\int_m \left[ {\bom K}^{a,b}(x) \cdot
d\sigma_b^{B}(xp) \right]
+ \sum_b \int_0^1 dx \;
\int_m \left[ {\bom P}^{a,b}(xp,x;\mu_F^2) \cdot
d\sigma_b^{B}(xp) \right] \;\;.\nonumber
\eeeq
Note that all the insertion operators ${\bom I}(\ep), {\bom K}^{a,b}(x),
{\bom P}^{a,b}(xp,x;\mu_F^2)$ depend on the colour charges and flavours of the
QCD partons. However, while this dependence is fully symmetric in
${\bom I}(\ep)$, the operators ${\bom K}^{a,b}(x)$ and
${\bom P}^{a,b}(xp,x;\mu_F^2)$ do depend
asymmetrically on the flavour and colour charge of the incoming parton $p$.
In addition, ${\bom I}(\ep)$ depends on the parton momenta,
${\bom K}^{a,b}(x)$ depends on the momentum fraction $x$ (but not on the
parton momenta) and
${\bom P}^{a,b}(xp,x;\mu_F^2)$ depends on $x$, parton momenta and factorization
scale.

As in the case of processes with no initial-state hadrons, the term
$d\sigma_a^{B}(p) \cdot {\bom I}(\ep)$ in Eq.~(\ref{dsadis}) cancels all the
$\ep$-poles in the virtual contribution $d\sigma_a^{V}(p)$, thus making
the NLO cross section in Eq.~(\ref{dsradis}) finite. The other two terms on the
right-hand side of Eq.~(\ref{dsadis}) are finite remainders that are left
after factorization of the initial-state collinear singularities into the
parton densities.

The operator ${\bom P}^{a,b}(xp,x;\mu_F^2)$ is directly related to the
Altarelli-Parisi probabilities. In particular, from Eq.~(\ref{pdef}) we
can check that it fulfils the relation:
\beq
\frac{\partial {\bom P}^{a,b}(p_1,...,p_m;xp_a,x;\mu_F^2)}{\partial \ln \mu_F^2}
=  \frac{\as}{2\pi} \;P^{ab}(x) \;\frac{1}{{\bom T}_{b}^2}
\sum_{i} {\bom T}_i \cdot {\bom T}_b = - \frac{\as}{2\pi} \;P^{ab}(x) \;\;,
\eeq
where we have used colour-charge conservation
($\sum_{i} {\bom T}_i = - {\bom T}_b$). It follows that it cancels the similar
(and with opposite sign) factorization-scale dependence of the parton densities
$f_a(x,\mu_F^2)$ thus making the hadronic cross section (\ref{1hxs})
$\mu_F$-independent to NLO accuracy.

The operator ${\bom K}^{a,b}(x)$ contains $( \;\;)_+$ and $\delta$
distributions with coefficients $\gamma_i$ of the same type as those appearing
in the
singular operator ${\bom I}(\ep)$ (see Eqs.~(\ref{calvexp},\ref{i0dis})). These
terms, due to colour correlations between the incoming parton and the
final-state partons, are the heritage of the initial-state collinear
divergences originally present in the real contribution $d\sigma^R_a(p)$.
Moreover, ${\bom K}^{a,b}(x)$ depends on the flavour (and colour diagonal)
kernels $\Kab(x)$
and ${\overline K}^{ab}(x)$. The kernel $\Kab(x)$ is related to the definition
of the factorization scheme (cfr.~Eq.~(\ref{saC})) while ${\overline K}^{ab}(x)$
has a close relationship with the parton splitting functions. As a matter of
fact, if we define ${\hat P}^{ \;\prime}_{ab}(x)$ by expanding in $\ep$
the $d$-dimensional Altarelli-Parisi splitting functions in
Eqs.~(\ref{avhpqq}--\ref{avhpgg}):
\beq
\label{avhpep}
< {\hat P}_{ab}(x;\ep) > \;= \;< {\hat P}_{ab}(x;\ep=0) >
- \,\ep \;{\hat P}^{ \;\prime}_{ab}(x) + {\cal O}(\ep^2) \;\;,
\eeq
we can rewrite Eqs.~(\ref{kbarqg}--\ref{kbargg}) as follows
\beeq
\label{kkernel}
&&{\overline K}^{ab}(x) =
{\hat P}^{ \;\prime}_{ab}(x) + P^{ab}_{{\rm reg}}(x) \;\ln\frac{1-x}{x}
\nonumber \\
&& + \;\delta^{ab} \left[ {\bom T}_{a}^2
\left( \frac{2}{1-x} \ln\frac{1-x}{x} \right)_+
- \delta(1-x) \left( \gamma_a + K_a - \frac{5}{6}\pi^2 \;{\bom T}_{a}^2 \right)
\right] \;\;,
\eeeq
where $P^{ab}_{{\rm reg}}(x)$ is the non-singular part of the (four-dimensional)
Altarelli-Parisi probabilities (see Eq.~(\ref{preg}))
and the coefficients $K_a$ are defined in Eq.~(\ref{kcondef}). The contribution
${\hat P}^{ \;\prime}_{ab}(x)$ on the right-hand side of Eq.~(\ref{kkernel}),
is directly related to the dimensional regularization of the initial-state
collinear singularities. Indeed, it comes from the interference of the
$1/\epsilon$ collinear pole and the ${\cal O}(\epsilon)$-contribution to the
$d$-dimensional splitting functions. Also the other terms on the right-hand
side of Eq.~(\ref{kkernel}) have a simple interpretation. After having
factorized the initial-state collinear divergences, the finite remainder
${\overline K}^{ab}(x)$ is proportional to the phase space available for the
emission from the incoming parton. The factor $\ln(1-x)/x$ multiplying
$P^{ab}_{{\rm reg}}(x)$ has this kinematic origin. The same factor controls
radiation that is both collinear and soft and thus it enters into the
$(\; )_+$-distribution in the square bracket.

\subsection{Final formulae}
\label{findis}

Summarizing the results derived in the previous Subsection, we obtain the
following final formulae for jet cross sections involving one initial-state
hadron.

The partonic cross section on the right-hand side of Eq.~(\ref{1hxs}) consists
of a LO and a NLO component. In the case of an incoming parton of flavour $a$
and momentum $p_a$, the explicit expression for the LO component is:
\beq
\label{LOdisfin}
\sigma^{LO}_a(p_a) = \int_m d\sigma^{B}_a(p_a) = \int d\Phi^{(m)}(p_a)
\; \frac{1}{n_c(a)} \;| \cm_{m,a}(p_1, ...,p_m;p_a)|^2
\;F_J^{(m)}(p_1, ...,p_m;p_a) \;\;,
\eeq
where $\cm_{m,a}$ is the tree-level matrix element to produce $m$ final-state
partons, $F_J^{(m)}$ is the jet defining function (it fulfils the properties in
Eqs.~(\ref{fjdiscoll},\ref{fjdis}) in addition to those in
Eqs.~(\ref{fjsoft}--\ref{fjLO})), the factor $1/n_c(a)$ comes from the average
over the colours of the initial-state parton and $d\Phi^{(m)}(p_a)$ collects
all the remaining factors (phase space, flux, spin average)
on the right-hand side of Eq.~(\ref{dsab}). The evaluation of the LO cross
section (\ref{LOdisfin}) is carried out in four space-time dimensions.

The NLO partonic cross section is split into three terms, as in
Eq.~(\ref{sNLO4}):
\beq
\label{sNLOdis}
\sigma^{NLO}_a(p_a;\mu_F^2) =
\sigma^{NLO\,\{m+1\}}_a(p_a) + \sigma^{NLO\,\{m\}}_a(p_a) + \int_0^1 dx \;
{\hat \sigma}^{NLO\,\{m\}}_a(x;xp_a,\mu_F^2) \;\;.
\eeq
The term with $m+1$-parton kinematics is given by
\beeq
\label{NLOdismp}
&&\sigma^{NLO\,\{m+1\}}_a(p_a) =
\int_{m+1} \left[ \left( d\sigma^{R}_a(p_a) \right)_{\ep=0} -
\left( \sum_{{\rm dipoles}} \;d\sigma^{B}_a(p_a) \otimes
\left( \;dV_{{\rm dipole}}
+  dV^{\prime}_{{\rm dipole}} \right)
\right)_{\ep=0} \;\right] \nonumber \\
&&= \int d\Phi^{(m+1)}(p_a) \left\{
\; \frac{1}{n_c(a)} \;| \cm_{m+1,a}(p_1, ...,p_{m+1};p_a)|^2
\;F_J^{(m+1)}(p_1, ...,p_{m+1};p_a) \right. \nonumber \\
&& \left. - \sum_{{\rm dipoles}}
\left( {\cal D} \cdot F^{(m)} \right)(p_1, ...,p_{m+1};p_a) \right\} \;\;,
\eeeq
where $\cm_{m+1,a}$ is the tree-level matrix element with $m+1$ partons in the
final state and
$\sum_{{\rm dipoles}}
\left( {\cal D} \cdot F^{(m)} \right)(p_1, ...,p_{m+1};p_a)$
is the sum of the dipole factors contained into the curly bracket on the
right-hand side of Eq.~(\ref{dsapa}). Note that the $m+1$-parton matrix element
is multiplied by $F_J^{(m+1)}$, the jet function for $m+1$ final-state partons,
while the dipole contributions involve the $m$-parton jet function $F_J^{(m)}$.
All the terms in Eq.~(\ref{NLOdismp}) are
evaluated and integrated in four dimensions.

The NLO contribution with $m$-parton kinematics is exactly like that
in Eq.~(\ref{meefin}) for \ee-type processes, apart from the additional
dependence on the colour and momentum of the incoming parton. Indeed,
combining the virtual cross section with the first term on the right-hand side
of Eq.~(\ref{dsadis}), we obtain
\beeq
\label{NLOdism}
&&\sigma^{NLO\,\{m\}}_a(p_a) = \int_m \left[ d\sigma^{V}_a(p_a) +
d\sigma^{B}_a(p_a) \otimes {\bom I}
\right]_{\ep=0} \nonumber \\
&&=  \int d\Phi^{(m)}(p_a) \left\{  \frac{1}{n_c(a)}
\;| \cm_{m,a}(p_1, ...,p_{m};p_a)|^2_{(1-loop)} \right. \nonumber \\
&& \left. + \;\;{}_{m,a}\!\!<{1, ...., m;a}| \;{\bom I}(\ep)
\;|{1, ...., m;a}>_{m,a} \frac{}{}
\right\}_{\ep=0} F_J^{(m)}(p_1, ...,p_{m};p_a) \;\;,
\eeeq
where $| \cm_{m,a}|^2_{(1-loop)}$ is the one-loop matrix element square and
the colour charge operator ${\bom I}(\ep)$ is given in Eq.~(\ref{i0dis})
(see also Appendix~C). The two terms in the curly bracket have to be separately
computed in $d=4-2\ep$ dimensions and the limit $\ep \to 0$ performed after
having cancelled analytically the $\ep$ poles. At this point the phase space
integration is carried out in four dimensions.

The NLO component involving the one-dimensional convolution with respect to
the longitudinal-momentum fraction $x$ is given by the last two terms on the
right-hand side of Eq.~(\ref{dsadis}):
\beeq
\label{NLOdisx}
&&\!\!\!\!\!\int_0^1 dx \;{\hat \sigma}^{NLO\,\{m\}}_a(x;xp_a,\mu_F^2) = \sum_b
\int_0^1 dx \;
\int_m \left[ d\sigma^{B}_b(xp_a) \otimes \left( {\bom K}
+ {\bom P} \right)^{a,b}(x) \right]_{\ep=0} \nonumber \\
&&\!\!\!\!\!= \sum_b \int_0^1 dx \int d\Phi^{(m)}(xp_a) \;
F_J^{(m)}(p_1, ...,p_{m};xp_a) \nonumber \\
&&\!\!\!\!\!\cdot \;{}_{m,b}\!\!<{1, ...., m;xp_a}| \;\left({\bom K}^{a,b}(x)
+ {\bom P}^{a,b}(xp_a, x;\mu_F^2) \right) \;
|{1, ...., m;xp_a}>_{m,b} \;\;.
\eeeq
The colour-charge operators ${\bom K}$ and  ${\bom P}$ are respectively defined
in Eqs.~(\ref{kdef}) and (\ref{pdef}). Their explicit expressions, as well as
those
of the related flavour kernels $P^{ab}(x), {\overline K}^{ab}(x)$ and
$\Kab(x)$, are recalled in Appendix C. The calculation of Eq.~(\ref{NLOdisx})
is directly performed in four space-time dimensions.

The partonic cross sections in Eqs.~(\ref{LOdisfin}--\ref{NLOdisx}) have to
be convoluted with the parton densities as in Eq.~(\ref{1hxs}), in order to
compute the corresponding hadronic cross sections.

Note that, because of the $x$-dependence of the operators $\bom K$ and
$\bom P$ in Eqs.~(\ref{kdef},\ref{pdef}), the cross section component in
Eq.~(\ref{NLOdisx}) is boost-invariant with respect to the direction of the
incoming momentum $p_a$. Therefore, in the evaluation of the hadronic cross
section, this contribution enters in the form of multiple convolution of
a Born-type partonic cross section with the kernel $\bom K$ or $\bom P$ and
with the parton densities.

\newpage

\setcounter{equation}{0}

\section{Jet cross sections with one final-state identified hadron}
\label{sifra}

Let us now consider fragmentation processes, starting with the simplest case,
which does not involve initial-state hadrons.
According to the definition in Sect.~\ref{NLOxs}, the inclusive cross section
to produce a parton of flavour $a$ and momentum $p_a$ has the following
expression at the Born level
\beeq
\label{dsabinc}
d\sigma_{(incl) a}^{B}(p_a) &=& {\cal N}_{in}
\sum_{\{ m \} } \, d\phi_m(p_1, ...,p_m;Q-p_a)
\;\frac{1}{S_{\{ m \} }} \nonumber \\
&\cdot&|{\cal{M}}_{m+a}(p_a,p_1, ...,p_m) |^2
\;F_J^{(m)}([p_a],p_1, ...,p_m) \;\;.
\eeeq
The notation in Eq.~(\ref{dsabinc}) is similar to that in Eqs.~(\ref{dsmb}) and
(\ref{dsab}). The only relevant difference regards the jet
defining function $F_J^{(m)}([p_a],p_1, ...,p_m)$. Besides being infrared
and collinear safe (see Eqs.~(\ref{fjsoft}--\ref{fjLO})), it should guarantee
the factorizability of final-state collinear singularities. This implies the
following general (i.e.\ for any number $n$ of partons) property
\beq
\label{fjfracol}
F_J^{(n+1)}([p_a],p_1,..,p_i,..,p_{n+1}) \to
F_J^{(n)}([p_a/z],p_1,...,p_{n+1}) \;\;, \;\;\;\;{\rm if} \;\;\;
p_i \to (1/z - 1) p_a \;\;,
\eeq
and the leading-order constraint
\beq
\label{fjfra}
F_J^{(m)}([p_a],p_1,...,p_{m}) \to 0 \;\;, \;\;\;\;{\rm if} \;\;\;
p_i \cdot p_a \to 0 \;\;.
\eeq

\subsection{Implementation of the subtraction procedure}
\label{impfra}

As usual, in order to evaluate the NLO partonic cross section, we rewrite
Eq.~(\ref{afNLO}) in the following form
\beeq
\label{afsub}
\sigma_{(incl) \,a}^{NLO}(p_a) &=& \int_{m+1} \left(
d\sigma_{(incl) \,a}^{R}(p_a) - d\sigma_{(incl) \,a}^{A}(p_a) \right) \\
&+& \left[ \;\int_{m+1} d\sigma_{(incl) \,a}^{A}(p_a) +
\int_m d\sigma_{(incl) \,a}^{V}(p_a) +
\int_m d\sigma_{(incl) \,a}^{C}(p_a;\mu_F^2) \right] \;, \nonumber
\eeeq
where, using the dipole formulae in Eqs.~(\ref{ffa}) and (\ref{aff}),
we define the local counterterm:
\beeq
\label{dsapainc}
d\sigma_{(incl) a}^{A}(p_a) &=& {\cal N}_{in}
\sum_{\{ m+1 \} } \,
d\phi_{m+1}(p_1, ...,p_{m+1};Q-p_a)
\frac{1}{S_{\{ m+1 \} }}  \nonumber \\
&\cdot&\!\!\! \left\{ \sum_{\mathrm{pairs}\atop i,j}
\;\sum_{k\not=i,j} {\cal D}_{ij,k}(p_a,p_1, ...,p_{m+1}) \;
F_J^{(m)}([p_a],p_1, .. {\widetilde p}_{ij},..{\widetilde p}_k, ..,p_{m+1})
\right. \nonumber \\
&+&\!\!\! \left. \sum_{\mathrm{pairs}\atop i,j} \;
{\cal D}_{ij,a}(p_a,p_1, ...,p_{m+1}) \;
F_J^{(m)}([{\widetilde p}_a],p_1, .. {\widetilde p}_{ij}, ..,p_{m+1})
\right.  \nonumber \\
&+&\!\!\! \left. \sum_i \;\sum_{k\not=i}
{\cal D}_{ai,k}(p_a,p_1, ...,p_{m+1}) \;
F_J^{(m)}([{\widetilde p}_{ai}],p_1, .. {\widetilde p}_{k}, ..,p_{m+1})
\right\} \;.
\eeeq
The four-dimensional integrability of $( d\sigma_{(incl) \,a}^{R}(p_a)
- d\sigma_{(incl) \,a}^{A}(p_a) )$ follows in exactly the same way as for
Eqs.~(\ref{xsid}) and (\ref{dsradis}).

To compute the term in the square bracket of
Eq.~(\ref{afsub}),
we write $d\sigma_{(incl) a}^{A}(p_a)$ as follows
\beq
\label{ds123inc}
d\sigma_{(incl) a}^{A}(p_a)= d\sigma_{(incl) a}^{A \prime}(p_a) +
d\sigma_{(incl) a}^{A \prime \prime}(p_a) +
d\sigma_{(incl) a}^{A \prime \prime \prime}(p_a) \;\;,
\eeq
where the three terms on the right-hand side are in
one-to-one correspondence with those in the curly bracket on the
right-hand side of Eq.~(\ref{dsapainc}).

The integration of $d\sigma_{(incl) a}^{A \prime}(p_a)$ can be carried out
analogously to that of $d\sigma^{A}$ in Sect.~\ref{siee} and, thus, we obtain
\beeq
\label{dsap1inc}
\int_{m+1} d\sigma_{(incl) a}^{A \prime}(p_a) &=& - \int_m {\cal N}_{in}
\sum_{\{ m \} }
d\phi_{m}(p_1,..,p_m;Q-p_a)
\; \frac{1}{S_{\{ m \} }} \nonumber \\
&\cdot&
F_J^{(m)}([p_a],p_1, ...,p_m) \sum_i \sum_{k \neq i}
|{\cal{M}}_{m+a}^{i,k}(p_a,p_1,...,p_m) |^2  \nonumber \\
&\cdot&
\frac{\as}{2\pi}
\frac{1}{\Gamma(1-\ep)} \left(
\frac{4\pi \mu^2}{2 p_i\cdot p_k} \right)^{\ep}
\;\frac{1}{{\bom T}_{i}^2} \;{\cal V}_i(\ep) \;\;,
\eeeq
where the functions ${\cal V}_i(\ep)$ are defined in
Eqs.~(\ref{vfep},\ref{vgep}).

The integration of $d\sigma_{(incl) a}^{A \prime \prime}(p_a)$ is similar
to that of $d\sigma_a^{A \prime \prime}(p_a)$ in Sect.~\ref{sidis}. The main
difference is
that the phase space convolution in Eq.~(\ref{psconvfr}) replaces that in
Eq.~(\ref{psconv}). Thus, we find:
\beeq
\label{dsap2inc}
\int_{m+1} d\sigma_{(incl) a}^{A \prime \prime}(p_a)
&=& - \int_m {\cal N}_{in} \int_0^1
\frac{dz}{z^{2-2\ep}} \; \sum_{\{ m \} }
d\phi_{m}(p_1,..,p_m;Q-p_a/z)  \nonumber \\
&\cdot& \frac{1}{S_{\{ m \} }} \;
F_J^{(m)}( \,[p_a/z],p_1, ..,p_m)
\sum_i \; |{\cal{M}}_{m+a}^{i,a}(p_a/z,p_1,
..,p_m) |^2 \nonumber \\
&\cdot&
\frac{\as}{2\pi}
\frac{1}{\Gamma(1-\ep)} \left(
\frac{4\pi \mu^2}{2 p_i\cdot p_a} \right)^{\ep}
\;\frac{1}{{\bom T}_i^2} \;{\overline {\cal V}}_{i}(z;\ep) \;\;,
\eeeq
where we have defined
\beeq
\label{vizep}
{\overline {\cal V}}_i(z;\ep) &\equiv& {\overline {\cal V}}_{qg}(z;\ep) \;,
\;\;\;\;\;\;\;\;\;\;\;\;\;\;\;\;\;\;\;\;\;\;\;\;\;\;\;\; {\rm if}
\;\;i=q,{\bar q}
\;\;, \\
\label{vgzep}
{\overline {\cal V}}_i(z;\ep) &\equiv& \frac{1}{2}
\;{\overline {\cal V}}_{gg}(z;\ep) + N_f \,
{\overline {\cal V}}_{q{\bar q}}(z;\ep)\;, \;\;\; {\rm if} \;\; i=g \;\;.
\eeeq

The integration of $d\sigma_{(incl) a}^{A \prime \prime \prime}(p_a)$
is again analogous to that of $d\sigma_a^{A \prime \prime \prime}(p_a)$
in Sect.~\ref{sidis}, apart from the different phase space convolution in
Eq.~(\ref{psconfrag}) for the dipole $\{ai,k\}$. We find:
\beeq
\label{dsap3inc}
\int_{m+1} d\sigma_{(incl) a}^{A \prime \prime \prime}(p_a) &=&
- \int_m {\cal N}_{in} \sum_{\{ m \} } \int_0^1
\frac{dz}{z^{2-2\ep}} \;\;
d\phi_{m}(p_1,..,p_m;Q-p_a/z) \; \frac{1}{S_{\{ m \} }} \nonumber \\
&\cdot& F_J^{(m)}(\, [p_a/z],p_1,..,p_m)
\; \sum_{k} \sum_{b}
\; |{\cal{M}}_{m+b}^{b,k}(p_a/z,p_1,
...,p_m) |^2 \nonumber \\
&\cdot&
\frac{\as}{2\pi}
\frac{1}{\Gamma(1-\ep)} \left(
\frac{4\pi \mu^2}{2 p_k\cdot p_a} \right)^{\ep}
\;\frac{1}{{\bom T}_{b}^2} \;{\cal V}_{b,a}(z;\ep) \;\;,
\eeeq
where the functions ${\cal V}_{a,ai}(z;\ep)$ are given in
Eqs.~(\ref{cvqgzep}--\ref{cvggzep}).

Combining the results in Eqs.~(\ref{dsap1inc},\ref{dsap2inc},\ref{dsap3inc})
and adding the collinear counterterm in Eq.~(\ref{saCinc}), we have
\beeq
\label{dsAin}
&&\!\!\!\!\!\!\!\!\int_{m+1} d\sigma_{(incl) a}^{A}(p_a) +
\int_m d\sigma_{(incl) a}^{C}(p_a) =
- \sum_{b} \; \int_0^1 \frac{dz}{z^{2-2\ep}} \;\;
\int_m {\cal N}_{in} \nonumber \\
&\cdot& \sum_{\{ m \} }
d\phi_{m}(p_1,..,p_m;Q-p_a/z) \frac{1}{S_{\{ m \} }}
\; F_J^{(m)}(\,[p_a/z],p_1,..,p_m)
\; \frac{\as}{2\pi}
\frac{1}{\Gamma(1-\ep)}
\nonumber \\
&\cdot& \left\{ \delta_{ab} \;\delta(1-z) \sum_i
\sum_{k \neq i}
\; |{\cal{M}}_{m+b}^{i,k}(p_a/z,p_1,
...,p_m) |^2 \;\left(
\frac{4\pi \mu^2}{2 p_i\cdot p_k} \right)^{\ep}
\;\frac{1}{{\bom T}_{i}^2} \;{\cal V}_{i}(\ep) \right.
\nonumber \\
&+& \left. \delta_{ab} \sum_{i}
\; |{\cal{M}}_{m+b}^{i,b}(p_a/z,p_1,
...,p_m) |^2 \;\left(
\frac{4\pi \mu^2}{2 p_i\cdot p_a} \right)^{\ep}
\;\frac{1}{{\bom T}_{i}^2} \;{\overline {\cal V}}_{i}(z;\ep) \right.
\nonumber \\
&+& \left. \sum_{i}
\; |{\cal{M}}_{m+b}^{i,b}(p_a/z,p_1,
...,p_m) |^2 \;\left(
\frac{4\pi \mu^2}{2 p_i\cdot p_a} \right)^{\ep}
\;\frac{1}{{\bom T}_{b}^2} \;{\cal V}_{b,a}(z;\ep) \right.
\nonumber \\
&+& \left.
\; |{\cal{M}}_{m+b}(p_a/z,p_1,
...,p_m) |^2 \;\left[ - \frac{1}{\ep}
\left(
\frac{4\pi \mu^2}{\mu_F^2} \right)^{\ep}
\;P_{ba}(z)  + \Hba(z) \right] \right\} \;\;. \nonumber\\
\!\!\!\!\!\!&\!\!\!\!\!\!& \!\!\!\!\!\!\;
\eeeq
Hence, $\int_{m+1} d\sigma_{(incl) a}^{A}(p_a) +
\int_m d\sigma_{(incl) a}^{C}(p_a)$
is obtained
from the leading order expression $\int_m d\sigma_{(incl) a}^{B}(p_a)$ by
replacing the Born-level matrix element squared
\beq
{}_{m+a}\!\!<{p_a,...}|{p_a,...}>_{m+a} \;,
\eeq
by
\beeq
\sum_b \;\;
{}_{m+b}\!\!<{p_a/z,....}| \;{\bom I}_{b,a}(z;\ep) \;|{p_a/z,...}>_{m+b} \;,
\eeeq
where the insertion operator ${\bom I}(z;\ep)$ depends on the colour charges,
momenta and flavours of the QCD partons:
\beeq
\label{ifrag1}
\!\!\!\!\!\!\!\!\!\!&&{\bom I}_{b,a}(p_a,p_1,...,p_m;z;\ep) = -
\frac{\as}{2\pi}
\frac{1}{\Gamma(1-\ep)} \nonumber \\
\!\!\!\!\!\!\!\!\!\!&&\cdot \left\{ \delta_{ab} \;\delta(1-z)
\sum_i \sum_{k \neq i} {\bom T}_i \cdot {\bom T}_k
\;\left(
\frac{4\pi \mu^2}{2 p_i\cdot p_k} \right)^{\ep}
\;\frac{1}{{\bom T}_{i}^2} \;{\cal V}_{i}(\ep)
+ \delta_{ab} \sum_{i} {\bom T}_i \cdot {\bom T}_b
\;\left(
\frac{4\pi \mu^2}{2 p_i\cdot p_a} \right)^{\ep}
\;\frac{1}{{\bom T}_{i}^2} \;{\overline {\cal V}}_{i}(z;\ep)
\right. \nonumber \\
\!\!\!\!\!\!\!\!\!\!&&+ \left.
\sum_{i} {\bom T}_i \cdot {\bom T}_b
\left(
\frac{4\pi \mu^2}{2 p_i\cdot p_a} \right)^{\ep}
\;\frac{1}{{\bom T}_{b}^2} \;{\cal V}_{b,a}(z;\ep)
- \;\frac{1}{\ep}
\left(
\frac{4\pi \mu^2}{\mu_F^2} \right)^{\ep}
\;P_{ba}(z)  + \Hba(z)  \right\} \;\;.
\eeeq

This insertion operator is similar to that in Eq.~(\ref{idis1}) for the
cross section with a single incoming parton, apart from the
replacements ${\cal V}_{i}(x;\ep) \to {\overline {\cal V}}_{i}(z;\ep), \;
{\cal V}^{a,b}(x;\ep) \to {\cal V}_{b,a}(z;\ep), \; P_{ab}(x) \to P_{ba}(z),
\; \Kab(x) \to \Hba(z)$ (note, in particular, the transposition of the
flavour indices that is involved in these replacements).
Therefore, using the analogues of Eqs.~(\ref{vixmvi},\ref{vabpab}),
namely
\beeq
\label{bvixmvi}
{\overline {\cal V}}_{i}(z;\ep) - \delta(1-z)
\; {\cal V}_{i}(\ep) &=& \gamma_i + {\bom T}_i^2 \;
\left( \frac{2}{1-z} \ln\frac{1}{1-z} \right)_+
\nonumber \\
&-& \gamma_i \left[ \left( \frac{1}{1-z} \right)_+ + \delta(1-z) \right]
+ {\cal O}(\ep) \;\;,
\eeeq
\beeq
\label{bvabpab}
&&\!\!\!\!\!\!\!\!{\cal V}_{b,a}(z;\ep)
+ \frac{1}{\ep} \;P_{ba}(z)
- \delta_{ba} \delta(1-z) \;{\cal V}_{a}(\ep) =
{\overline K}^{ba}(z)
+ 2 P_{ba}(z) \;\ln z \nonumber \\
&-& \delta^{ba} \; {\bom T}_{a}^2
\left( \frac{2}{1-z} \ln\frac{1}{1-z} \right)_+
+ {\cal O}(\ep) \;\;,
\eeeq
and performing the same algebraic manipulations as in Sect.~\ref{sidis},
we end up with the final result:
\beeq
\label{dsainc}
&&\!\!\!\!\!\!\!\!\int_{m+1} d\sigma_{(incl)a}^{A}(p) +
\int_m d\sigma_{(incl)a}^{C}(p;\mu_F^2)
= \int_m \left[ d\sigma_{(incl)a}^{B}(p) \cdot {\bom I}(\ep) \right]  \\
&&\!\!\!\!\!\!\!\!+ \sum_b \int_0^1 \frac{dz}{z^2} \;
\int_m \left[ d\sigma_{(incl)b}^{B}(p/z) \cdot {\bom H}_{b,a}(z)
\right]
+ \sum_b \int_0^1 \frac{dz}{z^2} \;
\int_m \left[ d\sigma_{(incl)b}^{B}(p/z) \cdot {\bom P}_{b,a}(p/z, z;\mu_F^2)
\right] \;\;,\nonumber
\eeeq
where the insertion operator ${\bom I}(\ep)$ is exactly the same as
in Eq.~(\ref{i0dis})
and the insertion operators
${\bom H}_{b,a}(z)$ and
${\bom P}_{b,a}(p/z, z;\mu_F^2)$
are defined as follows
\beeq
\label{hdef}
&&{\bom H}_{b,a}(z)
= \frac{\as}{2\pi}
\left\{ \frac{}{} {\overline K}^{ba}(z) + 3 P_{ba}(z) \;\ln z - \Hba(z)
\right. \nonumber \\
&&\left. +  \; \delta_{ab} \;\sum_{i} {\bom T}_i \cdot {\bom T}_b
\frac{\gamma_i}{{\bom T}_i^2}
\left[ \left( \frac{1}{1-z} \right)_+ + \delta(1-z) - 1 \right]
\right\} \;\;,
\eeeq
\beeq
\label{pfindef}
{\bom P}_{b,a}(p_1,...,p_m;p_a/z,z;\mu_F^2) =
\frac{\as}{2\pi} \;P_{ba}(z) \;\frac{1}{{\bom T}_{b}^2}
\sum_{i} {\bom T}_i \cdot {\bom T}_b
\;\ln \frac{z \mu_F^2}{2 p_a \cdot p_i} \;\;.
\eeeq

Equation (\ref{dsainc}) is the time-like (a single identified parton in the
final state) analogue of Eq.~(\ref{dsadis}) for the space-like (a single parton
in the initial state) case. The contribution
$d\sigma_{(incl)a}^{B}(p) \cdot {\bom I}(\ep)$ cancels all the $\ep$-poles
in $d\sigma_{(incl)a}^{V}(p)$ thus making the NLO cross section (\ref{afsub})
finite in the four-dimensional limit.

The operators ${\bom H}_{b,a}(z)$ and ${\bom P}_{b,a}(p/z, z;\mu_F^2)$
are instead finite for $\ep \to 0$ (for this reason, in Eq.~(\ref{dsainc})
we have replaced the phase space factor $dz/z^{2-2\ep}$ of Eq.~(\ref{dsAin})
with $dz/z^2$) and are similar to the operators
${\bom K}^{a,b}(x)$ and ${\bom P}^{a,b}(xp,x;\mu_F^2)$  in Eq.~(\ref{dsadis}).
Actually, apart from the momentum rescaling $xp \to p/z$,
the only other difference between the two ${\bom P}$ operators is in the
transposition of the flavour indices $a$ and $b$. Therefore, in spite of the
identity $P_{ab}(z) = P^{ab}(z)$ of the Altarelli-Parisi probabilities
for time-like and space-like splittings, in the case of the operators ${\bom P}$
we have ${\bom P}_{b,a}(p,z;\mu_F^2) \neq {\bom P}^{b,a}(p,z;\mu_F^2)$. This
difference is due to the colour correlations ${\bom T}_i \cdot {\bom T}_b$ or,
more precisely, to the fact that momentum fraction and colour flow in opposite
direction in the time-like and space-like cases (Fig.~4).
\begin{figure}
\centerline{\epsfbox{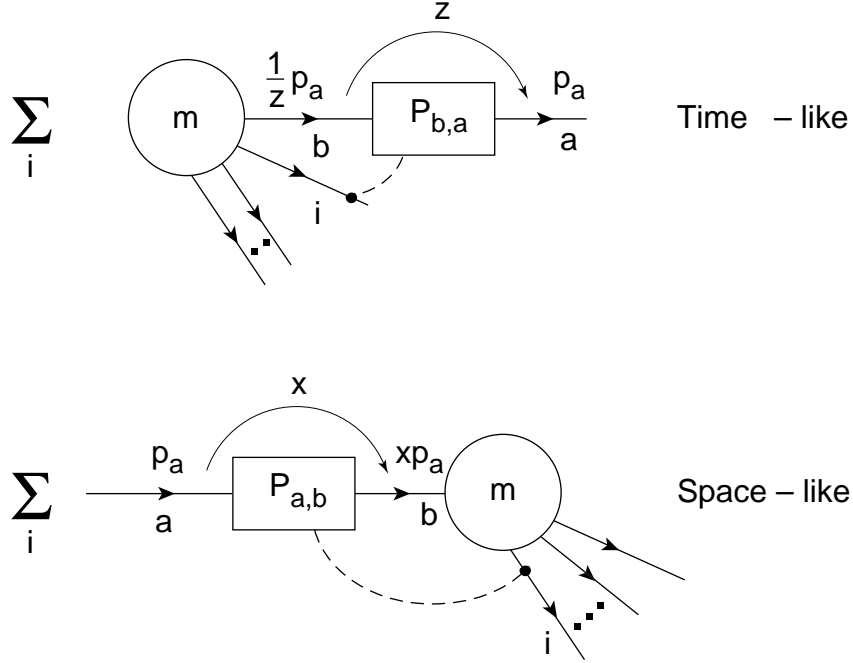}}
\caption{Altarelli-Parisi insertion operators for the time-like
(Eq.~(\protect\ref{pfindef}))
and the space-like
(Eq.~(\protect\ref{pdef}))
cases.}
\end{figure}

Comparing Eq.~(\ref{kdef}) and (\ref{hdef}), we see that this transposition of
the
flavour indices also affects the difference between ${\bom K}$ and ${\bom H}$.
Moreover, ${\bom H}$ differs from ${\bom K}$ by an extra term
$3 P_{ba}(z) \;\ln z - \delta_{ab} \;
\sum_{i} {\bom T}_i \cdot {\bom T}_b \;\gamma_i/{\bom T}_i^2$,
which can be attributed to the kinematic crossing from initial- to
final-state partons.
The appearance of the two
different kernels $\Kab$ and $\Hba$ is trivially related to the choice
of the factorization scheme of collinear singularities (cfr. Eqs.(\ref{saC}) and
(\ref{saCinc})).

\subsection{Final formulae}
\label{finfra}

The calculations carried out in Subsection~\ref{impfra} for jet cross sections
with
a single identified hadron in the final state lead to results that are very
similar to those described in Subsection~\ref{findis} for the
kinematically-crossed process with a single hadron in the initial state.

In summary, the LO parton-level cross section (\ref{afLO}), to be convoluted
with the non-perturbative fragmentation function as in Eq.~(\ref{1fxs}),
can be written as follows
\beq
\label{LOfrafin}
\sigma_{(incl) \,a}^{LO}(p_a) = \int_m d\sigma_{(incl) \,a}^{B}(p_a)
= \int d\Phi^{(m)}(p_a)
\; |{\cal{M}}_{m+a}(p_a,p_1, ...,p_m) |^2
\;F_J^{(m)}([p_a],p_1, ...,p_m) \;\;.
\eeq
Here $a$ and $p_a$ respectively denote the flavour and momentum of the
identified
parton, ${\cal{M}}_{m+a}$ is the tree-level matrix element to produce
$m$ unidentified partons in the final state, $F_J^{(m)}$ is the jet defining
function (its general properties are listed in
Eqs.~(\ref{fjfracol},\ref{fjfra}) and
Eqs.~(\ref{fjsoft}--\ref{fjLO})) and $d\Phi^{(m)}(p_a)$ stands for all the
remaining
phase-space factors on the right-hand side of Eq.~(\ref{dsabinc}).

The NLO partonic cross section is decomposed into three terms. Following the
symbolic notation in Eq.~(\ref{sNLO4}), we write:
\beq
\label{NLOfrafin}
\sigma^{NLO}_{(incl) \,a}(p_a;\mu_F^2) =
\sigma^{NLO\,\{m+1\}}_{(incl) \,a}(p_a) +
\sigma^{NLO\,\{m\}}_{(incl) \,a}(p_a) + \int_0^1 \frac{dz}{z^2} \;
{\hat \sigma}^{NLO\,\{m\}}_{(incl) \,a}(z;p_a/z,\mu_F^2) \;\;.
\eeq

The NLO contribution with $m+1$-parton kinematics has the following explicit
expression
\beeq
\label{NLOframp}
&&\!\!\!\!\sigma^{NLO\,\{m+1\}}_{(incl) \,a}(p_a) =
\int_{m+1} \Biggl[ \left( d\sigma^{R}_{(incl) \,a}(p_a) \right)_{\ep=0}
\Biggr. \nonumber \\
&&- \left.
\left( \sum_{{\rm dipoles}} \;d\sigma^{B}_{(incl) \,a}(p_a) \otimes
\left( \;dV_{{\rm dipole}}
+  dV^{\prime}_{{\rm dipole}} \right)
\right)_{\ep=0} \;\right] \nonumber \\
&&\!\!\!\!= \int d\Phi^{(m+1)}(p_a) \;\Bigl\{
\;| \cm_{m+1+a}(p_a,p_1, ...,p_{m+1})|^2
\;F_J^{(m+1)}([p_a],p_1, ...,p_{m+1})  \nonumber \\
&&\!\!\!\!  - \sum_{{\rm dipoles}}
\left( {\cal D} \cdot F^{(m)} \right)(p_a,p_1, ...,p_{m+1}) \Bigr\} \;\;,
\eeeq
where $\cm_{m+1+a}$ is the tree-level matrix element with $m+1$ unidentified
partons in the final state and
$\sum_{{\rm dipoles}}
\left( {\cal D} \cdot F^{(m)} \right)(p_a,p_1, ...,p_{m+1})$
is the sum of the dipole factors contained into the curly bracket on the
right-hand side of Eq.~(\ref{dsapainc}).

The NLO term with $m$-parton kinematics is obtained by adding the virtual
cross section and the first contribution on the right-hand side of
Eq.~(\ref{dsainc}). This term is completely analogous to that in
Eq.~(\ref{NLOdism}) and is given by:
\beeq
\label{NLOfram}
&&\sigma^{NLO\,\{m\}}_{(incl) \,a}(p_a) = \int_m \left[
d\sigma^{V}_{(incl) \,a}(p_a) +
d\sigma^{B}_{(incl) \,a}(p_a) \otimes {\bom I}
\right]_{\ep=0} \nonumber \\
&&=  \int d\Phi^{(m)}(p_a) \left\{ \frac{}{}
\;| \cm_{m+a}(p_a,p_1, ...,p_{m})|^2_{(1-loop)} \right. \nonumber \\
&& \left. + \;\;{}_{m+a}\!\!<{a,1, ...., m}| \;{\bom I}(\ep)
\;|{a,1, ...., m}>_{m+a} \frac{}{}
\right\}_{\ep=0} \;F_J^{(m)}([p_a],p_1, ...,p_{m}) \;\;.
\eeeq
where the colour-charge operator ${\bom I}(\ep)$ is defined in Eq.~(\ref{i0dis})
(see also Appendix~C).

The third contribution on the right-hand side of Eq.~(\ref{NLOfrafin}) involves
the integration of an $m$-parton cross section with respect to the
fraction $z$ of the longitudinal momentum carried by the identified parton. This
contribution is given by the last two terms on the right-hand side of
Eq.~(\ref{dsainc}):
\beeq
\label{NLOfrax}
&&\!\!\!\!\!\!\!\!\int_0^1 \frac{dz}{z^2}
\;{\hat \sigma}^{NLO\,\{m\}}_{(incl) \,a}(z;p_a/z,\mu_F^2)
= \sum_b
\int_0^1 \frac{dz}{z^2} \;
\int_m \left[ d\sigma^{B}_{(incl) \,b}(p_a/z) \otimes \left( {\bom H}
+ {\bom P} \right)_{b,a}(z) \right]_{\ep=0} \nonumber \\
&&\!\!\!\!\!\!\!\!= \sum_b \int_0^1  \frac{dz}{z^2}
\int d\Phi^{(m)}(p_a/z) \;F_J^{(m)}([p_a/z],p_1, ...,p_{m}) \\
&&\!\!\!\!\!\!\!\!\cdot \;{}_{m+b}\!\!<{p_a/z,1,..., m}|
\;\left({\bom H}_{b,a}(z)
+ {\bom P}_{b,a}(p_a/z, z;\mu_F^2) \right) \;
|{p_a/z,1,..., m}>_{m+b} \;\;, \nonumber
\eeeq
where the colour-charge operators ${\bom H}$ and  ${\bom P}$ are respectively
defined in Eqs.~(\ref{hdef}) and (\ref{pfindef}) (see also Appendix C).

The actual evaluation of Eqs.~(\ref{LOfrafin}), (\ref{NLOframp}) and
(\ref{NLOfrax}) is directly performed in four space-time dimensions.
As for Eq.~(\ref{NLOfram}), one should first cancel analytically
the $\ep$ poles of the one-loop matrix element with those of the insertion
operator ${\bom I}$, perform the limit $\ep \to 0$ and then carry out
the phase-space integration in four space-time dimensions.

\newpage

\setcounter{equation}{0}

\section{Jet cross sections with two initial-state hadrons}
\label{sipp}

In the case of unpolarized scattering, the Born-level cross section with two
incoming partons of flavours $a$ and $b$ and momenta $p_a$ and $p_b$ is the
following
\beeq
\label{dsabb}
d\sigma_{ab}^{B}(p_a,p_b) &=& {\cal N}_{in}
\frac{1}{n_s(a) n_s(b) n_c(a) n_c(b)
\Phi(p_a \cdot p_b)}
\sum_{\{ m \} } \, d\phi_m(p_1, ...,p_m;p_a+p_b+Q)
\nonumber \\
&\cdot& \frac{1}{S_{\{ m \} }} \;|{\cal{M}}_{m,ab}(p_1, ...,p_m;p_a,p_b) |^2
\;F_J^{(m)}(p_1, ...,p_m;p_a,p_b) \;\;.
\eeeq
Here the factor $1/(n_s(a) n_s(b) n_c(a) n_c(b))$  accounts for the average
over the number of initial-state polarizations and colours
and $\Phi(p_a\cdot p_b)$ is the flux factor.
The flux factor fulfils the following scaling property
\beq
\label{fluxeab}
\Phi( \eta p_a\cdot p_b) = \eta \,\Phi(p_a\cdot p_b) \;\;.
\eeq
The function $F_J^{(m)}(p_1, ...,p_m;p_a,p_b)$ defines the jet observable and
has the same properties as the function $F_J^{(m)}(p_1, ...,p_m;p_a)$
in Sect.~\ref{sidis} (more precisely, the
factorizability of initial-state collinear singularities has to be valid with
respect to both $p_a$ and $p_b$).
All the other factors in Eq.~(\ref{dsabb}) are analogous to those in
Eq.~(\ref{dsab}).

\subsection{Implementation of the subtraction procedure}
\label{imppp}

In order to compute the NLO cross section, we can write it as follows
\beeq
\sigma_{ab}^{NLO}(p_a,p_b;\mu_F^2)
&=&  \int_{m+1} \left( d\sigma_{ab}^{R}(p_a,p_b) -
d\sigma_{ab}^{A}(p_a,p_b) \right)  \nonumber \\
&+& \left[ \;\int_{m+1} d\sigma_{ab}^{A}(p_a,p_b) +
\int_m d\sigma_{ab}^{V}(p_a,p_b) +
\int_m d\sigma_{ab}^{C}(p_a,p_b;\mu_F^2) \right] \;\;, \nonumber \\
\;\;
\eeeq
where the local counterterm
$d\sigma_{ab}^{A}(p_a,p_b)$ is given by
\beeq
\label{dsapapb}
d\sigma_{ab}^{A}(p_a,p_b) \!&\!=\!& {\cal N}_{in} \frac{1}{n_s(a) n_s(b)
\Phi(p_ap_b)}
\sum_{\{ m+1 \} } \,
d\phi_{m+1}(p_1, ...,p_{m+1};p_a+p_b+Q)
\frac{1}{S_{\{ m+1 \} }}  \nonumber \\
\!&\!\cdot\!& \left\{ \sum_{\mathrm{pairs}\atop i,j}
\;\sum_{k\not=i,j} {\cal D}_{ij,k}(p_1, ...,p_{m+1};p_a,p_b) \;
F_J^{(m)}(p_1, .. {\widetilde p}_{ij}, {\widetilde p}_k, ..,p_{m+1};p_a,p_b)
\right. \;\; \nonumber \\
\!&\!+\!& \left. \sum_{\mathrm{pairs}\atop i,j} \;
\left[ {\cal D}_{ij}^{a}(p_1, ...,p_{m+1};p_a,p_b) \;
F_J^{(m)}(p_1, .. {\widetilde p}_{ij}, ..,p_{m+1};{\widetilde p}_a,p_b)
+ ( a \leftrightarrow b ) \right]
\right. \;\; \nonumber \\
\!&\!+\!& \left. \sum_i \;\sum_{k\not=i}
\left[ {\cal D}_{k}^{ai}(p_1, ...,p_{m+1};p_a,p_b) \;
F_J^{(m)}(p_1, .. {\widetilde p}_{k}, ..,p_{m+1};{\widetilde p}_{ai},p_b)
+ ( a \leftrightarrow b ) \right] \right. \;\; \nonumber \\
\!&\!+\!&  \left. \sum_i
\left[ {\cal D}^{ai,b}(p_1, ...,p_{m+1};p_a,p_b) \;
F_J^{(m)}({\widetilde p}_1, ...,{\widetilde p}_{m+1};{\widetilde p}_{ai},p_b)
+ ( a \leftrightarrow b ) \right] \right\} \;.
\eeeq
While the first three terms in the curly bracket exactly correspond to those in
the curly bracket of Eq.~(\ref{dsapa}), the last term is the new dipole
contribution introduced in Eq.~(\ref{iif}).

In order to compute the integral of $d\sigma_{ab}^{A}(p_a,p_b)$, we write it
as follows
\beq
\label{ds1234}
d\sigma_{ab}^{A}(p_a,p_b)= d\sigma_{ab}^{A \prime}(p_a,p_b) +
d\sigma_{ab}^{A \prime \prime}(p_a,p_b) +
d\sigma_{ab}^{A \prime \prime \prime}(p_a,p_b) +
d\sigma_{ab}^{A \prime \prime \prime \prime}(p_a,p_b) \;\;,
\eeq
where the four terms on the right-hand side are in
one-to-one correspondence with those in the curly bracket of
Eq.~(\ref{dsapapb}).

The integration over $p_i$ in the first three terms on the right-hand side
of Eq.~(\ref{ds1234}) is completely analogous to that carried out in the
Sect.~\ref{sidis}. Thus we obtain:
\beeq
\label{dsabp1}
\int_{m+1} d\sigma_{ab}^{A \prime}(p_a,p_b) &=& - \int_m {\cal N}_{in}
\frac{1}{n_s(a) n_s(b) \Phi(p_ap_b)}
\sum_{\{ m \} }
d\phi_{m}(p_1,..,p_m;p_a+p_b+Q) \nonumber \\
&\cdot& \frac{1}{S_{\{ m \} }}
F_J^{(m)}(p_1, ...,p_m;p_a,p_b) \sum_i \sum_{k\not=i}
|{\cal{M}}_{m,ab}^{i,k}(p_1,...,p_m;p_a,p_b) |^2  \nonumber \\
&\cdot&
\frac{\as}{2\pi}
\frac{1}{\Gamma(1-\ep)} \left(
\frac{4\pi \mu^2}{2 p_i\cdot p_k} \right)^{\ep}
\;\frac{1}{{\bom T}_{i}^2} \;{\cal V}_i(\ep) \;\;,
\eeeq
\beeq
\label{dsabp2}
\int_{m+1} d\sigma_{ab}^{A \prime \prime}(p_a,p_b) &=& - \int_m {\cal N}_{in}
\int_0^1 dx \;
\frac{1}{n_s(a) n_s(b) \Phi(xp_ap_b)} \sum_{\{ m \} }
d\phi_{m}(p_1,..,p_m;xp_a +p_b+ Q)  \nonumber \\
&\cdot& \frac{1}{S_{\{ m \} }} \;
F_J^{(m)}(p_1, ..,p_m;xp_a,p_b)
\sum_i \; |{\cal{M}}_{m,ab}^{i,a}(p_1,
..,p_m;xp_a,p_b) |^2 \nonumber \\
&\cdot&
\frac{\as}{2\pi}
\frac{1}{\Gamma(1-\ep)} \left(
\frac{4\pi \mu^2}{2 p_i\cdot p_a} \right)^{\ep}
\;\frac{1}{{\bom T}_i^2} \;{\cal V}_{i}(x;\ep) + (a \leftrightarrow b) \;\;,
\eeeq
\beeq
\label{dsabp3}
\int_{m+1} d\sigma_{ab}^{A \prime \prime \prime}(p_a,p_b) &=&
- \int_m {\cal N}_{in} \sum_{\{ m \} } \int_0^1 dx \;\;
d\phi_{m}(p_1,..,p_m;xp_a +p_b+ Q) \nonumber \\
&\cdot& \frac{1}{S_{\{ m \} }} \; F_J^{(m)}(p_1,..,p_m;xp_a,p_b)
\nonumber \\
&\cdot& \sum_{k} \sum_{c}
\frac{1}{n_s(c) n_s(b) \Phi(xp_ap_b)}
\; |{\cal{M}}_{m,cb}^{c,k}(p_1,
...,p_m;xp_a,p_b) |^2 \nonumber \\
&\cdot&
\frac{\as}{2\pi}
\frac{1}{\Gamma(1-\ep)} \left(
\frac{4\pi \mu^2}{2 p_k\cdot p_a} \right)^{\ep}
\;\frac{1}{{\bom T}_{c}^2} \;{\cal V}^{a,c}(x;\ep) +
(a \leftrightarrow b) \;\;,
\eeeq
where the functions ${\cal V}_i(\ep)$, ${\cal V}_i(x;\ep)$,
and ${\cal V}^{a,b}(x;\ep)$
are respectively defined in
Eqs.~(\ref{vfep},\ref{vgep}), Eqs.~(\ref{vixep},\ref{vgxep})
and Eqs.~(\ref{cvqgxep}--\ref{cvggxep}).

Let us now consider the $p_i$-integration of
$d\sigma_{ab}^{A \prime \prime \prime \prime}(p_a,p_b)$.
We first use
the phase space convolution in Eqs.~(\ref{psconvab},\ref{dpiabx}) in order to
factorize the $p_i$ integration. Then we can integrate the splitting function
${\bom V}^{ai,c}$ over $p_i$ and we find:
\beeq
\label{dsab3}
\int_{m+1} d\sigma_{ab}^{A \prime \prime \prime \prime}(p_a,p_b) &=&
- \int_m {\cal N}_{in} \int_0^1 dx \;
\sum_{\{ m+1 \} }
\sum_i \frac{1}{n_s({\widetilde {ai}}) n_s(b) \Phi(xp_ap_b)}
\nonumber \\
&\cdot& d\phi_{m}(p_1, ..,p_m;xp_a +p_b+ Q)
\; \frac{1}{S_{\{ m+1 \} }}
\nonumber \\
&\cdot&
F_J^{(m)}(p_1, ..,p_m;xp_a,p_b)
\; |{\cal{M}}_{m,{\widetilde {ai}}b}^{ai,b}(p_1,
...,p_m;xp_a,p_b) |^2 \nonumber \\
&\cdot&
\frac{\as}{2\pi}
\frac{1}{\Gamma(1-\ep)} \left(
\frac{4\pi \mu^2}{2 p_a \cdot p_b} \right)^{\ep}
\;\frac{1}{{\bom T}_{ai}^2} \;{\widetilde {\cal V}}^{a,ai}(x;\ep)
+ (a \leftrightarrow b) \;\;,
\eeeq
where the functions ${\widetilde {\cal V}}^{a,ai}(x;\ep)$ are given in
Eq.~(\ref{wtcalv}). In order to rewrite Eq.~(\ref{dsab3}) in terms of a sum
over $m$-parton configurations, we have to perform the corresponding counting
of symmetry factors. However this counting is exactly the same as that already
considered in Sect.~\ref{sidis} (see Eqs.~(\ref{sm1},\ref{sfi}))
for the case in which the spectator is a final-state
parton $k$. Thus we obtain:
\beeq
\label{dsabp4}
\int_{m+1} d\sigma_{ab}^{A \prime \prime \prime \prime}(p_a,p_b) &=&
- \int_m {\cal N}_{in} \sum_{\{ m \} } \int_0^1 dx \;\;
d\phi_{m}(p_1,..,p_m;xp_a +p_b+ Q) \; \frac{1}{S_{\{ m \} }} \nonumber \\
&\cdot& F_J^{(m)}(p_1,..,p_m;xp_a,p_b)
\nonumber \\
&\cdot& \sum_{c}
\frac{1}{n_s(c) n_s(b) \Phi(xp_ap_b)}
\; |{\cal{M}}_{m,cb}^{c,b}(p_1,
...,p_m;xp_a,p_b) |^2 \nonumber \\
&\cdot&
\frac{\as}{2\pi}
\frac{1}{\Gamma(1-\ep)} \left(
\frac{4\pi \mu^2}{2 p_a\cdot p_b} \right)^{\ep}
\;\frac{1}{{\bom T}_{c}^2} \;{\widetilde {\cal V}}^{a,c}(x;\ep)
+ (a \leftrightarrow b)\;\;.
\eeeq

Collecting Eqs.~(\ref{dsabp1},\ref{dsabp2},\ref{dsabp3},\ref{dsabp4})
and adding Eq.~(\ref{sabC}), we can write the following expression
\beeq
\label{dsabac}
&&\!\!\!\!\!\!\!\!\int_{m+1} d\sigma_{ab}^{A}(p_a,p_b) +  \int_m
d\sigma_{ab}^{C}(p_a,p_b;\mu_F^2) =
\sum_{c,d} \; \int_0^1 dx \;\;\int_0^1 dy \;\;
\int_m {\cal N}_{in}
\frac{1}{n_s(c) n_s(d) \Phi(xyp_ap_b)} \nonumber \\
&\cdot& \sum_{\{ m \} }
d\phi_{m}(p_1,..,p_m;xp_a +yp_b+ Q) \;\frac{1}{S_{\{ m \} }}
\; F_J^{(m)}(p_1,..,p_m;xp_a,yp_b)
\nonumber \\
&\cdot& {}_{m,cd}\!\!\bra{1,..,m;a,b}
{\bom I}^{ab,cd}(x,y;\ep)
\ket{1,..,m;a,b}_{m,cd} \;\;.
\eeeq
Here
$\int_{m+1} d\sigma_{ab}^{A}(p_a,p_b)
+ \int_m d\sigma_{ab}^{C}(p_a,p_b;\mu_F^2)$ is obtained from the leading-order
contribution $d\sigma_{ab}^{B}(xp_a,yp_b)$ by
replacing the corresponding Born-level matrix element squared
\beq
\frac{1}{n_s(a) n_s(b)} \;{}_{m,ab}\!\!<{....}|{....}>_{m,ab} \;,
\eeq
by
\beeq
\sum_{c,d} \frac{1}{n_s(c) n_s(d)}
\;\;{}_{m,cd}\!\!<{....}| \;{\bom I}^{ab,cd}(x,y;\ep) \;|{....}>_{m,cd} \;,
\eeeq
and performing the $x$ and $y$ integrations.
The insertion operator ${\bom I}(x,y;\ep)$ depends on the colour charges,
momenta and flavours of the QCD partons. Its explicit expression can be
written as follows
\beeq
\label{ipp1}
\!\!&&\!\!\!\!\!\!\!\!{\bom I}^{ab,cd}(p_1,...,p_m;p_a,x;p_b,y;\ep;\mu_F^2) =
\delta^{ac} \;\delta^{bd} \;\delta(1-x) \;\delta(1-y)
\; {\bom I}(p_1,...,p_m,p_a,p_b;\ep) \nonumber \\
\!\!&&\!\!\!\!\!\!\!\!+ \;\delta^{ac} \;\delta^{bd} \;
\left [ \delta(1-y) \;{\bom I}_{(1)}(p_1,...,p_m;p_a,x;\ep) +
\delta(1-x) \;{\bom I}_{(1)}(p_1,...,p_m;p_b,y;\ep) \right]
\nonumber \\
\!\!&&\!\!\!\!\!\!\!\!+ \left[ \delta^{bd} \;\delta(1-y)
{\bom I}_{(2)}^{a,c}(p_1,...,p_m;p_a,x;p_b;\ep;\mu_F^2)
+\delta^{ac} \;\delta(1-x)
{\bom I}_{(2)}^{b,d}(p_1,...,p_m;p_a;p_b,y;\ep;\mu_F^2)
\right] \;\;, \nonumber \\
&~&
\eeeq
where
\beeq
\label{i0pp}
&&{\bom I}(p_1,...,p_m,p_a,p_b;\ep) =  - \;\frac{\as}{2\pi}
\frac{1}{\Gamma(1-\ep)} \left\{ \sum_i
\frac{1}{{\bom T}_{i}^2} \;{\cal V}_{i}(\ep)
\left[ \sum_{k \neq i} {\bom T}_i \cdot {\bom T}_k
\;\left(
\frac{4\pi \mu^2}{2 p_i\cdot p_k} \right)^{\ep} \right. \right.
\nonumber \\
&& + \left. \left. {\bom T}_i \cdot {\bom T}_a
\;\left(
\frac{4\pi \mu^2}{2 p_i\cdot p_a} \right)^{\ep}
+ {\bom T}_i \cdot {\bom T}_b
\;\left(
\frac{4\pi \mu^2}{2 p_i\cdot p_b} \right)^{\ep}
\; \right] \right. \nonumber \\
&& \left.  +
\frac{1}{{\bom T}_{a}^2} \;{\cal V}_{a}(\ep)
\left[ \sum_{i} {\bom T}_i \cdot {\bom T}_a
\left(
\frac{4\pi \mu^2}{2 p_i\cdot p_a} \right)^{\ep}
+ {\bom T}_b \cdot {\bom T}_a
\left(
\frac{4\pi \mu^2}{2 p_b\cdot p_a} \right)^{\ep} \; \right] \right.
\nonumber \\
&& \left.
+ \frac{1}{{\bom T}_{b}^2} \;{\cal V}_{b}(\ep)
\left[ \sum_{i} {\bom T}_i \cdot {\bom T}_b
\left(
\frac{4\pi \mu^2}{2 p_i\cdot p_b} \right)^{\ep}
+ {\bom T}_b \cdot {\bom T}_a
\left(
\frac{4\pi \mu^2}{2 p_b\cdot p_a} \right)^{\ep} \; \right]
\right\} \;\;,
\eeeq
\beeq
\label{i1pp}
{\bom I}_{(1)}(p_1,...,p_m;p_a,x;\ep) &=&
- \;\frac{\as}{2\pi} \frac{1}{\Gamma(1-\ep)}
\sum_{i} {\bom T}_i \cdot {\bom T}_a
\;\left(
\frac{4\pi \mu^2}{2 p_i\cdot p_a} \right)^{\ep}
\frac{1}{{\bom T}_{i}^2} \nonumber \\
&\cdot& \Bigl[ {\cal V}_{i}(x;\ep) - \delta(1-x)
\; {\cal V}_{i}(\ep) \Bigr] \;\;,
\eeeq
\beeq
\label{i2pp}
\!\!\!\!\!\!\!\!\!\!\!\!\!\!\!&&\!\!\!\!\!\!{\bom I}_{(2)}^{a,c}
(p_1,...,p_m;p_a,x;p_b;\ep;\mu_F^2) =
- \;\frac{\as}{2\pi} \frac{1}{\Gamma(1-\ep)}
\left\{ \sum_{i} {\bom T}_i \cdot {\bom T}_c
\;\left(
\frac{4\pi \mu^2}{2 p_i\cdot p_a} \right)^{\ep}
\;\frac{1}{{\bom T}_{c}^2} \right. \nonumber \\
\!\!\!\!\!\!\!\!\!\!\!\!\!\!\!&\cdot&\!\!\!\! \left.
\Bigl[ \;{\cal V}^{a,c}(x;\ep)
- \delta^{ac} \;\delta(1-x) \;{\cal V}_{a}(\ep) \Bigr]
+ {\bom T}_b \cdot {\bom T}_c
\;\left(
\frac{4\pi \mu^2}{2 p_b\cdot p_a} \right)^{\ep}
\;\frac{1}{{\bom T}_{c}^2} \right. \nonumber \\
\!\!\!\!\!\!\!\!\!\!\!\!\!\!\!&\cdot&\!\!\!\! \left.
\Bigl[ \;{\widetilde {\cal V}}^{a,c}(x;\ep)
- \delta^{ac} \;\delta(1-x) \;{\cal V}_{a}(\ep) \Bigr]
+ \left[ \; - \frac{1}{\ep}
\;\left(
\frac{4\pi \mu^2}{\mu_F^2} \right)^{\ep}
\;P^{ac}(x)  + \Kac(x) \right]  \right\} \;\;. \nonumber \\
&~&
\eeeq
All the contributions that are not proportional to $\delta(1-x)$
in Eqs.~(\ref{i1pp}) and (\ref{i2pp}) respectively come from Eqs.~(\ref{dsabp2})
and (\ref{sabC},\ref{dsabp3},\ref{dsabp4}). The operator
${\bom I}$ in Eq.~(\ref{i0pp}) instead contains all the terms coming from
Eq.~(\ref{dsabp1}) plus those proportional to $\delta(1-x)$ that have been
subtracted in Eqs.~(\ref{i1pp}) and (\ref{i2pp}).

We see that ${\bom I}(p_1,...,p_m,p_a,p_b;\ep)$ in Eq.~(\ref{i0pp}) is exactly
like that in Eq.~(\ref{iee}), apart from depending on the additional
initial-state partons $a$ and $b$. Therefore it cancels all the
singularities in
the virtual contribution $\int_m d\sigma_{ab}^V(p_a,p_b)$.

The other
two operators ${\bom I}_{(1)}$ and ${\bom I}_{(2)}$ contribute as finite
counterterms.
As a matter of fact, the insertion operator ${\bom I}_{(1)}$
in Eq.~(\ref{i1pp}) is exactly the same as the insertion operator in
Eqs.~(\ref{i1dis}, \ref{i1dis1}) for the case of cross sections with a single
incoming
parton (note, however, that in Eq.~(\ref{i1pp}) colour-charge conservation
reads $\sum_i {\bom T}_i = - ({\bom T}_a + {\bom T}_b )$ ).
As for the operator ${\bom I}_{(2)}$,
in order to show that it is finite itself for $\ep \to 0$,
we rewrite Eq.~(\ref{i2pp}) as follows
\beeq
\label{i2pp1}
&&\!\!\!\!{\bom I}_{(2)}^{a,c}(p_1,...,p_m;p_a,x;p_b;\ep;\mu_F^2) =
- \;\frac{\as}{2\pi} \frac{1}{\Gamma(1-\ep)}
\left\{ \left[ \sum_{i} {\bom T}_i \cdot {\bom T}_c
\;\left(
\frac{4\pi \mu^2}{2 p_i\cdot p_a} \right)^{\ep} \right. \right. \nonumber \\
&+& \left. \left. {\bom T}_b \cdot {\bom T}_c
\;\left(
\frac{4\pi \mu^2}{2 p_b\cdot p_a} \right)^{\ep} \;\right]
\;\frac{1}{{\bom T}_{c}^2}
\left[ \;{\cal V}^{a,c}(x;\ep)
+ \frac{1}{\ep} \;P^{ac}(x)
- \delta^{ac} \;\delta(1-x) \;{\cal V}_{a}(\ep) \right] \right.
\nonumber \\
&-& \left.
\left[ \sum_{i} {\bom T}_i \cdot {\bom T}_c
\;\left(
\frac{4\pi \mu^2}{2 p_i\cdot p_a} \right)^{\ep}
\;\frac{1}{{\bom T}_{c}^2}
+ {\bom T}_b \cdot {\bom T}_c
\;\left(
\frac{4\pi \mu^2}{2 p_b\cdot p_a} \right)^{\ep}
\;\frac{1}{{\bom T}_{c}^2}
+ \;\left(
\frac{4\pi \mu^2}{\mu_F^2} \right)^{\ep} \;\right]
\frac{1}{\ep}
\;P^{ac}(x) \right. \nonumber \\
&+& \left. {\bom T}_b \cdot {\bom T}_c
\;\left(
\frac{4\pi \mu^2}{2 p_b\cdot p_a} \right)^{\ep}
\;\frac{1}{{\bom T}_{c}^2}
\left[ \;{\widetilde {\cal V}}^{a,c}(x;\ep)
- {\cal V}^{a,c}(x;\ep) \right]
+ \Kab(x)  \right\} \;\;.
\eeeq
Then, the first term in the curly bracket of Eq.~(\ref{i2pp1}) gives:
\beeq
- {\overline K}^{ac}(x) - P^{ac}(x) \ln x
+ \delta^{ac} \; {\bom T}_{a}^2
\left[
\left( \frac{2}{1-x} \ln\frac{1}{1-x} \right)_+ + \frac{2}{1-x} \ln(2-x) \right]
+ {\cal O}(\ep) \;\;,
\eeeq
where we have used Eq.~(\ref{vabpab}) and charge conservation ($\sum_i
{\bom T}_{i} + {\bom T}_{b} = - {\bom T}_{c}$). From the
second term in the curly bracket of Eq.~(\ref{i2pp1}) we obtain:
\beeq
&-& \left[ \sum_{i} {\bom T}_i \cdot {\bom T}_c
\; \frac{1}{{\bom T}_{c}^2} \;\ln \frac{\mu^2_F}{2 p_i\cdot p_a}
+ {\bom T}_b \cdot {\bom T}_c
\; \frac{1}{{\bom T}_{c}^2}
\;\ln \frac{\mu^2_F}{2 p_b\cdot p_a} \right] \;P^{ac}(x)
+ {\cal O}(\ep) \;\;, \nonumber \\
&~&
\eeeq
where we have used charge conservation and performed the $\ep$ expansion
as in Eq.~(\ref{oep}).
Finally, the third term in the curly bracket of Eq.~(\ref{i2pp1}) gives:
\beeq
&&{\bom T}_b \cdot {\bom T}_c
\; \frac{1}{{\bom T}_{c}^2} {\widetilde K}^{ac}(x)
+ \delta^{ac} \;{\bom T}_b \cdot {\bom T}_a
\left[ \left( \frac{2}{1-x} \ln\frac{1}{1-x} \right)_+ +
\frac{2}{1-x} \ln(2-x) \right]
+ {\cal O}(\ep) \;\;, \nonumber \\
&&
\eeeq
where we have used Eq.~(\ref{wtcalv}).
Collecting these results we find:
\beeq
\!\!\!\!&&{\bom I}_{(2)}^{a,c}(p_1,...,p_m;p_a,x;p_b;\ep;\mu_F^2) =
- \frac{\as}{2\pi} \left\{ - {\overline K}^{ac}(x) - P^{ac}(x) \ln x
+ \Kac(x)
\right.
\nonumber \\
\!\!\!\!&+& \left. \delta^{ac} \;{\bom T}_{a}^2 \left[
\left( \frac{2}{1-x} \ln\frac{1}{1-x} \right)_+ + \frac{2}{1-x}
\ln (2-x) \right]
\right. \nonumber \\
\!\!\!\!&-& \left. \left[ \sum_{i} {\bom T}_i \cdot {\bom T}_c
\;\frac{1}{{\bom T}_{c}^2}
\ln \frac{\mu_F^2}{2 p_i \cdot p_a} +
{\bom T}_b \cdot {\bom T}_c
\;\frac{1}{{\bom T}_{c}^2}
\ln \frac{\mu_F^2}{2 p_b \cdot p_a} \right]
\;P^{ac}(x) \right.  \\
\!\!\!\!&+& \left.
{\bom T}_b \cdot {\bom T}_c
\; \frac{1}{{\bom T}_{c}^2} {\widetilde K}^{ac}(x)
+ \delta^{ac} \;{\bom T}_b \cdot {\bom T}_a
\left[ \left( \frac{2}{1-x} \ln\frac{1}{1-x} \right)_+ +
\frac{2}{1-x} \ln(2-x) \right] \right\}
+ {\cal O}(\ep) \;\;, \nonumber
\eeeq
and, adding ${\bom I}_{(1)}$,
we can write the following final expression
\beeq
\label{dsahad}
&&\int_{m+1} d\sigma_{ab}^{A}(p,{\bar p}) +
\int_m d\sigma_{ab}^{C}(p,{\bar p};\mu_F^2) =
\int_m \left[ d\sigma_{ab}^{B}(p,{\bar p}) \cdot {\bom I}(\ep)
\right]  \\
&&+ \sum_{a'} \int_0^1 dx \;
\int_m \left[ {\bom K}^{a,a'}(x) \cdot
d\sigma_{a'b}^{B}(xp,{\bar p}) \right]
+ \sum_{a'} \int_0^1 dx \;
\int_m \left[ {\bom P}^{a,a'}(xp,x;\mu_F^2) \cdot
d\sigma_{a'b}^{B}(xp,{\bar p}) \right] \nonumber \\
&&+ \sum_{b'} \int_0^1 dx \;
\int_m \left[ {\bom K}^{b,b'}(x) \cdot
d\sigma_{ab'}^{B}(p,x{\bar p}) \right]
+ \sum_{b'} \int_0^1 dx \;
\int_m \left[ {\bom P}^{b,b'}(x{\bar p},x;\mu_F^2) \cdot
d\sigma_{ab'}^{B}(p,x{\bar p}) \right]
\;\;, \nonumber
\eeeq
where ${\bom I}(\ep)$ is given by Eq.~(\ref{i0pp}) and the insertion operators
${\bom K}^{a,b}(x)$ and ${\bom P}^{a,b}(xp,x;\mu_F^2)$ are:
\beeq
\label{kdefhad}
\!\!\!\!\!\!&&{\bom K}^{a,a'}(x)
= \frac{\as}{2\pi}
\left\{ \frac{}{} {\overline K}^{aa'}(x) - \KFS{aa'}(x) \right.  \\
\!\!\!\!\!\!&& \left.
+ \;\delta^{aa'} \; \sum_{i} {\bom T}_i \cdot {\bom T}_a
\;\frac{\gamma_i}{{\bom T}_i^2} \left[
\left( \frac{1}{1-x} \right)_+ + \delta(1-x) \right]
\right\} -
\frac{\as}{2\pi} {\bom T}_b \cdot {\bom T}_{a'} \frac{1}{{\bom T}_{a'}^2}
{\widetilde K}^{aa'}(x) \;\;, \nonumber
\eeeq
\beeq
\label{pdefpp}
&&{\bom P}^{a,a'}(p_1,...,p_m,p_b;xp_a,x;\mu_F^2)
\nonumber \\ && \hspace{2em} =
\frac{\as}{2\pi} \;P^{aa'}(x) \;\frac{1}{{\bom T}_{a'}^2}
\left[
\sum_{i} {\bom T}_i \cdot {\bom T}_{a'}
\;\ln \frac{\mu_F^2}{2 xp_a \cdot p_i}
+ {\bom T}_b \cdot {\bom T}_{a'}
\;\ln \frac{\mu_F^2}{2 xp_a \cdot p_b} \right] .
\eeeq

The operators ${\bom I}(\ep)$ and ${\bom P}^{a,a'}(xp,x;\mu_F^2)$ are completely
analogous to those in Eqs.~(\ref{i0dis}) and (\ref{pdef}) for the case with a
single
incoming parton, apart from the trivial dependence on the additional
initial-state parton. Note, instead, a new feature of the operator ${\bom K}$.
While the term in the curly bracket on the right-hand side of
Eq.~(\ref{kdefhad}) is equal to that in Eq.~(\ref{kdef}), in the present case
there is an additional contribution to ${\bom K}$, namely
${\widetilde K}^{aa'}(x)$,
due to {\em parton-parton correlations\/} in the initial state.

\subsection{Final formulae}
\label{finpp}

The results of the previous Subsection can be combined into the following
final expressions for the jet cross sections in hadron-hadron
scattering processes.

The hadron-level cross section is obtained by convoluting the partonic
cross sections on the right-hand side of Eq.~(\ref{2hxs}) with the parton
densities. The LO parton-level cross section is given by
\beeq
\label{LOppfin}
&&\sigma^{LO}_{ab}(p_a,p_b) = \int_m d\sigma^{B}_{ab}(p_a,p_b)  \\
&&= \int d\Phi^{(m)}(p_a,p_b)
\; \frac{1}{n_c(a) n_c(b)} \;| \cm_{m,ab}(p_1, ...,p_m;p_a,p_b)|^2
\;F_J^{(m)}(p_1, ...,p_m;p_a,p_b) \;\;, \nonumber
\eeeq
where $a$ and $b$ denote the flavours of the incoming partons, $p_a$ and
$p_b$ are their momenta and $n_c(a), n_c(b)$ are their number of colours.
The matrix element $| \cm_{m,ab}|^2$ is the square of the tree-level amplitude
to produce $m$ final-state partons and $F_J^{(m)}$ is the function that
defines the jet observable we want to compute (the properties that $F_J$ has to
fulfil are given in Eqs.~(\ref{fjsoft}--\ref{fjLO})) and
Eqs.~(\ref{fjdiscoll},\ref{fjdis})). All the other phase space factors on
the right-hand side of Eq.~(\ref{dsabb}) are collected into the factor
$d\Phi^{(m)}(p_a,p_b)$.

According to the general notation in Eq.~(\ref{sNLO4}), the full
NLO partonic cross section is obtained by adding three different types of
contribution:
\beeq
\label{sNLOppfin}
\sigma^{NLO}_{ab}(p_a,p_b;\mu_F^2) &=&
\sigma^{NLO\,\{m+1\}}_{ab}(p_a,p_b) + \sigma^{NLO\,\{m\}}_{ab}(p_a,p_b) \\
&+&
\int_0^1 dx \;
\left[ \;{\hat \sigma}^{NLO\,\{m\}}_{ab}(x;xp_a,p_b,\mu_F^2)
+ {\hat \sigma}^{NLO\,\{m\}}_{ab}(x;p_a,xp_b,\mu_F^2) \;\right] \;\;. \nonumber
\eeeq

The first contribution has $m+1$-parton kinematics and is given by the
following expression
\beeq
\label{NLOppmp}
\!\!\!\!\!\!\!\!\!&&\sigma^{NLO\,\{m+1\}}_{ab}(p_a,p_b) =
\int_{m+1} \left[ \frac{}{} \left( d\sigma^{R}_{ab}(p_a,p_b) \right)_{\ep=0}
\right. \nonumber \\
\!\!\!\!\!\!\!\!\!&&- \left.
\left( \frac{}{} \right.
\sum_{{\rm dipoles}}  \;d\sigma^{B}_{ab}(p_a,p_b) \otimes
\left( \;dV_{{\rm dipole}}
+  dV^{\prime}_{{\rm dipole}} \right) \left. \frac{}{}
\right)_{\!\ep=0} \right. \left. \frac{}{}  \;\right] \nonumber \\
\!\!\!\!\!\!\!\!\!&&= \int d\Phi^{(m+1)}(p_a,p_b) \left\{
\; \frac{1}{n_c(a) n_c(b)} \;| \cm_{m+1,ab}(p_1, ...,p_{m+1};p_a,p_b)|^2
\;F_J^{(m+1)}(p_1, ...,p_{m+1};p_a,p_b) \right. \nonumber \\
\!\!\!\!\!\!\!\!\!&&- \left.  \sum_{{\rm dipoles}}
\left( {\cal D} \cdot F^{(m)} \right)(p_1, ...,p_{m+1};p_a,p_b) \right\} \;\;,
\eeeq
where $\cm_{m+1,ab}$ is the tree-level matrix element with $m+1$ partons in the
final state and $\sum_{{\rm dipoles}}
\left( {\cal D} \cdot F^{(m)} \right)(p_1, ...,p_{m+1};p_a,p_b)$
is the sum of the dipole factors contained into the curly bracket on the
right-hand side of Eq.~(\ref{dsapapb}).

The NLO contribution with $m$-parton kinematics is obtained by adding the
virtual cross section and the first term on the right-hand side
of Eq.~(\ref{dsahad}). As in all the other scattering processes, its explicit
expression is given in terms of the square of the one-loop matrix element
$| \cm_{m,ab}|^2_{(1-loop)}$ and of the insertion operator ${\bom I}(\ep)$:
\beeq
\label{NLOppm}
&&\sigma^{NLO\,\{m\}}_{ab}(p_a,p_b) = \int_m \left[ d\sigma^{V}_{ab}(p_a,p_b) +
d\sigma^{B}_{ab}(p_a,p_b) \otimes {\bom I}
\right]_{\ep=0} \nonumber \\
&&=  \int d\Phi^{(m)}(p_a,p_b) \left\{  \frac{1}{n_c(a) n_c(b)}
\;| \cm_{m,ab}(p_1, ...,p_{m};p_a,p_b)|^2_{(1-loop)} \right.  \\
&& \left. + \;\;{}_{m,ab}\!\!<{1, ...., m;a,b}\,| \;{\bom I}(\ep)
\;|{1, ...., m;a,b}>_{m,ab} \frac{}{}
\right\}_{\ep=0} \;F_J^{(m)}(p_1, ...,p_{m};p_a,p_b) \;\;. \nonumber
\eeeq
In the present case, the colour-charge operator ${\bom I}(\ep)$ is explicitly
written down in Eq.~(\ref{i0pp}) (see also Appendix~C).

The third term on the right-hand side of Eq.~(\ref{sNLOppfin}) comes from the
second and third lines on the right-hand side of Eq.~(\ref{dsahad}) and
contains two
contributions that are similar to that involved in the processes with a single
incoming hadron (cfr. Eq.~(\ref{sNLOdis})). Each of these contributions is
obtained
by integrating a cross section with $m$-parton kinematics with respect to the
fraction $x$ of the longitudinal momentum carried by one of the incoming
partons. When this parton is the parton $a$, we explicitly have:
\beeq
\label{NLOppx}
&&\!\!\!\int_0^1 dx \;{\hat \sigma}^{NLO\,\{m\}}_{ab}(x;xp_a,p_b,\mu_F^2)
= \sum_{a'}
\int_0^1 dx \;
\int_m \left[ d\sigma^{B}_{a'b}(xp_a,p_b) \otimes \left( {\bom K}
+ {\bom P} \right)^{a,a'}(x) \right]_{\ep=0} \nonumber \\
&&\!\!\!= \sum_{a'} \int_0^1 dx \int d\Phi^{(m)}(xp_a,p_b) \;
F_J^{(m)}(p_1, ...,p_{m};xp_a,p_b) \\
&&\!\!\!\cdot \;
{}_{m,a'b}\!\!<{1, ...., m;xp_a,p_b}| \;\left({\bom K}^{a,a'}(x)
+ {\bom P}^{a,a'}(xp_a, x;\mu_F^2) \right) \;
|{1, ...., m;xp_a,p_b}>_{m,a'b} \;\;, \nonumber
\eeeq
where the colour-charge operators ${\bom K}$ and  ${\bom P}$ are respectively
defined in Eqs.~(\ref{kdefhad}) and (\ref{pdefpp}) (see also Appendix C).
The expression
for ${\hat \sigma}^{NLO\,\{m\}}_{ab}(x;p_a,xp_b,\mu_F^2)$ is completely
analogous to Eq.~(\ref{NLOppx}), apart from the replacements $xp_a \to p_a,
p_b \to xp_b$ and $\sum_{a'} \to \sum_{b'}$ (as in Eq.~(\ref{dsahad})). Note
that the
right-hand side of Eq.~(\ref{NLOppx}) has exactly the same structure as in
Eq.~(\ref{NLOdisx}) for the case with a single incoming parton. However,
we should
recall that the colour-charge operator ${\bom K}^{a,a'}$ entering into
Eq.~(\ref{NLOppx}) differs from that appearing into Eq.~(\ref{NLOdisx}) by
the additional
correlation term ${\widetilde K}^{a,a'}$ (see Eq.~(\ref{kdefhad})), which
is due to the presence of the other incoming parton $b$.

Equations (\ref{LOppfin}), (\ref{NLOppmp}) and (\ref{NLOppx}) are directly
evaluated in four space-time dimensions.
As for Eq.~(\ref{NLOppm}), one should first cancel analytically
the $\ep$ poles of the one-loop matrix element with those of the insertion
operator ${\bom I}$, perform the limit $\ep \to 0$ and then carry out
the phase-space integration in four space-time dimensions.

\newpage

\setcounter{equation}{0}

\section{Multi-particle correlations}
\label{eemp}

In the case of processes involving multi-particle correlations
(see Eq.~(\ref{mpkin})), the partonic cross section at the
Born level is given by
\beeq
\label{dsbmp}
&&d\sigma_{ab,(incl) a_1,..,a_n}^{B}(p,{\bar p};q_1,...,q_n)
= {\cal N}_{in} \frac{1}{n_s(a) n_s(b) n_c(a) n_c(b) \Phi(p \cdot {\bar p})}
\nonumber \\
&&\cdot \;
\sum_{\{ m \} } \, d\phi_m(p_1, ...,p_m;p+{\bar p}+Q-q_1-...-q_n)
\;\frac{1}{S_{\{ m \} }} \nonumber \\
&&\cdot \;|{\cal{M}}_{m+a_1+a_2...,ab}(q_1,...,q_n,p_1, ...,p_m;p,{\bar p}) |^2
\;F_J^{(m)}([q_1],...[q_n],p_1, ...,p_m;p,{\bar p}) \;\;.
\eeeq
Here, we denote by $a$ and $b$ the flavour indices of the two incoming partons
with momenta $p$ and ${\bar p}$, while $a_1, ...,a_n$ are the flavour indices
of the final-state identified partons with momenta $q_1, ...,q_n$. In addition,
the leading-order cross section in Eq.~(\ref{dsbmp}) has $m$ final-state
unidentified partons with momenta $p_1, ...,p_m$ (non-QCD partons are
understood). The jet defining function $F_J$ has the properties already
discussed in Sects.~\ref{siee}--\ref{sipp}, namely, infrared and collinear safety
and factorizability of initial- and final-state collinear singularities.
Remember that, by definition, the momenta $p,{\bar p},q_1, ...,q_n$ are supposed
not to be parallel to each other.

\subsection{Implementation of the subtraction procedure}

According to our general procedure, we write the NLO partonic cross section
(\ref{mpNLO}) in the following form
\beeq
\label{mprw}
\!\!\!&&\sigma_{ab,(incl) a_1,..,a_n}^{NLO}(p,{\bar p};
q_1,...,q_n;\mu_F^2, \mu_1^2,..,\mu_n^2) \nonumber \\
\!\!\!&&=
\int_{m+1} \left( d\sigma_{ab,(incl) a_1,..,a_n}^{R}(p,{\bar p};q_1,...,q_n)
- d\sigma_{ab,(incl) a_1,..,a_n}^{A}(p,{\bar p};q_1,...,q_n) \right)
\nonumber \\
\!\!\!&&+ \left[ \;\int_{m+1}
d\sigma_{ab,(incl) a_1,..,a_n}^{A}(p,{\bar p};q_1,...,q_n) +
\int_m d\sigma_{ab,(incl) a_1,..,a_n}^{V}(p,{\bar p};q_1,...,q_n) \right.
\nonumber \\
\!\!\!&&+ \left. \int_m d\sigma_{ab,(incl)
a_1,..,a_n}^{C}(p,{\bar p};q_1,...,q_n;\mu_F^2, \mu_1^2,..,\mu_n^2) \right]
\;\;,
\eeeq
where the subtraction term is defined by
\beeq
\label{dsamp}
&&d\sigma_{ab,(incl) a_1,..,a_n}^{A}(p,{\bar p};q_1,...,q_n)
= {\cal N}_{in} \frac{1}{n_s(a) n_s(b) \Phi(p \cdot {\bar p})} \nonumber \\
&&\cdot \;
\sum_{\{ m+1 \} } \,
d\phi_{m+1}(p_1, ...,p_{m+1};p+{\bar p}+Q-q_1-...-q_n) \;
\frac{1}{S_{\{ m+1 \} }}  \\
&&\cdot \left\{
\sum_{\mathrm{pairs}\atop i,j} \left( {\cal D} \cdot F_J^{(m)} \right)_{ij} +
\sum_{l=1}^n  \sum_i \left( {\cal D} \cdot F_J^{(m)} \right)_{a_li} +
\sum_i \left[ \left( {\cal D} \cdot F_J^{(m)} \right)^{ai} +
\left( {\cal D} \cdot F_J^{(m)} \right)^{bi} \right] \right\} \;\;. \nonumber
\eeeq
Here we have introduced the shorthand notation ${\cal D} \cdot F_J^{(m)}$
to denote
the different dipole contributions in which the emitter is a final-state
unidentified parton, a final-state identified parton or
an initial-state parton. Their explicit expressions, according to the dipole
formulae in Sect.~\ref{dff}, are respectively the following
\beeq
\label{DFij}
\!\!\!\!\!&&\!\!\left( {\cal D} \cdot F_J^{(m)} \right)_{ij} = \sum_{k\neq i,j}
{\cal D}_{ij,k}(q_1,..,q_n,p_1, ...,p_{m+1};p,{\bar p}) \;
F_J^{(m)}([q_1],..,[q_n],p_1, .. {\widetilde p}_{ij},
{\widetilde p}_k, ..,p_{m+1};p,{\bar p}) \nonumber \\
\!\!\!\!\!&&\!\!+
\sum_{l=1}^n {\cal D}_{ij,a_l}(q_1,..,q_n,p_1, ...,p_{m+1};p,{\bar p}) \;
F_J^{(m)}([q_1],..,[{\widetilde q}_l],..[q_n],p_1, .. {\widetilde p}_{ij},
..,p_{m+1};p,{\bar p}) \nonumber \\
\!\!\!\!\!&&+\!\! \;
{\cal D}_{ij}^a(q_1,..,q_n,p_1, ...,p_{m+1};p,{\bar p}) \;
F_J^{(m)}([q_1],..,[q_n],p_1, .. {\widetilde p}_{ij},
..,p_{m+1};{\widetilde p},{\bar p}) \nonumber \\
\!\!\!\!\!&&\!\!+ \;
{\cal D}_{ij}^b(q_1,..,q_n,p_1, ...,p_{m+1};p,{\bar p}) \;
F_J^{(m)}([q_1],..,[q_n],p_1, .. {\widetilde p}_{ij},
..,p_{m+1};p,{\widetilde {\bar p}}) \;\;,
\eeeq
\beeq
\label{DFli}
\!\!\!\!\!\!\!\!&&\left( {\cal D} \cdot F_J^{(m)} \right)_{a_li} =
\sum_{k\neq i}
{\cal D}_{a_li,k}(q_1,..,q_n,p_1, ...,p_{m+1};p,{\bar p}) \;
\nonumber \\
\!\!\!\!\!\!\!\!&&\cdot \left.\right.
F_J^{(m)}([q_1],..,[{\widetilde q}_l],..[q_n],p_1,...,
{\widetilde p}_k, ..,p_{m+1};p,{\bar p}) \nonumber \\
\!\!\!\!\!\!\!\!&&+ \Bigl[ \;\sum_{r=1 \atop r \neq l}^n
{\cal D}_{a_li,a_r}^{(n)}(q_1,..,q_n,p_1, ...,p_{m+1};p,{\bar p})
+ {\cal D}_{a_li}^{(n) \,a}(q_1,..,q_n,p_1, ...,p_{m+1};p,{\bar p}) \Bigr.
\nonumber \\
\!\!\!\!\!\!\!\!&& + \Bigl.
{\cal D}_{a_li}^{(n) \,b}(q_1,..,q_n,p_1, ...,p_{m+1};p,{\bar p})
\left. \frac{}{} \right.
\Bigr] F_J^{(m)}([q_1],..,[{\widetilde q}_l],..[q_n],{\widetilde p}_1,..,
{\widetilde p}_{m+1};p,{\bar p}) \;\;,
\eeeq
\beeq
\label{DFai}
\!\!\!\!\!\!\!\!\!\!&&\left( {\cal D} \cdot F_J^{(m)} \right)^{ai} +
(a \leftrightarrow b) =
\Bigl\{ \sum_{k \neq i}
{\cal D}^{ai}_{k}(q_1,..,q_n,p_1, ...,p_{m+1};p,{\bar p})
\; \Bigr. \nonumber \\
\!\!\!\!\!\!\!\!\!\!&&\cdot \left.
F_J^{(m)}([q_1],..,[q_n],p_1,...,
{\widetilde p}_k, ..,p_{m+1};{\widetilde p},{\bar p}) \right. \nonumber \\
\!\!\!\!\!\!\!\!\!\!&&+ \left. \left[ \;\sum_{l=1}^n
{\cal D}^{(n) \,ai}_{a_l}(q_1,..,q_n,p_1, ...,p_{m+1};p,{\bar p}) +
{\cal D}^{(n) \,ai,b}(q_1,..,q_n,p_1, ...,p_{m+1};p,{\bar p}) \right] \right.
\nonumber \\
\!\!\!\!\!\!\!\!\!\!&&\cdot \Bigl. \frac{}{}
F_J^{(m)}([q_1],..,[q_n],{\widetilde p}_1,
..,{\widetilde p}_{m+1};{\widetilde p},{\bar p}) \Bigr\} +
(a \leftrightarrow b)
\;\;.
\eeeq
Comparing Eqs.~(\ref{dsamp}--\ref{DFai}) with the form of the subtraction terms
introduced in Sects.~\ref{siee}--\ref{sipp}, we see that the only new feature
of the present case is due to the `pseudodipoles' ${\cal D}_{a_li,a_r}^{(n)},
{\cal D}_{a_li}^{(n) \,a}, {\cal D}_{a_li}^{(n) \,b}, {\cal D}^{(n) \,ai}_{a_l},
{\cal D}^{(n) \,ai,b}$ where both the emitter and the spectator are
incoming partons or identified final-state partons. These dipoles are defined
in Sect.~\ref{dcorr}. Note, in particular, that no distinction is made between
initial- and final-state spectator (i.e.\ ${\cal D}_{a_li}^{(n) \,a} =
{\cal D}_{a_li,a}^{(n)}$ in Eq.~(\ref{DFli}) and ${\cal D}^{(n) \,ai,b}
= {\cal D}^{(n) \,ai}_{b}$ in Eq.~(\ref{DFai})) and that the dipole partonic
states $\{q_1,..,{\widetilde q}_l,..q_n,{\widetilde p}_1,..,
{\widetilde p}_{m+1},p,{\bar p}\} ,
\{q_1,..,q_n,{\widetilde p}_1, ..,{\widetilde p}_{m+1},{\widetilde p},{\bar p}
\} , \{q_1,..,q_n,{\widetilde p}_1, ..,{\widetilde p}_{m+1},p,
{\widetilde {\bar p}}\}$ depend only on the emitter (the corresponding jet
functions $F_J$ appear as common factors in Eqs.~(\ref{DFli},\ref{DFai})).

The subtracted contribution $( d\sigma^R - d\sigma^A )$ in Eq.~(\ref{mprw}) is
integrable in four dimensions by explicit construction.

In order to evaluate
the $d$-dimensional integral of  $d\sigma^A$, we decompose it as follows
\beq
\label{mp1234}
\int_{m+1} d\sigma^{A}= \int_{m+1} \left( d\sigma^{A \prime} +
d\sigma^{A \prime \prime} +
d\sigma^{A \prime \prime \prime} +
d\sigma^{A \prime \prime \prime \prime} \right)\;\;.
\eeq
The first three terms on the right-hand side of Eq.~(\ref{mp1234}) respectively
correspond to the integral of $({\cal D} \cdot F_J)_{ij}, {\cal D}_{a_li,k}$
and ${\cal D}^{ai}_{k}$. Their treatment has been already considered in
Sects.~\ref{siee}--\ref{sifra}. The fourth term contains the integral of all the
pseudodipoles and can be handled as $d\sigma^{A \prime \prime \prime \prime}$
in Sect.~\ref{sipp}. As a result of this integration procedure we find:
\beeq
\label{mpAC}
&&\!\!\!\!\!\!\!\!\!\!\int_{m+1}
d\sigma_{ab,(incl) a_1,..,a_n}^{A}(p,{\bar p};q_1,...,q_n)
+ \int_m d\sigma_{ab,(incl)
a_1,..,a_n}^{C}(p,{\bar p};q_1,...,q_n;\mu_F^2, \mu_1^2,..,\mu_n^2) \nonumber \\
&&\!\!\!\!\!\!\!\!\!\!= \int_m
\left[ d\sigma_{ab,(incl) a_1,..,a_n}^{B}(p,{\bar p};q_1,...,q_n)
\cdot {\bom I}(\ep) \right] \nonumber \\
&&\!\!\!\!\!\!\!\!\!\!+ \sum_{a'} \int_0^1 dx \int_m
\left[ \left( {\bom K}^{a,a'}(x)
+ {\bom P}^{a,a'}(xp,x;\mu_F^2) \right) \cdot
d\sigma_{a'b,(incl) a_1,..,a_n}^{B}(xp,{\bar p};q_1,...,q_n) \right]
\\
&&\!\!\!\!\!\!\!\!\!\!+ \sum_{b'} \int_0^1 dx \int_m
\left[ \left( {\bom K}^{b,b'}(x)
+ {\bom P}^{b,b'}(x{\bar p},x;\mu_F^2) \right) \cdot
d\sigma_{ab',(incl) a_1,..,a_n}^{B}(p,x{\bar p};q_1,...,q_n) \right]
\nonumber \\
&&\!\!\!\!\!\!\!\!\!\!+ \sum_{l=1}^n \sum_{a_l^\prime} \int_0^1 \frac{dz}{z^2}
\int_m \left[
d\sigma_{ab,(incl) a_1,..,a_l^\prime,..a_n}^{B}(p,{\bar p};q_1,..,q_l/z,..,q_n)
\cdot \left( {\bom H}_{a_l^\prime,a_l}(z)
+ {\bom P}_{a_l^\prime,a_l}(q_l/z,z;\mu_l^2) \right) \right] \nonumber \;.
\eeeq
The factor ${\bom I}(\ep)$ comes from the integration of ${\cal D}_{ij,k}$ and,
in addition, collects all the $\ep$-poles due to the other dipole factors. The
finite parts of the dipoles ${\cal D}_{ij}^{a},$ ${\cal D}_{ij}^{b},$
${\cal D}^{ai}_{k},$
${\cal D}_{a_l}^{(n) \,ai},$ ${\cal D}^{(n) \,ai,b},$ ${\cal D}^{(n) \,bi,a}$
contribute to the initial-state operators ${\bom K}(x)+ {\bom P}(xp,x;\mu_F^2)$,
and those of the dipoles ${\cal D}_{ij,a_l}, {\cal D}_{a_li,k},
{\cal D}_{a_li,a_r}^{(n)}, {\cal D}_{a_li}^{(n) \,a}, {\cal D}_{a_li}^{(n) \,b}$
contribute to the final-state operators
${\bom H}(z)+ {\bom P}(q_l/z,z;\mu_l^2)$.

The insertion operator ${\bom I}(\ep)$ in Eq.~(\ref{mpAC}) is fully symmetric
with respect to all the QCD partons. Therefore, denoting by $I$ a generic QCD
parton $(I=\{i,a_l,a,b\})$, we can write (the singular factors
${\cal V}_{I}(\ep)$ are given in Eq.~(\ref{calvexp})):
\beq
\label{mpI}
{\bom I}(q_1,...,q_n,p_1,...,p_m,p,{\bar p};\ep) =  - \;\frac{\as}{2\pi}
\frac{1}{\Gamma(1-\ep)} \; \sum_I
\frac{1}{{\bom T}_{I}^2} \;{\cal V}_{I}(\ep)
 \sum_{J \neq I} {\bom T}_I \cdot {\bom T}_J
\;\left(
\frac{4\pi \mu^2}{2 p_I\cdot p_J} \right)^{\ep}  \;.
\eeq

The initial-state operator ${\bom P}^{a,a'}(xp,x;\mu_F^2)$ is instead symmetric
with respect to all the partons except $p$. Its explicit expression is:
\beq
\label{mpIin}
{\bom P}^{a,a'}(q_1,..,q_n,p_1,...,p_m,{\bar p};xp,x;\mu_F^2) =
\frac{\as}{2\pi} \;P^{aa'}(x) \;\frac{1}{{\bom T}_{a'}^2}
\sum_{I \neq a'} {\bom T}_I \cdot {\bom T}_{a'}
\;\ln \frac{\mu_F^2}{2 x p \cdot p_I}
\;\;.
\eeq
Similarly, for the final-state operator
${\bom P}_{a_l^\prime,a_l}(q_l/z,z;\mu_l^2)$ we find
\beq
\label{mpIfin}
{\bom P}_{a_l^\prime,a_l}(q_1,..,q_n,p_1,...,p_m,p,{\bar p};q_l/z,z;\mu_l^2) =
\frac{\as}{2\pi} \;P_{a_l^\prime,a_l}(z) \;\frac{1}{{\bom T}_{a_l^\prime}^2}
\sum_{I \neq a_l^\prime} {\bom T}_I \cdot {\bom T}_{a_l^\prime}
\;\ln \frac{z \mu_l^2}{2 q_l \cdot p_I}
\;\;.
\eeq

The insertion operators ${\bom K}^{a,a'}(x)$ and ${\bom H}_{a_l^\prime,a_l}(z)$
are separately symmetric with respect to the sets of the unidentified and
identified (or initial-state) partons. They are given by the
following equations
\beeq
\label{mpK}
\!\!\!\!&&{\bom K}^{a,a'}(x)
= \frac{\as}{2\pi}
\left\{ \frac{}{} {\overline K}^{aa'}(x) - \KFS{aa'}(x) \right. \nonumber \\
\!\!\!\!&&+ \left. \; \delta^{aa'} \sum_{i} {\bom T}_i \cdot {\bom T}_a
\;\frac{\gamma_i}{{\bom T}_i^2} \left[ \left( \frac{1}{1-x} \right)_+
+ \delta(1-x) \right]
-  \;\frac{1}{{\bom T}_{a'}^2}
 \left( \sum_{l=1}^n  {\bom T}_{a_l} \cdot {\bom T}_{a'}
+ {\bom T}_{b} \cdot {\bom T}_{a'} \right)
{\widetilde K}^{aa'}(x)
  \right. \nonumber \\
\!\!\!\!&&- \left.  \frac{1}{{\bom T}_{a'}^2} \left[ \;
\sum_{l=1}^n  {\bom T}_{a_l} \cdot {\bom T}_{a'} \;{\cal L}^{a,a'}(x;p,q_l,n)
+ {\bom T}_{b} \cdot {\bom T}_{a'} \;{\cal L}^{a,a'}(x;p,{\bar p},n)
\right] \right\} \;\;,
\eeeq
\beeq
\label{mpH}
\!\!\!\!\!&&{\bom H}_{a_l^\prime,a_l}(z) =
\frac{\as}{2\pi}
\left\{ \frac{}{} {\overline K}^{a_l^\prime a_l}(z) +
3 P_{a_l^\prime a_l}(z) \;\ln z - \HFS{a_l^\prime a_l}(z)
\right. \nonumber \\
\!\!\!\!\!&& + \;\delta_{a_l^\prime a_l} \sum_i
{\bom T}_i \cdot {\bom T}_{a_l^\prime} \frac{\gamma_i}{{\bom T}_i^2}
\left[ \left( \frac{1}{1-z} \right)_+ + \delta(1-z) - 1 \right] \nonumber \\
\!\!\!\!\!&& +\left. \frac{1}{{\bom T}_{a_l^\prime}^2}
\left( \sum_{r=1 \atop r \neq l}^n {\bom T}_{a_r} \cdot {\bom T}_{a_l^\prime}
+ {\bom T}_{a} \cdot {\bom T}_{a_l^\prime}
+ {\bom T}_{b} \cdot {\bom T}_{a_l^\prime} \right)
\left[ \;P_{a_l^\prime a_l}(z) \;\ln z -{\widetilde K}^{a_l^\prime a_l}(z)
\frac{}{} \right] \right. \nonumber \\
\!\!\!\!\!&&- \left.  \frac{1}{{\bom T}_{a_l^\prime}^2} \left[
\sum_{r=1 \atop r \neq l}^n
{\bom T}_{a_r} \cdot {\bom T}_{a_l^\prime,a_l}
\;{\cal L}^{a_l^\prime,a_l}(z;q_l,q_r,n)
+ {\bom T}_{a} \cdot {\bom T}_{a_l^\prime}
\;{\cal L}^{a_l^\prime,a_l}(z;q_l,p,n) \right. \right. \nonumber \\
\!\!\!\!\!&& \left. \left.
+ {\bom T}_{b} \cdot {\bom T}_{a_l^\prime}
\;{\cal L}^{a_l^\prime,a_l}(z;q_l,{\bar p},n) \frac{}{}
\right] \right\}  \;\;,
\eeeq
where the flavour kernels ${\overline K}^{ab}, \Kab, {\widetilde K}^{ab},
{\cal L}^{a,b}, \Hba$ are defined respectively in
Eqs.~(\ref{kbarqg}--\ref{kbargg}), (\ref{saC}),
(\ref{wkdef}), (\ref{calldef}) and (\ref{saCinc}).

Note that the operators ${\bom I}$ and ${\bom P}$ in
Eqs.~(\ref{mpI}--\ref{mpIfin}) are completely analogous to the corresponding
operators defined in Sects.~\ref{siee}--\ref{sipp}. Some new features instead
appear in Eqs.~(\ref{mpK}) and (\ref{mpH}). Comparing Eq.~(\ref{mpK}) with
Eq.~(\ref{kdefhad}), we find a new contribution (that in the last square
bracket of
(\ref{mpK})) due to {\em correlations\/} between the incoming parton $a$ and
the identified final-state partons.
As a result, the insertion operator ${\bom K}$ explicitly depends on the
momenta of the identified partons unlike the previous cases.
This contribution is indeed non-vanishing
only if there are final-state identified partons.
A similar term is also present in the
expression (\ref{mpH}) for ${\bom H}$. However, in this case also the term in
the second square bracket on the right-hand side of Eq.~(\ref{mpH}) has no
analogue in the corresponding Eq.~(\ref{hdef}). This term is due to
{\em final-state\/} parton correlations and is similar to that proportional to
${\widetilde K}^{aa'}(x)$
on the right-hand side of Eq.~(\ref{mpK}). In this respect the
difference between initial- and final-state parton correlations simply amounts
to the replacement ${\widetilde K}^{a^\prime,a}(z) \to
{\widetilde K}^{a^\prime,a}(z) - P^{a^\prime,a}(z) \ln z$.

\subsection{Final formulae}
\label{finmpc}

The results of the previous Subsection can be summarized by the following final
formulae.

The LO parton-level cross section, which enters into the calculation of the
multi-particle hadronic cross section of Eq.~(\ref{1mxs}), is obtained as
follows
\beeq
\label{LOmpcfin}
&&\sigma_{ab,(incl) a_1,..,a_n}^{LO}(p,{\bar p};q_1,...,q_n) =
\int_m d\sigma_{ab,(incl) a_1,..,a_n}^{B}(p,{\bar p};q_1,...,q_n) \nonumber \\
&&= \int d\Phi^{(m)}(p,{\bar p};q_1,...,q_n)
\; \frac{1}{n_c(a) n_c(b)}
\;|{\cal{M}}_{m+a_1+a_2...,ab}(q_1,...,q_n,p_1, ...,p_m;p,{\bar p}) |^2
\nonumber \\
&&\cdot \;F_J^{(m)}([q_1],...[q_n],p_1, ...,p_m;p,{\bar p}) \;\;.
\eeeq
Here $n_c(a)$ and $n_c(b)$ are the number of colours of the incoming partons,
${\cal{M}}_{m+a_1+a_2...,ab}$ is the tree-level matrix element to produce
$m$ final-state partons in addition to the identified partons, and
$F_J^{(m)}$ is the most general jet defining function (it fulfils
Eqs.~(\ref{fjsoft}--\ref{fjLO}), Eqs.~(\ref{fjdiscoll},\ref{fjdis}) and
Eqs.~(\ref{fjfracol},\ref{fjfra}) with respect to the dependence on the momenta
of the unidentified partons, of the incoming partons and of the
final-state identified partons). The term
$d\Phi^{(m)}(p,{\bar p};q_1,...,q_n)$ in Eq.~(\ref{LOmpcfin}) collects
all the other kinematic
factors on the right-hand side of Eq.~(\ref{dsbmp}).

As in the symbolic notation of Eq.~(\ref{sNLO4}),
the NLO partonic cross section contains a contribution with $m+1$-parton
kinematics, a contribution with $m$-parton kinematics and $n+2$ terms
obtained by a one-dimensional convolution of cross sections with $m$-parton
kinematics. We have:
\beeq
\label{sNLOmpc}
&&\!\!\sigma^{NLO}_{ab,(incl) a_1,..,a_n}(p,{\bar p};
q_1,...,q_n;\mu_F^2, \mu_1^2,..,\mu_n^2) =
\sigma^{NLO\,\{m+1\}}_{ab,(incl) a_1,..,a_n}(p,{\bar p};
q_1,...,q_n) \nonumber \\
&&\!\!+\;\;
\sigma^{NLO\,\{m\}}_{ab,(incl) a_1,..,a_n}(p,{\bar p};
q_1,...,q_n) \nonumber \\
&&\!\!+
\int_0^1 dx \;
\left[ \;{\hat \sigma}^{NLO\,\{m\}}_{ab,(incl) a_1,..,a_n}(x;xp,{\bar p};
q_1,...,q_n;\mu_F^2) +
{\hat \sigma}^{NLO\,\{m\}}_{ab,(incl) a_1,..,a_n}(x;p,x{\bar p};
q_1,...,q_n;\mu_F^2)  \;\right]  \nonumber \\
&&\!\!+ \sum_{l=1}^n \int_0^1 \frac{dz}{z^2}
\;{\hat \sigma}^{NLO\,\{m\}}_{ab,(incl) a_1,..,a_n}(z;p,{\bar p};
q_1,..,q_l/z,.,q_n;\mu_l^2)  \;\;.
\eeeq

The NLO contribution with $m+1$-parton kinematics is given by the  following
expression
\beeq
\label{NLOmpcmp}
&&\sigma^{NLO\,\{m+1\}}_{ab,(incl) a_1,..,a_n}(p,{\bar p};
q_1,...,q_n)
= \int_{m+1} \left[ \frac{}{} \right.
\left( d\sigma^{R}_{ab,(incl) a_1,..,a_n}(p,{\bar p};
q_1,...,q_n) \right)_{\ep=0} \nonumber \\
&&- \left( \frac{}{} \right.
\sum_{{\rm dipoles}} \;d\sigma^{B}_{ab,(incl) a_1,..,a_n}(p,{\bar p};
q_1,...,q_n) \otimes
\left( \;dV_{{\rm dipole}}
+  dV^{\prime}_{{\rm dipole}} \right) \left. \frac{}{}
\right)_{\ep=0}  \left. \frac{}{} \;\right] \nonumber \\
&&= \int d\Phi^{(m+1)}(p,{\bar p};q_1,...,q_n)
 \left\{
\; \frac{1}{n_c(a) n_c(b)}
\;|{\cal{M}}_{m+1+a_1+a_2...,ab}(q_1,...,q_n,p_1, ...,p_{m+1};p,{\bar p}) |^2
\right. \nonumber \\
&&\left. \cdot \;F_J^{(m+1)}([q_1],...[q_n],p_1, ...,p_{m+1};p,{\bar p})
\right.
\left. - \sum_{{\rm dipoles}}
\left( {\cal D} \cdot F^{(m)} \right)(q_1,..,q_n,p_1, ...,p_{m+1};p,{\bar p})
\right\} \;\;, \nonumber \\
&&
\eeeq
where ${\cal{M}}_{m+1+a_1+a_2...,ab}$ is the tree-level matrix element with
$m+1$ unidentified partons in the final state and
$\sum_{{\rm dipoles}} \left( {\cal D} \cdot F^{(m)}
\right)(q_1,..,q_n,p_1, ...,p_{m+1};p,{\bar p})$
is the sum of the dipole factors in the curly bracket on the
right-hand side of Eq.~(\ref{dsamp}).

The NLO term with $m$-parton kinematics is obtained by adding the virtual
cross section and the first contribution on the right-hand side of
Eq.~(\ref{mpAC}). Its explicit form is given by:
\beeq
\label{NLOmpcm}
&&\!\!\!\sigma^{NLO\,\{m\}}_{ab,(incl) a_1,..,a_n}(p,{\bar p};
q_1,...,q_n) \nonumber \\
&&\!\!\!= \int_m \left[ d\sigma^{V}_{ab,(incl) a_1,..,a_n}(p,{\bar p};
q_1,...,q_n) +
d\sigma^{B}_{ab,(incl) a_1,..,a_n}(p,{\bar p};q_1,...,q_n) \otimes {\bom I}
\right]_{\ep=0} \nonumber \\
&&\!\!\!=  \int d\Phi^{(m)}(p,{\bar p};q_1,...,q_n)
\left\{  \frac{1}{n_c(a) n_c(b)}
\;| \cm_{m+a_1+a_2...,ab}(q_1,...,q_n,p_1, ...,p_{m};p,{\bar p})|^2_{(1-loop)}
\right.  \nonumber \\
&&\!\!\! \left. + \;\;{}_{m+a_1+a_2...,ab}\!\!<{a_1,..,a_n,1, ....,m;a,b\;}|
\;{\bom I}(\ep)
\;|{a_1,..,a_n,1,...,m;a,b}>_{m+a_1+a_2...,ab} \frac{}{}
\right\}_{\ep=0} \nonumber \\
&&\!\!\!\cdot \;F_J^{(m)}([q_1],...[q_n],p_1, ...,p_{m};p,{\bar p})
\;\;,
\eeeq
where $| \cm_{m+a_1+a_2...,ab}|^2_{(1-loop)}$ is the one-loop matrix element
squared and the colour-charge operator ${\bom I}(\ep)$ is defined in
Eq.~(\ref{mpI}) (see also Appendix~C).

The last three terms on the right-hand side of Eq.~(\ref{sNLOmpc})
are in one-to-one correspondence with the last three terms on the
right-hand side of Eq.~(\ref{mpAC}). Their expressions as a function of the QCD
matrix elements are the following
\beeq
\label{NLOmpcx}
&&\!\!\!\!\!\!\!\!\!\int_0^1 dx
\;{\hat \sigma}^{NLO\,\{m\}}_{ab,(incl) a_1,..,a_n}(x;xp,{\bar p};
q_1,...,q_n;\mu_F^2) \nonumber \\
&&\!\!\!\!\!\!\!\!\!= \sum_{a'}
\int_0^1 dx \;
\int_m \left[ d\sigma^{B}_{a'b,(incl) a_1,..,a_n}(x;xp,{\bar p};
q_1,...,q_n) \otimes \left( {\bom K}
+ {\bom P} \right)^{a,a'}(x) \right]_{\ep=0} \nonumber \\
&&\!\!\!\!\!\!\!\!\!= \sum_{a'}
\int_0^1 dx \int d\Phi^{(m)}(xp,{\bar p};q_1,...,q_n)
\;F_J^{(m)}([q_1],...[q_n],p_1, ...,p_{m};xp,{\bar p}) \\
&&\!\!\!\!\!\!\!\!\!\cdot \;
{}_{m+a_1..,a'b}\!\!<{q_1,.,q_n,1,..,m;xp,{\bar p}\;}|
\;\left({\bom K}^{a,a'}(x)
\right. \nonumber \\ &&\!\!\!\!\!\!\!\!\! \left.
+ \;{\bom P}^{a,a'}(xp, x;\mu_F^2) \right) \;
|{q_1,.,q_n,1,..,m;xp,{\bar p}}>_{m+a_1..,a'b} \;\;, \nonumber
\eeeq
\beeq
\label{NLOmpcz}
&&\!\!\!\!\!\!\!\!\int_0^1 \frac{dz}{z^2}
\;{\hat \sigma}^{NLO\,\{m\}}_{ab,(incl) a_1,..,a_n}(z;p,{\bar p};
q_1,..,q_l/z,.,q_n;\mu_l^2)   \nonumber \\
&&\!\!\!\!\!\!\!\!= \sum_{a_l^\prime}
\int_0^1 \frac{dz}{z^2} \;
\int_m \left[ d\sigma^{B}_{ab,(incl) a_1,.,a_l^\prime,.,a_n}(z;p,{\bar p};
q_1,..,q_l/z,.,q_n) \otimes \left( {\bom H}
+ {\bom P} \right)_{a_l^\prime,a_l}(z) \right]_{\ep=0} \nonumber \\
&&\!\!\!\!\!\!\!\!= \sum_{a_l^\prime} \int_0^1  \frac{dz}{z^2}
\int d\Phi^{(m)}(p,{\bar p};q_1,..,q_l/z,..,q_n)
\;F_J^{(m)}([q_1],..,[q_l/z],..[q_n],p_1, ...,p_{m};p,{\bar p})
\phantom{(99.99)}\\
&&\!\!\!\!\!\!\!\!\cdot \;
{}_{m+..+a_l^\prime+..,ab}\!\!<{q_1,.,q_l/z,..,q_n,1,..,m;p,{\bar p}}\,|
\;\left({\bom H}_{a_l^\prime,a_l}(z)
\right. \nonumber \\ &&\!\!\!\!\!\!\!\!
+ \;\left. {\bom P}_{a_l^\prime,a_l}(q_l/z, z;\mu_l^2) \right) \;
|{q_1,.,q_l/z,..,q_n,1,..,m;p,{\bar p}}>_{m+..+a_l^\prime+..,ab} \;\;.
\nonumber
\eeeq
The contribution
${\hat \sigma}^{NLO\,\{m\}}_{ab,(incl) a_1,..,a_n}(x;p,x{\bar p};
q_1,...,q_n;\mu_F^2)$ is obtained from Eq.~(\ref{NLOmpcx}) by the replacements
$xp \to p, \;{\bar p} \to x{\bar p}, \;\sum_{a'} \to \sum_{b'}$
(see Eq.~(\ref{mpAC})). Note that each of the contributions
${\hat \sigma}^{NLO\,\{m\}}_{ab,(incl) a_1,..,a_n}$ in
Eqs.~(\ref{sNLOmpc},\ref{NLOmpcx},\ref{NLOmpcz}) depends on a single
factorization scale.

The colour-charge operators ${\bom K}$ and  ${\bom P}$ of Eq.~(\ref{NLOmpcx})
are defined in Eqs.~(\ref{mpK}) and (\ref{mpIin}) respectively.
The colour-charge operators ${\bom H}$ and  ${\bom P}$ of Eq.~(\ref{NLOmpcz})
are respectively given in Eqs.~(\ref{mpH}) and (\ref{mpIfin}). The definitions
of the
related flavour kernels $P^{aa'}(x),$ ${\overline K}^{aa'}(x),$ $\KFS{aa'}(x),$
${\widetilde K}^{aa'}(x),$ $\HFS{a'a}(x)$ and of the functions ${\cal L}^{a,a'}$
are also recalled in Appendix C.

The actual evaluation of Eqs.~(\ref{LOmpcfin}),
(\ref{NLOmpcmp}),  (\ref{NLOmpcx}) and
(\ref{NLOmpcz}) is directly performed in four space-time dimensions.
As for Eq.~(\ref{NLOmpcm}), one should first cancel analytically
the $\ep$ poles of the one-loop matrix element with those of the insertion
operator ${\bom I}$, perform the limit $\ep \to 0$ and then carry out
the phase-space integration in four space-time dimensions.

Note that the formulae presented in this Subsection can be applied also to the
simplified cases of multi-particle correlations with a single incoming
hadron or with no hadrons in the initial state. For this purpose, it is
sufficient
to remove one or both of the contributions in the square bracket on the
right-hand side of Eq.~(\ref{sNLOmpc}) and to set equal to zero the colour charge
of the removed incoming parton both in the dipole factors of
Eq.~(\ref{NLOmpcmp}) and in the colour-charge operators of
Eqs.~(\ref{NLOmpcm}--\ref{NLOmpcz}).

The results obtained in this Section show that, in the most general case of
multi-particle correlations, the NLO partonic cross section in
Eq.~(\ref{sNLOmpc}) is finite, i.e.\ free from soft and collinear singularities.
We should point out that its finiteness directly follows from the primitive
definition in Eq.~(\ref{mpNLO}), which is nothing other than the formal
restatement,
to the lowest non-trivial order,  of the factorization theorem of mass
singularities [\ref{CSS}]. Doubts have been raised in the past about the
validity of the theorem for processes involving more than one identified
parton, for instance, the Drell-Yan process. Although those doubts have been
proved to be unfounded,
a Cartesian (and fully general) proof of the factorization theorem in the
context of QCD is still missing. Our check, by means of the explicit
calculations in this Section, that the partonic cross section in
Eq.~(\ref{mpNLO}) is finite is,
to our knowledge, the first proof (at NLO) of the factorization theorem in the
most general case of QCD jet cross sections.

\newpage

\setcounter{equation}{0}

\section{Summary and discussion}
\label{summ}

\subsection{Summary}

In this paper, we have introduced a new general algorithm for calculating
arbitrary jet cross sections to NLO in arbitrary scattering processes,
based on the subtraction method.  The key ingredients are the dipole
factorization formulae, which implement both the usual soft and collinear
factorization formulae, smoothly interpolating the two.  The corresponding
dipole phase space obeys exact factorization, so that the dipole
contributions to the cross section can be exactly integrated analytically
over the whole of phase space.

The steps that are necessary to set up a general method for evaluating
NLO QCD cross sections are recalled in Sect.~\ref{int}.
In order to compute a jet quantity to NLO, one must calculate the
contributions from one-loop corrections to the Born-level cross section and
from the cross section for processes in which one additional parton is
present in the final state.  Each leads to singularities, which are regularized
by working in a number of dimensions $d=4-2\ep$ other than four.
Analytical integration must then be used to extract the singular terms as
poles in $\ep$ and combine them with one another to yield a finite result.
However, almost all experimentally important jet quantities are
sufficiently complicated that analytical integration is impossible and one
must resort to numerical techniques.  Thus one must somehow extract the
singular parts of the cross section in a way that is independent of the
exact details of the observable and treat them analytically.  This leaves a
remainder that depends on the full complications of the jet quantity, but
which is finite so can be treated numerically.

One way of doing this is provided by the subtraction method,
described in Sect.~\ref{subproc}.  This works
by introducing an approximate cross section, which
is defined in such a way as to match all
the singularities of the real cross section to produce one additional
parton.  Thus the difference between the two is guaranteed to be finite and
numerical integration can be used in 4 dimensions.  However, the
approximate cross section is required to be simple enough to be integrated
analytically without knowing the details of the definition of the jet
observable.

This is achieved by considering, for every $m+1$-parton configuration, a
mapping to a `similar' $m$-parton configuration (or in general, several of
them), defined such that in the singular regions of phase space the two
configurations become indistinguishable.  The approximate cross section is
then defined to be proportional to the jet quantity calculated from this
similar $m$-parton configuration. Thus, the analytical integration can be
performed without knowing the definition of the jet quantity, by holding
the $m$-parton configuration fixed and integrating over all $m+1$-parton
configurations that map onto it.  This gives rise to $\ep$ poles, which
cancel those from the one-loop cross section to produce that same
$m$-parton configuration.  The result is a finite $m$-parton cross section
that can then be integrated numerically.

Note that, using the subtraction method, no approximation is actually performed
in the evaluation of the NLO cross section. The approximate cross section is
subtracted from the real cross section and then added back to the one-loop
cross section. Moreover, and most importantly, no approximation is performed
in the analytical integration of the approximate cross section. In that respect,
the adjective `approximate' may sound misleading. Rather than approximating the
true cross section, the subtracted contribution defines a fake cross section
that has the same dynamical singularities as the real cross section and whose
kinematics are sufficiently simple to allow its analytic integration.

To define this fake cross section, one must find an approximation
(in $d$ dimensions) to the matrix element (or its square) that matches the
real matrix element in all the singular regions of phase space.  While this
can be done by calculating the full matrix element in $d$ dimensions and
taking all the relevant limits, this is an extremely laborious and
ungeneralizable procedure.  Instead, one can use an approach based on the
soft and collinear factorization theorems (see Sect.~\ref{facsc}),
which guarantee that the
singular terms are process independent.  Thus one can define the
subtraction cross section as a sum of soft and collinear pieces in a
completely process-independent way and integrate it once-and-for-all.

In fact, we go one step further even than this.  By introducing a {\em dipole\/}
factorization theorem (see Sect.~\ref{dff}), we are able to construct a single
approximation to
the matrix element that matches all of its soft and collinear
singularities.  This approximation is given in terms of the Born-level
matrix element times universal (process independent) dipole factors.
Thus we can provide a {\em single\/} formula that
approximates the real matrix element (squared) for an arbitrary process, in {\em
  all\/} of its singular limits.  This avoids problems that can arise from
using the separate soft and collinear approximations, where one must either
define an explicit arbitrary cutoff to separate the two regions, or add
back on another term to compensate for the double-counting of the region of
overlapping singularities that gives rise to double poles.

While the soft and collinear factorization theorems dictate the form the
dipole factor must take in the exactly singular limits, we are free to
choose the extrapolation away from those limits arbitrarily.  Clearly we
should use this freedom to make the necessary analytical integrals as
straightforward as possible.  Physical processes in which some external
momenta are fixed (by incoming hadrons or by measurement of outgoing
hadrons) impose additional constraints on the phase space.  In order to
ensure that the phase-space integrals are still analytically tractable, we
make different choices for the extrapolation away from these limits,
depending on which of the participating parton momenta are fixed by
external hadrons.  Thus we end up with a finite set of different dipole
formulae (see Sects.~\ref{dfss}--\ref{dcorr}),
applicable to different physical processes.

Closely related to the choice of dipole formulae is the definition of the
`similar' $m$-parton configuration. Each dipole factor smoothly describes
the merging of three partons into two new partons (emitter and spectator)
while one of the three partons approaches a singular region. Thus,
for each of the different constraints
on the $m+1$-parton phase space, we are able to define a {\em one-to-one\/}
mapping to a set of $m$ momenta plus a single-parton momentum in such a way that
the single-parton subspace
obeys {\em exact\/} phase space factorization.  That is, the
$m+1$-parton phase space can be written as the product of a {\em physical\/}
$m$-parton phase space times the dipole phase space. By physical, we
mean that all partons are on-shell and energy-momentum conservation is
implemented exactly, as are all the phase-space constraints.
The fact that each dipole factor depends on more than two parton momenta is
essential in order to implement these kinematic features.

Owing to the exact phase-space factorization and the convenient choice of
dipole formulae, it is possible to integrate all of the dipole
contributions analytically over the full dipole phase space in $d$
dimensions.  These result in a set of $\ep$ poles that cancel those in the
one-loop cross section, as well as a set of finite corrections
(see Sects.~\ref{intsub},\ref{1lc}).  It is
worth noting that because of the simple definitions of the dipole formulae,
the origin of all the finite terms that arise can be simply traced, and all
are well-known constants, allowing a powerful check that the integrals have
been performed correctly.

In addition to the poles that cancel with the one-loop contribution, when
there are identified external hadrons the integration of the dipole factors
leads to additional poles that must
be subtracted into the process-independent parton distribution functions.
Practically, this means cancellation against universal (but factorization
scheme- and scale-dependent) collinear counter-terms (see Sect.~\ref{NLOxs}).
The scheme- and
scale-dependences resurface in the finite remainder left over after the
cancellation (see Sect.~\ref{subpp} and Sects.~\ref{sidis}--\ref{eemp}).
Once again, the finite terms can be easily checked, because
they have simple physical origins related to different integrals and
projections of the Altarelli-Parisi splitting functions.

One feature of our algorithm is that it does not require the convolution
with the parton distribution function to be made during Monte Carlo
integration.  One is free to choose either to calculate a hadron-level
cross section, including the convolution, or a parton-level cross section
as a function of the partonic momentum fraction.  The latter can then be
convoluted with the distribution function after Monte Carlo integration.
This can be extremely useful in many respects. One can produce cross sections
with a wide
variety of parton distribution functions, or study the scheme- and
scale-dependence of the results without having to reintegrate for each new
scheme or scale. Moreover one can check whether, in extreme phase space regions,
the NLO partonic cross section contains enhanced (typically,
logarithmically-enhanced) contributions that may spoil the convergence of the
fixed-order perturbative expansion, thus requiring all-order summations.
Finally, the parton-level calculation
can also be important for comparing different
computations, since all the dependence on non-perturbative input can be
removed, and the result is a purely perturbative, parameter-free,
quantity.

Starting from the dipole formulae and the integrals of the dipole factors, in
Sects.~\ref{siee}--\ref{eemp} we have explicitly carried out all the
$d$-dimensional analytic work that is necessary for a straightforward
numerical implementation of NLO calculations in any scattering process. The
results are collected in effective final formulae, which are recalled in the
last Subsection of Sects.~\ref{siee}--\ref{eemp} for each different class of
scattering processes. Using these final formulae, any NLO calculation
requires only the corresponding matrix elements as input. More precisely,
the only additional ingredients needed to
construct a numerical program to calculate the NLO corrections to arbitrary
jet quantities in a given process are:
\vspace{-\parskip}
\begin{itemize}
\addtolength{\itemsep}{-\parskip}
\item a set of independent colour projections of the matrix element squared
  at the Born level, summed over parton polarizations, in $d$ dimensions
  (if the total number of QCD partons involved in the LO matrix element is
  less than or equal to three this is unnecessary, because the colour
  structure exactly factorizes, see Appendix A);
\item the one-loop matrix element in $d$ dimensions;
\item an additional projection of the Born level matrix element over the
  helicity of each external gluon in four dimensions;
\item the tree-level NLO matrix element in four dimensions.
\end{itemize}

We should emphasize that, independently of the actual set-up of our algorithm,
the dipole formalism is fully general and, hence, highly {\em flexible}. The
main point is that it provides an explicit, universal and simple
(in $d$ dimensions) expression for the approximate cross section $d\sigma^A$.
Starting from it and
having the process-independent dipole splitting functions and
their integrals to hand, by direct inspection one can try to modify the
subtraction term (for instance, introducing finite weighting factors or
cut-offs for the dipole terms)
to simplify its treatment in any particular
scattering process. This may be useful for improving the convergence of the
numerical program  and can always be done at the expense of some extra analytic
work in four dimensions on the finite difference between the two approximate
cross sections. In other words, using the dipole formalism one can set up ones
own sub-algorithm.

Generalizing the procedure for constructing NLO Monte Carlo programs for
arbitrary quantities has several advantages.  These are principally because
of the reduction in the number and complexity of ingredients that have to
be calculated for each new process, and because the $d$-dimensional
integrals only need be done once and can be easily checked independently,
rather than being buried inside a specific calculation.

Perhaps the single biggest advantage is the fact that the NLO matrix
element for additional real emission can be calculated in four dimensions.
In calculations involving several partons this can result in great savings
in computation time and in size of the final expressions, because helicity
amplitude techniques can be used.  On the other hand, specific calculations
that construct the approximate cross section directly from the real one
must work in $d$ dimensions, producing very cumbersome formulae at
intermediate stages even though the final result is simple, since it is
just the soft and collinear limits, which factorize.

It is often said that the bottleneck in producing new NLO calculations is
the calculation of the necessary one-loop amplitudes.  While this is
partially true, one can cite many examples where the relevant matrix
elements have been available for a long time, yet no Monte Carlo programs
for arbitrary jet quantities have been available.  Most notorious has been
the case of jet production in deep inelastic scattering, where the matrix
elements can be simply obtained by crossing the $e^+e^-$ ones, which have
been known for many years, but only recently have any NLO jet calculations
been produced [\ref{Mirkes},\ref{CSdis}].  This has severely hampered jet
studies by the HERA experiments, which have been forced to compare data
with parton shower models, or partial calculations.  Other examples include
the longitudinal fragmentation function in $e^+e^-$ annihilation, which was,
until very recently [\ref{eefrag}], perhaps the single simplest uncalculated
QCD quantity.

It is thus clear that the numerical implementation of NLO calculations
forms a second bottleneck.  General algorithms such as ours will certainly
help to reduce the amount of work needed to make these implementations, so
will help to reduce this bottleneck and hence increase the number of
processes in which perturbative QCD can be compared with data in a
quantitative way.

As a final comment on the Monte Carlo implementation of the subtraction
method, we should note that the cancellation of soft and collinear
singularities does not completely solve all the numerical problems.  Although
the subtracted cross section is finite, integrable square-root singularities
may arise in the $m+1$ parton integral.  Square-root singularities cannot be
integrated by \naive\ Monte Carlo methods because they have infinite variance
and the integration procedure would never converge.  The presence of these
singularities can depend on the behaviour of the particular jet observable, so
no subtraction procedure can universally overcome them.  However, it is
straightforward to control the square-root singularities in a completely
general way, without any knowledge of the jet quantity being calculated, using
the standard technique of multi-channel Monte Carlo integration [\ref{Kleiss}]
during the phase-space generation.  In the Monte Carlo implementations of our
algorithm a user-supplied routine analyses the generated momenta, and any
infrared safe jet observable can be used [\ref{CSlett},\ref{CSdis}].

\subsection{Comparison with other general approaches}

The first general algorithm for constructing NLO Monte Carlo programs for
jet cross sections in arbitrary processes was proposed in
Refs.~[\ref{GG},\ref{GGK}], using the phase-space slicing method.  Several
subsequent calculations have been based on it.  In the simple case in which
there are no identified hadrons, discussed in Ref.~[\ref{GG}],
colour-ordered amplitudes are used to derive the soft and collinear
factorization formulae.  An arbitrary, but small, parameter is introduced
to separate the phase space into `resolved' and `unresolved' regions.  In
the latter the cross section and jet quantity are really approximated by their
soft and collinear limits.  The resulting formulae can then be
integrated analytically over the unresolved regions, to cancel the poles in
the virtual cross section.

Although this is an approximate calculation, it becomes exact in the limit
that the cutoff parameter goes to zero.  Unfortunately the errors on the
numerical
integral over the resolved region of phase space then diverge and one must
always make a compromise in the choice of cutoff.  While rules of thumb
have developed to give a rough idea of how small it needs to be, there is
no substitute for explicitly checking that there is no dependence on it.
However, this is not always feasible owing to constraints on computer time.
A particularly poignant example is that of the energy-energy correlation in
\ee\ annihilation.  Since discrepancies exist between dedicated analytical
calculations and general-purpose Monte Carlo programs, a very
high-statistics comparison between three different Monte Carlo programs was
made~[\ref{CGNS}], to check that they really were in agreement.  The result
was a small, but statistically significant, discrepancy between the program
of Ref.~[\ref{GG}] and those of Refs.~[\ref{KN},\ref{CSlett}].  While this
was attributed to a residual dependence on the cutoff, it was not possible
within the available computer time to make a sufficiently accurate
calculation to confirm this, or to extrapolate to zero cutoff.

These problems become increasingly severe as one approaches the edges of
phase space because the approximation performed
within the slicing procedure can strongly interfere with the actual definition
of certain jet observables. Indeed, at the edges of phase space there are
physical parameters, namely ratios of physical scales, that become large.
These large parameters can produce logarithmic and even power-like (!)
enhancement of the cutoff-dependence, thus
requiring such small cutoffs that numerical stability is never reached,
in practice.

However, we should stress that these problems do not imply that, in some case,
the phase-space slicing method is unable to provide reliable QCD predictions.
On one hand, in the evaluation of jet quantities far from the phase-space
boundary, these problems are all obviated if proper
care is taken to explicitly check the cutoff-dependence for every result
quoted and to work in a region in which it is negligible. On the other hand,
in the case of quantities at the edges of phase space, the kinematic region
where one can lose control of the numerical stability is that in which
the perturbative expansion is affected by very large coefficients: in this
region the convergence of the fixed-order expansion is spoiled, the NLO
calculation is insufficient to provide reliable QCD predictions and the latter
require, anyhow, analytic summation techniques to all orders.

A possible disadvantage of the phase-space slicing method
regards precisely this last point. For most jet observables
the only analytical NLO results available
are for the coefficients of logarithmically enhanced terms near the
kinematic limits. Thus, it is a useful test of a complete NLO numerical
calculation to compare it with
these analytical results. At the same time, the numerical calculation can be
used to check these analytical results, which are the first step in the
resummation procedures. The approximation embodied in the phase-space slicing
method, and the ensuing problems of numerical stability in extreme phase-space
regions, may reduce the amount of information that, otherwise, can be provided
by a NLO calculation.

The slicing method of Ref.~[\ref{GG}] was extended to include identified
hadrons in Ref.~[\ref{GGK}].
The approach is to first consider the case in which all partons are
outgoing and then cross some to the initial state as necessary.  As we have
seen in our approach, the kinematic constraints imposed by identifying
external hadrons do not act symmetrically on initial- and final-state
partons.  Therefore `crossing kernels' have to be introduced.

These crossing kernels are similar to the insertion operators $\bom K$
and $\bom P$ (and $\bom H$, in the case of fragmentation processes) that, in
our approach, arise as finite remainders left over after the subtraction
of the universal collinear counter-terms. The main difference between the
crossing kernels of Ref.~[\ref{GGK}] and our kernels $\bom K$
and $\bom P$ is that the latter do depend on the colour charges of the partons
involved in the scattering process. In order to compute the full parton-level
cross sections in NLO, the crossing kernels as well as the insertion operators
$\bom K$ and $\bom P$ must be convoluted with $m$-parton cross sections
(see Sect.~\ref{subpp}).

In Ref.~[\ref{GGK}], the crossing kernels are preconvoluted with the
parton density set to provide a new effective set of `crossing functions',
which are
used in the main integration stage of the Monte Carlo program.  This
procedure can be a considerable problem in one of the main applications of
NLO jet calculations, the extraction of parton distribution functions by
fitting to the data, since this process is usually iterative, with the
distribution functions gradually converging on the best fit.  The use of the
crossing functions would require a
new Monte Carlo integration for each iteration.  In our method on the other
hand, it is straightforward to calculate parton-level cross sections.
These can then be convoluted with any parton distribution functions after
the Monte Carlo integration is completed.  Thus each iteration would then
only require a simple one-dimensional numerical integral.

This disadvantage of the usual implementation of the method of
Ref.~[\ref{GGK}] can obviously be avoided if the crossing kernels are first
convoluted with partonic cross section contributions rather than with
parton densities.

The general properties of soft and collinear radiation were used to
construct a subtraction algorithm for the first time in Ref.~[\ref{KS}],
for one- and two-jet production in hadron-hadron collisions.  This has
recently been modified to deal with three-jet production [\ref{Frix},\ref{NT}]
and, in general, with $n$-jet cross sections. Although this formalism
is based on the subtraction method, so looks superficially similar to
our algorithm, there are in fact a great many differences.

The subtraction procedures used in Refs.~[\ref{Frix}] and [\ref{NT}] differ in
many details but share some common features. Firstly, they select
energy and angle variables by working in a definite reference frame and thus
breaking Lorentz invariance at
intermediate steps of the calculation (although of course it is restored in
the final results). The definite choice of a reference frame then unambiguously
specify the integration variables that can lead to singularities in the
integration of the real matrix element. Thus, one can introduce a partition of
the $m+1$-parton phase space in such a way that in each subregion only one
energy or angle variable can kinematically vanish. Having done that, the
approximate cross section $d\sigma^A$ is defined by means of a double (soft and
collinear) subtraction procedure as the product of the singular variable times
its residue. The residue is evaluated by exactly performing the soft and
collinear limits and thus it can be computed in a process independent manner.

The two main differences between this `residue approach' and our formalism
regard the treatment of the soft and collinear limits and the related definition
of the subtracted $m$-parton configuration. While the dipole formulae provide
a single and smooth approximation of the real matrix element in all of its
singular limits, in the residue approach the soft and collinear regions are
treated separately. Correspondingly, while in the dipole formalism the
subtracted parton configuration is obtained by a one-to-one mapping from the
original $m+1$-parton configuration, in the residue approach the mapping
is achieved by a projection procedure. This
projection of the $m+1$-parton phase space onto $m$-parton phase space does
not allow for an exact phase space factorization. For this reason, one
has to introduce arbitrary cutoffs to define the upper limits on the
integration variables, although the dependence on these cutoffs should
cancel, to within the numerical accuracy, in the final result.

The residue approach can be recovered as a particular set-up of the dipole
formalism. Starting from our general expression for the subtracted cross
section $d\sigma^A$ one can project it onto any properly defined soft and
collinear subspace.

\subsection{Future outlook}

Looking to the future, there are several avenues along which the present
work could be continued.  The first is obviously to apply it to as many
processes as possible, in particular those for which no other calculations
exist.  We have already constructed Monte Carlo programs for jets in \ee\
annihilation [\ref{CSlett}] and deep inelastic scattering [\ref{CSdis}] and
several more applications are in progress.

In terms of extending the algorithm itself, the most obvious missing
feature at present is the treatment of heavy quark effects.  It is
straightforward to extend the dipole formalism to incorporate massive
partons, either in jet calculations or with fragmentation
functions [\ref{CS1}].
This is clearly an important extension, as many heavy quark
processes are good probes of perturbative QCD, but many of these that are
being experimentally measured are not yet predicted to NLO\@.

Another potentially important extension is the generalization to polarized
partons, which is also straightforward in the dipole formalism.

Looking further ahead, it is to be hoped that at some stage NNLO
calculations of jet observables will be attempted.  Even once the necessary
two-loop matrix elements are calculated, the amount of work needed to
provide a numerical implementation will be enormous.  Clearly any progress
that can be made in the meantime to set up a general-purpose NNLO
subtraction algorithm will speed up the process of bringing the new
calculation to the marketplace.  The dipole formalism seems particularly
suited to this task.

When starting this research project, we had in mind a main final goal:
a method for carrying out NNLO QCD calculations.
Having set up a completely general NLO algorithm, we are
confident that this main goal can indeed be achieved.

\section*{Acknowledgements}
\addcontentsline{toc}{section}{Acknowledgements}

We are grateful to Willy van Neerven for useful discussions, and for pointing
out the reference in which the integral in Eq.~(\ref{intapp}) was performed.
One of
us (SC) would like to thank the CERN TH Division for hospitality and partial
support during the course of this work.  We would also like to thank the
organisers of the QCD Summer Institute at Gran Sasso Nat.\ Lab.\ in August 1994
where this project was begun.

We are grateful to Keith Ellis for pointing out an error in the Lorentz
transformations originally used in sections~\ref{dissa} and~\ref{dcorr}
(Eqs.~(\ref{ktilde}--\ref{klor2}) and~(\ref{klorn}--\ref{qdef}) respectively),
which has been fixed in the present version.  No other results are
affected, because they only rely on the existence of such a transformation,
and not on its exact definition.

\newpage

\renewcommand{\theequation}{A.\arabic{equation}}
\setcounter{equation}{0}

\section*{Appendix A:~~Colour Algebra}
\addcontentsline{toc}{section}{Appendix A:~~Colour Algebra}

In order to practice with the colour algebra we can consider some simple
examples. In the simplest cases with two or three partons
(regardless of whether they are final- or initial-state partons),
the colour algebra can be performed in factorized (closed) form.

For the case with two partons, using colour conservation,
we have:
\beq
\label{colour1}
{\bom T}_1 \cdot {\bom T}_2 \; \ket{1,2} =
- {\bom T}_1 \cdot {\bom T}_1 \; \ket{1,2} =
- {\bom T}_1^2 \; \ket{1,2} = - {\bom T}_2^2 \; \ket{1,2} \;,
\eeq
so that all the charge operators $\{ {\bom T}_1^2, {\bom T}_2^2, \;
- {\bom T}_1 \cdot {\bom T}_2 \}$ are factorizable in terms of the (scalar)
Casimir operator $C_1 = C_2$.

Using colour conservation for the three-parton case
we have:
\beeq
0 &=& \left(\sum_{i=1}^3 {\bom T}_i \right)^2 \ket{1,2,3}
\nonumber \\
&=& \left({\bom T}_1^2 + {\bom T}_2^2 + {\bom T}_3^2 +
2 \,{\bom T}_1 \cdot {\bom T}_2 + 2 \,{\bom T}_1 \cdot {\bom T}_3
+ 2 \,{\bom T}_2 \cdot {\bom T}_3 \right) \; \ket{1,2,3} \;\;,
\eeeq
and
\beq
( {\bom T}_1 \cdot {\bom T}_2 + {\bom T}_1 \cdot {\bom T}_3 )
\; \ket{1,2,3} = - {\bom T}_1^2 \;\ket{1,2,3} \;.
\eeq
Combining these two equations we get:
\beq
\label{colour2}
2 \,{\bom T}_2 \cdot {\bom T}_3 \;\ket{1,2,3} =
\left( {\bom T}_1^2 - {\bom T}_2^2 - {\bom T}_3^2 \right) \;\ket{1,2,3}
\eeq
and similarly for ${\bom T}_1 \cdot {\bom T}_3$ and
${\bom T}_1 \cdot {\bom T}_2$.
Therefore, all the charge operators are factorizable in terms of linear
combinations of the Casimir invariants $C_1, C_2, C_3$ of the three partons.

When the total number $n$ of partons is $n \geq 4$ the colour algebra does not
factorize any longer.
For instance, if $n = 4$ we have four trivial relations, namely
\beq
{\bom T}_i^2 \ket{1,2,3,4} = C_i \ket{1,2,3,4} \;\;, \;\;\;\;i=1,...,4 \;\;.
\eeq
As for the remaining six charge operators
${\bom T}_i \cdot {\bom T}_j (i \neq j)$, we can use the following four
identities (due to charge conservation)
\beq
{\bom T}_i \cdot \sum_{j=1}^4 {\bom T}_j \ket{1,2,3,4} = 0
\;\;, \;\;\;\;i=1,...,4 \;\;,
\eeq
in order to single out two independent charge operators. For instance we can
write:
\beeq
{\bom T}_3 \cdot {\bom T}_4 \ket{1,2,3,4} &=&
\left[ \frac{1}{2} \left( C_1 + C_2 - C_3 - C_4 \right)
+ {\bom T}_1 \cdot {\bom T}_2 \right] \ket{1,2,3,4} \;\;, \nonumber \\
{\bom T}_2 \cdot {\bom T}_4 \ket{1,2,3,4} &=&
\left[ \frac{1}{2} \left( C_1 + C_3 - C_2 - C_4 \right)
+ {\bom T}_1 \cdot {\bom T}_3 \right] \ket{1,2,3,4} \;\;,\nonumber\\
{\bom T}_2 \cdot {\bom T}_3 \ket{1,2,3,4} &=&
\left[ \frac{1}{2} \left( C_4 - C_1 - C_2 - C_3 \right)
- {\bom T}_1 \cdot {\bom T}_2 - {\bom T}_1 \cdot {\bom T}_3 \right]
\ket{1,2,3,4} \;\;,\nonumber\\
{\bom T}_1 \cdot {\bom T}_4 \ket{1,2,3,4} &=&
- \left( C_1 + {\bom T}_1 \cdot {\bom T}_2 + {\bom T}_1 \cdot {\bom T}_3
\right) \ket{1,2,3,4} \;\;,
\eeeq
and express all the charge operators in terms of Casimir invariants and
${\bom T}_1 \cdot {\bom T}_2, {\bom T}_1 \cdot {\bom T}_3$. The actual values
of the latter depend on the detailed colour configuration of the four-parton
state, namely\footnote{To be precise, if some of the partons in
  $\ket{1,2,3,4}$ are initial-state partons, the right-hand sides of
  Eq.~(\ref{n4colour}) should contain an additional normalization factor as in
  Eq.~(\ref{ccamp}).}
\beeq
\label{n4colour}
\bra{1,2,3,4} {\bom T}_1 \cdot {\bom T}_2 \ket{1,2,3,4} \!\!&\!\!=\!\!&\!\!
\left[ {\cal M}_4^{b_1 b_2 a_3 a_4}(p_1,p_2,p_3,p_4) \right]^*
T_{b_1a_1}^c T_{b_2a_2}^c {\cal M}_4^{a_1 a_2 a_3 a_4}(p_1,p_2,p_3,p_4) \;,
\nonumber \\
\bra{1,2,3,4} {\bom T}_1 \cdot {\bom T}_3 \ket{1,2,3,4} \!\!&\!\!=\!\!&\!\!
\left[ {\cal M}_4^{b_1 a_2 b_3 a_4}(p_1,p_2,p_3,p_4) \right]^*
T_{b_1a_1}^c T_{b_3a_3}^c {\cal M}_4^{a_1 a_2 a_3 a_4}(p_1,p_2,p_3,p_4) \;.
\nonumber \\
&~&
\eeeq

In the general case with $n > 4$ partons, one should consider $n(n+1)/2$
charge operators ${\bom T}_i \cdot {\bom T}_j$. Among them there are $n$ trivial
factorizable contributions, i.e.\ ${\bom T}_i^2 \ket{...}_n = C_i \ket{...}_n$.
In addition, one can use $n$ charge conservation constraints (i.e.\
${\bom T}_i \cdot \sum_{j=1}^n  {\bom T}_j  \ket{...}_n = 0$) to end up
with $n(n-3)/2$ independent charge operators whose action onto the matrix
elements has to be evaluated explicitly.

\newpage

\renewcommand{\theequation}{B.\arabic{equation}}
\setcounter{equation}{0}

\section*{Appendix B:~~Soft Integrals}
\addcontentsline{toc}{section}{Appendix B:~~Soft Integrals}

In this Appendix we explicitly perform the integration of the soft terms
$v_{i,ab}$ in Eqs.~(\ref{calvtavn}) and (\ref{calvtavnf}) over
the phase space volumes in Eqs.~(\ref{dpinx}) and (\ref{dpinz}) respectively.

Let us first discuss the space-like case (see Eq.~(\ref{intVainin})). We have
to carry out the following integration
\beq
\label{Is}
I_{soft}(x) = 8 \pi \as  \mu^{2\ep} \int \left[ dp_i(n,p_a, x) \right]
\frac{1}{2p_ap_i} \;v_{i,ab} \;\;.
\eeq
Since the vector $n^\mu$ in Eq.~(\ref{ndef}) is time-like, we can work in its
rest frame and consider the integration variables $E_i,\theta,\phi$ defined
as follows
\beeq
\label{kin}
&&n^\mu = E (1, \dots) \;\;, \;\;\;\;\;\;p_a^\mu = E_a (1, \dots, 1)\;\;,
\;\;\;\;\;\;
p_a^\mu + p_b^\mu  = E_{ab} (1, \dots, v \sin \chi, v \cos \chi) \;\;,
\nonumber \\
&&p_i^\mu  =
E_{i} (1, ..\mbox{`angles'}.., \sin \theta \cos \phi, \cos \theta) \;\;.
\eeeq
In Eq.~(\ref{kin}) the dots stand for vanishing components, while the notation
`angles' denotes the dependence of $p_i$ on the $d-4$ angular variables that
can be trivially integrated in Eq.~(\ref{Is}).

Using Eqs.~(\ref{xaindef},\ref{dpinx}) for the phase space
$\left[ dp_i(n,p_a, x) \right]$ and Eq.~(\ref{viabdef}) for $v_{i,ab}$ and
introducing the variables in Eq.~(\ref{kin}), the integral (\ref{Is}) becomes:
\beeq
\label{Is1}
I_{soft}(x) \!\!&\!=\!& \Theta(x) \;\Theta(1-x) \;\frac{\as}{2\pi}
\;\frac{\Gamma(1-\ep)}{\Gamma(1-2\ep)} \int_0^\infty \frac{dE_i}{E_i}
\left( \frac{\pi \mu^2}{E_i^2} \right)^\ep \delta\!\left(x - \frac{E_a-E_i}{E_a}
\right) \\
\!\!&\!\cdot\!&\!\!\! \int_0^\pi d\theta (\sin \theta)^{1-2\ep} \int_0^\pi
\frac{d\phi}{\pi}
(\sin \phi)^{-2\ep} \;\frac{1- v \cos \chi}{(1- \cos \theta)
(1 - v \sin \chi \sin \theta \cos \phi - v \cos \chi \cos \theta)} \;\;.
\nonumber
\eeeq
The integration over the energy $E_i$ in Eq.~(\ref{Is1}) is straightforward:
\beq
\label{Ein}
\int_0^\infty \frac{dE_i}{E_i} \left( \frac{\pi \mu^2}{E_i^2} \right)^\ep
\;\delta\!\left(x - \frac{E_a-E_i}{E_a} \right) = \frac{1}{1-x}
\left( \frac{\pi \mu^2}{(1-x)^2 E_a^2} \right)^\ep \;\;,
\eeq
while for the angular integrals we can use the result of Ref.~[\ref{Been}],
namely
\beeq
\label{intapp}
&&\int_0^\pi d\theta (\sin \theta)^{1-2\ep} \int_0^\pi \frac{d\phi}{\pi}
(\sin \phi)^{-2\ep} \;\frac{1- v \cos \chi}{(1- \cos \theta)
(1 - v \sin \chi \sin \theta \cos \phi - v \cos \chi \cos \theta)}
\nonumber \\
&&= - \frac{1}{\ep} \left[ \frac{1-v^2}{(1-v \cos \chi)^2} \right]^\ep
\; - \ep \left[ 2 {\rm Li}_2\left( - \frac{v(1-\cos \chi)}{1-v} \right) -
2 {\rm Li}_2\left( - \frac{v(1+\cos \chi)}{1-v \cos \chi} \right) \right.
\nonumber \\
&&+ \left.
\ln^2 \frac{1-v}{1-v \cos \chi} - \frac{1}{2} \;\ln^2 \frac{1+v}{1-v}
- \frac{1}{2} \;\ln^2 \frac{1-v^2}{(1-v \cos \chi)^2} \right] + {\cal O}(\ep^2)
\;\;,
\eeeq
which, by means of the identity ${\rm Li}_2(-v(1+\cos \chi)/(1-v \cos \chi)) =
- {\rm Li}_2(v(1+\cos \chi)/(1+v)) - (1/2) \ln^2((1-v \cos \chi)/(1+v))$
(i.e.\ ${\rm Li}_2(1-1/x) = - {\rm Li}_2(1-x) -(1/2) \ln^2x$,
for $0\leq x \leq 1$), can be rewritten as follows
\beeq
\!\!\!\!\!&&\int_0^\pi d\theta (\sin \theta)^{1-2\ep} \int_0^\pi \frac{d\phi}{\pi}
(\sin \phi)^{-2\ep} \;\frac{1- v \cos \chi}{(1- \cos \theta)
(1 - v \sin \chi \sin \theta \cos \phi - v \cos \chi \cos \theta)}
\nonumber \\
\!\!\!\!\!&&= - \left[ \frac{1-v^2}{(1-v \cos \chi)^2} \right]^\ep
\left\{ \frac{1}{\ep}
+ \ep \left[ 2 {\rm Li}_2\!\left( - \frac{v(1-\cos \chi)}{1-v} \right) +
2 {\rm Li}_2\!\left( \frac{v(1+\cos \chi)}{1+v} \right) \right]
+ {\cal O}(\ep^2) \right\} \;\;. \nonumber \\
\!\!\!\!\!&~&
\eeeq
Thus, we find
\beeq
I_{soft}(x) \!\!&\!=\!& \Theta(x) \;\Theta(1-x) \;\frac{\as}{2\pi}
\;\frac{\Gamma(1-\ep)}{\Gamma(1-2\ep)} \;(1-x)^{-1 -2\ep}
\left[ \frac{(1-v^2) \pi \mu^2}{E_a^2 (1-v \cos \chi)^2} \right]^\ep
\nonumber \\
\!\!&\!\cdot\!& \left\{ - \frac{1}{\ep}
- \ep \left[ 2 {\rm Li}_2\!\left( - \frac{v(1-\cos \chi)}{1-v} \right) +
2 {\rm Li}_2\!\left( \frac{v(1+\cos \chi)}{1+v} \right) \right]
+ {\cal O}(\ep^2) \right\} \;\;.
\eeeq
We can then perform the $\ep$-expansion of the factor $(1-x)^{-1 -2\ep}$
according to Eq.~(\ref{vpd}):
\beq
(1-x)^{-1 -2\ep} = - \frac{1}{2\ep} \delta(1-x) + \left( \frac{1}{1-x}\right)_+
+ 2 \ep \left(\frac{1}{1-x} \ln\frac{1}{1-x} \right)_+ + {\cal O}(\ep^2) \;\;,
\eeq
and, relating the variables $v, \cos \chi, E_a$  in Eq.~(\ref{kin}) to the
Lorentz invariants:
\beeq
&&\frac{4E_a^2 (1- v \cos \chi)^2}{1-v^2} = 2 p_a \cdot p_b \;\;,
\;\;\;\;\frac{2 (1- v \cos \chi)}{1-v^2} =
\frac{(p_a + p_b) \cdot n}{p_a \cdot n} \;\;, \nonumber \\
&&v= \sqrt { 1 - \frac{ n^2 (p_a+p_b)^2 }{ [ (p_a+p_b)\cdot n ]^2}} \;\;,
\eeeq
we end up with the final result (note, $\Gamma^2(1-\ep)/\Gamma(1-2\ep) =
1 - \ep^2 \pi^2/6 + {\cal O}(\ep^3)$)
\beeq
I_{soft}(x) &=& \Theta(x) \;\Theta(1-x) \frac{\as}{2\pi}
\;\frac{1}{\Gamma(1-\ep)}
\left( \frac{4 \pi \mu^2}{2 p_ap_b} \right)^\ep \left\{ \frac{1}{2\ep^2}
\delta(1-x) - \frac{1}{\ep} \left( \frac{1}{1-x}\right)_+ \right.
\nonumber \\
&-& \left. \left( \frac{2}{1-x} \ln\frac{1}{1-x} \right)_+ + \delta(1-x)
\left[ 2
{\rm Li}_2\!\left(1- \frac{(1+v)}{2}\frac{(p_a+p_b)\cdot n}{p_a \cdot n} \right)
\right. \right. \nonumber \\
&+& \left. \left. 2 {\rm Li}_2\!\left(1- \frac{(1-v)}{2}
\frac{(p_a+p_b)\cdot n}{p_a \cdot n} \right) - \frac{\pi^2}{12} \right]
+ {\cal O}(\ep) \right\} \;\;.
\eeeq

Let us now consider the time-like case (see Eq.~(\ref{intVainf})) and
compute the integral
\beq
\label{Js}
J_{soft}(z) = 8 \pi \as  \mu^{2\ep} \int \left[ dp_i(n;p_a, z) \right]
\frac{1}{2p_ap_i} \;\frac{v_{i,ab}}{z} \;\;.
\eeq
As in the previous case we work in the rest frame of  $n^\mu$.
Using Eqs.~(\ref{zaindef},\ref{dpinz}) for $\left[ dp_i(n;p_a, z) \right]$
and Eq.~(\ref{viabdef}) for $v_{i,ab}$ and introducing the kinematic variables
in Eq.~(\ref{kin}), we obtain an expression for $J_{soft}(z)$ that is equal
to that in Eq.~(\ref{Is1}) apart from the replacement:
\beq
\delta\!\left(x - \frac{E_a-E_i}{E_a} \right) \to
z^{1-2\ep} \;\delta\!\left(z - \frac{E_a}{E_a+E_i} \right) \;.
\eeq
Therefore the $E_i$ integration in $J_{soft}(z)$, namely
\beq
\int_0^\infty \frac{dE_i}{E_i} \left( \frac{\pi \mu^2}{E_i^2} \right)^\ep
z^{1-2\ep} \delta\left(z - \frac{E_a}{E_a+E_i} \right) = \frac{1}{1-z}
\left( \frac{\pi \mu^2}{(1-z)^2 E_a^2} \right)^\ep \;\;,
\eeq
gives exactly the same factor as in Eq.~(\ref{Ein}) and we immediately obtain
\beq
J_{soft}(z) = I_{soft}(z) \;\;.
\eeq

\newpage

\renewcommand{\theequation}{C.\arabic{equation}}
\setcounter{equation}{0}

\section*{Appendix C:~~Collection of Main Formulae}
\addcontentsline{toc}{section}{Appendix C:~~Collection of Main
  Formulae}

In this Appendix we collect together the main formulae that are needed to
implement our algorithm for calculating jet cross sections.

\begin{table}[b]
\begin{center}
\begin{tabular}{|l|l|l|l|l|}
\hline
 dipole                  &               & dipole     & phase &
 dipole splitting    \\
 factor                  & definition    & kinematics & space &
 functions {\bom V}  \\
\hline \hline \normalsize
                         &               &            &       &              \\
 ${\cal D}_{ij,k}$       & (\ref{dipff}) & (\ref{pk}),
                                           (\ref{yijk}),
                                         (\ref{tilz}) &
                                               (\ref{exdpi}) &
                                               (\ref{Vqigjk}--\ref{vggk})    \\
                         &               &            &      &               \\
 ${\cal D}_{ij}^a$       & (\ref{dipfi}) & (\ref{wpa}),
                                           (\ref{wpax}),
                                        (\ref{tilzi}) &
                                              (\ref{expx}) &
                                              (\ref{Vqigja}--\ref{Vqibqja})  \\
                         &               &            &      &               \\
 ${\cal D}^{ai}_k$       & (\ref{dipif}) & (\ref{wpaia}--\ref{wpk}),
                                          (\ref{dui}) &
                                               (\ref{exdpix}) &
                                               (\ref{Vqagik}--\ref{Vgagik})  \\
                         &               &            &      &               \\
 ${\cal D}_{ij,a}$       & (\ref{dipffa})& (\ref{dzija}--\ref{tzia}),
                                                      &
                                             (\ref{exdpiz}) &
                                             (\ref{Vqigjaa}--\ref{Vgigjaa})  \\
                         &               &            &      &               \\
 ${\cal D}_{ai,k}$       & (\ref{dipaff})& (\ref{wpaimu}),
                                           (\ref{wpaimuz}),
                                          (\ref{duia})&
                                             (\ref{expsaik}) &
                                             (\ref{Vqagika}--\ref{Vgagika})  \\
                         &               &            &      &               \\
 ${\cal D}^{ai,b}$       & (\ref{dipiif})& (\ref{wpab}--\ref{dqmu}),
                                                      &
                                               (\ref{exdpiabx}) &
                                               (\ref{Vqagib}--\ref{Vgagib})  \\
                         &               &            &      &               \\
 ${\cal D}^{(n)\, ai}_b\,,$
 ${\cal D}^{(n)\, ai,b}$ &
                         (\ref{dipainin})& (\ref{ndef}),
                                           (\ref{wpai}--\ref{qdef}),
                                           (\ref{viabdef})
                                                      &
                                               (\ref{exdpinx}) &
                                               (\ref{Vqagin}--\ref{Vgagin})  \\
                         &               &            &      &               \\
 ${\cal D}^{(n)}_{ai,b}$,
${\cal D}^{(n)\, b}_{ai}$&
                          (\ref{dipainf})& (\ref{ndef}),
                                           (\ref{wpaif}),
                                           (\ref{zaindef}),
                                                      &
                                             (\ref{exdpinz}) &
                                             (\ref{Vqaginb}--\ref{Vgaginb})  \\
&&\hspace{1em}                             (\ref{klorn}--\ref{qdef}),
                                           (\ref{viabdef})&&\\
                         &               &            &      &               \\
\hline
\end{tabular}
\end{center}
\caption{List of the dipole factorization formulae used to construct the
universal subtraction term $d\sigma^A$. The numbering refers to the Equations in
Sect.~\protect\ref{dff}.}
\end{table}
According to our method, the final expressions for the NLO cross sections
are given in terms of contributions with $m+1$-parton and $m$-parton kinematics,
denoted as $\sigma^{NLO \,\{m+1\}}$, $\sigma^{NLO \,\{m\}}$,
${\hat \sigma}^{NLO \,\{m\}}$. The cross section $\sigma^{NLO \,\{m+1\}}$ is
obtained by subtracting the counter-term $d\sigma^A$ from the real contribution
$d\sigma^R$. The counter-term $d\sigma^A$ is
constructed by using the dipole factors introduced in Sect.~\ref{dff}. These
dipole factors are collected in Table~1, where we list the Equations with the
dipole definition, the corresponding kinematics and the related splitting
functions. Remember that, when computing $\sigma^{NLO \,\{m+1\}}$,
all the dipole factors have to be directly evaluated in four space-time
dimensions.

The cross sections $\sigma^{NLO \,\{m\}}$ and ${\hat \sigma}^{NLO \,\{m\}}$ are
obtained by adding the virtual contribution $d\sigma^V$ and the collinear
counter-term $d\sigma^C$ to the integral of the subtraction counter-term
$d\sigma^A$. In the rest of this Appendix we concentrate
on the formulae needed to construct the final integral of the subtraction
counter-term.

\subsection*{Flavour kernels}
The majority of the functions and constants we encounter are derived from
the spin-averaged Altarelli-Parisi splitting functions in
$d = 4 - 2 \ep$ dimensions:
\beeq
< {\hat P}_{qq}(x;\ep) > \;\,=\;\, < {\hat P}_{{\bar q}{\bar q}}(x;\ep) >
&=& C_F
\;\left[ \frac{1 + x^2}{1-x} - \ep (1-x) \right] \;\;,
\\
< {\hat P}_{qg}(x;\ep) > \;\,=\;\, < {\hat P}_{{\bar q}g}(x;\ep) > &=& C_F
\;\left[ \frac{1 + (1-x)^2}{x} - \ep x \right] \;\;,
\\
< {\hat P}_{gq}(x;\ep) > \;\,=\;\, < {\hat P}_{g{\bar q}}(x;\ep) >&=& T_R
\left[ 1 - \frac{2 x(1-x)}{1-\ep} \right] \;\;,
\\
< {\hat P}_{gg}(x;\ep) > &=& 2C_A
\;\left[ \frac{x}{1-x} + \frac{1-x}{x}
+ x(1-x) \right] \;\;,
\eeeq
and $< {\hat P}_{q{\bar q}}(x;\ep) > \,=\,
< {\hat P}_{{\bar q}q}(x;\ep) > \,=\, 0$.
It is sometimes useful to express these as a power series in $\ep,$
\beq
< {\hat P}_{ab}(x;\ep) > \;=\; < {\hat P}_{ab}(x;\ep=0) >
- \,\ep \;{\hat P}^{ \;\prime}_{ab}(x) + {\cal O}(\ep^2) \;\;.
\eeq
The regularized Altarelli-Parisi probabilities are the sum of these and the
virtual corrections, evaluated in four dimensions,
\beeq
\label{pqgapp}
P^{qg}(x) \;\,=\;\, P^{{\bar q}g}(x) &=& C_F \;\frac{1 + (1-x)^2}{x} \;\;,
\\
\label{pgqapp}
P^{gq}(x) \;\,=\;\, P^{g{\bar q}}(x) &=& T_R \left[ x^2 + (1-x)^2 \right] \;\;,
\\
\label{pqqapp}
P^{qq}(x) \;\,=\;\, P^{{\bar q}{\bar q}}(x)
&=& C_F \;\left( \frac{1 + x^2}{1-x} \right)_+ \;\;,
\eeeq
\beq
\label{pggapp}
P^{gg}(x) = 2C_A \;\left[ \left( \frac{1}{1-x} \right)_+ + \frac{1-x}{x}
-1 + x(1-x) \right]
+ \delta(1-x) \left( \frac{11}{6} C_A - \frac{2}{3}
N_f T_R \right) \;\;.
\eeq
Note that the final-state Altarelli-Parisi probabilities $P_{ab}(x)$ are
identical to these initial-state ones, $P_{ab}(x) = P^{ab}(x),$ so we do
not usually make any distinction.  Their regular parts are denoted as follows
\beeq
&&P^{ab}_{{\rm reg}}(x) = P^{ab}(x) \;\; \;\;\;\;\;\;\;\;\;\;\;\;\;{\rm if}
\;\;a \neq b \;\;,
\nonumber \\
&&P^{qq}_{{\rm reg}}(x) = - C_F \,(1 + x) \;\;, \;\;\;
P^{gg}_{{\rm reg}}(x) = 2\, C_A \left[ \frac{1-x}{x}  - 1 + x(1-x) \right] \;\;.
\eeeq
After integrating the dipole formulae for final state emission, we obtain
the constants
\beq
\label{gaapp}
\gamma_q = \gamma_{\bar q} = \frac{3}{2} C_F \;, \;\;\;\;
\gamma_{g} = \frac{11}{6} C_A - \frac{2}{3} T_R N_f \;,
\eeq
and
\beq
\label{kcapp}
K_q = K_{\bar q} = \left( \frac{7}{2} - \frac{\pi^2}{6} \right) C_F \;,
\;\;\;\;
K_{g} = \left( \frac{67}{18} - \frac{\pi^2}{6} \right) C_A - \frac{10}{9}
T_R N_f \;,
\eeq
which are related to various integrals of the Altarelli-Parisi splitting
functions. As a matter of fact, we have
\beq
- \frac{1}{2} \sum_b \int_0^1 dz \;\left( z (1-z) \right)^{-\ep}
\;< {\hat P}_{ab}(z;\ep) > = 2 {\bom T}_a^2 \frac{1}{\ep} + \gamma_a
+ \left( K_a - \frac{\pi^2}{6}{\bom T}_a^2 \right) \;\ep + {\cal O}(\ep^2) \;\;.
\eeq

After integrating the expressions for emission from a dipole in which one
identified hadron participates, we obtain the functions:
\beeq
\label{bkabapp}
&&{\overline K}^{ab}(x) =
{\hat P}^{ \;\prime}_{ab}(x) + P^{ab}_{{\rm reg}}(x) \;\ln\frac{1-x}{x}
\nonumber \\
&& + \;\delta^{ab} \left[ {\bom T}_{a}^2
\left( \frac{2}{1-x} \ln\frac{1-x}{x} \right)_+
- \delta(1-x) \left( \gamma_a + K_a - \frac{5}{6}\pi^2 \;{\bom T}_{a}^2 \right)
\right] \;\;,
\eeeq
or, explicitly,
\beeq
{\overline K}^{qg}(x) \;\,= \;\,{\overline K}^{{\bar q}g}(x)
&=& P^{qg}(x) \ln\frac{1-x}{x} + C_F \;x \;\;, \\
\label{okgq}
{\overline K}^{gq}(x) \;\,=\;\, {\overline K}^{g{\bar q}}(x)
&=& P^{gq}(x) \ln\frac{1-x}{x} + T_R \;2x(1-x) \;\;, \\
\label{okqq}
{\overline K}^{qq}(x) \;\,=\;\, {\overline K}^{{\bar q}{\bar q}}(x)
&=& C_F \left[
\left( \frac{2}{1-x} \ln\frac{1-x}{x} \right)_+
- (1+x) \ln\frac{1-x}{x} + (1-x) \right] \nonumber\\
&-& \delta(1-x) \left( 5 - \pi^2 \right) C_F\;\;, \\
{\overline K}^{gg}(x) &=& 2 C_A \left[
\left( \frac{1}{1-x} \ln\frac{1-x}{x} \right)_+
+ \left( \frac{1-x}{x} - 1 + x(1-x) \right) \ln\frac{1-x}{x} \right] \nonumber\\
&-& \delta(1-x) \left[ \left( \frac{50}{9} -\pi^2 \right) C_A -
\frac{16}{9} T_R N_f \right]
\;\;,
\eeeq
and ${\overline K}^{{\bar q}q}(x) = {\overline K}^{q{\bar q}}(x) = 0$.

Finally, when two identified hadrons participate, we obtain the following
additional functions
\beeq
\label{wkabapp}
&&\!\!\!\!\!{\widetilde K}^{ab}(x) = P^{ab}_{{\rm reg}}(x) \;\ln(1-x)
+ \;\delta^{ab} \,{\bom T}_a^2 \left[ \left( \frac{2}{1-x} \ln (1-x)
\right)_+ - \frac{\pi^2}{3} \delta(1-x) \right] \;\;.
\eeeq

Although it is not as closely related to the Altarelli-Parisi splitting
functions, we include in this Section the integral of the
{\em pseudo}dipole splitting function encountered in multiparticle
correlations,
\beeq
\label{clabapp}
\!\!{\cal L}^{a,b}(x;p_a,p_b,n) &=& \delta^{ab} \;
\delta(1-x) \,2 \,{\bom T}_a^2 \left[ \,
{\rm Li}_2\!\left(1- \frac{(1+v)}{2} \frac{(p_a+p_b)\cdot n}{p_a\cdot n}
\right) \right.
\nonumber \\
\!\!&+& \left. {\rm Li}_2\!\left(1- \frac{(1-v)}{2}
\frac{(p_a+p_b)\cdot n}{p_a \cdot n} \right)
\right]
- P^{ab}_{{\rm reg}}(x) \ln \frac{n^2(p_a \cdot p_b)}{2 (p_a \cdot n)^2} \;\;,
\eeeq
\beq
v= \sqrt { 1 - \frac{ n^2 (p_a+p_b)^2 }{ [ (p_a+p_b)\cdot n ]^2}} \;\;,
\eeq
\beq
{\rm Li}_2(x) = - \int_0^x \frac{dz}{z} \;\ln (1-z) \;\;.
\eeq
Remember that the four-momentum $n^\mu$ is defined as follows
\beq
n^\mu = p^\mu_{\rm {in}} - \sum_{a
\in {\rm {final \,state}}} p_a^\mu \;\;,
\eeq
where $p^\mu_{\rm {in}}$ is the total incoming momentum in the scattering
process and the second term on
the right-hand side is the sum of all the momenta of the identified partons
in the final-state.

In addition, partonic cross sections for processes involving identified
hadrons also depend on the scheme-dependent flavour functions $\Kab(x)$
and $\Hba(x)$.  In the ${\overline {\rm MS}}$ scheme, all are zero,
\beq
\label{kabapp}
  K^{ab}_{\overline {\rm MS}} = H_{ba}^{\overline {\rm MS}} = 0.
\eeq
In the DIS scheme, the initial-state functions are given by (see [\ref{AEM}]
and Appendix~D),
\beeq\label{kqqscheme}
  K^{qq}_{\rm DIS}(x) &=& K^{{\bar q}{\bar q}}_{\rm DIS}(x) \;=\; C_F\left[
    \frac{1+x^2}{1-x}\left(\log\frac{1-x}{x}
      -\frac34\right)+\frac14(9+5x)\right]_+, \\
\label{kgqscheme}
  K^{gq}_{\rm DIS}(x) &=& K^{g{\bar q}}_{\rm DIS}(x) \;=\; T_R\left[
    (x^2+(1-x)^2)\log\frac{1-x}{x}+8x(1-x)-1
    \right] \;\;,
\eeeq
and $K^{qg}_{\rm DIS}(x) = K^{{\bar q}g}_{\rm DIS}(x) = - K^{qq}_{\rm DIS}(x)$,
$K^{gg}_{\rm DIS}(x) = - 2 N_f \,K^{gq}_{\rm DIS}(x)$,
$K^{{\bar q}q}_{\rm DIS}(x) = K^{q{\bar q}}_{\rm DIS}(x) = 0$.

\noindent In the case of fragmentation processes, a factorization scheme
conceptually analogous to the DIS scheme has been introduced in
Ref.~[\ref{AEMP}].

\subsection*{Insertion Operators}

The final result for singular part of the integral of the dipole splitting
functions is the same in all processes, and depends on the universal insertion
operator $\bom I$.
In order to write the result in a uniform way for all dipole contributions,
we use the notation $\{p\}$ to denote a set of parton momenta, without
specifying which are identified and whether they are in the initial or
final state.  $I$ and $J$ are indices over all these momenta, and we obtain
\beeq
\label{Iapp}
{\bom I}(\{p\};\ep) = -
\frac{\as}{2\pi}
\frac{1}{\Gamma(1-\ep)} \sum_I \;\frac{1}{{\bom T}_{I}^2} \;{\cal V}_I(\ep)
\; \sum_{J \neq I} {\bom T}_I \cdot {\bom T}_J
\; \left( \frac{4\pi \mu^2}{2 p_I\cdot p_J} \right)^{\ep}
 \;\;. \phantom{9.99,9.99,9.99}
\eeeq
Note that the scalar products $p_I\cdot p_J$ are always positive.  If either
momentum is crossed between the initial and final states, our uniform notation
ensures that $p_I\cdot p_J$ retains the same sign.
The universal singular function ${\cal V}_I(\ep)$ in Eq.~(\ref{Iapp}) depends
only on the parton flavour and has the following $\ep$-expansion
\beq
\label{VIapp}
{\cal V}_{I}(\ep) = {\bom T}_{I}^2 \left( \frac{1}{\ep^2} -
\frac{\pi^2}{3} \right) + \gamma_I \;\frac{1}{\ep}
+ \gamma_I + K_I + {\cal O}(\ep) \;\;,
\eeq
where the constants $\gamma_I$ and  $K_I$ are given in
Eqs.~(\ref{gaapp},\ref{kcapp}). The insertion operator $\bom I$ enters into the
calculation of the cross sections $\sigma^{NLO \,\{m\}}$ with $m$-parton
kinematics.

When there are identified external hadrons, we obtain finite remainders
from the infinite subtraction into the parton distribution functions.
These finite remainders, which contribute to the one-dimensional convolution
of the cross sections ${\hat \sigma}^{NLO \,\{m\}}$ with $m$-parton
kinematics, contain two terms which respectively depend on the factorization
scale and on the factorization scheme.

The factorization-scale dependent term is proportional to the insertion operator
$\bom P$, which is again universal.  Using the $\{p\}$ and $I$ notation again,
it is given by
\beeq
\label{pinapp}
{\bom P}^{a,b}(\{p\};xp_a,x;\mu_F^2) =
\frac{\as}{2\pi} \;P^{ab}(x) \;\frac{1}{{\bom T}_{b}^2}
\sum_{I \neq b} {\bom T}_I \cdot {\bom T}_b
\;\ln \frac{\mu_F^2}{2 x p_a \cdot p_I} \;\;,
\eeeq
for initial-state partons and
\beeq
\label{pfinapp}
{\bom P}_{b,a}(\{p\};p_a/z,z;\mu_F^2) =
\frac{\as}{2\pi} \;P_{ba}(z) \;\frac{1}{{\bom T}_{b}^2}
\sum_{I \neq b} {\bom T}_I \cdot {\bom T}_b
\;\ln \frac{z \mu_F^2}{2 p_a \cdot p_I} \;\;,
\eeeq
for final-state partons. Here $P^{ab}= P_{ab}$ are the Altarelli-Parisi
probabilities in Eqs.~(\ref{pqgapp}--\ref{pggapp}). Note that if we denote by
$p_b$ the rescaled momentum in the insertion operator $\bom P$ (i.e.\ $p_b=xp_a$
in Eq.~(\ref{pinapp}) and $p_b=p_a/z$ in Eq.~(\ref{pfinapp})), the right-hand
sides of Eqs.~(\ref{pinapp}) and (\ref{pfinapp}) only differ by the
transposition $ab \to ba$ of the flavour indices in the Altarelli-Parisi
probabilities. Nonetheless, as discussed at the end of Sect.~\ref{impfra},
we have ${\bom P}^{a,b}(\{p\};p_b,x;\mu_F^2) \neq
{\bom P}_{a,b}(\{p\};p_b,x;\mu_F^2)$.

As for the term that depends on the factorization scheme, it is
proportional to the initial-state insertion operator $\bom K$ and/or to the
final-state insertion operator $\bom H$. When there is only one initial-state
hadron, the insertion operator $\bom K$ is:
\beeq
\label{Kdisapp}
{\bom K}^{a,b}(x)
= \frac{\as}{2\pi}
\left\{ \frac{}{} {\overline K}^{ab}(x) - \Kab(x)
+ \delta^{ab} \; \sum_{i} {\bom T}_i \cdot {\bom T}_b
\frac{\gamma_i}{{\bom T}_i^2}
\left[ \left( \frac{1}{1-x} \right)_+ + \delta(1-x) \right] \right\}
\;.
\eeeq
Similarly, when there is only one final-state identified hadron,
the insertion operator $\bom H$ is:
\beeq
&&{\bom H}_{b,a}(z)
= \frac{\as}{2\pi}
\left\{ \frac{}{} {\overline K}^{ba}(z) + 3 P_{ba}(z) \;\ln z - \Hba(z)
\right. \nonumber \\
&&\left. +  \; \delta_{ab} \;\sum_{i} {\bom T}_i \cdot {\bom T}_b
\frac{\gamma_i}{{\bom T}_i^2}
\left[ \left( \frac{1}{1-z} \right)_+ + \delta(1-z) - 1 \right]
\right\} \;\;.
\eeeq
Likewise, when there are two initial-state hadrons, we obtain
\beeq
\!\!\!\!\!\!&&{\bom K}^{a,a'}(x)
= \frac{\as}{2\pi}
\left\{ \frac{}{} {\overline K}^{aa'}(x) - \KFS{aa'}(x) \right.  \\
\!\!\!\!\!\!&& \left.
+ \;\delta^{aa'} \; \sum_{i} {\bom T}_i \cdot {\bom T}_a
\;\frac{\gamma_i}{{\bom T}_i^2} \left[
\left( \frac{1}{1-x} \right)_+ + \delta(1-x) \right]
\right\} -
\frac{\as}{2\pi} {\bom T}_b \cdot {\bom T}_{a'} \frac{1}{{\bom T}_{a'}^2}
{\widetilde K}^{aa'}(x) \;\;. \nonumber
\eeeq
Finally, in the most general case of multiparton correlations, we obtain
\beeq
\label{Kmpapp}
\!\!\!\!&&{\bom K}^{a,a'}(x)
= \frac{\as}{2\pi}
\left\{ \frac{}{} {\overline K}^{aa'}(x) - \KFS{aa'}(x) \right. \nonumber \\
\!\!\!\!&&+ \left. \; \delta^{aa'} \sum_{i} {\bom T}_i \cdot {\bom T}_a
\;\frac{\gamma_i}{{\bom T}_i^2} \left[ \left( \frac{1}{1-x} \right)_+
+ \delta(1-x) \right]
-  \;\frac{1}{{\bom T}_{a'}^2}
 \left( \sum_{l=1}^n  {\bom T}_{a_l} \cdot {\bom T}_{a'}
+ {\bom T}_{b} \cdot {\bom T}_{a'} \right)
{\widetilde K}^{aa'}(x)
  \right. \nonumber \\
\!\!\!\!&&- \left.  \frac{1}{{\bom T}_{a'}^2} \left[ \;
\sum_{l=1}^n  {\bom T}_{a_l} \cdot {\bom T}_{a'} \;{\cal L}^{a,a'}(x;p,q_l,n)
+ {\bom T}_{b} \cdot {\bom T}_{a'} \;{\cal L}^{a,a'}(x;p,{\bar p},n)
\right] \right\} \;\;,
\eeeq
and
\beeq
\label{Hmpapp}
\!\!\!\!\!&&{\bom H}_{a_l^\prime,a_l}(z) =
\frac{\as}{2\pi}
\left\{ \frac{}{} {\overline K}^{a_l^\prime a_l}(z) +
3 P_{a_l^\prime a_l}(z) \;\ln z - \HFS{a_l^\prime a_l}(z)
\right. \nonumber \\
\!\!\!\!\!&& + \;\delta_{a_l^\prime a_l} \sum_i
{\bom T}_i \cdot {\bom T}_{a_l^\prime} \frac{\gamma_i}{{\bom T}_i^2}
\left[ \left( \frac{1}{1-z} \right)_+ + \delta(1-z) - 1 \right] \nonumber \\
\!\!\!\!\!&& +\left. \frac{1}{{\bom T}_{a_l^\prime}^2}
\left( \sum_{r=1 \atop r \neq l}^n {\bom T}_{a_r} \cdot {\bom T}_{a_l^\prime}
+ {\bom T}_{a} \cdot {\bom T}_{a_l^\prime}
+ {\bom T}_{b} \cdot {\bom T}_{a_l^\prime} \right)
\left[ \;P_{a_l^\prime a_l}(z) \;\ln z -{\widetilde K}^{a_l^\prime a_l}(z)
\frac{}{} \right] \right. \nonumber \\
\!\!\!\!\!&&- \left.  \frac{1}{{\bom T}_{a_l^\prime}^2} \left[
\sum_{r=1 \atop r \neq l}^n
{\bom T}_{a_r} \cdot {\bom T}_{a_l^\prime,a_l}
\;{\cal L}^{a_l^\prime,a_l}(z;q_l,q_r,n)
+ {\bom T}_{a} \cdot {\bom T}_{a_l^\prime}
\;{\cal L}^{a_l^\prime,a_l}(z;q_l,p,n) \right. \right. \nonumber \\
\!\!\!\!\!&& \left. \left.
+ {\bom T}_{b} \cdot {\bom T}_{a_l^\prime}
\;{\cal L}^{a_l^\prime,a_l}(z;q_l,{\bar p},n) \frac{}{}
\right] \right\}  \;\;.
\eeeq
Note that setting ${\bom T}_{b} = 0$ in Eqs.~(\ref{Kmpapp}) and (\ref{Hmpapp})
we respectively obtain the operators $\bom K$ and $\bom H$ for the case of
multiparton correlations with a single incoming parton. Likewise, setting
${\bom T}_{a} = {\bom T}_{b} = 0$ in Eq.~(\ref{Hmpapp}) we obtain the operator
$\bom H$ for multiparton correlations in processes with no hadrons in the initial
state.

The definition of the flavour kernels ${\overline K}^{ab}(x)$,
${\widetilde K}^{ab}(x)$, ${\cal L}^{a,b}$, $\Kab(x)$ and $\Hba(x)$
that appear in Eqs.~(\ref{Kdisapp}--\ref{Hmpapp}) is recalled in
Eqs.~(\ref{bkabapp}), (\ref{wkabapp}),
(\ref{clabapp}) and (\ref{kabapp}--\ref{kgqscheme}).

In order to evaluate the NLO cross sections with $m$-parton kinematics,
the colour-charge operators $\bom I$, $\bom P$, $\bom K$ and $\bom H$ have
to be inserted into the tree-level matrix elements. This leads to the
computation of colour-correlated tree-amplitudes. We conclude this Appendix by
recalling their definition. As above
we denote by
$\{p \}$ a generic set of $N$ parton momenta. The square $| \cm^{I,J} |^2$ of
the colour-correlated amplitude has the following expression in terms of the
coloured tree-level amplitude $\cm^{a_1...a_N}(\{p \})$:
\beeq
\label{ccapp}
| \cm^{I,J}(\{p \}) |^2 &\equiv& \bra{\{p \}} \;{\bom T_I} \cdot {\bom T_J}\;
\ket{\{p \}}  \\
&=& \frac{1}{n_c(a) n_c(b)}
\left[ {\cal M}^{a_1..b_I..b_J..a_N}(\{p \}) \right]^*
T_{b_Ia_I}^c T_{b_Ja_J}^c {\cal M}^{a_1..a_I..a_J..a_N}(\{p \}) \;\;. \nonumber
\eeeq
Here the labels $a$ and $b$ refer to the initial-state partons in $\ket{\{p \}}$
and $n_c(a)$ and $n_c(b)$ is their number of colours. The factor
$1/(n_c(a) n_c(b))$ on the right-hand side of Eq.~(\ref{ccapp}) comes from the
definition of the state vector $\ket{\{p \}}$
(see Eqs.~(\ref{cmmdef},\ref{ketin})) used
throughout this paper.  If there is only one initial state parton, or none,
then this factor becomes $1/n_c(a)$ or $1$ respectively, as in
Eqs.~(\ref{ccamp}) and (\ref{colam}).

\newpage

\renewcommand{\theequation}{D.\arabic{equation}}
\setcounter{equation}{0}

\section*{Appendix D:~~Examples}
\addcontentsline{toc}{section}{Appendix D:~~Examples}

In this appendix we give some simple applications of our method.

\subsection*{\boldmath $e^+e^-\to2$ jets}

We begin with a trivial example: two-jet observables in $e^+e^-$ annihilation.
We use customary notation for the kinematic variables: $Q^2$ is the square of
the centre-of-mass energy, $y_{ij}=2p_i\cdot p_j/Q^2$ and $x_i=2p_i\cdot Q/Q^2$,
where $p_i$ is the momentum of any QCD parton in the final state.

The LO contribution is the parton model process
$e^+e^- \to q(p_1)+\bar{q}(p_2),$ with matrix element
$\cm_2$.  We average over event orientation, so $\cm_2$ has no
dependence on the parton momenta. Moreover, we choose the overall normalization
of $\cm_2$ such that the two-parton phase space is:
\beq
d\Phi^{(2)} =
dy_{12} \;\delta(1 - y_{12}) \;\;,
\eeq
and the LO cross section in Eq.~(\ref{LOeefin}) is given by
\beq
\sigma^{LO} = |\cm_2|^2 \int dy_{12} \;\delta(1 - y_{12}) \;F_J^{(2)}(p_1,p_2)
\;\;.
\eeq

The NLO real-emission process is $e^+e^- \to q(p_1)+{\bar q}(p_2)+g(p_3),$
with matrix element $\cm_3(p_1,p_2,p_3)$. In four dimensions, the matrix element
is:
\beq\label{cm3}
  |\cm_3(p_1,p_2,p_3)|^2 =
  C_F\frac{8\pi\as}{Q^2}\frac{x_1^2+x_2^2}{(1-x_1)(1-x_2)} |\cm_2|^2 \;\;,
\eeq
and the phase-space is given by
\beq\label{dphi3}
d\Phi^{(3)} =
\frac{Q^2}{16\pi^2} dx_1 dx_2 \;\; \Theta(1-x_1)\Theta(1-x_2)\Theta(x_1+x_2-1).
\eeq

The calculation of the subtracted cross section involves the evaluation of
two dipole contributions: ${\cal D}_{13,2}$ and ${\cal D}_{23,1}$. Their
definition is given in Eqs.~(\ref{dipff},\ref{Vqigjk}). The
associated colour algebra is trivial, as shown in Eq.~(\ref{colour1}), and
we find
\beq\label{d132}
  {\cal D}_{13,2}(p_1,p_2,p_3) = \frac1{2p_1p_3} V_{q_1g_3,2} \;|\cm_2|^2 \;\;,
\eeq
with the following dipole kinematics
\beq\label{d132kin}
{\widetilde p}_2^{\; \mu} = \frac{1}{x_2} \;p_2^{\mu} \;, \;\;\;\;\;\;
{\widetilde p}_{13}^{\; \mu} = Q^{\mu} - \frac{1}{x_2} \;p_2^{\mu} \;\;.
\eeq
The dipole contribution
${\cal D}_{23,1}$ is obtained from Eqs.~(\ref{d132},\ref{d132kin})
by the replacement $p_1 \leftrightarrow p_2$. Inserting the (four-dimensional)
definitions of
$V_{q_1g_3,2}$ and $x_i,$ we obtain the final expression for the
three-parton cross section in Eq.~(\ref{m1eefin}):
\beeq
\label{ee3}
&&\sigma^{NLO \,\{3\}} =
  \int_3 \left[ d\sigma^R_{\ep=0} - d\sigma^A_{\ep=0} \right] \nonumber \\
&=& |\cm_2|^2
  \frac{C_F\as}{2\pi} \int_0^1 dx_1 \; dx_2 \;\Theta(x_1+x_2-1)
  \left\{ \frac{x_1^2+x_2^2}{(1-x_1)(1-x_2)}F_J^{(3)}(p_1,p_2,p_3)
\right.\nonumber\\&&\left.
- \left[ \frac{1}{1-x_2} \left(\frac{2}{2-x_1-x_2} - (1+x_1) \right)
 + \frac{1-x_1}{x_2} \right]
\;F_J^{(2)}({\widetilde p}_{13},{\widetilde p}_2) \right. \nonumber \\
&&\left.
- \left[ \frac{1}{1-x_1} \left(\frac{2}{2-x_1-x_2} - (1+x_2) \right)
 +\frac{1-x_2}{x_1} \right]
\;F_J^{(2)}({\widetilde p}_{23},{\widetilde p}_1) \right\} \;\;.
\eeeq
Since the three-parton matrix element can be written as follows
\beq
\frac{x_1^2+x_2^2}{(1-x_1)(1-x_2)} =
\frac{1}{1-x_2} \left( \frac{2}{2-x_1-x_2} - (1+x_1) \right) +
\left( x_1 \leftrightarrow x_2 \right) \;\;,
\eeq
it is straightforward to see that for any infrared safe observable
(implying that $F_J^{(3)} \to F_J^{(2)}$ as $x_i\to1$), Eq.~(\ref{ee3}) is
finite.

Next we have to evaluate the insertion operator ${\bom I}(\ep)$ and combine
it with the virtual cross section. The one-loop matrix element in the
${\overline {\rm MS}}$ renormalization scheme is given by
\beq\label{m21loop}
  |\cm_2|^2_{(1-loop)} = |\cm_2|^2 \; \frac{C_F\as}{2\pi}
  \frac{1}{\Gamma(1-\epsilon)}\left(\frac{4\pi\mu^2}{Q^2}\right)^\epsilon
  \left\{ -\frac{2}{\ep^2}-\frac{3}{\ep}-8+\pi^2+{\cal O}(\ep)\right\} \;\;,
\eeq
while the insertion operator in Eqs.~(\ref{Iapp},\ref{VIapp}) gives
\beq\label{I2ee}
  {}_2\!\!<{1,2}| \;{\bom I}(\ep) \;|{1,2}>_2 =
  |\cm_2|^2 \; \frac{C_F\as}{2\pi}
  \frac{1}{\Gamma(1-\epsilon)}\left(\frac{4\pi\mu^2}{Q^2}\right)^\epsilon
  \left\{ \frac{2}{\ep^2}+\frac{3}{\ep}+10-\pi^2+{\cal O}(\ep)\right\}.
\eeq
Note that in Eqs.~(\ref{m21loop},\ref{I2ee}) the two-parton matrix element
$\cm_2$ is consistently evaluated in $d=4-2\ep$ dimensions.
Combining these two contributions as in Eq.~(\ref{meefin}), we obtain a
finite (for $\ep \to 0$) final expression for the two-parton cross section:
\beq
\label{ee2}
\sigma^{NLO \,\{2\}} =
  \int_2 \left[ d\sigma^{V}  + \int_1 d\sigma^{A} \right]_{\ep=0} =
|\cm_2|^2 \frac{C_F\as}{\pi} \int dy_{12} \;\delta(1 - y_{12})
\;F_J^{(2)}(p_1,p_2) \;\;.
\eeq

It is straightforward to check that the total cross section
($F_J^{(3)}=F_J^{(2)}=1$) agrees with the well-known result,
$\sigma^{NLO}_{{\rm tot}}=\frac{3}{4}C_F\frac{\as}{\pi}
\sigma^{LO}_{{\rm tot}}$.

\subsection*{\boldmath $e^+e^-\to3$ jets}

In Ref.~[\ref{CSlett}] we presented the simplest non-trivial case:
three-jet production in $e^+e^-$ annihilation.  For completeness we repeat
it here.  We again average over event orientation, so our formalism can be
directly compared with that in Ref.~[\ref{ERT}].

The LO partonic process to be considered is $e^+e^- \to q+{\bar q}+g,$ with
matrix element $\cm_3$ and kinematic variables as defined above for the case
of $e^+e^- \to  2$ jets.

At NLO, two different real-emission subprocesses contribute:
a) $e^+ e^- \to q(p_1) + {\bar q}(p_2)  + g (p_3) + g(p_4)$;
b) $e^+ e^- \to q(p_1) + {\bar q}(p_2)  + q(p_3) + {\bar q}(p_4)$.
The calculation of the subtracted cross section (\ref{m1eefin}) for the
subprocess
a) involves the evaluation of the following dipole contributions:
${\cal D}_{13,2}, {\cal D}_{13,4}, {\cal D}_{14,2}, {\cal D}_{14,3},
{\cal D}_{23,1}, {\cal D}_{23,4},
{\cal D}_{24,1},$ ${\cal D}_{24,3}, {\cal D}_{34,1}, {\cal D}_{34,2}$.
The associated colour algebra can again be easily performed because the
several colour projections of the three-parton matrix element fully
factorize (see Eq.(\ref{colour2})).  Thus we do not need to compute any
colour-correlated tree amplitudes and we find
\beeq
\label{dipa}
{\cal D}_{13,2}(p_1,p_2,p_3,p_4) &=& \frac{1}{2p_1p_3}
\left(1- \frac{C_A}{2C_F}\right)
V_{q_1g_3,2} \;|\cm_3({\widetilde p_{13}},{\widetilde p_2},p_4)|^2 \;\;,
\nonumber \\
{\cal D}_{13,4}(p_1,p_2,p_3,p_4) &=& \frac{1}{2p_1p_3} \,\frac{C_A}{2C_F} \;
V_{q_1g_3,4} \;|\cm_3({\widetilde p_{13}},p_2,{\widetilde p_4})|^2 \;\;, \\
{\cal D}_{34,1}(p_1,p_2,p_3,p_4) &=& \frac{1}{2p_3p_4} \,\frac{1}{2} \;
V_{g_3g_4,1}^{\mu \nu}
\;{\cal T}_{\mu \nu}({\widetilde p_{1}},p_2,{\widetilde p_{34}}) \;\;.
\nonumber
\eeeq
The dipole contributions ${\cal D}_{23,1}, {\cal D}_{23,4},
{\cal D}_{34,2}$ are obtained respectively
from ${\cal D}_{13,2},{\cal D}_{13,4}, {\cal D}_{34,1} $ by the replacement
$p_1 \leftrightarrow p_2$. Analogously, one can obtain
${\cal D}_{14,2}$ and ${\cal D}_{14,3}$ respectively from
${\cal D}_{13,2}$ and ${\cal D}_{13,4}$ by the replacement
$p_3 \leftrightarrow p_4$, and ${\cal D}_{24,1}$ and
${\cal D}_{24,3}$ respectively from ${\cal D}_{13,2}$ and
${\cal D}_{13,4}$ by the replacement $p_1 \leftrightarrow p_2, \;
p_3 \leftrightarrow p_4$.

In the case of the subprocess b) we have to consider the following
dipole contributions:
${\cal D}_{12,3}, {\cal D}_{12,4}, {\cal D}_{14,2}, {\cal D}_{14,3},
{\cal D}_{23,1}, {\cal D}_{23,4}, {\cal D}_{34,1}, {\cal D}_{34,2}$.
Performing the colour algebra we get
\beq
\label{dipb}
{\cal D}_{34,1}(p_1,p_2,p_3,p_4) = \frac{1}{2p_3p_4} \frac{1}{2} \;
V_{q_3{\bar q}_4,1}^{\mu \nu}
{\cal T}_{\mu \nu}({\widetilde p_{1}},p_2,{\widetilde p_{34}}) \;,
\eeq
and all the other dipoles are obtained by the corresponding permutation of
the parton momenta.

The splitting functions $V_{ij,k}$ of Eqs.~(\ref{dipa},\ref{dipb}) are
explicitly given in Eqs.~(\ref{Vqigjk}--\ref{vggk}).
The tensor ${\cal T}_{\mu \nu}$ is
the squared amplitude for the LO process $\mathrm{e^+e^-\to q\bar{q}g}$
not summed over the gluon polarizations ($\mu$ and $\nu$ are the gluon spin
indices and $-g^{\mu \nu} {\cal T}_{\mu \nu} = |\cm_3|^2 $).
This can be easily calculated. In the case of jet observables averaged over
the directions of the incoming leptons (un-oriented events) we find
(in $d=4$ dimensions)
\beq
{\cal T}^{\mu \nu}(p_1,p_2,p_3) = - \frac{1}{x_1^2 + x_2^2}
\; |\cm_3(p_1,p_2,p_3)|^2 \;T^{\mu \nu} \;,
\eeq
where
\beeq\label{tmunu}
  T^{\mu\nu} &=&
    +2                            \frac{p_1^\mu p_2^\nu}{Q^2}
    +2                            \frac{p_2^\mu p_1^\nu}{Q^2}
    -2\frac{1-x_1}{1-x_2}         \frac{p_1^\mu p_1^\nu}{Q^2}
    -2\frac{1-x_2}{1-x_1}         \frac{p_2^\mu p_2^\nu}{Q^2}
\nonumber\\&&
    -\frac{1-x_1-x_2+x_2^2}{1-x_2}\left[
                                  \frac{p_1^\mu p_3^\nu}{Q^2}
    +                             \frac{p_3^\mu p_1^\nu}{Q^2}
                                  \right]
    -\frac{1-x_2-x_1+x_1^2}{1-x_1}\left[
                                  \frac{p_2^\mu p_3^\nu}{Q^2}
    +                             \frac{p_3^\mu p_2^\nu}{Q^2}
                                  \right]
\nonumber\\&&
    +\left(1+\hf x_1^2+\hf x_2^2-x_1-x_2\right)g^{\mu\nu} \;.
\eeeq

The few ingredients listed in Eqs.~(\ref{dipa}--\ref{tmunu}) have to be combined
with the four-parton matrix elements $\cm_4$ for evaluating the four-parton
cross section $\sigma^{NLO \,\{4\}}$ in Eq.~(\ref{m1eefin}). Obviously, due
to the very long expressions for the matrix elements [\ref{ERT}], we do not
report here the explicit formula for $\sigma^{NLO \,\{4\}}$.

To complete the NLO calculation we also need the virtual cross section.
In the case of un-oriented events, we take the one-loop matrix element
in the ${\overline {\rm MS}}$ renormalization scheme from Ref.~[\ref{ERT}]
(we use slightly different notation):
\beeq
\label{ert1loop}
&&|\cm_3(p_1,p_2,p_3)|^2_{(1-loop)} =
|\cm_3(p_1,p_2,p_3)|^2 \; \frac{\as}{2\pi}\frac{1}{\Gamma(1-\epsilon)}
\left(\frac{4\pi\mu^2}{Q^2}\right)^\epsilon
\nonumber \\
&&\cdot
\left\{ \,-\frac{1}{\ep^2} \left[ (2C_F -C_A) y_{12}^{-\ep} +
C_A \left(y_{13}^{-\ep} + y_{23}^{-\ep} \right) \right]
-\frac{1}{\ep} \left( 3C_F + \frac{11}{6}C_A -\frac{2}{3}T_R N_f \right)
\right. \nonumber \\
&&+ \left. \frac{\pi^2}{2}(2C_F+C_A) - 8 C_F \right\}
+ \frac{\as}{2\pi}\left[ F(y_{12},y_{13},y_{23}) +{\cal O}(\ep) \right] \;\;,
\eeeq
where $F(y_{12},y_{13},y_{23})$ is defined in Eq.~(2.21) of Ref.~[\ref{ERT}].

The explicit evaluation of the insertion operator ${\bom I}(\ep)$ in
Eqs.~(\ref{Iapp},\ref{VIapp}) gives:
\beeq
\label{ertinsop}
&&{}_3\!\!<{1,2,3}| \;{\bom I}(\ep) \;|{1,2,3}>_3 =
|\cm_3(p_1,p_2,p_3)|^2 \; \frac{\as}{2\pi}\frac{1}{\Gamma(1-\epsilon)}
\left(\frac{4\pi\mu^2}{Q^2}\right)^\epsilon
\nonumber \\
&&
\left\{ \,\frac{1}{\ep^2} \left[ (2C_F -C_A) y_{12}^{-\ep} +
C_A \left(y_{13}^{-\ep} + y_{23}^{-\ep} \right) \right]
+\frac{1}{\ep} \left( 2\gamma_q + \gamma_g \right) \right.
\nonumber \\
&&- \left. \gamma_q \frac{1}{C_F} \left[ (2C_F - C_A) \ln y_{12} + \frac{1}{2}
C_A \ln (y_{13}y_{23}) \right] - \frac{1}{2} \gamma_g \ln (y_{13}y_{23})
\right. \nonumber \\
&&\left. -\frac{\pi^2}{3}(2C_F+C_A) + 2(\gamma_q + K_q) + \gamma_g + K_g
+{\cal O}(\ep) \right\} \;\;.
\eeeq
Combining the one-loop  matrix element (\ref{ert1loop}) with the result
(\ref{ertinsop}), and using the explicit
expressions (\ref{gaapp},\ref{kcapp}) for $\gamma_I$ and $K_I$,
all the pole terms cancel.
Note that as well as the pole terms, the closely related $\pi^2$ and
$\ln^2$ terms cancel:
\beeq\label{ertsub}
&&|\cm_3(p_1,p_2,p_3)|^2_{(1-loop)} +
{}_3\!\!<{1,2,3}| \;{\bom I}(\ep) \;|{1,2,3}>_3 = |\cm_3(p_1,p_2,p_3)|^2
\nonumber \\
&&\cdot \;\frac{\as}{2\pi}
\left [ - \frac{3}{2} (2C_F - C_A) \ln y_{12}
-\frac{1}{3} (5C_A - T_R N_f) \ln (y_{13}y_{23}) \right. \nonumber \\
&&\left. + 2 C_F +\frac{50}{9} C_A
- \frac{16}{9} T_R N_f \right]
+ \frac{\as}{2\pi}\left[ F(y_{12},y_{13},y_{23}) +{\cal O}(\ep) \right] \;\;.
\eeeq
The integration of the expression (\ref{ertsub}) (with the matrix element
$\cm_3$ given in Eq.~(\ref{cm3})) over the phase space in Eq.~(\ref{dphi3})
provides the three-parton cross section $\sigma^{NLO \,\{3\}}$
in Eq.~(\ref{meefin}).

The results presented here form the basis for a Monte Carlo program
[\ref{CSlett}]
that can calculate
the NLO corrections to arbitrary three-jet-like observables in \ee\
annihilation: it will be described in more detail elsewhere.

\subsection*{{\boldmath $1+1$ jets} in DIS and the Structure
Function {\boldmath$F_2$}}

We next discuss the simplest case in which there is an incoming hadron:
1+1-jet observables in deep inelastic lepton-hadron scattering (DIS).
We limit ourselves to considering the case of virtual-photon exchange and, in
particular, we compute $\sigma = \sigma_T + \sigma_L$, $\sigma_T$ and
$\sigma_L$ respectively being the scattering cross sections off transversely and
longitudinally polarized photons. In the fully inclusive limit (i.e.\ when no
final-state jet observable is defined), our cross section is simply
proportional to the customary structure function $F_2$. We use standard notation
for the kinematic variables: $p$ is the incoming momentum, $q$ is the off-shell
photon momentum $(q^2 = - Q^2 < 0)$, $x=Q^2/(2pq)$ is the Bjorken variable and
$z_i= p_ip/pq$, $p_i$ being any parton (hadron) momentum in the final state.
The relevant matrix elements are obtained from the hadronic tensor
$W_{\mu\nu}$ by applying the following ($d$-dimensional) projection operator
\beq
  P_2^{\mu\nu} = \frac{Q^2}{2pq}
  \left[-g^{\mu\nu}+\frac{(d-1) Q^2}{(pq)^2}p^\mu p^\nu\right].
\eeq
Note that to implement the final formulae of our algorithm,
the $d$-dimensional definition of $P_2^{\mu\nu}$ is relevant only for a
consistent calculation of the one-loop matrix element.

The hadronic cross section $\sigma(p)$ in Eq.~(\ref{1hxs}) is obtained by
convoluting partonic cross sections $\sigma_a = \sigma_q , \sigma_{\bar q},
\sigma_g$ with parton densities $f_a$. Since we are considering only photon
exchange, by charge conjugation invariance we have $\sigma_{\bar q}=\sigma_q$.
Thus, we explicitly compute only $\sigma_q$ and $\sigma_g$. At LO,
$\sigma_g^{LO}=0$
while $\sigma_q^{LO}$ is obtained by the parton model process $q(p) + \gamma^*(q)
\to q(p_1)$. By momentum conservation we have $p_1 = q + p$, so the LO matrix
element $\cm_{1,q}(q + p;p)$ has, in practice, no dependence on final-state
parton momenta. We choose the overall normalization of $\cm_{1,q}(q + p;p)$
such that the single-parton phase space is:
\beq\label{dphi1dis}
d\Phi^{(1)}(p) = \frac{Q^2}{2pq} \;\delta\!\left(\frac{2pq}{Q^2} - 1 \right)
= \delta(1-x) \;\;,
\eeq
and the LO cross section in Eq.~(\ref{LOdisfin}) is given by
\beq
\sigma^{LO}_q(p) = \frac{1}{N_c} \;|\cm_{1,q}(q + p;p)|^2 \;\delta(1-x)
\;F_J^{(1)}(q + p;p) \;\;.
\eeq

At NLO, we have to consider real emission processes with two final-state
partons with momenta $p_1$ and $p_2$. By momentum conservation we have $2p_1p_2
= (1-x)Q^2/x$ and the two-parton phase space is given by
\beq\label{dphi2dis}
d\Phi^{(2)}(p) = \frac{Q^2 \,x}{16 \pi^2} dz_1 \;dz_2 \;
\frac{d\phi_1}{2\pi} \; \frac{d\phi_2}{2\pi} \;
\Theta(z_1) \Theta(z_2) \;\delta(1-z_1-z_2) \;
2\pi \;\delta(\pi+\phi_1-\phi_2)\;\;.
\eeq
In the following, we consider jet quantities that are averaged over the
azimuthal angles~$\phi_{1,2},$ so they can be trivially integrated and $\int
d\phi_1 d\phi_2 \;\delta(\pi+\phi_1-\phi_2) \to 2\pi$.
The NLO cross sections $\sigma_q^{NLO}(p)$, $\sigma_g^{NLO}(p)$ respectively
receive
contributions from the real emission processes
$q(p) + \gamma^*(q) \to q(p_1) + g(p_2)$ and
$g(p) + \gamma^*(q) \to q(p_1)+\bar{q}(p_2)$. The corresponding matrix elements
in $d=4$ dimensions are:
\beeq\label{m2dis}
\frac1{N_c}| \cm_{2,q}(p_1,p_2;p) |^2 &=& C_F \;\frac{8\pi \as}{Q^2}
\left[ \frac{x^2+z_1^2}{(1-x)(1-z_1)} + 2 (1+3x z_1) \right] \nonumber\\
&\cdot& \frac1{N_c}|\cm_{1,q}(q + xp;xp)|^2 \;\;, \nonumber\\
\frac1{N_c^2 -1}| \cm_{2,g}(p_1,p_2;p) |^2 &=& T_R \;\frac{8\pi \as}{Q^2}
\left[ \frac{(z_1^2+(1-z_1)^2) (x^2+(1-x)^2)}{z_1(1-z_1)}
+ 8x (1-x)\right] \nonumber \\
&\cdot& \frac1{N_c}|\cm_{1,q}(q + xp;xp)|^2 \;\;.
\eeeq

According to Eq.~(\ref{NLOdismp}), the calculation of the two-parton
cross sections
$\sigma_q^{NLO \,\{2\}}$ and $\sigma_g^{NLO \,\{2\}}$ is performed by
subtracting, in the first case, the final-state dipole ${\cal D}_{12}^{q}$
and the initial-state dipole ${\cal D}^{q2}_{1}$ and, in the second case,
the two initial-state dipoles ${\cal D}^{g1}_{2}$, ${\cal D}^{g2}_{1}$.
The associated colour algebra is trivial
(as in the example of \ee $\to$ 2 jets) and we find
\beeq
\label{didis}
{\cal D}_{12}^q(p_1,p_2;p) &=& \frac{1}{2p_1p_2} \;\frac{1}{x}
\;V_{q_1g_2}^{q} \;\frac{1}{N_c} \;
|\cm_{1,q}(q + {\widetilde p};{\widetilde p})|^2\;\;,
\nonumber \\
{\cal D}^{q2}_1(p_1,p_2;p) &=& \,\frac{1}{2pp_2} \;\,\frac{1}{x}
\;V^{qg_2}_1 \;\frac{1}{N_c} \;
|\cm_{1,q}(q + {\widetilde p};{\widetilde p})|^2\;\;, \\
{\cal D}^{g1}_2(p_1,p_2;p) &=& \,\frac{1}{2pp_1} \;\,\frac{1}{x}
\;V^{gq_1}_2 \;\frac{1}{N_c} \;
|\cm_{1,q}(q + {\widetilde p};{\widetilde p})|^2\;\;, \nonumber
\eeeq
with ${\cal D}^{g2}_1$ being obtained from ${\cal D}^{g1}_2$ by the replacement
$p_1 \leftrightarrow p_2$. Note that the dipole kinematics turns out to be the
same for all these dipole contributions: the incoming and outgoing parton
momenta in each dipole respectively are ${\widetilde p}^\mu = x p^\mu $ and (by
momentum conservation) $q^\mu  +  x p^\mu$.

Inserting into Eq.~(\ref{didis}) the definitions
(\ref{Vqigja},\ref{Vqagik},\ref{Vqagik1})
of the splitting functions $V$ and combining
Eqs.~(\ref{dphi2dis},\ref{m2dis},\ref{didis}) as in Eq.~(\ref{NLOdismp}),
we obtain the
following expressions for the NLO cross sections contributions
$\sigma_a^{NLO \,\{2\}}$
\beeq\label{si2qdis}
\sigma_q^{NLO \,\{2\}}(p) &=& \int_0^1 dz_1 \;dz_2
\;\delta(1-z_1-z_2) \;\frac{1}{N_c} \;
|\cm_{1,q}(q + xp;xp)|^2 \;\frac{C_F \as}{2\pi} \; x \nonumber \\
&\cdot& \left\{ \left[ \frac{x^2+z_1^2}{(1-x)(1-z_1)} + 2 (1+3x z_1) \right]
F_J^{(2)}(p_1,p_2;p) \right. \nonumber \\
&~& - \left. \frac{x^2+z_1^2}{(1-x)(1-z_1)} \;F_J^{(1)}(q+xp;xp) \right\} \;\;,
\eeeq
\beeq\label{si2gdis}
\sigma_g^{NLO \,\{2\}}(p) &=& \int_0^1 dz_1 \;dz_2
\;\delta(1-z_1-z_2) \;\frac{1}{N_c} \;
|\cm_{1,q}(q + xp;xp)|^2 \;\frac{T_R \as}{2\pi} \; x \nonumber \\
&\cdot& \left\{ \left[ \frac{(z_1^2+(1-z_1)^2) (x^2+(1-x)^2)}{z_1(1-z_1)}
+ 8x (1-x)\right] F_J^{(2)}(p_1,p_2;p) \right. \nonumber \\
&~& - \left. \frac{x^2+(1-x)^2}{z_1(1-z_1)} \;F_J^{(1)}(q+xp;xp) \right\} \;\;.
\eeeq
Clearly, for any jet observable (implying that $F_J^{(2)} \to F_J^{(1)}$ as
$z_1 \to 1, 0$ or $x \to 1$) the integrals in Eqs.~(\ref{si2qdis},\ref{si2gdis})
are finite.

In order to compute the NLO cross sections with $1 \to 1$ parton kinematics, we
have to evaluate the insertion operators $\bom I$, $\bom P$ and $\bom K$ in
Eqs.~(\ref{Iapp},\ref{pinapp},\ref{Kdisapp}). We find
\beeq\label{I2dis}
  {}_{1,q}\!\!<{1;p}| \;{\bom I}(\ep) \;|{1;p}>_{1,q} &=& \frac{1}{N_c} \;
  |\cm_{1,q}(q + p;p)|^2 \\
  &\cdot& \frac{C_F\as}{2\pi}
  \frac{1}{\Gamma(1-\epsilon)}\left(\frac{4\pi\mu^2}{Q^2}\right)^\epsilon
  \left\{ \frac{2}{\ep^2}+\frac{3}{\ep}+10-\pi^2+{\cal O}(\ep)\right\}
  \nonumber \;\;,
\eeeq
\beq\label{pin2dis}
\sum_b {}_{1,b}\!\!<{1;zp}| {\bom P}^{q,b}(zp,z;\mu_F^2) \;|{1;zp}>_{1,b}
\; = - \frac{1}{N_c} \;|\cm_{1,q}(q + zp;zp)|^2 \;\;
\frac{\as}{2\pi}P^{qq}(z) \;\ln\frac{x\mu_F^2}{zQ^2} \;\;,
\eeq
\beq
\sum_b {}_{1,b}\!\!<{1;zp}| {\bom P}^{g,b}(zp,z;\mu_F^2) \;|{1;zp}>_{1,b}
\; = - \frac{1}{N_c} \;|\cm_{1,q}(q + zp;zp)|^2 \;\;
2 \;\frac{\as}{2\pi}P^{gq}(z) \;\ln\frac{x\mu_F^2}{zQ^2} \;\;,
\eeq
\beeq
\sum_b {}_{1,b}\!\!<{1;zp}| {\bom K}^{q,b}(z) \;|{1;zp}>_{1,b}
&=& \frac{1}{N_c} \;|\cm_{1,q}(q + zp;zp)|^2 \;\;\frac{\as}{2\pi}\\
&\cdot&
\left\{ \overline{K}^{qq}(z)-\KFS{qq}(z)
    -\frac32C_F\left[\left(\frac1{1-z}\right)_++\delta(1-z)\right]\right\}
\nonumber  \;\;,
\eeeq
\beq\label{k2dis}
\sum_b {}_{1,b}\!\!<{1;zp}| {\bom K}^{g,b}(z) \;|{1;zp}>_{1,b}
\; = \frac{1}{N_c} \;|\cm_{1,q}(q + zp;zp)|^2 \;
2 \;\frac{\as}{2\pi}\left\{
    \overline{K}^{gq}(z)-\KFS{gq}(z)\right\} \;\;.
\eeq

According to Eq.~(\ref{NLOdism}), the result in Eq.~(\ref{I2dis}) has to be
combined with the following one-loop matrix element
\beeq
  \frac1{N_c}|\cm_{1,q}(q + p;p)|^2_{(1-loop)} &=&
  \frac1{N_c}|\cm_{1,q}(q + p;p)|^2
\nonumber\\&\cdot&
  \frac{C_F\as}{2\pi}
  \frac{1}{\Gamma(1-\epsilon)} \left(\frac{4\pi\mu^2}{Q^2}\right)^\epsilon
  \left\{ -\frac{2}{\ep^2}-\frac{3}{\ep}-8+{\cal O}(\ep)\right\} \;\;.
\eeeq
Obviously, the $\ep$ poles cancel and we obtain
\beq\label{si21qdis}
\sigma_q^{NLO \,\{1\}}(p) = \frac{1}{N_c} \;|\cm_{1,q}(q + p;p)|^2
\;\frac{C_F \as}{2\pi}  \;(2 - \pi^2) \;\delta(1-x) \;F_J^{(1)}(q+p;p) \;\;.
\eeq

{}From Eqs.~(\ref{pin2dis}--\ref{k2dis}) we can evaluate the NLO cross
section contributions in Eq.~(\ref{NLOdisx}). Using the phase space factor in
Eq.~(\ref{dphi1dis}),
we find
\beeq\label{hs2q}
&&\int_0^1 dz \;{\hat \sigma}_q^{NLO \,\{1\}}(z;zp,\mu_F^2) = \int_0^1 dz
\;\delta(z/x -1) \;F_J^{(1)}(q+zp;zp) \;\frac{1}{N_c} \;|\cm_{1,q}(q + zp;zp)|^2
\nonumber \\
&&\cdot \frac{\as}{2\pi}
\left\{ \overline{K}^{qq}(z)-\KFS{qq}(z)
    -\frac32C_F\left[\left(\frac1{1-z}\right)_++\delta(1-z)\right]
    - P^{qq}(z) \ln\frac{x\mu_F^2}{zQ^2} \right\}
\nonumber  \\
&&= \frac{1}{N_c} \;|\cm_{1,q}(q + xp;xp)|^2 \;F_J^{(1)}(q+xp;xp)
\;\frac{\as}{2\pi}\;C_F \; x  \\
&&\cdot \left\{ \left[ \frac{1+x^2}{1-x}
\left( \ln\frac{(1-x)Q^2}{x\;\mu_F^2}
 -\frac34\right) \right]_+ + \frac{1-7x}{4} + \delta(1-x) \left(\pi^2
 -\frac{39}{8} \right) -\frac{1}{C_F} \KFS{qq}(x) \right\} \;\;, \nonumber
\eeeq
\beeq\label{hs2g}
&&\int_0^1 dz \;{\hat \sigma}_g^{NLO \,\{1\}}(z;zp,\mu_F^2) = \int_0^1 dz
\;\delta(z/x -1) \;F_J^{(1)}(q+zp;zp) \;\frac{1}{N_c} \;|\cm_{1,q}(q + zp;zp)|^2
\nonumber \\
&&\cdot \frac{\as}{2\pi} 2
\left\{ \overline{K}^{gq}(z)-\KFS{gq}(z)
- P^{gq}(z) \ln\frac{x\mu_F^2}{zQ^2} \right\}
\nonumber  \\
&&= \frac{1}{N_c} \;|\cm_{1,q}(q + xp;xp)|^2 \;F_J^{(1)}(q+xp;xp)
\;\frac{\as}{2\pi}\; T_R \;2x \nonumber \\
&&\cdot \left\{ [ x^2+(1-x)^2 ] \ln\frac{(1-x)Q^2}{x\;\mu_F^2}+  2x(1-x)
    -\frac{1}{T_R} \KFS{gq}(x) \right\} \;\;,
\eeeq
where, in the last expression on the right-hand side of
Eqs.~(\ref{hs2q},\ref{hs2g})
we have performed the $z$-integration and introduced the explicit formulae
(\ref{okgq},\ref{okqq})
for the kernels $\overline{K}^{ab}$ and (\ref{pgqapp},\ref{pqqapp}) for
the splitting probabilities $P^{ab}$.

We can explicitly check that our calculation correctly reproduces the known
NLO results~[\ref{AEM}] for the structure function $F_2$. The partonic
coefficient functions $F_{2a}^{NLO}$ for $F_2$ are obtained from our cross
sections by integrating over all possible final states (i.e.\ by setting
$F_J^{(1)} = F_J^{(2)} = 1$) and by fixing the overall normalization
with $|\cm_{1,q}(q + xp;xp)|^2 = N_c$. Gathering together all the terms
in Eqs.~(\ref{si2qdis},\ref{si21qdis},\ref{hs2q}) and those in
Eqs.~(\ref{si2gdis},\ref{hs2g}), we
obtain
\beq\label{f2qnlo}
F_{2q}^{NLO}(p,\mu_F^2) = \frac{\as}{2\pi}C_F\;x \left\{ \left[
    \frac{1+x^2}{1-x}\left(\ln\frac{(1-x)Q^2}{x\;\mu_F^2}
      -\frac34\right)
    +\frac14(9+5x)
    \right]_+ - \frac{1}{C_F} \KFS{qq}(x)\right\} \;\;,
\eeq
\beq\label{f2gnlo}
F_{2g}^{NLO}(p,\mu_F^2) = \frac{\as}{2\pi}T_R\;2x \left\{
 [ x^2+(1-x)^2 ] \ln\frac{(1-x)Q^2}{x\;\mu_F^2}+8x(1-x)-1
    - \frac{1}{T_R} \KFS{gq}(x) \right\} \;\;.
\eeq

The NLO expressions in Eqs.~(\ref{f2qnlo},\ref{f2gnlo}) are factorization-scheme
dependent and this dependence is accounted for by the kernels $\KFS{qq}$ and
$\KFS{gq}$. The DIS factorization scheme is defined in such a way that
$F_{2q}^{NLO}(p,Q^2) = F_{2g}^{NLO}(p,Q^2) = 0$: thus
Eqs.~(\ref{kqqscheme},\ref{kgqscheme}) follow. The definition of the gluon
kernels $K^{qg}_{\rm DIS}$ and $K^{gg}_{\rm DIS}$ is chosen in order to define
parton densities that fulfil momentum conservation [\ref{AEM}].

The next simplest process involving an incoming hadron is the 2+1-jet rate
in DIS\@.  Conceptually, there are no additional problems that are not dealt
with either in the example of 1+1 jets in DIS or in $e^+e^-\to3$ jets, and
it is straightforward to implement a general purpose Monte Carlo program
for arbitrary 2+1-jet-like quantities in DIS\@.  Specific details of the
algorithm and numerical results will be presented elsewhere.

\newpage

\section*{References}
\addcontentsline{toc}{section}{References}
\begin{enumerate}

\item \label{QCDrev}
S.\ Catani, in Proc.\ of the Int.\ Europhysics Conf.\ on High Energy Physics,
HEP 93,
eds.\ J.\ Carr and M.\ Perrottet (Editions
Frontieres, Gif-sur-Yvette, 1994), p.~771; B.R.\ Webber,
in Proc.\ of the 27th Int.\ Conf.\ on High Energy Physics,
eds. P.J.\ Bussey and I.G.\ Knowles (Institute of Physics, Bristol,
1995), p.~213.

\item \label{moscow}
F.V.\ Tkachov, \pl{100}{65}{81}; K.G.\ Chetyrkin and F.V.\ Tkachov,
\np{192}{159}{81}.

\item \label{loopy}
Z.\ Bern, L.\ Dixon, D.C.\ Dunbar and D.A.\ Kosower, \np{425}{217}{94};
Z.\ Kunszt, A.\ Signer and Z.\ Tr\'{o}cs\'{a}nyi, \pl{336}{529}{94};
Z.\ Bern, L.\ Dixon and D.A.\ Kosower, preprint SLAC-PUB-7111 (hep-ph/9602280).

\item \label{KS}
Z.\ Kunszt and D.E.\ Soper, \pr{46}{192}{92}.

\item \label{KL}
K.\ Fabricius, G.\ Kramer, G.\ Schierholz and I.\ Schmitt, \zp{11}{315}{81};
G.\ Kramer and B.\ Lampe, Fortschr. Phys. 37 (1989) 161.

\item \label{ERT}
R.K.\ Ellis, D.A.\ Ross and A.E.\ Terrano,
\np{178}{421}{81}.

\item  \label{BCM}
See, for instance: A.\ Bassetto, M.\ Ciafaloni and G.\ Marchesini,
\prep{100}{201}{83}; Yu.L.\ Dokshitzer, V.A.\ Khoze, A.H.\ Mueller and
S.I.\ Troyan, {\em Basics of Perturbative QCD}\/, Editions Frontieres,
Paris, 1991.

\item \label{GG}
W.T.\ Giele and E.W.N.\ Glover, \pr{46}{1980}{92}.

\item \label{GGK}
W.T.\ Giele, E.W.N.\ Glover and D.A.\ Kosower, \np{403}{633}{93}.

\item \label{BOO}
H.\ Baer, J.\ Ohnemus and J.F.\ Owens, \pr{42}{61}{90}; B.\ Bailey,
J.F.\ Owens and J.\ Ohnemus, \pr{46}{2018}{92}.

\item \label{Aver}
F.\ Aversa, P.\ Chiappetta, M.\ Greco and J.P.\ Guillet, \np{327}{105}{89};
F.\ Aversa, L.\ Gonzales, M.\ Greco, P.\ Chiappetta and J.P.\ Guillet,
\zp{49}{459}{91}.

\item \label{Chiap}
P.\ Chiappetta, R.\ Fergani and J.P.\ Guillet, \zp{69}{443}{96}.

\item \label{GKL}
W.T.\ Giele, S.\ Keller and E.\ Laenen, \pl{372}{141}{96};
S.\ Keller and E.\ Laenen, preprint in preparation.

\item  \label{KN}
Z.\ Kunszt and P.\ Nason, in `Z Physics at LEP 1', CERN 89-08, vol.~1, p.~373.

\item \label{EKS}
S.D.\ Ellis, Z.\ Kunszt and D.E.\ Soper, \pr{40}{2188}{89}, \prl{69}{1496}{92}.

\item \label{Frix}
S.\ Frixione, Z.\ Kunszt and A.\ Signer, preprint SLAC-PUB-95-7073
(hep-ph/9512328).

\item \label{NT}
Z.\ Nagy and  Z.\ Tr\'{o}cs\'{a}nyi, preprint KLTE-DTP/96-1 (February 1996).

\item \label{MNR}
M.L.\ Manga\-no, P.\ Nason and G.\ Ridolfi, \np{373}{295}{92}.

\item \label{CS1}
S.\ Catani and M.H.\ Seymour, in preparation.

\item \label{CSlett}
S.\ Catani and M.H.\ Seymour, preprint CERN-TH/96-28 (hep-ph/9602277), to
appear in Phys.\ Lett.\ B.

\item \label{CSdis}
S.\ Catani and M.H.\ Seymour, in preparation.

\item \label{HA}
P.\ de Causmaecker, R.\ Gastmans, W.\ Troost and T.T.\ Wu, \pl{105}{215}{81};
R.\ Kleiss, \np{241}{61}{84}; F.A.\ Berends, P.H.\ Daverveldt and R.\ Kleiss,
\np{253}{441}{85}; J.F.\ Gunion and Z.\ Kunszt, \pl{161}{333}{85}; Z.\ Xu,
D.H.\ Zhang and L.\ Chang, \np{291}{392}{87}.

\item \label{MP}
M.L.\ Mangano and S.J.\ Parke, \prep{200}{301}{91}, and references therein.

\item \label{KST}
Z.\ Kunszt, A.\ Signer and Z.\ Tr\'{o}cs\'{a}nyi, \np{411}{397}{94}.

\item \label{AP}
G.\ Altarelli and G.\ Parisi, \np{126}{298}{77}.

\item \label{CFP}
E.G.\ Floratos, D.A.\ Ross and C.T.\ Sachrajda, \np{129}{66}{77};
A.\ Gonzalez-Arroyo, C.\ Lopez and F.J.\ Yndurain, \np{153}{161}{79};
G.\ Curci, W.\ Furmanski, R.\ Petronzio, \np{175}{27}{80};
W.\ Furmanski, R.\ Petronzio, \pl{97}{437}{80};
A.\ Gonzalez-Arroyo and C.\ Lopez, \np{166}{429}{80};
E.G.\ Floratos, P.\ Lacaze and K.\ Kounnas, \pl{98}{225}{81}.

\item \label{Sud}
J.\ Kodaira and L.\ Trentadue, \pl{123}{335}{83};
S.\ Catani and L.\ Trentadue, \np{327}{323}{89};
S.\ Catani, E. d'Emilio and L.\ Trentadue, \pl{211}{335}{88};
S.\ Catani, L.\ Trenta\-due, G.\ Turnock and B.R.\ Webber, \np{407}{3}{93}.

\item \label{CMW}
S.\ Catani, G.\ Marchesini and B.R.\ Webber, \np{349}{635}{91}.

\item \label{KST2}
Z.\ Kunszt, A.\ Signer and Z.\ Tr\'{o}cs\'{a}nyi, \np{420}{550}{94}.

\item \label{CSS}
J.C.\ Collins, D.E.\ Soper and G.\ Sterman, in {\it Perturbative Quantum
Chromodynamics}, ed. A.H.\ Mueller (World Scientific, Singapore, 1989), p.~1
and references therein.

\item \label{Mirkes}
E.\ Mirkes and D.\ Zeppenfeld, preprint MADPH-95-916 (hep-ph/9511448).

\item \label{eefrag}
P.J.\ Rijken and W.L.\ van Neerven, preprint INLO-PUB-4-96 (hep-ph/9604436).

\item \label{Kleiss}
R.\ Kleiss and R.\ Pittau, Comp.\ Phys.\ Comm.\ 83 (1994) 141.

\item \label{CGNS}
K.\ Clay, E.W.N.\ Glover, P.\ Nason and M.H.\ Seymour, unpublished;\\
P.\ Nason and B.R.\ Webber, convenors, {\em QCD}, to appear in the
proceedings of the workshop on physics at LEP2,
preprint CERN-96-01 (hep-ph/9602288).

\item \label{Been}
W.\ Beenakker, H.\ Kuijf, W.L.\ van Neerven and J.\ Smith, \pr{40}{54}{89}.

\item \label{AEM}
G.\ Altarelli, R.K.\ Ellis and G.\ Martinelli, \np{157}{461}{79}.

\item \label{AEMP}
G.\ Altarelli, R.K.\ Ellis, G.\ Martinelli and So-Young Pi, \np{160}{301}{79}.

\end{enumerate}


\newpage

\renewcommand{\theequation}{E.\arabic{equation}}
\setcounter{equation}{0}

\section*{Note added}
\addcontentsline{toc}{section}{Note added}

All the detailed calculations in this paper have been performed using the 
conventional dimensional-regularization scheme (see Sect.~\ref{not}). 
As discussed in Sect.~\ref{oneloop}, 
the unphysical dependence on the regularization scheme can
be parametrized in terms of simple coefficients ${\tilde \gamma}_i$ 
(see Eq.~(\ref{mscheme})) that enter in the one-loop contribution. 
However, it is worth emphasizing some peculiar features\footnote{We thank 
Z.\ Tr\'{o}cs\'{a}nyi for pointing out these features to us.} of one particular
regularization scheme, namely, the dimensional reduction (or 
four-dimensional helicity) scheme\footnote{See: 
W.\ Siegel, \pl{84}{193}{79}; 
D.M.\ Capper, D.R.T.\ Jones and P.\ van Nieuwenhuizen, \np{167}{479}{80};
Z.\ Bern and D.A.\ Kosower, \np{379}{451}{92}; Ref.~[\ref{KST}].}.

All the final results of our algorithm, summarised in Sects.~\ref{finfo}, 
\ref{findis}, \ref{finfra}, \ref{finpp} and \ref{finmpc}, 
can be directly translated into the
dimensional-reduction scheme by simply modifying the explicit expression of the 
cross section component $\sigma^{NLO \,\{m\}}(p)$. This is the contribution to 
the NLO cross section that involves $m$-parton kinematics and no additional
convolutions with respect to longitudinal momentum fractions. It is obtained by
integrating the following combination of the one-loop matrix element and the
insertion operator $\bf I$
\beeq 
\!\!\!\!&&\!\!\!\! \left\{  \frac{1}{n_c(a) n_c(b)}
\;| \cm_{m+a_1+a_2...,ab}(q_1,...,q_n,p_1, ...,p_{m};p,{\bar p})|^2_{(1-loop)}
\right.  \\
\!\!\!\!&&\!\!\!\! \left.  + \;\;{}_{m+a_1+a_2...,ab}\!\!<{a_1,..,a_n,1, ....,m;a,b\;}|
\;{\bom I}(\ep)
\;|{a_1,..,a_n,1,...,m;a,b}>_{m+a_1+a_2...,ab} \frac{}{}
\right\}_{\ep=0} \;\;. \nonumber
\eeeq
The corresponding expression in the dimensional-reduction scheme is obtained by
the following replacements
\beeq
| \cm |^2_{(1-loop)} &\to&  | \cm^{\scriptscriptstyle\rm D\!.R\!.} |^2_{(1-loop)}
\;\;, \\
<...| \;{\bom I}(\ep) \;|...> &\to& 
<...| \;{\bom I}^{\scriptscriptstyle\rm D\!.R\!.}(\ep) 
\;|...>_{\scriptscriptstyle\rm D\!.R\!} \;\;,
\eeeq
where $| \cm^{\scriptscriptstyle\rm D\!.R\!.} |^2_{(1-loop)}$ and
$|...>_{\scriptscriptstyle\rm D\!.R\!.}$ respectively denote the one-loop
and tree-level matrix elements evaluated in the dimensional-reduction
regularization
and the colour-charge operator ${\bf I}^{\scriptscriptstyle\rm D\!.R\!.}(\ep)$
has the same general expression as in Eq.~(\ref{Iapp}) apart from the
replacement:
\beq
{\cal V}_I(\ep) \to {\cal V}_I^{\scriptscriptstyle\rm D\!.R\!.}(\ep)
= {\cal V}_I(\ep) - {\tilde \gamma}_I + {\cal O}(\ep) \;\;,
\eeq
\beq
{\tilde \gamma}_q = {\tilde \gamma}_{\bar q} = \frac{1}{2} \,C_F
\;, \;\;\;\;\;\;\; {\tilde \gamma}_g = \frac{1}{6} \,C_A \;\;.
\eeq

The dimensional reduction scheme is particularly relevant because most of the
one-loop matrix elements [\ref{loopy}] have been computed in precisely this 
scheme.

Moreover in dimensional reduction, just because of its definition, the 
$d$-dimensional and four-dimensional tree-level matrix elements exactly
coincide.
Thus, no calculation of $d$-dimensional Born-level matrix elements is necessary.
Actually, owing to the general structure of the one-loop corrections discussed 
in Sect.~\ref{1lc} and Ref.~[\ref{KST2}], it follows that from the 
$\ep$-expansion
(see Eq.~(\ref{1lep})) of 
$| \cm^{\scriptscriptstyle\rm D\!.R\!.} |^2_{(1-loop)}$  
one can directly extract all the   
independent colour projections of the matrix element squared
at the Born level.

Thus, for our purposes, the main feature of dimensional reduction is that,
provided the one-loop matrix elements are computed in this regularization 
scheme,  the shopping list (see Sects.~\ref{subee} and \ref{summ}) to
construct a numerical program to calculate the NLO corrections to arbitrary
jet quantities in a given process can be shortened as follows
\vspace{-\parskip}
\begin{itemize}
\addtolength{\itemsep}{-\parskip}
\item the one-loop matrix element in the dimensional-reduction scheme 
in $d$ dimensions;
\item an additional projection of the Born level matrix element over the
  helicity of each external gluon in four dimensions;
\item the tree-level NLO matrix element in four dimensions.
\end{itemize}

The computation of one-loop matrix elements in the dimensional-reduction scheme
is greatly simplified by directly evaluating helicity amplitudes [\ref{loopy}].
Using these calculations, the above shopping list can be further shortened by
eliminating the second item. 

\end{document}